\documentclass[11pt,a4paper,preprintnumbers]{book}

\usepackage{vmargin}
\setpapersize{A4}
\setmarg{40mm}{45mm}{130mm}{210mm}
 \usepackage[utf8]{inputenc}       

\usepackage[T1]{fontenc}

\usepackage[english]{babel}

\usepackage{graphicx}
\usepackage{xspace,colortbl}
\usepackage{epsfig}
\usepackage[small]{caption}

\usepackage{amstext}
\usepackage{amsmath}
\usepackage{amssymb}

\usepackage{fancyhdr}
\fancypagestyle{plain}{
\fancyhead{}

}

 \makeatletter
  \def\cleardoublepage{\clearpage\if@twoside \ifodd\c@page\else%
    \hbox{}
    \thispagestyle{empty}               
       \newpage
    \if@twocolumn\hbox{}\newpage\fi\fi\fi}
\makeatother

\addto\captionsspanish{}

\usepackage[numbers,square,comma,sort&compress]{natbib}
\usepackage{hyperref}
\usepackage{rotating}
\usepackage{subfigure}
\usepackage{dcolumn}
\usepackage{bm}
\usepackage{tabularx}
\usepackage{colortbl}
\usepackage{color}
\usepackage{array}
\usepackage{amssymb}
\usepackage{amsmath}
\usepackage{graphicx}
\usepackage{tikz}
\usetikzlibrary{shapes,arrows}
\usetikzlibrary{positioning}
\usetikzlibrary{shadows}
\usepackage{sidecap}
\usepackage{float}
\usepackage{wrapfig}
\usepackage{enumerate}
\usepackage{textcomp}

\begin{document}



\pagenumbering{roman}


\thispagestyle{empty}\
\mbox{}
\newpage


\thispagestyle{empty}
\mbox{}
\newpage


\thispagestyle{empty}

\vspace*{30mm}

\hrule\hrule\hrule\hrule
\begin{flushleft}
{\textsc{\textbf{Material Laws and numerical Methods in applied
superconductivity\\}}}
\end{flushleft}
\hrule\hrule\hrule\hrule

\newpage


\thispagestyle{empty} 

\vspace*{15mm} 

\noindent {\Large \bf
ZAGUAN: Theses Digital Repository\\ University of Zaragoza (Spain)}
%


\bigskip
\bigskip
\bigskip
\bigskip

\begin{quotation}

{\large \em Thesis submitted for the Degree of Doctor in Physics.\\
University of Zaragoza, Spain (2012).\\[10mm]}

{Author: Harold Steven Ruiz Rondan\\[3mm]}

{Advisor: Dr. Antonio Bad\'ia Maj\'os\\[5mm]}
\end{quotation}

\vspace{5.5cm}

\hspace{3cm}\includegraphics[scale=0.15]{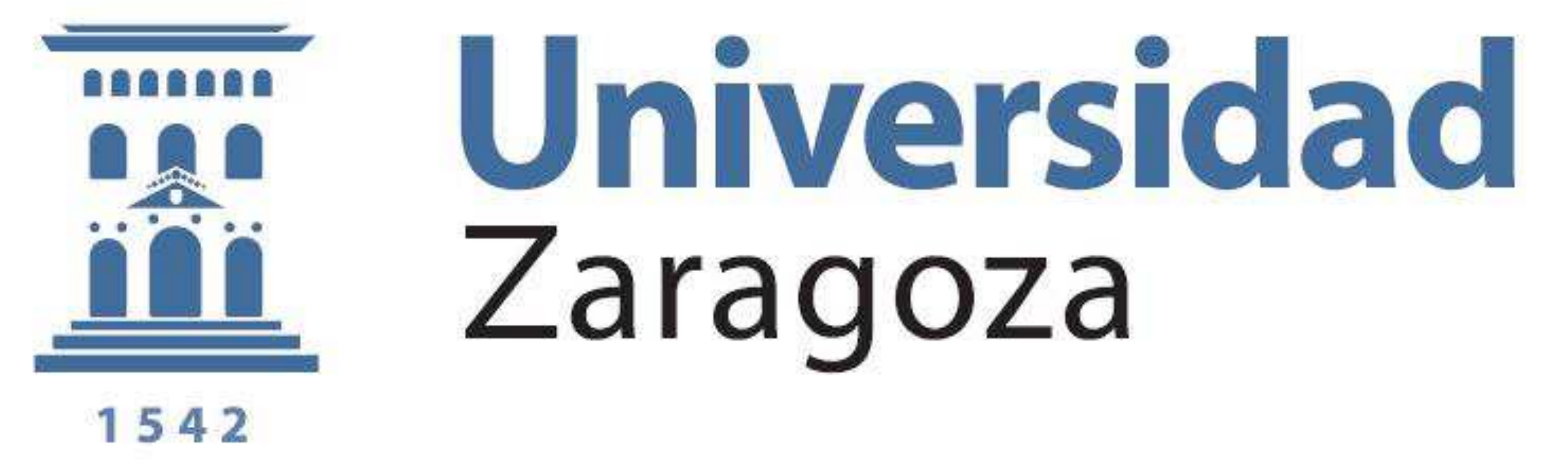}

\newpage


\thispagestyle{empty}

\hrule\hrule\hrule
\begin{flushleft}
\vspace*{-2.5mm}
\begin{center}
{\Large \bf Material Laws And Numerical Methods In Applied
Superconductivity\\}
\end{center}
\vspace*{-0.5mm}
\end{flushleft}
\hrule\hrule\hrule

\vspace{5cm}%

\begin{center}
{\Large \bf Harold Steven Ruiz Rondan} 
\end{center}

\vspace{5cm}%

\begin{center}
\includegraphics[scale=0.22]{logo_uz_new.pdf} \hspace*{0.2cm}
\includegraphics[scale=0.5]{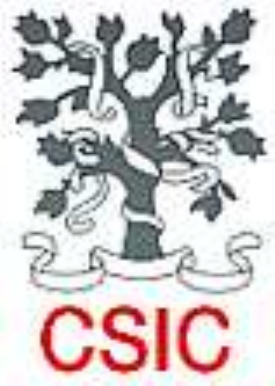} \hspace*{0.2cm}
\includegraphics[scale=0.62]{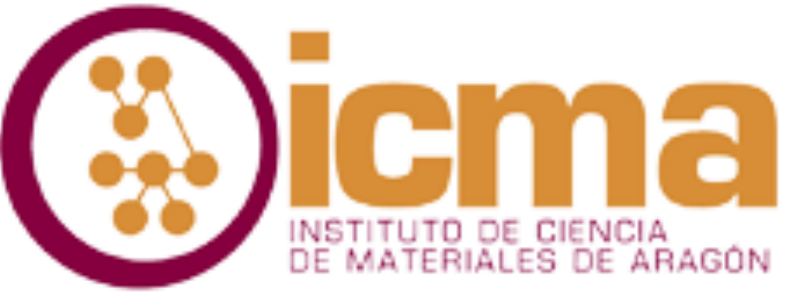}
\end{center}

\vspace{0.1cm}

\begin{center}
{\sc \bf Department of Condensed Matter Physics, University of Zaragoza.\\
Spanish National Research Council (CSIC), Institute of Materials Science of
Arag\'on (ICMA)} 
\end{center}

\newpage


\thispagestyle{empty}\
\mbox{}
\newpage


\thispagestyle{plain}

\vspace*{15mm}
\begin{quotation}
\begin{flushright}
{\bf \em Every step of my career\\ makes me feel proudest \\of my family. \\To
them.}
\end{flushright}
\end{quotation}

\newpage


\thispagestyle{empty}\
\mbox{}
\newpage


\parskip 0cm
\tableofcontents
\markboth{Contents}{Contents}
\parskip 0.2cm

%



\chapter*{\sc \textbf{Preface}}
\markboth{\sc \textbf{Preface}}{\sc \textbf{Preface}}
\addcontentsline{toc}{chapter}{\sc \textbf{Preface}}
\vspace{2cm}
%
%
One century has elapsed since the discovery of superconductivity
by Heike Kamerlingh Onnes, opening a new world of significant
applications in technologies ranging from electric power devices such as
motors and generators, large magnet systems such as those needed in
storage rings for particle accelerators, and electricity transmission in power
lines. As it is well-known, the technological usage of any superconducting
material is based upon its ability
to carry and maintain a current with no applied voltage whatsoever, i.e., with
an almost negligible loss of energy even in those cases when the superconductor
is subjected to strong enough applied magnetic fields. Although electrical
currents can flow with a negligible loss of energy maintaining the
superconductor in an appropriate temperature environment, superconductivity
can be destroyed by the effect of a sufficiently intense magnetic field or the
flow of a current density exceeding a critical value. Indeed, most of the
technological applications of the superconductors are directly linked to their
magnetic properties, and in particular in the way that they expell the magnetic
fields. This fact leads to the classification of superconductors in two
different kinds. On the one hand, \textit{Type-I} superconductors are mainly
characterized by a unique curve for the maximal applied magnetic field which a
superconductor is able to expell before the sudden transition to the normal
state occurs. On the other hand, \textit{Type-II} superconductors are
characterized by a new phase or ``mixed-state'' where the transition from the
superconducting state to the normal state allows the existence of bundles of
magnetic flux penetrating the sample (vortices), before reaching the sudden
transition to the normal state. This remarkable property allows to preserve the
superconducting state with the advantage of sustaining much higher magnetic
fields, and therefore carrying much higher current densities. However, this
ability is directly related to the pinning efficiency of a given material as the
motion of vortices produces a high dissipation of energy which in turn can lead
to the \textit{quench} of the superconducting state.  It is worth noting that
all the superconductors, from metal-alloys to cuprates, fullerenes, $MgB_{2}$,
iron-based systems that have been discovered  along the last 60 years, are
\textit{Type-II} superconductors, and consequently almost all the actual
superconducting technology is based on these kind of materials.
Thus, since the vortices must be pinned by the underlying crystallographic
structure and the presence of different kind of defects, the knowledge of the
electromagnetic properties and laws governing the pinning of vortices becomes a
crucial but not trivial issue for the understanding and developing of devices in
the framework of applied superconductivity.

In spite of significant theoretical and practical interest, from the
macroscopical point of view, the material laws and the electromagnetic
properties of type-II superconductors still deserve attention,
and currently no book exists that covers all the aspects about this
topic in full depth. This thesis attempts to contribute with some novel
numerical methods in applied superconductivity, including a comprehensive
discussion of the different mechanisms involved in the vortex dynamics.

The book has been structured in three parts with sequential chapters increasing
the level of complexity, both from the mathematical point of view and as
concerns to the underlying phenomena. On the one hand, a general critical
state theory for type-II superconductors with magnetic anisotropy, its
computation, implications, and consequently some applications for particular
problems, is what the first and second part try to convey. 
On the other hand, some microscopical aspects of the superconductivity have been
also considered and the attained results have been compiled along the third part
of this thesis.

In detail, the first part of this book is devoted to the study of the
electromagnetic properties of type-II superconductors in the critical state
regime.
After an introductory Chapter 1, which reviews the classical statements of the
critical state theory and derived approaches, Chapter 2 focuses on the
variational theory for critical state problems and the establishment of a 
general material law for 3D critical states with an associated magnetic
anisotropy and the underlying physics for the mechanism of flux depinning and
cutting. Then, a technical but important issue arises and is covered in Chapter
3: how to deal with large-scale nonlinear minimization problems such as those
presented in the general critical state theory for applied superconductivity,
but in a personal computer. A well defined structure for the minimization
functional, constraints, bounds, and preconditioners, based upon a set of
FORTRAN packages, solve this problem.

Hopefully, at the end of the first part, the reader will feel either attracted
or at least intrigued by the scope of our theory and methods. In
this sense, the second part of this book is devoted to sketch some of the main
results obtained along this line, i.e., we show some examples where we
have implemented our general critical state theory whose impact affects not
only the understanding of the physical properties of a superconducting system
but also at its potential applications. 

In Chapter 4, the advantages of the variational method are emphasized focusing
on its numerical performance that allows to explain a wide number of physical
scenarios. In particular, we present a thorough analysis of the underlying
effects derived of the three dimensional magnetic anisotropy and different
material laws (\textit{or models}) which allow us to treat with the flux
depinning and cutting mechanisms.

Chapter 5 deals with the study of the longitudinal transport problem
(the current is applied parallel to some bias magnetic field) in type II
superconductors. In particular, for the slab geometry with three dimensional
components of the local electromagnetic quantities, the complex interaction
between shielding and transport is solved. On the one hand, based on a
simplified analytical method for 2D configurations, and on the other hand,
based on a wide set of numerical studies for general scenarios (3D), it is shown
that an outstanding inversion of the current flow in a surface layer, and the
remarkable enhancement of the current density by their compression towards the
center of the sample, are straightforwardly predicted when the physical
mechanisms of flux depinning and consumption (via line cutting) are recalled. In
addition, a number of striking collateral effects, such as local and global
paramagnetic behavior, are predicted. 

Chapter 6 addresses to a comprehensive study of the electromagnetic response
of superconducting wires subjected to diverse configurations of transverse
magnetic field and/or longitudinal transport current. 
In particular, we have performed a wide set of numerical experiments dealing
with the local and global effects underlying to the distribution of field and
current for a straight, infinite, type II superconducting wire, it immersed in
an oscillating magnetic field applied perpendicular to its surface
($\textbf{B}_{0}$), and the simultaneous action of an AC transport current
($I_{tr}$). Thus, in a first part we have introduced the theoretical framework
of this problem focusing on the numerical advantages of our variational method.
Likewise, we provide a thorough discussion about some of the main macroscopic
quantities which may be experimentally measured, such as the magnetization curve
and the hysteretic AC loss, as well as on the local behavior of the
electromagnetic quantities \textbf{E}, \textbf{B}, and \textbf{J}. Three
different regimes of excitation have been considered: (\textit{i}) Isolated
electromagnetic excitations, where only the action of $\textbf{B}_{0}$ or
$I_{tr}$ is considered, (\textit{ii}) Synchronous electromagnetic sources,
where the concomitant action of $\textbf{B}_{0}$ and $I_{tr}$ shows a
unique oscillating phase and frequency, and (\textit{iii}) Asynchronous
electromagnetic sources, where $\textbf{B}_{0}$ and $I_{tr}$ do not show the
same oscillating frequency and therefore are out-phase. The underlying
effects of considering premagnetized wires under the above mentioned regimes are
also considered. Thus, several striking effects as the strong localization of
the local density of power loss, a distinct low-pass filtering effect intrinsic
to the wire's magnetic response, exotic magnetization loops, increases and
decreases of the hysteretic AC loss by power supplies with double frequency
effects, and significant differences between the widely used approximate
formulae and the actual AC loss numerically calculated, have been detected
and explained.

The last part of this dissertation concerns our contribution to another aspect
of superconductivity. By means of a specific integral method applied to
spectroscopic data, we have been able to draw some conclusions on the influence
of the Electron-Phonon (E-Ph) coupling mechanism in cuprate superconductors.
More specifically, we have focused on the analysis of high-resolution angle
resolved photoemission spectroscopies (ARPES) in several families of cuprate
superconductors. Although relying on solid (and sophisticated) techniques in
the realm of quantum theory, we describe a phenomenological procedure that
allows to obtain relevant physical parameters concerning the E-Ph interaction. 

Thus, in chapter~\ref{ch-7}, we introduce a novel theoretical model which allows
a quite general explanation of the so-called nodal \textit{kink effect} observed
in ARPES, for several doping levels in the cuprate
families $La_{2-x}Sr_{x}CuO_{4}$, $Bi_{2}Sr_{2}CaCu_{2}O_{8+x}$, and
$YBa_{2}Cu_{3}O_{6+x}$. 

Finally, in an effort to clarify the influence of
the E-Ph coupling mechanism to the boson mechanism which causes the pair
formation in the superconducting state, chapter~\ref{ch-8} addresses the study
of the superconducting thermodynamical quantities, $T_{c}$, the ratio gap
$2\Delta_{0}/k_{B}T_{c}$, and the zero temperature gap $\Delta_{0}$, for a wide
set of natural and empirical equations. 

In reading this book, we want to remark that each one of its parts have its
own introduction and concluding sections, and also the list of references to the
literature have been placed forward. In addition, a small glossary can be found
at the end of this book.

\break

Hopefully, this thesis may serve to bring a bigger community interested in
the world of superconductivity, either in the application of their macroscopical
properties or the understanding of their microscopical ground.

%
\vspace*{2cm}

\hspace{7cm} February 2012, Zaragoza - Spain.
%
\cleardoublepage

\pagenumbering{arabic}



\part{\label{Part_1}\textsc{\textbf{Electromagnetism of type II
superconductors}}}


\chapter*{}
\vspace*{-3cm}
\section*{\label{Intro-P1}\hspace*{0.55cm} \sc \textbf{Introduction}}
\markboth{\sc \textbf{Introduction}}{\hspace*{0.55cm} \sc \textbf{Introduction}}
\addcontentsline{toc}{chapter}{\hspace*{0.55cm} \sc \textbf{Introduction}}

The high interest concerning the investigation of the macroscopic
magnetic properties of type-II superconductors in the mixed state is markedly
associated with its relevance to technological and industrial applications
achieving elevated transport currents with no discernible energy dissipation.
It relates to a wide list of physical phenomena concerning the physics of
vortices, which may be basically analyzed in terms of
interactions between the flux lines themselves (lattice elasticity and line
cutting), and interactions with the underlying crystal structure averaged
by the so-called flux pinning mechanism.

In a mesoscopic description of real type-II superconductors, the distribution
of vortices may be simplified through a mean-field approach for a volume
containing a big enough number of vortices and making use of an appropriate
material law incorporating the intrinsic properties of the material. This
picture of  coarse-grained fields, i.e.: magnetic induction ${\bf B}\equiv
\langle{\bf b}\rangle$, electric current density ${\bf J\equiv
\langle{\bf j}\rangle}$ and electric field ${\bf E\equiv \langle{\bf
e}\rangle}$, allows to state the problem of the driving force due to the
currents circulating in the superconducting sample and their balance with the
limiting pinning force acting on the vortex lattice so as to
prevent destabilization and the consequent propagation of dissipative states. 
Per unit volume, this reads ${\bf J}\times{\bf B}={\bf F}_p$ (or $J_{\perp}B =
F_p$). The underlying concept behind this balance condition is already a
classical subject well known as the critical state model by Charles P.
Bean~\cite{P1-Bean_1964}. In this simple, but brilliant model, the response of
the superconducting sample is provided by assuming that the electrical current
density vector $\textbf{J}$ (oriented perpendicular to the direction of
the local magnetic field vector ${\bf B}$) compensates with the pinning force,
and then, it is constrained by a threshold value $\textbf{J}_{c}$ which defines
a local critical state for the array of magnetic flux lines. Thus, in
view of Faraday's law, external field variations are opposed by the maximum
current density $J_{c}$ within the material, and after the changes occur,
$J_{c}$ persists in those regions which have been affected by an electric field.
Although, such a model allows to capture the main features of the magnetic
response of superconductors with pinning at low frequencies and temperatures,
through the minimal mathematical complication, the stronger limitation
of Bean's ansatz is that one can just apply it to vortex lattices composed by
parallel flux lines perpendicular to the current flow, and unless for highly
symmetric situations \textbf{J} does not necessarily satisfy the condition
$\textbf{J}=\textbf{J}_{\perp}$. In fact, a proper theory for the critical state
must allow the coexistence of nonparallel flux lines. Thus, rotations of
\textbf{B} can lead to entanglement and recombination of neighboring flux lines
which brings a component of the current density along the local magnetic field,
$\textbf{J}_{\parallel}$. This component generates distortions which also become
unstable when a threshold value $J_{c\parallel}$ is exceeded, giving way to the
so-called flux cutting phenomenon. 

When the conditions $J_{\parallel}=J_{c\parallel}$ and $J_{\perp}=J_{c\perp}$
become active, the so-called double critical state appears~\cite{P1-Clem_DCSM}.
In simple words, this upgraded theory ({\em double critical state model} or
DCSM) generalises the one-dimensional concept introduced by
Bean~\cite{P1-Bean_1964} to anisotropic scenarios for the material law in terms
of the natural concepts $E_{\parallel}(J_{\parallel})$ and
$E_{\perp}(J_{\perp})$~\cite{P1-Ruiz_PRB_2009,P1-Ruiz_SUST_2010,
P1-Ruiz_SUST_2011}. 
From the mathematical point of view, the critical state problem can be
understood as finding the equilibrium distribution for the circulating
current density $\textbf{J}(\textbf{r})$ defined by the conditions
$J_{\parallel}\leq J_{c\parallel}$ and $J_{\perp}\leq J_{c\perp}$, both
consistent with the Maxwell equations in quasistationary form, and under
continuity boundary conditions that incorporate the influence of the sources.
Being a quasi-stationary approach, the critical state is customarily stated
without an explicit role for the transient electric field. Thus, Faraday's law
is implicitly used through Lenz' law by selecting the actual value $\pm
J_{c}$ or $0$ that minimizes flux variations when solving Ampere's law
$\nabla\times \textbf{B}=\mu_{0}\textbf{J}$ along the process. Customarily, one
also considers situations
where the local components of the magnetic field $\textbf{H}(\textbf{r})$ along
the superconductor (SC) are much higher than the lower critical field
$H_{c1}$ and well below $H_{c2}$ to allow the use of the linear relation
$\textbf{B}=\mu_{0}\textbf{H}$.

Within this picture of the electromagnetic problem, in this first \textit{part}
of the book we introduce the important definitions and concepts of those topics
behind the critical state theory, extending its scope for three dimensional
cases with help of numerical methods in the
framework of the variational formalism for optimal control problems. We want to
emphasize that although it is not our intent to develop a comprehensive study of
these mathematical topics, we will show common mathematical techniques which are
found to be particularly useful in applied superconductivity.  The reader is
referred to the
references~\cite{P1-Badia_PRL_2001,P1-Badia_PRB_2002,P1-Jackson,P1-Mayergoyz,
P1-Arfken,P1-Pontryagin,P1-Leitmann,P1-Knowles} for a more thorough discussion
of this
material. 

Chapter 1 is devoted to introduce the theoretical background that
justifies the critical state concept as a valid
constitutive law for superconducting materials. First, the critical state is
described by the classical differential formalism of the Maxwell equations, and
then, the prescribed magnetoquasistationary approximation is thoroughly
discussed. 

In chapter 2, our proposed general critical state theory is developed in two
parts. Firstly, the critical state problem is posed in terms of an equivalent
optimal control problem with variational statements, i.e., the classical Maxwell
equations are translated to the variational formalism where a simpler set of
integral equations with boundary conditions is to be solved by a minimization
procedure. Is to be noticed, that despite of the fact that the reader can feel
more familiar to the differential formalism, the numerical solution of the
differential set of equations is much more cumbersome than minimizing an
integral functional. Secondly, the underlying vortex physics is posed in terms
of a quite general material law for type-II
superconductors with magnetic anisotropy, which characterizes
the conducting behavior in terms of the threshold values for the current
density and the physical mechanisms of flux depinning and cutting.

Finally, chapter 3 covers the basic facts related to the computational method 
adopted for the solution of general critical state problems such
as those tackled in the following part of this book. Here, no attempt
is made to scrutinize through the FORTRAN packages for large scale nonlinear
optimization. Instead, the presentation of this chapter must be
understood as a schematic tool for dealing with a wide variety of problems in
applied superconductivity. 








\chapter{\label{ch-1} \sc \textbf{General Statements Of The Critical State}}


\subsection*{\label{ch-1-1}
\hspace*{0.1cm} \textsl{1.1 The Critical State In The Maxwell Equations
Formalism}}
\vspace*{1cm}
\markboth{\hspace*{0.1cm}\textsl{1.1 The Critical State In The Maxwell
Equations Formalism}} 
{\hspace*{0.1cm} \textsl{1.1 The Critical State In The Maxwell Equations
Formalism}}
\addcontentsline{toc}{chapter}
{\hspace*{0.1cm} \textsl{1.1 The Critical State In The Maxwell Equations
Formalism}}

The fundamental concept on which the critical state theory relies is that, in
many cases, the experimental conditions allow to analyze the evolution of the
system in a magnetoquasisteady (MQS) regime of the time-dependent Maxwell
equations accompanied by material constitutive laws,  ${\bf H}({\bf
B})$, ${\bf D}({\bf E})$, and ${\bf J}({\bf E})$. Thus, Faraday's and Ampere's
laws represent a coupled system of time evolution field equations
\begin{eqnarray}\label{Eq.1.1}
\partial_{t}{\bf B}=-\nabla\times{\bf E}\quad , \quad
\partial_{t}{\bf D}=\nabla\times{\bf H}-{\bf J} \, ,
\end{eqnarray}
which together determine the distribution of supercurrents within the sample.
Here, the induced transient electric field is determined through an appropriate
material relation ${\bf J}({\bf E})$, and is used to update the profile of $\bf
J$.

\footnotetext[1]{If there were some average surface currents, then the actual
density of magnetic flux $\textbf{B}$ would differ from
$\mu_{0}\textbf{H}$.}

\begin{figure}[t]
\centering
\includegraphics[width=1.0\textwidth]{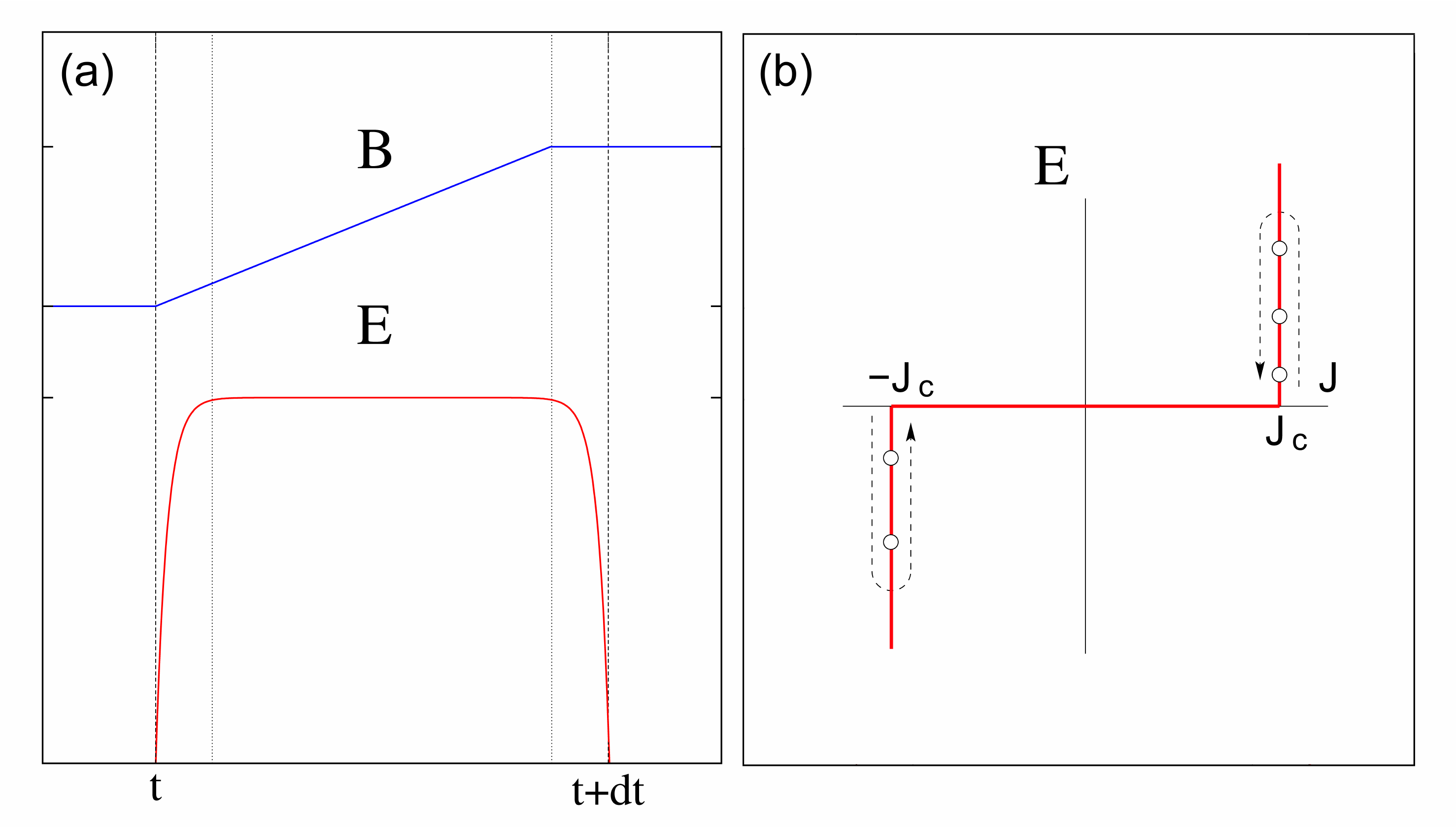}
\caption{\label{Figure_1_1} (a) Schematic representation the time
dependence of the electromagnetic fields within the MQS
regime. (b) Pictorial drawing of the critical state model in terms of a one
dimensional ${E}({J})$ law.}
\end{figure}

Notice that, as equilibrium magnetization is usually neglected
in the critical-state regime, one is enabled to use the
relation $\textbf{B}=\mu_{0}\textbf{H}$, so that there are no average surface
currents.\footnotemark[1] Furthermore, as the magnetic fields of interest are
some fraction of the critical transition magnetic field $H_{c2}$ that is much
greater than the penetration field $H_{c1}$, the distribution of vortices and
the corresponding supercurrents will be thermodynamically favoured to
go into the superconductor, such that a ramp in the magnetic field is induced
by the external excitation within the interval $[t,t+dt]$,  see
Fig.~\ref{Figure_1_1}(a). Thus, as a consequence of a very fast diffusion
(elevated flux flow resistivity), the electric field quickly adjusts to a
constant value along the excitation interval, and once the
magnetic field ramp stops, $E$ goes back to zero again. 

The {\em readjusting}
vertical bands are considered a second order effect and allow for charge
separation and recombination, according to the specific ${\bf E}({\bf J})$ model
[see Fig.~\ref{Figure_1_1}(b)]. Therefore,  we are allowed to model the
flux as entering the superconductor at zero field cooling, where the electric
field arises when some {\em critical} condition for the volume current density
is reached ($J_c$ in the 1D representation). Then, corresponding to the MQS
limit, the electric field instantaneously increases to a certain value
determined by the rate of variation of the magnetic field and then goes back to
zero. 

By taking divergence in both sides
of the Faraday's and Ampere's laws, and recalling integrability
(permutation of space and time derivatives) it leads to
the additional conditions
\begin{eqnarray}\label{Eq.1.2}
\partial_{t}(\nabla\cdot{\bf B})=0 \quad ,\quad
 \partial_{t} (\nabla\cdot{\bf D}) + \nabla\cdot{\bf J} = 0 \, .
\end{eqnarray}

Within this picture, the remaining
Maxwell equations can be interpreted as ``\textit{spatial
initial conditions}'' for Eq.~(\ref{Eq.1.2}) which are defining the existence of
conserved electric charges, i.e.,
\begin{eqnarray}\label{Eq.1.3}
 \nabla\cdot{\bf B}(t=0) = 0 \quad ,\quad \nabla\cdot {\bf D} (t=0) = \rho
(t=0)\, .
\end{eqnarray}

In this sense, the set of equations~(\ref{Eq.1.1}), upon substitution of $\bf
H$,  $\bf D$ and $\bf J$ through the constitutive laws, and with appropriate
initial conditions, uniquely determine the evolution profiles ${\bf
B}({\bf r},t)$ and ${\bf E}({\bf r},t)$.

%
%

Notice that, for {\em slow} and {\em uniform} sweep rates of the external
excitations (magnetic field sources and/or transport current), the
transient variables $\bf E$, $\bf D$ and $\rho$ are small, and proportional to
 $\dot{\bf B}$, whereas $\ddot {\bf B}$, $\dot {\bf E}$ and
$\dot {\rho}$ are negligible. Thus, the main hypothesis within the MQS regime is
that the {\em displacement} current densities $\partial _t{\bf D}$ are much
smaller than $\bf J$ in the bulk and vanish in a first order treatment. This
causes a crucial change in the
mathematical structure of the Maxwell equations: Ampere's law is no
longer a time evolution equation, but becomes a purely spatial
condition. It reads as 
\begin{eqnarray}\label{Eq.1.4}
\nabla\times {\bf B} \simeq \mu_{0}{\bf J} \, ,
\end{eqnarray}
with approximate integrability condition $\nabla\cdot{\bf J}\simeq 0$.

In the MQS limit, Faraday's law is the unique time evolution equation. Then, one
can find the evolution profile ${\bf B}({\bf r},t)$ from
\begin{eqnarray}\label{Eq.1.5}
\partial_{t}{\bf B}=-\nabla\times{\bf
E}=-\nabla\times\left[\rho(\textbf{J})~\mu_{0}\nabla\times{\bf
B}\right] \, .
\end{eqnarray}
Here, $\rho ({\bf J})$ plays the role of a nonlinear and possibly nonscalar
resistivity
that should properly incorporate the physics of the threshold and dissipation
mechanisms associated to the flux depinning and flux cutting mechanisms.

We want to mention that, although the \textit{B-formulation} in
Eq.~(\ref{Eq.1.5}) is definitely the most extended one, the possibilities of
\textit{E-formulations}~\cite{P1-Barret_2006},
\textit{J-formulations}~\cite{P1-Wolsky_2008}, or a vector potential oriented
theory (\textit{A-formulation})~\cite{P1-Campbell_2009}, in which the dependent
variables are the fields ${\bf E}$, ${\bf J}$, or $\textbf{A}$ respectively, 
have also been exploited by several authors.
%


\vspace*{0.5cm}
\subsection*{\label{ch-1-2}
\hspace*{0.1cm} \textsl{1.2 The Critical State Regime And The MQS
Limit}}
\vspace*{0.5cm}
\markboth{\hspace*{0.1cm}\textsl{1.2 The Critical State Regime And The MQS
Limit}} 
{\hspace*{0.1cm} \textsl{1.2 The Critical State Regime And The MQS
Limit}}
\addcontentsline{toc}{chapter}
{\hspace*{0.1cm} \textsl{1.2 The Critical State Regime And The MQS
Limit}}

In spite of the seeming simplicity of the MQS approach
($\partial_{t}\textbf{D}\approxeq0$), we want to emphasize that the numerical
procedure to solve a critical state problem is closely linked to the
consequences of having assumed this limit. Below, two of the most relevant
consequences of the MQS limit are highlighted.

\begin{enumerate}
\item  Notice that, making use of the conductivity law
through its inverse function ${\bf E}({\bf J})$, the successive field
penetration profiles within the superconductor
may be obtained by the finite-difference expression of Faraday's law,
\begin{eqnarray}\label{Eq.1.6}
\frac{B_{l+1}-B_{l}}{\delta t}=-\nabla\times{\bf
E}~\left(\mu_{0}\textbf{J}_{l+1}\approx\nabla\times\textbf{B}_{l+1}\right) \, .
\end{eqnarray}
Here we have assumed an evolutionary discretization scheme, where
$\textbf{B}_{l}$ stands for the local magnetic field induction at the time
layer $l\delta t$, and the current density profiles are related to some
magnetic diffusion process that takes place when the local condition for
critical state $\textbf{J}(\textbf{r})\leq \textbf{J}_{c}(\textbf{r})$ is
violated. On the other hand, the constitutive law ${\bf D}({\bf E})$ which is
not used in
Eq.~(\ref{Eq.1.6}), plays no role in the evolution of the magnetic variables
$\textbf{B}_{l+1}$ and $\textbf{J}_{l+1}$, which means that the magnetic
``\textit{sector}'' is decoupled from the charge density profile because the
coupling term (charge recombination) has disappeared. In this sense, notice that
the local profile $\textbf{B}_{l+1}$ can be solved in terms of the
previous field distribution $\textbf{B}_{l}$ and the boundary conditions at time
layer $(l+1)\delta t$.
\item As the initial conditions must fulfill the Ampere's
law $\nabla\times\textbf{B}_{l}=\mu_{0}\textbf{J}_{l}$ as well as
$\nabla\cdot\textbf{B}_{l}=0$ and $\nabla\cdot\textbf{J}_{l}=0$, only the
inductive component of $\textbf{E}$ (given by $\nabla\times {\bf E}_{\rm ind} =
- \dot {\bf B}$, $\nabla\cdot {\bf E}_{\rm ind} = 0$) determines the evolution
of $\bf B$ (Faraday's law). At this point, the conducting law in its inverse
formulation ${\bf E}({\bf J})$ seems show certain ambiguity, as far as
two different material laws related by ${\bf E}_2 ({\bf J}) = {\bf E}_1 ({\bf
J}) + \nabla\Phi ({\bf J})$ determine the same magnetic and current density
profiles. Going into some more detail, whereas for the complete Maxwell
equations  statement, the potential component of the electric field
($\nabla\times{\bf E}_{\rm pot} = 0$,
$\epsilon _0 \nabla\cdot {\bf E}_{\rm pot} = \rho$), is coupled to $\bf B$ and 
${\bf E}_{\rm ind}$ through the $\dot {\bf D}$
term (which contains both inductive and potential parts), within the MQS limit
it is irrelevant for the magnetic quantities. In
fact, one is enabled to include the presence of charge densities without
contradiction with the condition $\nabla\cdot {\bf J} \simeq 0$ by means of
inhomogeneity or nonlinearity in the $\textbf{E}(\textbf{J})$ relation. Then
one has that the condition $\nabla\cdot\textbf{J}=0$ does not imply 
$\nabla\cdot\textbf{E}=0$. The charge density ${\rho}$ can be
understood as a parametrized charge of {\it static} character as far as $\dot
{\rho}$ is
neglected. As indicated above, once the magnetic variables are computed, one
has the freedom to modify the ``\textit{electrostatic sector}''  if necessary
by the rule ${\bf E}({\bf J})+\nabla\Phi$ while still maintaining the values
of $\bf B$ and $\bf J$. This invariance can be of practical interest as far as
the ``electrostatic'' behavior in the critical state regime is still under
discussion, it because of the inherent difficulties in
the direct measurement of transient charge
densities~\cite{P1-Ruiz_SUST_2011,P1-Joos_2006,P1-Clem_2011_PRB,Clem_2011_SUST}.
\end{enumerate}
%


\chapter{\label{ch-2} \sc \textbf{Variational Theory for Critical State
Problems}}


\subsection*{\label{ch-2-1}
\hspace*{0.1cm} \textsl{2.1 General Principles Of The Variational
Method}}
\vspace*{1cm}
\markboth{\hspace*{0.1cm}\textsl{2.1 General Principles Of The Variational
Method}} 
{\hspace*{0.1cm} \textsl{2.1 General Principles Of The Variational
Method}}
\addcontentsline{toc}{chapter}
{\hspace*{0.1cm} \textsl{2.1 General Principles Of The Variational
Method}}

As we have mentioned before, the numerical solution of the critical state
problem from the differential formalism of the Maxwell equations may be
cumbersome. One
possibility for making the resolution of this system affordable is to find an
equivalent variational statement of Eq.~(\ref{Eq.1.6}). Then, one can avoid the
integration of these set of differential equations by \textit{inversion} of a
set of Euler-Lagrange equations
\begin{eqnarray}\label{Eq.2.1}
\mu_{0}\textbf{J}_{l+1}-\nabla\times\textbf{B}_{l+1}=0\, ,
\end{eqnarray} 
and
\begin{eqnarray}\label{Eq.2.2}
\mu_{0}\nabla\times\textbf{p}_{l}+\textbf{B}_{l+1}-\textbf{B}_{l}=0\, ,
\end{eqnarray} 
for arbitrary variations of the Lagrange multiplier
(i.e., $\delta\textbf{p}_{l}$), and the time-discretized local magnetic
induction field (i.e., $\delta\textbf{B}_{l+1}$).\footnotemark[1] Eventually,
the Lagrange multiplier, $\textbf{p}_{l}$, will be basically identified with the
electric field of the problem.
\footnotetext[1]{Recall that the magnetic field \textbf{H} is defined as
a modification of the induction field \textbf{B} due to magnetic fields produced
by material media. However, as in the critical state regime the use of the
linear relation $\textbf{B}=\mu_{0}\textbf{H}$ is allowed, henceforth, we
will refer to \textit{magnetic field} where either or both fields apply.}

Going into more detail, let us consider a small path step $\delta t$, from some
initial profile of the magnetic field $\textbf{B}_{l}(\textbf{r})$ to a
final profile $\textbf{B}_{l+1}(\textbf{r})$, and the
corresponding $\textbf{J}_{l}(\textbf{r})$ and $\textbf{J}_{l+1}(\textbf{r})$.
Defining $\Delta\textbf{B}=\textbf{B}_{l+1}-\textbf{B}_{l}$, both
configurations can be considered to be connected by a steady process
performing a small linear step, such that
$\textbf{B}_{l+1}=\textbf{B}_{l}+s\Delta\textbf{B}$ with $s\in[0,1]\delta t$.
Recalling that the initial condition fulfills Ampere's law
$\nabla\times\textbf{B}_{l}=\mu_{0}\textbf{J}_{l}$, as well as
$\nabla\cdot\textbf{B}_{l}=0$ and $\nabla\cdot\textbf{J}_{l}=0$, the
time-averaged Lagrange density (over whole space) is
\begin{eqnarray}\label{Eq.2.3}
{\cal L}
=\frac{1}{2}|\Delta\textbf{B}|^{2}+\textbf{E}\cdot(\nabla\times\textbf{B}_ { l+1
}
-\textbf{J}_{l+1})\delta t \, .
\end{eqnarray} 
Thus, the physically admissible Lagrangian multipliers in the critical state
regime must satisfy the condition
\begin{eqnarray}\label{Eq.2.4}
\textbf{p}=\textbf{E}_{cs}\delta t  \, ,
\end{eqnarray} 
where the critical state electric field $(\textbf{E}_{cs})$ must be properly
defined by the imposed material law $\textbf{E}(\textbf{J})$.

However, concerning the  ``{\it unknown parameter}'' $\textbf{J}_{l+1}$, as far
as it is not allowed to take arbitrary values, we cannot impose
arbitrary variations as it is customary for the typical steady condition
of the Euler-Lagrange equations. Instead, an Optimal Control-like Maximum
principle equivalent to a maximal projection rule
$\textbf{\^{E}}\cdot\textbf{J}$ must be
used (see Refs.~\cite{P1-Badia_PRL_2001,P1-Badia_PRB_1998}). For a
more comprehensive review
of
the optimal control theory which can be understood as a generalization of the
variational calculus, the interested reader is directed to see, for instance,
Refs.~\cite{P1-Pontryagin,P1-Leitmann,P1-Knowles}.

In simple terms, the optimal control concept introduces a geometrical picture of
the material law for the boundary conditions of the vector $\textbf{J}$ that may
be of much help when discussing the idea of a general
critical state theory.  Summarizing, it is necessary to declare that
there must be a region $\Delta_{\textbf{r}}$ within the \textbf{J}
space (possibly oriented according to the local magnetic field $\textbf{\^{B}}$,
and/or also depending on $|\textbf{B}|$ and \textbf{r}) such that nondissipative
current flow occurs when the condition $\textbf{J}\in\Delta_{\textbf{r}}$ is
verified. Thus, the minimum of the Lagrangian
must be sought within the set of current density vectors fulfilling
${\bf J}\in\Delta_{\textbf{r}}$, i.e.: $\textbf{J}_{l+1}$ is determined by
the condition
\begin{eqnarray}\label{Eq.2.5}
{\rm Min}\{ {\cal L} \}|_{\textbf{J}\in\Delta_{\textbf{\textbf{r}}}}\equiv{\rm
Max}\{\textbf{J}\cdot\textbf{p}\}|_{\textbf{J}\in\Delta_{\textbf{\textbf{r}}}}
\, .
\end{eqnarray}

Notice that an $\textbf{E}(\textbf{J})$ law is needed in addition to
Eq.~\ref{Eq.2.3}. Thus, together with the concept of a very high dissipation
when \textbf{J} is driven outside $\Delta_{\textbf{r}}$ by some nonvanishing
electric field, Eq.~\ref{Eq.2.5} suffices to determine the relation between
the directions of \textbf{J} and \textbf{E}. Notice that the maximal shielding
condition is equivalent to the maximum projection rule, it means that the
orthogonality condition of the electric field direction with the surface of
$\Delta_{\textbf{r}}$ is recalled, and the Lagrange multiplier can be
straightforwardly identified with the electric field of the problem, i.e., 
\begin{eqnarray}\label{Eq.2.6}
{\rm
Max}\{\textbf{J}\cdot\textbf{p}\}|_{\textbf{J}\in\Delta_{\textbf{r}}}\equiv{\rm
Max}\{\textbf{E}\cdot\textbf{J}\}|_{\textbf{J}\in\Delta_{\textbf{r}}} \, .
\end{eqnarray}

Notice also that Ampere's law is imposed [Eq.~(\ref{Eq.2.1})] through the
Lagrange multiplier, while the discretized version of Faraday's law
[Eq.~(\ref{Eq.2.2})] is derived as an Euler-Lagrange equation for the
variational problem, so that absolute consistency with the Maxwell equations is
obtained. In fact, maximal global (integral) shielding is achieved through a
maximal local shielding rule [Eq.~(\ref{Eq.2.6})] that reproduces the elementary
evolution of $\partial_{t}{\bf J}$ for a perfect conductor with restricted
currents. Thus, in practice, if one explicitly introduces Ampere's
law $(\nabla \textbf{B}_{l+1}=\mu_{0}\textbf{J}_{l+1})$, minimization is made in
terms of 
\begin{eqnarray}\label{Eq.2.7}
{\cal F}[\textbf{J}_{l+1}]={\rm
Min}\left\{ -\int_{\Re^{3}}\textbf{E}\cdot\textbf{J}_{l+1} \right\} \, ,
\end{eqnarray}
and the minimum is sought over $\textbf{J}_{l+1}\in\Delta_{r}$ for a fixed
$\textbf{E}$. However, as we have mentioned before, special attention must be
payed to the feasible ambiguity of the function $\textbf{E}(\textbf{J})$ as it
can lead to fake values of the variables $\textbf{J}$.

Likewise, the straightforward equivalence between the convex functionals for
Eqs.~(\ref{Eq.2.5}) and (\ref{Eq.2.7}) allows to establish an equivalent
minimization principle in terms of the general definitions
\begin{eqnarray}\label{Eq.2.8}
\textbf{B}=\nabla\times\textbf{A} \, ,
\end{eqnarray} 
and
\begin{eqnarray}\label{Eq.2.9}
\textbf{E}=-\partial_{t}\textbf{A}-\nabla\varPhi \, ,
\end{eqnarray} 
by imposing the material law $\textbf{E}(\textbf{J})$ through the Lagrange
multiplier $\textbf{p}_{l+1}$. Thus, the minimization problem turns to find out
the invariant gauge conditions $\nabla\varPhi_{l+1}$ and
$\textbf{J}_{l+1}\in\Delta_{\textbf{r}}$ for a given function
$\textbf{A}[\textbf{J}]$, in such manner that
\begin{eqnarray}\label{Eq.2.10}
{\cal F}[\textbf{B}_{l+1},\nabla \varPhi]
={\rm Min}\int_{\Re^{3}}&&
\frac{1}{2}|\Delta\textbf{B}|^{2}-\Delta\textbf{A}\cdot(\nabla\times\textbf{B}_{
l+1}-\textbf{J}_{l+1})\\
&&-\nabla\varPhi(\nabla\times\textbf{B}_{l+1}-\textbf{J}_{l
+1})\delta t \, . \nonumber
\end{eqnarray}

We call the readers' attention to notice that the uncoupling of the
electromagnetic potentials can be accomplished
by exploiting the arbitrariness involved in the definition of $\textbf{A}$. In
fact, since \textbf{B} is defined through Eq.~(\ref{Eq.2.8}) in terms of
\textbf{A}, the vector potential is arbitrary to the extent that the gradient of
some scalar function can be added. Therefore, the ``\textit{magnetic sector}''
could be decoupled of the ``\textit{electric sector}'' if the physical
admissible states in the time-averaged Lagrange density $L$ are invariant gauge
of the Lagrange multipliers \textbf{p}. As a consequence, if the problem is such
that there are no intrinsic electromagnetic sources, $\varPhi\equiv0$ (for
type-II superconductors it means absence of transport current), a proper choice
of \textbf{A} should satisfy the Coulomb gauge $(\nabla\cdot\textbf{A}\equiv0)$.
In this sense, by using the Laplace equation, the second term in
Eq.~(\ref{Eq.2.10}) is reduced to
$\Delta\textbf{A}\cdot\delta_{t}^{2}\textbf{A}$ meanwhile the third term
have vanished by assuming $\varPhi\equiv0$. Then, as the MQS approximation
relies in assume that the electric field quickly adjusts to a constant value
along the interval $[t+\delta t]$, for enough small time steps $\delta t$ (see
Fig.~\ref{Figure_1_1}) the action of 
$\Delta\textbf{A}\cdot\delta_{t}^{2}\textbf{A}$ may be neglected, and therefore 
the solution of the critical state problem can be also achieved from the
functional for the magnetic sector:
\begin{eqnarray}\label{Eq.2.11}
{\cal F}[\textbf{B}(\cdot)]={\rm Min}\int_{\Re^{3}}
\frac{1}{2}|\Delta\textbf{B}|^{2} \, .
\end{eqnarray}
Recall that, the minimization principle is based on a discretization of the path
followed by the external sources, meaning that it is an approximation to the
continuous evolution whose accuracy increases as the step diminishes. 

Moreover, we must emphasize that the derived functionals
[Eqs.~(\ref{Eq.2.7}) \& (\ref{Eq.2.11})] are in matter of fact fully
equivalents, as long as the minimization procedure accomplishes the boundary
conditions imposed by the prescribed sources and the material law
$\textbf{J}\in\Delta_{\textbf{r}}$. Thus, in those cases when an intrinsic
electromagnetic source must be considered, i.e., $\nabla\varPhi\neq0$, the
global set of variables must me constrained by the prescribed conditions. For
example, if the superconductor is carrying a transport
current $I_{tr}$ flowing through the surface $s$, one has to mandatorily
consider the external constraint
\begin{eqnarray}\label{Eq.2.12}
\int_{s}\textbf{J}\cdot\hat{\textbf{n}}ds=I_{tr} \, 
\end{eqnarray}
and further update the distribution of current to satisfy the physical condition
$\textbf{E}\cdot\textbf{J}=0$ (at those points where the magnetic flux does not
vary), by means the use of a \textit{calibrated} potential
$\textbf{\~{A}}$. Thus, one of the advantage
of the formulation in Eq.~(\ref{Eq.2.11}) is that the number of variables can be
reduced avoiding to include the intrinsic variables associated to $\varPhi$,
accordingly to the statement
$\textbf{E}=-\partial_{t}\textbf{A}-\nabla\varPhi\equiv\partial_{t}
\textbf{\~{A}}$

Concluding, for 3D problems, it must be emphasized that the introduced
minimization principle can be applied for any shape of the
superconducting volume $\Omega$ as well as for any general restriction (material
law) for the current density $\textbf{J}_{l+1}\in\Delta_{\textbf{r}}$.
Different possibilities for the material law are described in the following
chapter. It is
also to be noticed that the searching of the minimum for the allowed set of
current densities must fulfill the intrinsic condition
$\nabla\cdot\textbf{J}=0$ to be consistent with charge conservation
in the quasi-steady regime. Further, from the numerical point of view, the
advantage
of the variational formulation in  Eqs.~(\ref{Eq.2.7})~\&~(\ref{Eq.2.11}) is
that
one can avoid the integration of the equivalent partial differential equations
and straightforwardly minimize the discretized integral by using a numerical
algorithm for constrained minimization (see Chapter~\ref{ch-3}). This fact
represents an
important advantage in the performance and power of the numerical methods
applied to the design of superconducting devices, where symmetry arguments can
allow further simplifications and correspondingly faster numerical convergence.


\vspace*{1cm}
\subsection*{\label{ch-2-2}
\hspace*{0.1cm} \textsl{2.2 The Material Law: SCs with
magnetic anisotropy}}
\vspace*{1cm}
\markboth{\hspace*{0.1cm}\textsl{2.2 The Material Law: SCs with
magnetic anisotropy}} 
{\hspace*{0.1cm} \textsl{2.2 The Material Law: SCs with
magnetic anisotropy}}
\addcontentsline{toc}{chapter}
{\hspace*{0.1cm} \textsl{2.2 The Material Law: SCs with
magnetic anisotropy}}

In this section, we will continue our discussion of the critical state theory
which still needs the explicit inclusion of a material law
$\textbf{J}(\textbf{E})$ that dictates the magnetic response of a
superconducting sample for a given external excitation. For simplicity, we start
with an overview of the  material law for 1D systems, that will be gradually
generalized until a 3D formulation is reached.

\vspace*{0.5cm}
\hspace*{-0.3cm}
\textbf{\textit{2.2.1 ~Onto The 1D Critical States}}
\markboth{\hspace*{0.1cm} \textsl{2.2.1 ~1D Critical States}}{\hspace*{0.1cm}
\textsl{2.2.1 ~Onto The 1D Critical States}}
\addcontentsline{toc}{chapter}{\hspace*{0.1cm} \textsl{2.2.1 ~Onto the 1D
Critical States}}\label{ch-2.2.1}
\vspace*{0.5cm}

For our purposes, it is sufficient to recall that the basic structure of the
critical state problem (Fig.~\ref{Figure_1_1}) relates to an experimental graph
within the $\{V , I \}$ plane that basically contains two regions defined by
the critical current value $I_{c}$ as follows:
\begin{enumerate}
\item $-I_c \leq I \leq I_c$ with perfect conducting behavior, i.e.:
 $V=0$ and $\partial _t I = 0$.
\item For $I \gtrsim I_c$, the curve is characterized by a high  
${\partial_{I} V}$ slope (and antisymmetric for $I \lesssim - I_c$). Further
steps, with $I$ increasing above the critical value $I_c$, i.e., the eventual
transition to the normal state, may be neglected for slow sweep rates of the
external sources, which produce moderate electric fields.
\end{enumerate}

Within the local description of the electromagnetic quantities involved in the
superconducting response, different models have been used for the
corresponding $E \leftrightarrow J$ graph, the most popular being

\break
\begin{enumerate}
\item The {\em power law} model~\footnotemark[2]\\
$E=\alpha\;{\rm sgn}(J)\left({|J|}/{J_c}\right) ^n$, with $\alpha$ a constant
and $n$ high.
\item The {\em piecewise continuous linear} approximation~\footnotemark[2]\\
$E=0$ for $|J|\leq J_c$, and $E=\beta\;{\rm sgn}(J)(|J|-J_c)$ for $|J|> J_c$,
$\beta$ having a high value.
\item {\em Bean's model}~\footnotemark[3]\\
$J$ constant for $E=0$, and $J={\rm
sgn}(E)J_c$ for $E\neq 0$.
\footnotetext[2]{Is to be noted that these models (1 \& 2) present a small
dependence on the sweep rate, as far as different values of $E$ give way to a
slightly different $J$. The feasibility of the selected model can only be
justified by agreement of the free parameters with the experimental results.}
\footnotetext[3]{This is the simplest model, without sweep rate dependence
because only the sign of $E$ enters the theory (see Fig.~\ref{Figure_1_1}).}
\end{enumerate}

In some treatments, the first or second models are 
implemented, in order to transfer a full ${\bf E}({\bf J})$ law to the Maxwell
equations. Further, notice that Bean's model may be obtained from
the other representations through the limiting cases $n\to\infty$ and
$\beta\to\infty$ respectively.

The well known experimental evidence of a practical sweep rate independence for
magnetic moment measurements (unless for high frequency alternating sources or
at elevated temperatures) allows the use of the clearest Bean's model because
the critical state problem is no longer a time-dependent problem, but a
path-dependent one, meaning that the trajectory of the external sources
${\bf H}_{0}$ uniquely determines the magnetic evolution of the
sample. This makes an important difference when one compares to more
standard treatments, as far as Faraday's law is not completely determined from
the path~\cite{P1-Ruiz_PRB_2009}. Strictly speaking, one has
\begin{eqnarray}\label{Eq.2.13}
\Delta{\bf B} = - \nabla\times [{\bf E}\Delta t] \,\, ,
\end{eqnarray}
with $\Delta t$ (and therefore $|{\bf E}|$) depending on the external sources.
i.e., the absence of an intrinsic time constant gives way to the
arbitrariness in the time scale of the problem. 

Furthermore, in the actual 1D applications of Beans's model,
Faraday's law is not strictly solved and \textbf{E} is absent from the theory.
It is just the sign rule (the {\it
vectorial} part of the material law), that is used to integrate Ampere's law.
Notice that such sign rule corresponds to a maximal shielding response against
magnetic vector variations, and thus, determines the selection of $J= \pm J_c$.

Regarding the direction of \textbf{E}, in ``1D'' problems one has ${\bf
J}\parallel {\bf E}$ and both orthogonal to $\bf B$, such that the physical
threshold related to a maximum value of the force balancing the magnetostatic
term ${\bf J}\times{\bf B}$ gives place to the material law
\begin{eqnarray}\label{Eq.2.14}
J_{\perp}={\rm sgn}(E_{\perp})J_{c\perp} \qquad{\rm for}\qquad E_{\perp}\neq
0\,\, .
\end{eqnarray}
Here, $E_{\perp}$ stands for the component of ${\bf E}$ along the direction
${\bf B}\times ({\bf J}\times{\bf B})$, and the material law falls in a ``1D''
scalar condition which describes the physical mechanism of vortex depinning. 

At this point, the constitutive relation for the critical state describes the
underlying physics for the coarse-grained fields in homogeneous type-II
superconductors. However, it is well known that the coarse-grained behavior
approach straightforwardly depends on the manufacturing process of the
superconducting sample as far as inclusion of impurities, magnetic defects,
or deformation of their cristal structure imposes the local coupling of $J_{c}$
with the intrinsic variation of the magnetic field B. Thus, for practical
purposes we emphasize that the theoretical framework developed in this book is
fully general, with caution of suggest to the experimentalist the need of an
apriori measurement of the dependence $\textbf{J}_{c}[\textbf{r},\textbf{B},T]$
at least in those cases where the condition $\textbf{J}\perp \textbf{B}$ can be
asserted. Henceforth, the
implementation for a particular superconductor can be carried out.

\vspace*{0.5cm}
\hspace*{-0.3cm}
\textbf{\textit{2.2.2 ~Towards The 3D Critical States}}
\markboth{\hspace*{0.1cm} \textsl{2.2.2 ~Towards The 3D Critical
States}}{\hspace*{0.1cm} \textsl{2.2.2 ~Towards The 3D Critical States}}
\addcontentsline{toc}{chapter}{\hspace*{0.1cm} \textsl{2.2.2 ~Towards The 3D
Critical States}}\label{ch-2.2.2}
\vspace*{0.5cm}

In the case of superconductors with anisotropy of the critical current, the
description of their magnetic behavior requires the development of approaches
more sophisticated than 1D-Bean's model. The main issue is that, in general, the
parallelism of ${\bf E}$ and ${\bf J}$ and their perpendicularity to ${\bf B}$
are no longer warranted. Then, the {\em sign rule} of Eq.~(\ref{Eq.2.14})
does not suffice for determining the solution of Eq.~(\ref{Eq.2.6}), and the
optimal control condition with $\textbf{J}\in\Delta_{\textbf{r}}$
({\em a vectorial rule}) must be invoked.

The simplest assumption that translates the critical state problem to 3D
situations was already issued by Bean
in Ref.~\cite{P1-Bean_1970}. It has been called the \textbf{\textit{isotropic
critical state model}} (ICSM) and generalizes 1D Bean's law to
\begin{equation}\label{Eq.2.15}
{\bf J} =J_{c}\,\hat {\bf E}\quad{\rm if} \quad E \neq 0 \, ,
\end{equation}
i.e., the region $\Delta_{\textbf{r}}$ becomes a sphere. Noticeably, in spite
of lacking a solid physical basis, thanks to its mathematical simplicity, this
qualitative model has been widely used by several authors for reproducing a
number of experiments with rotating and crossed magnetic
fields~\cite{P1-Badia_PRL_2001,P1-Badia_PRB_2002,P1-ICSM_app}. In any case,
one could argue that statistical averaging over a system of entangled flux lines
within a random pinning structure might be responsible for the isotropization of
$\Delta_{\textbf{r}}$.
 
\begin{figure}[t]
\centering
\includegraphics[width=1.0\textwidth]{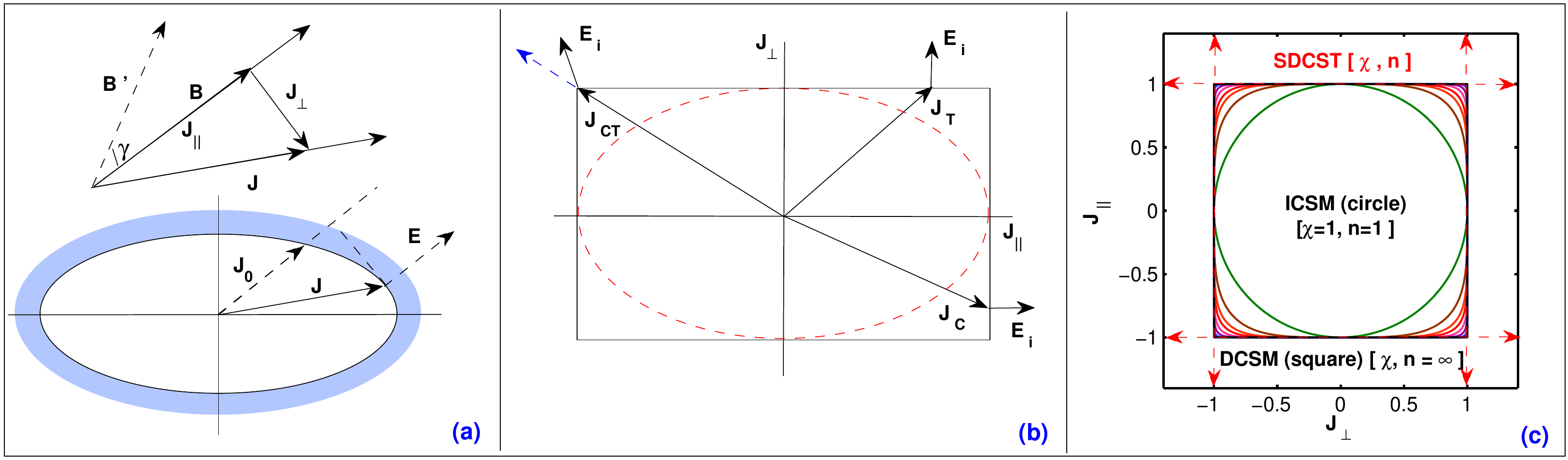}
\caption{\label{Figure_2_1} {\it Pane}~(a), Top: Schematic representation of the
local relative
orientations of {\bf B} and {\bf J}. Also sketched is the direction of
the magnetic field at some neighboring point, at an angle $\gamma$.
The vectors ${\bf B}$, ${\bf B}'$ and ${\bf J}$ do not
necessarily lie at the same plane. The current is decomposed into its parallel
and perpendicular components, i.e.: ${\bf J}={\bf
J}_{\parallel}+{\bf J}_{\perp}$.
Bottom: the {\em perfect conducting} region within the plane perpendicular to
$\textbf{B}$. An induced electric field is shown. Initially (${\bf
J}_{0}$), the high dissipation region is touched, but almost instantaneously
{\bf J} shifts along the boundary, reaching a point where the condition ${\bf
E}\perp\partial\Delta_{\textbf{r}}$ is fulfilled. Anisotropy within the plane is
allowed. {\it Pane}~(b): Geometric interpretation of the DCSM. {\bf J} is
constrained to the
boundary of a rectangular region. T, C and CT states are related to
the horizontal and vertical sides, and to the corners. Coupling between
the components $J_{c\parallel}$ and $J_{c\perp}$ is envisaged by the EDCSM
(dotted red ellipse). {\it Pane}~(c): Our generalization of the material law for
critical
state problems or SDCST. Several regions are shown from the degree of the
superelliptical functions ($n=1,2,3,4,6,10,20,40,\infty$) and
$\chi=J_{c\parallel}/J_{c\perp}=1$.}
\end{figure}

On the other hand, the general statement of the critical state in
terms of well accepted physical basis was firstly introduced by John R.
Clem~\cite{P1-Clem_DCSM}, and it is currently known as the
\textbf{\textit{double
critical
state model}} (DCSM). In particular, this theory
assumes two different critical parameters, $J_{c\parallel}$ and $J_{c\perp}$
acting as the thresholds for the components of ${\bf J}$ parallel and
perpendicular to ${\bf B}$ respectively. Notice
that, $J_{c\perp}$ relates to the flux depinning threshold induced by the
Lorentz force on flux tubes ($\textbf{J}\times\textbf{B}$),
while the additional $J_{c\parallel}$ is imposed by a maximum gradient in the
angle between adjacent vortices
($\textbf{B}\cdot\nabla\gamma=J_{\parallel}$) before mutual cutting
and recombination occurs  [see Fig.~\ref{Figure_2_1}~(a)]. In brief, the
DCSM may be expressed by the statement
\begin{eqnarray}\label{Eq.2.16}
\left\{
\begin{array}{ll}
{\bf J}_{\parallel} & =J_{c\parallel}\;\hat {\bf u}
\\
{\bf J}_{\perp} & =J_{c\perp}\,\hat {\bf v}
\end{array}
\right. \, ,
\end{eqnarray}
being $\textbf{\^{u}}$ the unit vector for the direction
of \textbf{B}, and $\textbf{\^{v}}$ a unit vector in the perpendicular plane
to \textbf{B}.

Within the DCSM, the region $\Delta_{\textbf{r}}$ is a cylinder with its axis
parallel to $\bf B$, and a rectangular longitudinal section in the plane defined
by the unit vectors $\hat{\bf B},\hat{\bf J}_{\perp}$ [see
Fig.~ \ref{Figure_2_1}~(b)]. The edges of the region $\Delta_{\textbf{r}}$
introduce a criterion for classifying the CS configurations into: 
\begin{enumerate}
\item T zones
or flux transport zones ($J_{\perp}=J_{c\perp}$;
$J_{\parallel}<J_{c\parallel}$) where the
flux depinning threshold has been reached (${\bf J}$ belongs to the
horizontal sides of the rectangle),
\item C-zones or flux cutting zones
($J_{\parallel}= J_{c \parallel}~;~J_{\perp}<J_{c \perp}$) where the
cutting threshold has been reached (${\bf J}$ belongs to the vertical sides
of the rectangle),
\item CT zones ($J_{\parallel} = J_{c \parallel}$ and
$J_{\perp} = J_{c \perp}$) where both $J_{\parallel}$
and $J_{\perp}$ have reached their critical values (corners of the
rectangle), and
\item O zones depicted the regions without energy dissipation (the current
density vector belongs to the interior of the rectangle). 
\end{enumerate}

%

Notice that $J_{c\parallel}$ and $J_{c\perp}$ are determined from different
physical phenomena, and their values may be very different (in general
$J_{c\parallel}>J_{c\perp}$ or even $J_{c\parallel}\gg J_{c\perp}$).
Nevertheless, the coupling of parallel and perpendicular effects has been longer
recognized by the experiments~\cite{Clem_2011_SUST,P1-Boyer_80} and, for
instance,
may be included in the
theory by the condition $J_{c\parallel}=K B J_{c\perp}$ with $K$ a
material dependent constant. 

Recalling that the mesoscopic parameters $\textbf{J}_{c}$
are related to averages over the flux line lattice, interacting activation
barriers for the mechanisms of flux depinning and cutting are expected and this
may give place to deformations of the boundary $\partial\Delta_{\textbf{r}}$
[see Fig.~ \ref{Figure_2_1}(b)]. Thus, validated in those cases where a good
agreement with the experiments is achieved, the theoretical scenario can be
enlarged by a number of alternative approaches that focus on different aspects
of the vast number of experimental activities in this field, e.g. one
can identify the so-called:

\begin{enumerate}
\item \textit{Isotropic critical state models
(ICSM)}~\cite{P1-Ruiz_PRB_2009,P1-Ruiz_SUST_2010,P1-ICSM_app}\\
\hspace*{3.4cm} $J^{2}=J_{\parallel}^2+J_{\perp}^2 \leq J_{c}^2$ 
 \item \textit{Elliptical double critical state
models (EDCSM)}~\cite{P1-Ruiz_PRB_2009,P1-EDCSM_app}\\
\hspace*{3.26cm}
$J_{\parallel}^2/J_{c\parallel}^2+J_{\perp}^2/J_{c\perp}^2\leq1$ 
 \item \textit{T critical state model
(TCSM)}~\cite{P1-Ruiz_PRB_2009,P1-Ruiz_SUST_2010,P1-Ruiz_SUST_2011,
P1-Brandt_2007,
P1-Ruiz_PRB_2011}\\
\hspace*{2.87cm} $J_{\parallel}$ unbounded $\forall$ $J_{\perp}\leq J_{c\perp}$
.
\end{enumerate}

Remarkably, the whole set of models have been recently unified by us in
Ref.~\cite{P1-Ruiz_SUST_2010} within a continuous two-parameter theory that
poses
the critical state problem in terms of geometrical concepts within the
$J_{\parallel}-J_{\perp}$ plane (see Fig.~ \ref{Figure_2_1}~(c)). To be
specific, in this framework, we have shown that by the application of our
variational statement~\cite{P1-Ruiz_PRB_2009}, one is able to specify almost any
critical state law by means of an integer index $n$, that accounts for the
smoothness of the $J_{\parallel}(J_{\perp})$ relation, and a certain {\em
bandwidth} characterizing the magnetic anisotropy ratio $\chi\equiv
J_{c\parallel}/J_{c\perp}$. This and the variational formalism introduced
above constitutes the so-called \textbf{\textit{Smooth Double
Critical State Theory}} (SDCST), which allows to elucidate the relation
between diverse physical processes and the actual material law.

Mathematically, the material law introduced in our general theory for the
critical state problem or SDCST is based upon the idea that either material or
extrinsic anisotropy can be easily incorporated by prescribing a region
$\Delta_{\textbf{r}}$ where the physically admissible states of $\textbf{J}$ are
hosted as limiting cases of a smooth expression defined
by the two-parameter family of superelliptic functions, 
\begin{equation}\label{Eq.2.17}
\left(\frac{J_{\parallel}}{J_{c\parallel}}\right)^{2n}+\left(\frac{J_{\perp}}{J_
{c\perp}}\right)^{2n}\leq 1.
\end{equation}

We call the readers' attention to the fact that an index $n=1$ and a bandwidth
defined by $\chi\equiv J_{c \parallel}/J_{c \perp}=1$ correspond to the standard
ICSM~\cite{P1-ICSM_app}. 
On the other hand, when one assumes enlarged bandwidth (i.e.: $\chi>1$), the
region $\Delta_{\textbf{r}}$ of the SDCST becomes the standard
EDCSM introduced by Romero-Salazar and
P\'erez-Rodr\'iguez~\cite{P1-EDCSM_app}. 
When the bandwidth $\chi$ is extremely large, i.e., $J_{c\parallel}\gg J_{c
\perp}$, one recovers the so-called $T-$zones treated by Brandt and
Mikitik~\cite{P1-Brandt_2007}. Rectangular regions strictly corresponding to the
DCSM~\cite{P1-Clem_DCSM} are
obtained for the limit $n\rightarrow\infty$ and arbitrary $\chi$. Finally,
allowing $n$ to take values over the positive integers, a wide scenario
describing anisotropy effects is envisioned [Fig.~\ref{Figure_2_1}(c)]. Such
regions will be named after superelliptical and their properties can be
understood in terms of the rounding (or smoothing) of the corners for the DCSM.


\chapter{\label{ch-3} \sc \textbf{Computational Method}}

In chapter~2.1 we have mentioned that the minimization functionals
[Eq.~(\ref{Eq.2.7}) or Eq.~(\ref{Eq.2.11})] may be transformed so as to get a
practical vector potential formulation. In turn, the resulting formulation can
be expressed in terms of the so-called magnetic inductance matrices which
allows a clearest identification of the set of elements playing some role in
the minimization procedure. In this chapter, we shall discuss how to implement
the above
statements for general critical states in the
framework of the computational methods for large scale nonlinear optimization
problems. 

Being more specific, in Eq.~(\ref{Eq.2.11}) the integrand $\frac{1}{2}
(\Delta{\bf B}) ^2$ can be rewritten as $\frac {1}{2} (\Delta{\bf B})\cdot
(\nabla \times \Delta{\bf A})$,
and manipulated to get $\frac {1}{2} (\Delta{\bf A})\cdot (\nabla \times
\Delta{\bf B})$ plus a divergence term, fixed by the external sources at a
distant surface. Now, the integral is restricted to the superconducting sample
volume $\Omega$, because $\nabla \times \Delta{\bf B} = \mu _0 \Delta{\bf J}$ is
only unknown within the superconductor. In addition, assuming that local
sources such as an injected transport current may be introduced as an external
constraint, and with the boundary condition that $\textbf{A}$ goes
to zero sufficiently fast as they approach infinity, the vector
potential can be expressed as:\footnotemark[1]
\footnotetext[1]{Recall that, \textbf{A} is determined by the Maxwell's
equations solution in the Lorenz gauge condition, i.e., the vector potential
must satisfy the condition $\partial_{\mu}A^{\mu}=0$ for any transformation
gauge $A^{\mu}\rightarrow A^{\mu}+\partial^{\mu}\psi$ with $\psi$ a scalar
function. Thus, as no local-sources are present into the minimization
principle of Eq.~\ref{Eq.2.11}, one is enabled to apriori assume $\psi=0$, and
thence simplify the vector potential in terms of the Coulomb gauge.}
\begin{equation}\label{Eq.3.1}
 \Delta{\bf A} = \Delta{\bf A}_{0}
+ \frac {\mu _0 }{4 \pi} \int _\Omega \frac {\Delta{\bf J}}{|{\bf r} - {\bf
r}'|}
d^3 {\bf r}' \, .
\end{equation}

This transforms ${\cal F}$ into a double integral over the body of the sample,
i.e.:
\begin{eqnarray}\label{Eq.3.2}
{\cal F}[\textbf{A}(\cdot),J\in\Delta_{\textbf{r}}]= &&\frac {8\pi}{\mu_0}\int
_\Omega \Delta{\bf A}_{0}\cdot {\bf J}_{l+1}({\bf r})d^3 {\bf r}
\nonumber\\&&
+\int\!\int _{\Omega \times \Omega}
\frac {{\bf J}_{l+1}({\bf r}')
\cdot [{\bf J}_{l+1}({\bf r}) - 2 {\bf J}_{l}({\bf r})]}{|{\bf r} - {\bf
r}'|}d^3 {\bf r}d^3 {\bf r}'\;
\end{eqnarray}

As a consequence, only the unknown current components within
the superconductor $(\textbf{J}_{l+1})$
appear in the computation so reducing the number of unknown variables. At this
point let me emphasize that Eq.~(\ref{Eq.3.2}) can be applied for any shape of
the superconducting volume $\Omega$ as well as for any physical constraint
(material law $\Delta_{\textbf{r}}$) for the local current density
$\textbf{J}_{l+1}$, and further for any condition defined by the external
sources ($\textbf{A}_{0}$). Above this, minimization must ensure
the charge conservation
condition by searching the minimum for the allowed set of current densities
fulfilling $\nabla\cdot\textbf{J}=0$.

On the other hand, also it may be noticed that the double
integral in Eq.~(\ref{Eq.3.2}) can be (\textit{eventually}) identified as the
Neumann formula once it has been transformed into filamentary closed circuits. A
noteworthy fact is that regarding the superconducting volume, the coefficients
of the intrinsic inductance matrices are straightforwardly independent of time
and consequently, they appear in the root problem before going to minimize the
functional. Indeed, the proper description of the inductance coefficients
directly depends on the geometry of the superconductor and the boundary
conditions defined by the dynamics of the external electromagnetic sources,
where any symmetry of the problem allows further simplifications and
correspondingly faster numerical convergence. To be specific, upon
discretization in current elements ($I_{i}=J_{i}s_{i}$), the minimization
functional for critical state problems bears
the algebraic structure
\begin{eqnarray}\label{Eq.3.3}
{\cal
F}[I_{l+1}]=\frac{1}{2}\sum_{i,j}I_{i,l+1}M_{ij}I_{j,l+1}-\sum_{i,j}I_{i,l}M_{
ij } I_ {j,l+1}+\sum_{i}I_{i,l+1}\Delta A_{0}(M_{0}) \, ,
\end{eqnarray}
with $\{I_{i,l+1}\}$ the set of unknown currents at specific
circuits for the problem of interest, $M_{ij}$ their intrinsic {\em inductance
coupling coefficients}, and $M_{0}$ the inductance matrices associated to the
external sources $A_{0}$. 

Corresponding to the critical state rule ${\bf J}\in\Delta_{\textbf{r}}$, in
order to minimize Eq.~(\ref{Eq.3.3}) each value $J_i$ must
be constrained. Thus, as it was described in chapter~\ref{ch-2}.2.2, we have
found that a number of constraints related to physically meaningful critical
state models may be expressed in the algebraic form  
\begin{eqnarray}\label{Eq.3.4}
{\rm
F}_{\alpha}\left(\sum_{i}I_{i}C_{ij}^{\alpha}I_{j}\right)\leq f_{0\alpha} \,
~\forall ~ j
\end{eqnarray}
with $f_{0}$ some constant representing the physical threshold, and 
${\rm F}_{\alpha}(\cdot)$ an algebraic function based upon a coupling matrix
$C_{ij}^{\alpha}$ whose elements depend on the physical model. 
For example, in the simpler cases (isotropic models), the constraints
correspond to assume the matricial elements $C_{ij}=\delta_{ij}$, and the
physical threshold $f_{0}={J}_{c}^{2}$.

For simplicity, most technical procedures related to the introduction of
intricate models and either depict the minimization functional in terms of the
inductance coefficients (including those for external sources) will be left as
matter of study of the following chapters (Part~\ref{Part_2}). In return, below
we present a thorough analysis of the computational tools handled for critical
state problems at large scale. 

\begin{figure}[t]
\centering
\includegraphics[height=13cm,width=13cm]{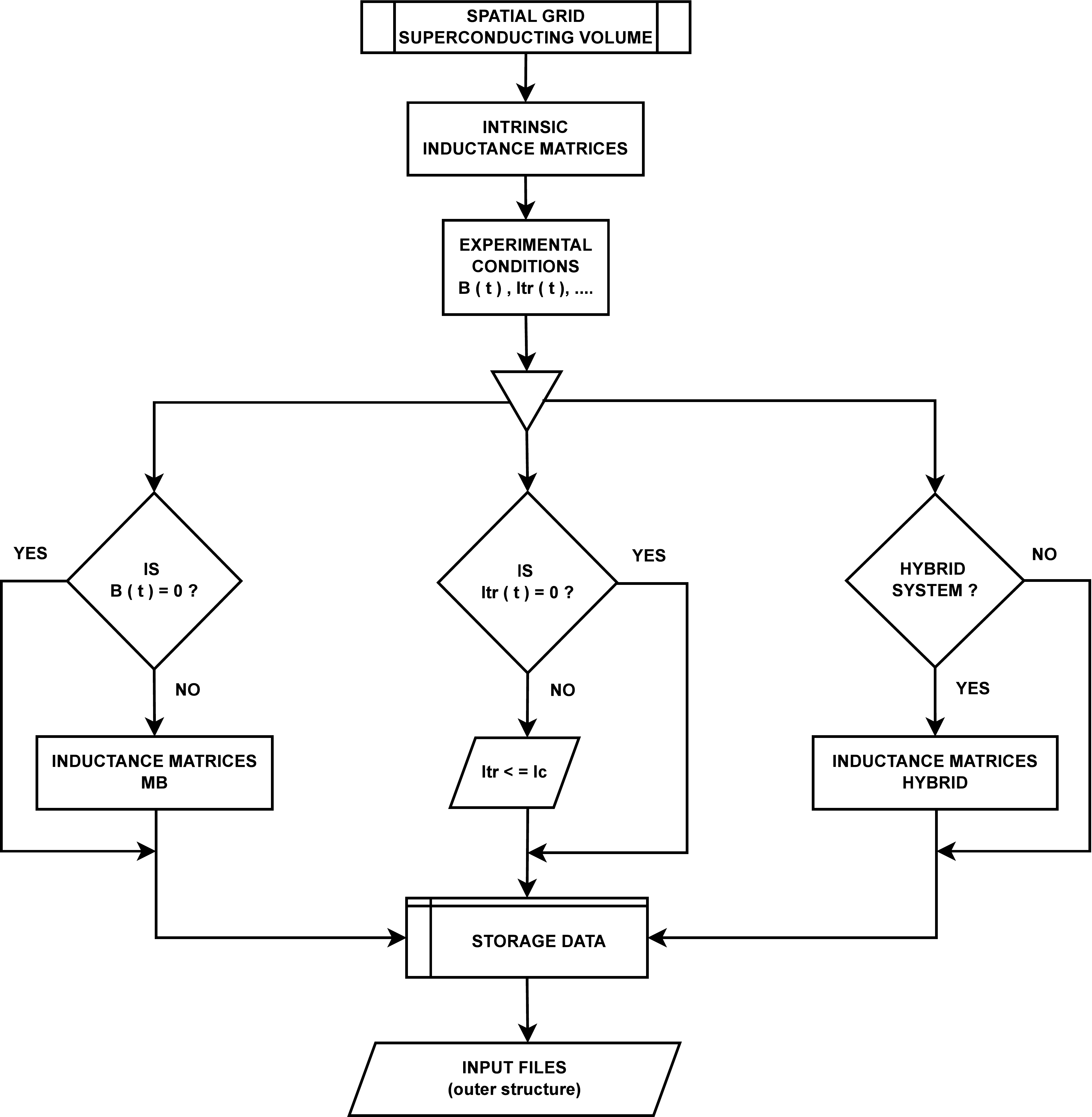}
\caption{\label{Figure_3_1} Flow chart describing  the preparation and
management of the input elements for the objective function.}
\end{figure}

\begin{figure}[t]
\centering
\includegraphics[height=13cm,width=13cm]{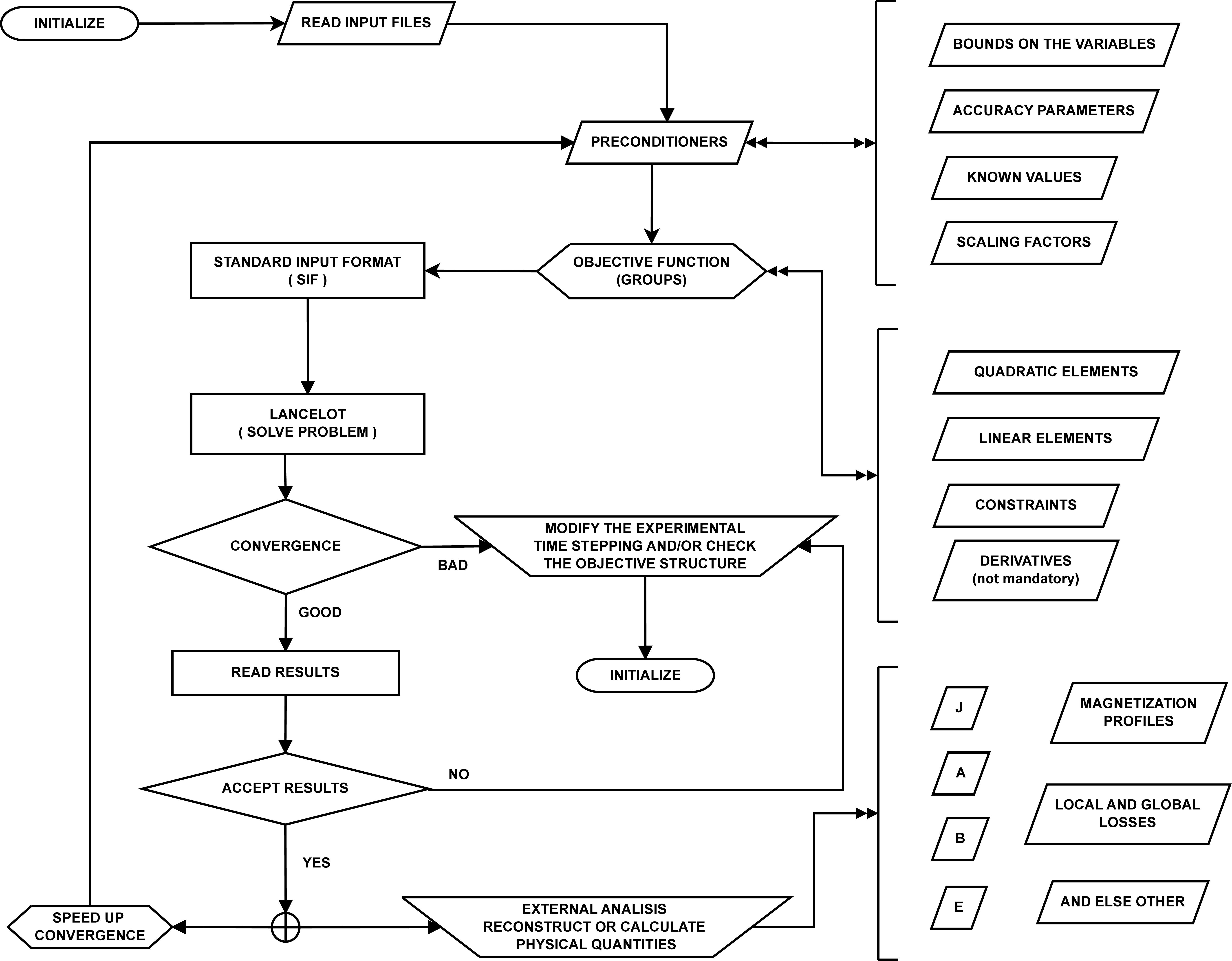}
\caption{\label{Figure_3_2} Flow chart describing the main structure of the
computational method implemented along this book.}
\end{figure}

With the purpose of obtaining a minimal understanding about how a critical state
problem can be tackled from the numerical point of view, the computational
method is sketched in the flow charts of
figures~\ref{Figure_3_1}~\&~\ref{Figure_3_2}.

The first step is designing a grid which will allows to describe the
superconducting volume $\Omega$ as a set of elements
$\delta\Omega_{i}$, each of them characterized by a well defined current
density flowing along the coordinates $\textbf{r}_{i}$. Then, the matrices for
the intrinsic inductance coefficients between the elements
$\textbf{J}(\textbf{r}_{i})$ and $\textbf{J}(\textbf{r}_{j})$ for all the set
of possible couples $(\textbf{r}_{i},\textbf{r}_{j})\in\Omega$ must be
calculated and stored on disk. As the mesh of points
$(\textbf{r}_{i},\textbf{r}_{j})$ can be considerably large, we suggest take
advantage on the matricial formalism provided by Matlab$^{\textregistered}$
and
their own language for storage data. Once the spatial elements playing some role
into the functional have been properly defined, the temporal \textit{sector}
must be introduced by means enough small path steps of the external
electromagnetic sources, i.e., the experimental conditions must be connected by
the finite difference expressions such a
$\Delta\textbf{B}_{0}=\textbf{B}_{l+1}-\textbf{B}_{l}$, where the
associated distribution of currents $\textbf{I}_{l+1}$ plays the role of
unknown. To be specific, in those cases where the superconductor is subjected
to an external magnetic field $\textbf{B}_{0}$, additional inductance matrices
($M_{0}$)
must be introduced according to the definition
\begin{equation}\label{Eq.3.5}
 \textbf{A}_{0}
(\textbf{B}_{0},\textbf{r}_{j})=\textbf{B}_{0}\times\textbf{r}_{ j } \, .
\end{equation}
On the other side, the vector potential $\textbf{A}_{0}$ not only allows to
define the contribution at the local potential $\textbf{A}$ produced by an
external magnetic field ($\textbf{A}_{{\rm B}}$), rather it also allows
consider the coupling with another materials such as ferromagnets. 

Before going within the minimization procedure, it has to be noticed that those
cases
considering a transport current along the superconducting sample must
be understood as a problem where the minimization variables are required to
satisfy a set of auxiliary constraints [see Eq.~(\ref{Eq.2.12})] under the
global critical state condition $I_{tr}\leq I_{c}$. Also, as the
values for the elements $\textbf{I}_{l}(\textbf{r})$ are
assumed to be known in advance, the linear elements into the argument of the
functional (objective function) can be calculated before minimizing. 

At this point, it is probably worthwhile to argue on what we mean by the
computational method for minimization of an objective function. Firstly, this
notion is clearly computer dependent, as the size of large scale problems can
require a substantial amount of memory and store. Moreover, what is large in a
personal computer can be significantly different from what is large on a super
computer. The  first machine just to have a smaller memory and storage than the
second one, and therefore has more difficulty handling problems involving a
large amount of data. Secondly, the size of the objective function strongly
depends on the structure and the mathematical formulation of the problem
and exploiting it is often crucial if one wants to obtain an answer
efficiently. The complexity of this structure is often a central key in
assessing the size of a problem\footnotemark[2]. For example, for linear
objective functions
(not our case) it is possible to solve pretty large size problems (say four
million variables). However, the objective function for problems in applied
superconductivity is in general highly nonlinear and, for instance, the
quadratic terms suggest to reduce the number of variables in a root square
factor (say two thousand variables). One advisable possibility for reducing the
number of elements in the objective function is subdividing the problem into
loosely connected subsystems, i.e., all the internal operations which do not
depend of the minimization variables must
be preallocated to a well structured data (Figure.~\ref{Figure_3_1}).
Lastly, an efficient algorithm for nonlinear optimization problems
must be either invoked or built. Fortunately, nowadays there is a significant
amount of available software with standard optimization tools which allow a
faster foray in this matter~\cite{P1-Matlab,P1-Lancelot}. 
\footnotetext[2]{Customarily memory access violations (segmentation faults)
appear on nonlinear large systems without a well designed structure.}

We must call reader's attention on the fact that efficient algorithms for
small-scale problems (in the sense that, assuming infinite precision,
quasi-Newton methods for unconstrained optimization are invariant under linear
transformations) do not necessarily translate into efficient algorithms
for large scale problems. Perhaps the main reason is that, in order to be able
to handle large problems with a high accuracy, the structure of the objective
function and the minimization algorithms have both of them to be enough simple
and tractable to avoid a wasting of time in the scaling of variables for the
inner iteration subproblem (the minimization itself) and the finding of an
optimum value for each one of the variables with respect to the remaining
variables sought. In this context, one  of the most powerful algorithms for
large and nonlinear constrained optimization problems, known as
LANCELOT, has been developed by the professors Andrew Conn (IBM corporation,
USA), Nick Gould (Rutherford Appleton Laboratory, UK), Philippe Toint
(Facult\'es Universitaries Notre-dame de la Paix, Belgium), and Dominique Orban
(Ecole Polytechnique de Montreal, Canada)~\cite{P1-Lancelot}. The wide number of
optimization techniques provided by this package and their flexibility in
handle and storage of large amounts of variables, make this program a
clever choice for tackle highly complicated systems as those described by
Eq.~(\ref{Eq.3.2}). 

A thorough study of the minimization techniques and the computational language
allocated in this package is far away of the purpose of this thesis. However,
the structure of a general problem can be understood via the flow chart in
figure~\ref{Figure_3_2}. In brief, the
minimization functional is translated into a suite of FORTRAN
procedures for minimizing an objective function, where the minimization
variables are required to satisfy a set of auxiliary constraints and possibly
internal bounds. Here, the major advantage of LANCELOT is the use of a Standard
Input Format (SIF) as a unified method for communicating numerical data and
FORTRAN subprograms with any optimization algorithm. Thus, when an optimization
problem (minimizing or maximizing a sought of variables) is specified in the SIF
decoder, one is required to write one or more files in ordered sections which
accomplish the role of introduce the set of preconditioners for the objective
function. 

Once the set of input data has been structured accordingly to the number of
variables and further on the temporal dependence of the experimental conditions
(see Fig.~\ref{Figure_3_1}), one is enabled to predefine a set of input cards
allowing the knowledgeable user to specify a priori known limits on the possible
values of the objective function, as well as on the specific optimization
variables, accuracy parameters, and scaling factors (see Fig.~\ref{Figure_3_2}).
Then, the minimization functional or so-called objective function is
subdivided in a set of groups, whose purpose is twofold: On the one hand, the
linear and nonlinear (\textit{quadratic}) elements for the minimization
procedure are identified in a fore. Likewise, the specification of analytical
first derivatives is optional, but recommended whenever possible. The SIF
decoder allows also include the Hessian matrix of the objective function, if the
second-order partial derivatives of the whole set of minimization variables are
known,\footnotemark[3] otherwise the derivatives of the nonlinear element
functions can be approximated by some finite difference method. 

Actually, the full Hessian matrix can be difficult to compute in practice; in
such situations, quasi-Newton algorithms\footnotemark[4] can be
straightforwardly called by LANCELOT where at least ten
different minimization algorithms have been already implemented and
coded according to the standard input format provided by the SIF
decoder~\cite{P1-Lancelot}. Notwithstanding, the solution obtained by LANCELOT
may be compromised if finite difference approximations are used, it has been our
experience that, once
understood, the programming language of SIF is in fact quite efficient for
problem specifications, in such manner that for objective functions
correctly written, and constraint functions well defined, any method can
efficiently reach to the solution sought. In this sense, additional groups may
be announced to make up the objective function by including an ``starting
point'' for the envisaged solution (if it is more or less known, or by defect
it is equals to zero) or, for introducing additional constraints (external
functions conditioning the system) as it is the case when the superconductor
implies a flow of transport current.

\footnotetext[3]{Given the real-valued function $f(x_{1},x_{2},...,x_{n}$, if
all second derivatives of $f$ exists, then the Hessian matrix of $f$ is the
matrix $H(f)_{ij}(x)=\partial_{i}\partial_{j} f(x)$. Hessian matrices are used
in large-scale optimization problems within Newton-type methods because they are
the coefficient of the quadratic term of the local Taylor expansion 
$f(x+\Delta x)\approx f(x)+\breve{J}(x)\Delta x + \frac{1}{2}\Delta x^{~
\textbf{T}} H(x)\Delta x$, where $\breve{J}$ is the Jacobian matrix, which is a
vector (the gradient for scalar-valued functions)}

\footnotetext[4]{Also known as variable metric methods, are algorithms for
finding local maxima and minima of a function, with the aim of find out the
stationary point of their local Taylor expansion where the gradient is 0. Thus,
these methods are so called quasi-Newton methods, because the Hessian matrix
does not need to be a priori computed, as the Hessian is straightforwardly
updated by analyzing successive gradient vectors instead.}

It may happen that a specific problem uses variables or general constraints
whose numerical values are widely different in magnitude, causing 
significant difficulties in the numerical convergence. However, LANCELOT also
gives the chance of incorporate a list of scaling factors which are applied to
the general constraints and variables separately before the optimization
commences, allowing a clearest handling of the group elements in highly
nonlinear problems as those herein considered. Thus, assuring a good
convergence, whether single or double precision,\footnotemark[5] is
implying  in turn to modify the experimental time stepping and either, the
accuracy
parameters such as, the number of iterations allowed, the constraint and
gradient accuracy, the penalty parameter and the trust region for the
optimizing.

\footnotetext[5]{For large scale programming in FORTRAN based languages, one
must care about exceeding the largest positive (or negative)
floating-point number defined by the FORTRAN distribution. By default, we have
defined an architecture of double precision in 64 bits conforming to the IEEE
standard 754 for the latest versions of Intel FORTRAN Compiler (see
Ref.~\cite{P1-IEEE}).}

Finally, in applied superconductivity, a set of complementary programs have to
be developed in an effort to provide a comprehensive understanding of the
temporal evolution of the electromagnetic quantities. For example, if the
set of minimization variables corresponds to the local profiles of current
density $\textbf{J}_{i}(\textbf{r}_{i})$, additional codes must to be used for
calculating $\textbf{A}_{i}(\textbf{r})$, $\textbf{B}_{i}(\textbf{r})$,
and $\textbf{E}_{i}(\textbf{r})$ in the whole $\Re^{3}$-space. Thus,
although integrated quantities such as the magnetic moment
$\textbf{M}(\textbf{J}_{i},\textbf{r}_{i})$ may be revealing a smooth trend
despite the use of a poor numerical accuracy, it is of utter importance testing
the numerical convergence by calculating the local profiles for the
electromagnetic quantities concerning to derived quantities. In this sense, the
following part of this book is devoted to the reliable solution of some
interesting problems in applied superconductivity, where the critical state
statement  falls into a large scale optimization problem.


 \chapter*{}
 \vspace*{-3cm}
 \section*{\Huge{Conclusions I}}
 \markboth{\sc \textbf{Conclusions I}}{\sc \textbf{Conclusions I}}
 \addcontentsline{toc}{chapter}{\sc \textbf{Conclusions I}}
 \vspace*{2cm}

In summary, in this part we have shown that the critical state theory for the
magnetic response of type-II superconductors may be built in a quite general
framework and in turn it may be solved by several means. As our interest is to
deal with highly nonlinear problems at large-scale, we
have emphasized in the performance of variational methods and computational
techniques for solving problems on personal computers.

We remark that the basic concepts underlying our generalization of the critical
state theory can be identified as follows:

\begin{enumerate}
\item The critical state theory bears a Magneto Quasi Steady (MQS) approximation
for the Maxwell equations in which $\ddot{\textbf{B}}$, $\dot{\bf E}$ and
$\dot{\rho}$
are second order quantities and consequently, the displacement current
densities $\dot{\textbf{D}}$ are much smaller than $\textbf{J}$ in the bulk and
vanish
in a first order treatment. This means that the {\em magnetic flux dynamics}
can be entirely described by the finite-difference expression of Faraday's law 
\begin{eqnarray}\label{Eq.3.6}
\Delta\textbf{B}=-\nabla\times(\textbf{E}\delta t) \, ,
\end{eqnarray}
where the physically admissible states must accomplish the MQS Ampere's law,
i.e., $\nabla\times\textbf{B}=\mu_{0}\textbf{J}$. Here, the inductive
part of ${\bf E}$ may be introduced through Faraday's law, whereas the role of
electrostatic quantities is irrelevant for the magnetic sector. In other
words, {\bf E} may be modified by a gradient function (${\bf E}\to{\bf
E}+\nabla\phi$) with no effect on the magnetic response.

\item In type-II superconductors, the law that characterizes the
\textit{conducting behavior} of the material may be written in terms of
thresholds values for the current density constrained to a geometrical region
$(\textbf{J}\in\Delta_{\textbf{r}})$ which suffices to determine the relation
between the directions of \textbf{E} and \textbf{J}. Thus, \textbf{E} is no
longer an unknown variable but rather plays the role of a parameter to be
adjusted in a direct algebraic minimization, i.e., 
\begin{eqnarray}\label{Eq.3.7}
{\rm Min}\{L\}|_{\textbf{J}\in\Delta_{\textbf{\textbf{r}}}}\equiv{\rm
Max}\{\textbf{J}\cdot\textbf{p}\}|_{\textbf{J}\in\Delta_{\textbf{\textbf{r}}}}
\equiv{\rm
Max}\{\textbf{E}\cdot\textbf{J}\}|_{\textbf{J}\in\Delta_{\textbf{\textbf{r}}}}
\, .
\end{eqnarray}
In physical terms, the material ``\textit{reacts}'' with a maximal shielding
rule when electric fields are induced, and a perfect conducting behavior
characterizes the magnetostatic equilibrium when external variations cease. In
fact, is to be noticed that the above representation can be
understood as the macroscopic counterpart of the underlying vortex physics.
Thus, recalling that, in type II superconductors an incomplete isotropy for the
limitations of the current density relative to the orientation of the local
magnetic field arises from the different physical conditions of current flow
either along or across the Abrikosov vortices, one may talk about
magnetically induced anisotropy where the physical barriers of flux depinning
and cutting are customarily depicted by the
condition $\textbf{J}\leq\textbf{J}_{c}\in\Delta_{\textbf{r}}$. 
The evolution
from one magnetostatic configuration to another
occurs through the local violation of this condition, i.e.: ${\bf
J}\notin\Delta_{\textbf{r}}$ ($\textbf{J}>\textbf{J}_{c}$). However, owing
to the high dissipation, an almost instantaneous response may be assumed,
represented by a {\em maximum shielding} rule in the form ${\rm Max}\{{\bf
J}\cdot\hat{\bf E}\}\left.\right|_{{\bf J}\in\Delta_{\textbf{r}}}$.
 
\item With the aim of offering a meaningful reduction of the number of
variables,
we have shown that the problem can be simplified by solving a minimization
functional with a underlying structure based upon inductance matrices [see
Eq.~(\ref{Eq.3.3})]. In
particular, the mutual inductance representation with ${\bf J}({\bf
r})$ as the unknown, offers two important advantages: 

(i) intricate boundary
conditions and infinite domains are avoided, and 

(ii) the transparency of the
numerical statement and its performance (stability) are outlined. 

Then, the quantities of interest (flux penetration profiles and magnetic moment)
are obtained by integration.

\item Most popular models for critical state problems have been
generalized in our so-called \textit{smooth double critical state theory}
(SDCST) for
anisotropic material laws~\cite{P1-Ruiz_SUST_2010}. This theory relies on our
variational framework for general critical state
problems~\cite{P1-Ruiz_PRB_2009} that allows us to
incorporate the above-mentioned physical structure in the form of mathematical
restrictions for the circulating current density. Two fundamental
material-dependent quantities play key roles in this theory
$(J_{c\parallel},J_{c\perp})$ related to the flux cutting and flux depinning
thresholds. Notoriously, the boundary condition for the material law
$\textbf{J}\in\Delta_{\textbf{r}}$ and the mutual interaction between the
critical thresholds have been described in a quite general picture, based upon
the relation between the coupling parameters
$\chi\equiv J_{c\parallel}/J_{c\perp}$
 and the smoothing index $n$ of the \textit{superelliptical} condition
$(J_{\parallel}/J_{c\parallel})^{2n}+(J_{\perp}/J_{c\perp})^{2n}\leq1$. 

Hence, our SDCST cover a wide range of laws: 

(i) the isotropic model ($\chi^{2}=1$, $n=1\Rightarrow \Delta_{\textbf{r}}$ is a
circle), 

(ii) the elliptical model ($\chi^{2}>1$, $n=1\Rightarrow \Delta_{\textbf{r}}$ is
an ellipse), 

(iii) the rectangular model ($\chi^{2}\geq 1$, $n\to\infty\Rightarrow
\Delta_{\textbf{r}}$ is a rectangle), 

(iv) the infinite band model ($\chi^{2}\rightarrow\infty$), and 

(v) else others with smooth magnetic anisotropy ($\chi^{2}\geq 1$, $n\in{\rm N}
\geqslant 1 \Rightarrow \Delta_{\textbf{r}}$ is a rectangle with smoothed
corners). 

\end{enumerate}

Finally, let me emphasize that the scope of our
theory is rather beyond the actual examples treated in the following part of
this thesis. On the one side, we have shown that the critical state concept
allows arbitrariness in the presence of electrostatic charge and potential, and
one could simply upgrade the models by the rule ${\bf E}\to{\bf E}+\nabla\phi$
if necessary. For instance, a scalar function  $\phi$ may be introduced if the
direction of ${\bf E}$ has to be modified respect to the maximum shielding rule
in the MQS limit. On the other side, the extension of the theory to arbitrary
sample geometries is intrinsically allowed by the mutual inductance
representation. Thus, this first part has laid necessary groundwork for
attacking general critical state problems in 3D geometry.


\renewcommand\bibname{References I}

%



\part{\label{Part_2}\textsc{\textbf{Critical State Problems:\newline \\ Effects
\& Applications}}}


\chapter*{}
\vspace*{-3cm}
\section*{\label{Intro-P2}\hspace*{0.55cm} \sc \textbf{Introduction}}
\markboth{\sc \textbf{Introduction}}{\hspace*{0.55cm} \sc \textbf{Introduction}}
\addcontentsline{toc}{chapter}{\hspace*{0.55cm} \sc \textbf{Introduction}}

In the first part of this book the magnetic flux dynamics of type-II
superconductors within the critical state regime has been posed in a
generalized framework, by using a variational theory supported by well
established physical principles and quite general numerical methods. The
equivalence between the variational statement and more conventional treatments,
based on the solution of the differential Maxwell equations together with
appropriate conductivity laws have been stated. On other side, in an effort to
explore new physical scenarios devoted to convey  the advantages of the
variational statement, in this part we present a thorough analysis of several
problems of recognized importance for the development and physical understanding
of intrinsic phenomena linked to the technological application of type-II
superconductors.

In particular, Chapter~\ref{ch-4} is devoted to present the
extensions of the so-called double critical state model to three dimensional
configurations in which either flux transport (T-states), cutting (C-states) or
both mechanisms (CT-states) occur. Firstly, we show the features
of the transition from T to CT states. Secondly, we focus on our generalized
expression for the flux cutting threshold in 3D systems and show its relevance
in the slab geometry. Recall that, our method has allowed us to unify a
number of conventional models describing the complex vortex configurations in
the critical state regime. Thus, in this chapter several material laws already
included in our generalized SDCST are compared to each other so as to weigh out
the inherent influence of the magnetic anisotropy and the coupling between
the flux depinning and cutting mechanisms. This is done by using different
initial configurations (diamagnetic and paramagnetic) of a superconducting slab
in 3D magnetic field, which allow to show that the predictions of the SDCST
range from the collapse to zero of transverse magnetic moment in the isotropic
model to nearly force-free configurations in which paramagnetic values can
arbitrarily increase with the applied field for magnetically anisotropic
current-voltage laws.

Chapter~\ref{ch-5} addresses the study of several intriguing phenomena for the
transport
current in type II superconductors. In particular, we present an exhaustive
study of the electromagnetic response for the so-called longitudinal transport
problem (current is applied parallel to the external magnetic field) in the
slab geometry. On the one hand, we will introduce a simplified analytical
model for a 2D configuration of the electromagnetic quantities. Then, based upon
numerical studies for general scenarios (3D) we will go beyond the analytical
models, and in general, it will shown that a remarkable inversion of the
current flow in a surface layer may be predicted under a wide set of
experimental conditions, including modulation of the applied magnetic field
either perpendicular or parallel (longitudinal) to the transport current
density. On the other hand, according to our SDCST where the magnetic
anisotropy of the superconducting material obeys a geometrical region enclosed
by a superelliptical function for the current density vector, a thorough
characterization of the underlying mechanism of flux cutting and depinning has
been performed. Thus, the intriguing occurrence of negative current patterns and
the enhancement of the transport current flow along the center of the
superconducting sample are reproduced as a straightforward consequence of the
magnetically induced internal anisotropy. Moreover, we establish that the
maximal transport current density allowed by the superconducting sample after
compression towards the center of the sample, is related to the
maximal projection of the current density vector onto the local magnetic field
or material law. Also, it will be shown that a high correlation exists between
the evolution of the transport current density and the appearance of striking
collateral effects, such as local and global paramagnetic
structures in terms of the applied longitudinal magnetic field. Finally,
the elusive measurement of the threshold value for the cutting current
component ($J_{c\parallel}$) is suggested on the basis of local measurements
of the transport current density.

Finally, chapter 6 is devoted to introduce a thorough study of the
electromagnetic response, either local or global, of straight infinite
superconducting wires in the critical state regime under the action of diverse
configurations of transverse magnetic field and/or longitudinal transport
current. A comprehensive theoretical framework for the physical concepts
underlying the temporal evolving of the electromagnetic quantities and the
production of hysteretic losses is in a fore. Thus,
along this line, and for the numerical implementation of our numerical
statement, we have considered three different excitation regimes which are
focused on the electromagnetic response of a superconducting wire with
cylindrical cross-section: (\textit{i}) Isolated electromagnetic
excitations, in which only the action of an external source of oscillating
transverse magnetic field, $\textbf{B}_{0}$, or an impressed AC transport
current, $I_{tr}$, is conceived. (\textit{ii}) Synchronous oscillating
excitations, which deals with the simultaneous action of $\textbf{B}_{0}$ and
$I_{tr}$ for experimental situations wherein both sources are showing the same
oscillating features (identical phase and frequency). Eventually, in (iii)
asynchronous excitation sources, we have addressed to most intricate
configurations where the oscillating sources are out of phase by assuming that
one of them sources is connected to a power supply with a double frequency than
the other. The temporal dynamics of the assorted electromagnetic quantities,
such as
the local profiles of current density $\textbf{J}_{i}$, the lines of magnetic
field (isolevels of the vector potential $\textbf{A}$), the vector components of
the magnetic flux density $\textbf{B}$, the local density of power dissipation
$\textbf{E}_{i}\cdot \textbf{J}_{i}$, the magnetic moment curves $\textbf{M}$,
and the hysteretic AC losses $L$, are shown for each one of the above mentioned
cases including a wide set of amplitudes for the oscillating excitations.
Striking differences between the actual hysteretic losses (predicted by
numerical methods) and the regular approximation formulas with the concomitant
action of both sources are highlighted. Also quite interesting magnetization
loops with exotic shapes non connected to Bean-like structures are outlined. An
outstanding low pass filtering effect intrinsic to the magnetic response of the
system, and a strongest localization of the heat release is envisioned for
systems subjected to synchronous excitations. Furthermore, contrary to the
generalized assumption that asynchronous sources may attain reductions in the
hysteretic losses, we show that as a consequence of considering double frequency
effects, noticeably increase of the hysteretic losses may be found.








\chapter{\label{ch-4} \sc \textbf{Type-II SCs With Intrinsic Magnetic
Anisotropy}}

\begin{figure}[t]
\centering
\includegraphics[width=1.0\textwidth]{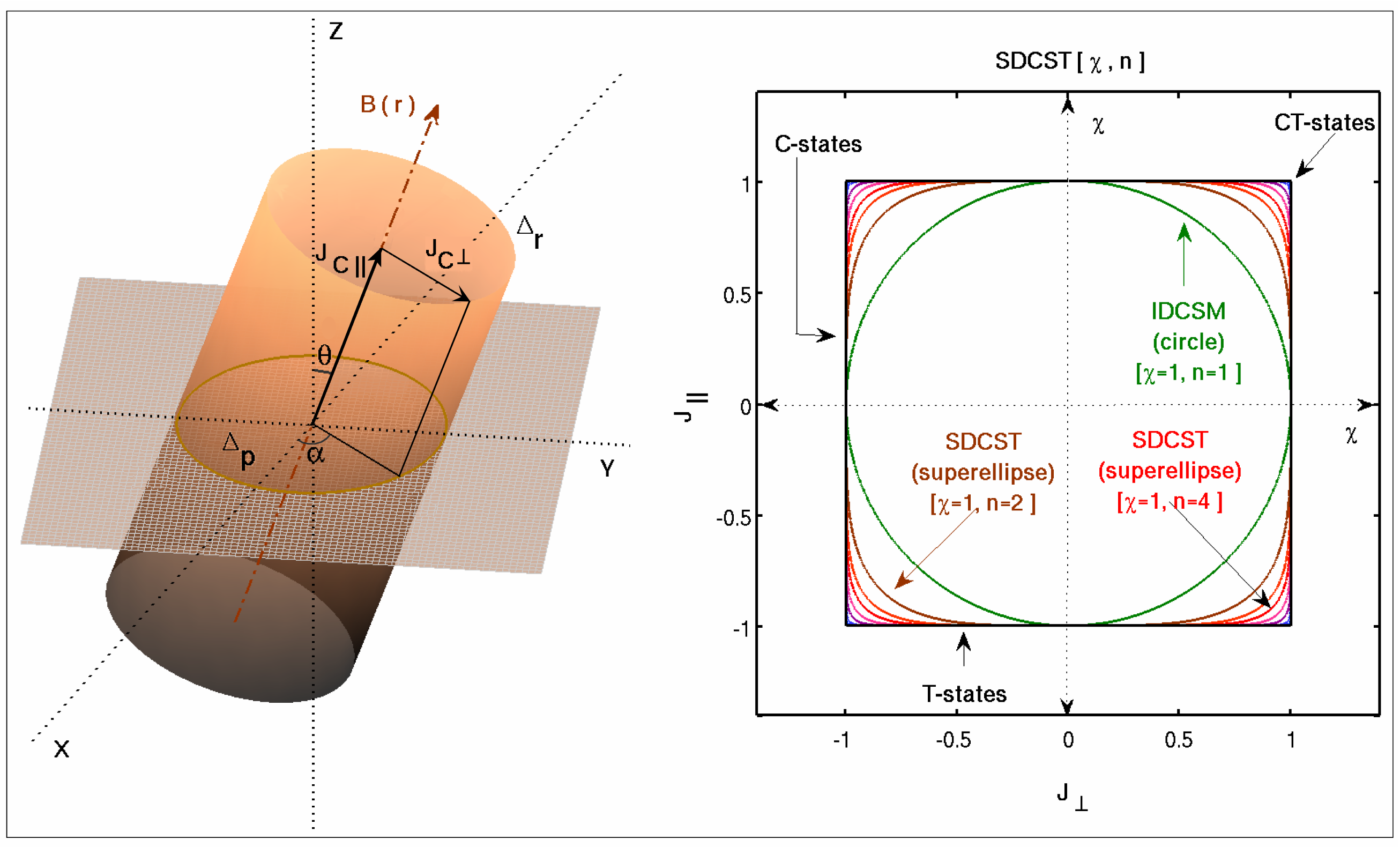}
\caption{\label{Figure_4_1} \textit{Left}: The critical current restriction is
represented by a cylindrical region $\Delta_{\textbf{r}}$ around the local
magnetic field axis (length $2J_{c\parallel}$ and diameter $2J_{c\perp}$),
$\Delta_{p}$ is the projection on the plane (x,y), $\alpha$ is the angle
between the in-plane field projection and the x-axis, and $\theta$ is the angle
between the field and the $z$ axis. \textit{Right}: Geometrical representation
of some of the material-law models depicted into the SDCST. The width and
height of the region is controlled by the anisotropy parameter
$\chi^{2}=J_{c\parallel}^{2}/J_{c\perp}^{2}$, and a smoothing index $n$.}
\end{figure}

As stated above, a rather complete description of irreversible phenomena in
type-II superconductors at a macroscopic level is done through the SDCST
framework by the application of our variational
statement~\cite{P2-Ruiz_PRB_2009} and further use of an appropriate material
law $\textbf{J}(\textbf{E})$~\cite{P2-Ruiz_SUST_2010}. Essentially,
our concept is to define the material law in terms of a geometrical region
$\Delta_{\textbf{r}}(\textbf{J})$ within the $J_{\parallel}-J_{\perp}$ plane,
such that nondissipative current flow occurs when the condition
$\textbf{J}=\textbf{J}_{\parallel}+\textbf{J}_{\perp}\in\Delta_{\textbf {r}}$ is
verified. In contrast, a very high dissipation is to be assumed when
$\textbf{J}$ is driven outside $\Delta_{\textbf{r}}$. Is of utter importance to
recall that the material law encodes the mechanism related to the breakdown of
magnetostatic equilibrium as well as the dissipation modes operating in the
transient from one state to the other. Thus, our scheme allows to translate the
DCSM physics~\cite{P2-Clem_DCSM} onto a region of currents defined in the
$\Re^{3}$--\textit{space} (3D) by a cylinder with its axis parallel to the local
magnetic field $\textbf{B}$, and a rectangular longitudinal section in the plane
defined by the vectors $\textbf{\textbf{J}}_{\parallel}=J_{c
\parallel}\textbf{\^{u}}$ and $\textbf{J}_{\perp}=J_{c
\perp}\textbf{\^{v}}$, being $\textbf{\^{u}}$ the unit vector for the direction
of \textbf{B}, and $\textbf{\^{v}}$ a unit vector in the perpendicular plane
to \textbf{B} (see Figure~\ref{Figure_4_1}). 

Is to be noticed
that in 2D problems with in-plane currents and magnetic field, the
current density region straightforwardly coincides with the above mentioned
longitudinal section ($\Delta_{\textbf{r}}=\Delta_{{\rm p}}$). We recall that,
in this scheme the parts of the sample where the local profiles of the current
density \textbf{J} have reached the boundary $J_{c\perp}$ (the flux depinning
threshold) are customarily called flux transport zones ($J_{\perp}
= J_{c \perp}~;~J_{\parallel}<J_{c \parallel}$), and the profiles satisfying
this condition are called T-states. They are represented by points in a
horizontal band. Physically, the flux lines are migrating while basically
retaining their orientation. On the other hand, regions where only the cutting
threshold is active are denoted as flux cutting zones ($J_{\parallel}
= J_{c \parallel}~;~J_{\perp}<J_{c \perp}$) or simply as C-states. They are
represented by points in a vertical band. In those regions where both mechanisms
have reached their critical values are defined as CT zones ($J_{\parallel} =
J_{c \parallel}$ and $J_{\perp} = J_{c \perp}$) or CT-states. The current
density vector belongs to the corners of
a rectangle. Finally, the regions without energy dissipation are called O zones,
and the current density vector belongs to the interior of the rectangle. 

In this chapter, and corresponding to the material laws depicted in the right
side of figure~\ref{Figure_4_1}, we will show that the variational statement
may be used to predict the magnetic response of type-II superconductors with
3D anisotropy. In a first part, we will give the details related to the
mathematical statement of the general critical state in a three
dimensional slab geometry, i.e., both in-plane and perpendicular magnetic field
components are applied to an infinite slab and varied in a given fashion. Then,
the second and third part are devoted to apply the theory and predict the
magnetic structure for the limiting cases, i.e., on one hand, the isotropic
model ($\chi^2=1,n=1$) and on the other hand, the infinite width-band model
($\chi^2\rightarrow\infty$) or model of \textit{T-states}.
Subchapter~\ref{ch-4}.4 is devoted to explore the set of effects associated to
the DCSM hypothesis ($\chi^2\geq1,n\rightarrow\infty$). A wide range of applied
fields will be considered, and our results
compared to those with possible analytical approaches. Finally, as we are
highly interested in knowing and understanding the role played by the physical
mechanisms of flux depinning and flux
cutting, the last part addresses different physical scenarios by means of
different smooth double critical state models paying special attention to
the influence of the smoothing index $n$ and the widthband $\chi$. In any case,
smooth models have to be considered as related to a number of experiments that
one could not explain within {\em piecewise continuous} models
\cite{P2-Voloshin_1997,P2-Fisher_2000,P2-Voloshin_2001,P2-Fisher_1997,
P2-Voloshin_2010,P2-Fisher_2000_b} or the previous ones. In addition, appendix
1 explores the concept of critical angle gradient in 3D systems, as an
alternative model to deal with anisotropic systems in the slab symmetry.


\vspace*{1cm}
\subsection*{\label{ch-4-1}
\hspace*{0.1cm} \textsl{4.1 3D variational statement in slab geometry}}
\vspace*{1cm}
\markboth{\hspace*{0.1cm}\textsl{4.1 3D variational statement in slab
geometry}} 
{\hspace*{0.1cm} \textsl{4.1 3D variational statement in slab geometry}}
\addcontentsline{toc}{chapter}
{\hspace*{0.1cm} \textsl{4.1 3D variational statement in slab geometry}}

\begin{figure}[t]
\centering
\includegraphics[width=1.0\textwidth]{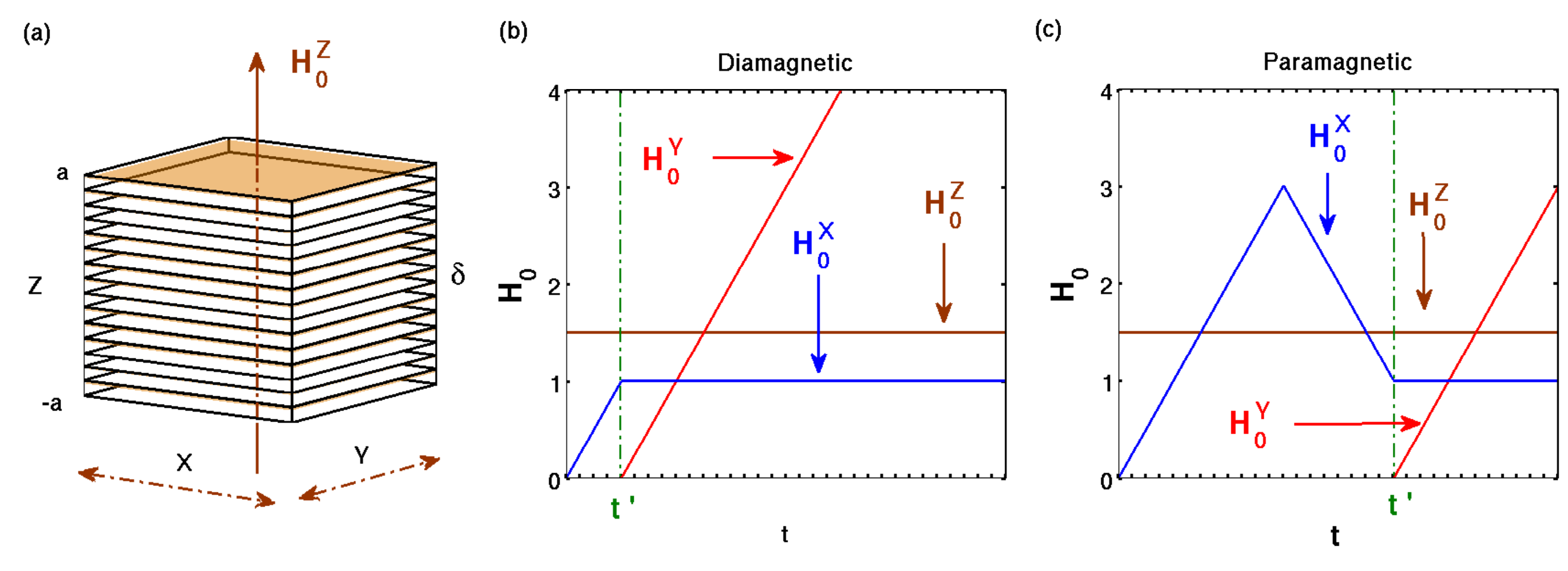}
\caption{\label{Figure_4_2} One the one hand, we show a pictorial illustration
of the slab geometry with a perpendicular magnetic component $H_{z0}$
[\textit{sketch (a)}]. On the other hand, schematics of the time dependence
of the applied magnetic fields in the diamagnetic and paramagnetic
configurations are depicted [\textit{sketches (b) and (c), respectively}] }
\end{figure}

In this subchapter, we derive a specific variational formulation in
superconducting slabs for eventual 3D local field configurations
(figure~\ref{Figure_4_1}), i.e., both in plane and transverse local magnetic
field components emerge as derived effects of the flux depinning and flux
cutting mechanisms. To be specific, we will consider an infinite slab, cooled
under the assumption of an initial state defined by a uniform vortex lattice
perpendicular to the external surfaces (i.e., a constant magnetic field
$H_{z0}$), and then subjected to
a certain process for the applied parallel field (i.e.,
$[H_{x0}(t),H_{y0}(t)]$) as is indicated in Fig.~\ref{Figure_4_2}. 

Recalling the symmetry properties of the electromagnetic quantities, one can
describe the problem as a stack of current layers parallel to the sample's
surface, in such manner that the slab occupies the space $|z|\leq a$. Thus, it
suffices to discretize the upper half, i.e.: $0\leq z_{i} \leq a$ as
symmetry (or antisymmetry) conditions may be applied, and the position
independence for a given value of $z_{i}$ ensures a divergenceless ${\bf J}$.
Notice also that, within this approximation, one has to include two components
of ${\bf J}$ within each layer, i.e.: $[J_{x}(z_{i}),J_{y}(z_{i})]$. At this
point it would be worth mentioning that in order to simplify the mathematical
statements we shall normalize the electrodynamic quantities by defining
$\textbf{h}\equiv\textbf{H}/J_{c\perp}a$,
$\textbf{j}\equiv\textbf{J}/J_{c\perp}$, and $\texttt{z}\equiv z/a$. Recall
that one may assume the numerical value $J_{c\perp}$ as known \textit{a priori}
or obtained from experiment. In turn,
our problem will be described in terms of $N_{s}$ discretized layers of
equal thickness $\delta$ ($z_{i}=\delta\, i\;,\; \delta\equiv a/N_{s}$), each
one
characterized by a current density function
$\textbf{j}(z_{i})=\textbf{j}_{x}(z_{i})+\textbf{j}_{y}(z_{i})$ distributed
along $|z_{i}|\leq N_{i} a/N_{s}$. Eventually,  for each layer, the unknown
variables entering the minimization procedure may be defined accordingly to
sheet currents, $I_{i,l+1}^{x}\equiv \delta j_{x}(z_{i},t=l+1)$ and
$I_{i,l+1}^{y}\equiv \delta j_{y}(z_{i},t=l+1)$. 

Then, a straightforward
application of Amp\`ere's law allows to express the penetrating
magnetic field along the $x-axis$ as the sums over the layers:
\begin{eqnarray}\label{Eq.4.1}
h_{x}(z_{i})\equiv h_{i}^{x}=-\sum_{j>i}I_{j}^{y}-I_{i}^{y}/2 \, .
\end{eqnarray}
Similarly, the local profiles for the longitudinal magnetic field component
$h_{y}(z_{i})$ can be evaluated from,
\begin{eqnarray}\label{Eq.4.2}
h_{y}(z_{i})\equiv h_{i}^{y}=\sum_{j>i}I_{j}^{x}+I_{i}^{x}/2 \, .
\end{eqnarray}

Following the concept introduced in the previous chapter [see
Eq.~(\ref{Eq.3.3})], the following form of the objective function over the
current sheets arises
\begin{eqnarray}\label{Eq.4.3} 
{\cal
F}[I_{l+1}]=&&~\frac{1}{2}\sum_{i,j}I_{i,l+1}^{x}{\rm
M}_{ij}^{x}I_{j,l+1}^{x}-\sum_{i,j}
{\rm I}_{i,l}^{x}{\rm M}_{ij}^{x}I_{j,l+1}^{x}\nonumber \\
&&+\frac{1}{2}\sum_{i,j}I_{i,l+1}^{y}{\rm M}_{ij}^{y}I_{j,l+1}^{y}
-\sum_{i,j} {\rm I}_{i,l}^{y}{\rm M}_{ij}^{y}I_{j,l+1}^{y}\nonumber \\
&&-\sum_{i}I_{i,l+1}^{y}(i-1/2)({\rm h}_{0,l+1}^{x}-{\rm h}_{0,l}^{x})\nonumber
\\
&&+\sum_{i}I_{i,l+1}^{x}(i-1/2)({\rm h}_{0,l+1}^{y}-{\rm h}_{0,l}^{y}) \, .
\end{eqnarray}

In an effort to provide an easier understanding of the above functional, we
stress that only the physical quantities playing the role of unknowns are shown
in italics. Then, we may straightforwardly identify the quadratic and linear
groups for the minimization procedure. In detail, the sheet currents
$I_{l+1}^{x}$ and $I_{l+1}^{y}$ represent the unknown variables to be
minimized as a given initial state $[{\rm I}_{l}^{x},{\rm I}_{l}^{y}]$ is
connected by the steady processes $\Delta {\rm h}_{0}^{x}={\rm
h}_{0,l+1}^{x}-{\rm h}_{0,l}^{x}$ and $\Delta {\rm h}_{0}^{y}={\rm
h}_{0,l+1}^{y}-{\rm h}_{0,l}^{y}$, and their mutual inductance matrices ${\rm
M}_{ij}^{x}$ and ${\rm M}_{ij}^{y}$. It is to be noticed that the index $l$ is
introduced to indicate time discretization, i.e., $I_{i}(l+\delta
t)-I_{i}(t)\equiv I_{i,l+1}-I_{i,l}$. When this index is omitted, it will be
meant that the element is time independent, i.e., it is valid for
any step $l$ and calculated as an external input for the objective function
(see Figs.~\ref{Figure_3_1}~\&~\ref{Figure_3_2}). 

On the other hand, recall that the inductance matrices are directly linked to
the design of a grid representing the \textit{location} of variables into the
superconducting volume. Thus, in the slab symmetry the circuits are just layers
made up of straight lines along the $x$ and $y$ axis, and
$\{I_{i}^{x},I_{i}^{y},\forall i\in\Omega\}$ is a compact notation for the whole
set. Then, for our discretized array of layers the reader can check that a
straightforward substitution of the squared components of the magnetic field
entering the expression in Eq.~(\ref{Eq.2.11}) in terms of
Eqs.~(\ref{Eq.4.1})~\&~(\ref{Eq.4.2}), leads to the following formulas for the
mutual inductance coupling elements:
\begin{eqnarray}\label{Eq.4.4} 
{\rm M}_{ij}^{x}={\rm M}_{ij}^{y} & \equiv & 1+2\left[{\rm
min}\left\{ i,j\right\}\right]\quad \forall\; i\neq j
\nonumber\\
{\rm M}_{ii}^{x}={\rm M}_{ii}^{y}& \equiv & 2\left(\frac{1}{4}+i-1\right)
\end{eqnarray}
Notice that inductive coupling only occurs between $x$ and $y$ layers
separately, and the corresponding coefficients are identical. 

Finally, we stress that minimization has to be
performed under a prescribed material law $\textbf{J}\in\Delta_{\textbf{r}}$
(i.e., some of the geometrical regions depicted in Figure~\ref{Figure_4_1}),
and ${\cal F}$ turns a new minimization functional for each different time step 
$(l=1,2,...)$. Specifically, the \textit{three dimensionality} of the local
magnetic field vector is controlled by the threshold values for the physical
mechanisms responsible of the depinning and cutting of the vortices, i.e., the
critical values $J_{c\parallel}$ and $J_{c\perp}$. Thus, in order to understand
the three dimensionality of the vector $\textbf{J}$ one has to
consider the polar decomposition 
\begin{eqnarray}\label{Eq.4.5}
{\bf J}_{i}={\bf J}_{i}^{\parallel}+{\bf J}_{i}^{\perp\alpha}+{\bf
J}_{i}^{\perp{\theta}}\, ,
\end{eqnarray}
with the parallel, azimuth and polar
components of ${\bf J}_{i}$ defined in terms of the magnetic field direction
$\textbf{\^{H}}_{i}$. After some simple algebraic operations in a Cartesian
coordinate system, the following expressions are obtained for such components:

\begin{enumerate}

 \item The current component parallel to $\textbf{\^{h}}_{i}$ or so-called
cutting
current component $I_{i}^{\parallel}$:
\begin{eqnarray}\label{Eq.4.6}
I_{i}^{\parallel}=\frac{h_{i}^{x}I_{i}^{x}+h_{i}^{y}I_{i}^{y}}{\left[(h_{i}^{x}
)^{2}+(h_{i}^{y})^{2}+(h_{i}^{z})^{2}\right]^{1/2}}
 \, .
\end{eqnarray}

 \item  The component of ${\bf I}$ perpendicular to the plane defined by the
vectors $\textbf{\^{z}}$ and $\textbf{\^{h}}$ or so-called azimuthal current
component
\begin{eqnarray}\label{Eq.4.7}
I_{i}^{\perp
\alpha}=\frac{-h_{i}^{y}I_{i}^{x}+h_{i}^{x}I_{i}^{y}}{\left[(h_{i}^{x})^{2}+(h_{
i}^{y})^{2}\right]^{1/2}}
 \, .
\end{eqnarray}

\item The component of ${\bf I}$ perpendicular to $\textbf{\^{h}}$ and contained
in the plane defined by the vectors $\textbf{\^{z}}$ and $\textbf{\^{h}}$ or
so-called polar current component $I_{\perp \theta}$:
\begin{eqnarray}\label{Eq.4.8}
I_{i}^{\perp
\theta}=\frac{h_{i}^{z}(h_{i}^{x}I_{i}^{x}+h_{i}^{y}I_{i}^{y})}{\left\{\left[(h_
{ i }^{x}
)^{2}+(h_{i}^{y})^{2}+(h_{i}^{z})^{2}\right]~
\left[(h_{i}^{x})^{2}+(h_{i}^{y})^{2}\right]\right\}^{1/2} } \, .
\end{eqnarray}
\end{enumerate}

Thus, as an example, within the framework of the DCSM hypothesis one has to
invoke the conditions
\begin{eqnarray}\label{Eq.4.9}
\left(1-(h_{i}^{x})^{2}\right)
(I_{i}^{x})^{2}+\left(1-(h_{i}^{y})^{2}\right)(I_{i}^{y})^{2}-2h_{i}^{x}h_{i}^{y
} I_{i}^{x}I_{i}^{y} \leq {\rm I}_{c\perp}^{2} \, ,
\end{eqnarray}
and
\begin{eqnarray}\label{Eq.4.10}
(h_{i}^{x} I_{i}^{x})^{2} + (h_{i}^{y} I_{i}^{y})^{2}+
2h_{i}^{x}h_{i}^{y}I_{i}^{x} I_{i}^{y} \leq {\rm I}_{c\parallel}^{2} \, . 
\end{eqnarray}

In summary, the objective function is constrained by the group of functions
defined by Eqs.~(\ref{Eq.4.9})~\&~(\ref{Eq.4.10}), and their minimization
provides
the magnetic response of the superconductor by means a collection of
discretized current elements for the planar sheets of current density
$[j_{i}^{x},j_{i}^{y}]$ at the time steps $l+1=1,2,3,...$.


\vspace*{1cm}
\subsection*{\label{ch-4-2}
\hspace*{0.1cm} \textsl{4.2 Isotropic predictions in ``3D''
configurations}}
\vspace*{1cm}
\markboth{\hspace*{0.1cm}\textsl{4.2 Isotropic predictions in ``3D''
configurations}} 
{\hspace*{0.1cm} \textsl{4.2 Isotropic predictions in ``3D''
configurations}}
\addcontentsline{toc}{chapter}
{\hspace*{0.1cm} \textsl{4.2 Isotropic predictions in ``3D''
configurations}}

\begin{figure}[t]
\centering
\includegraphics[height=5cm,width=13cm]{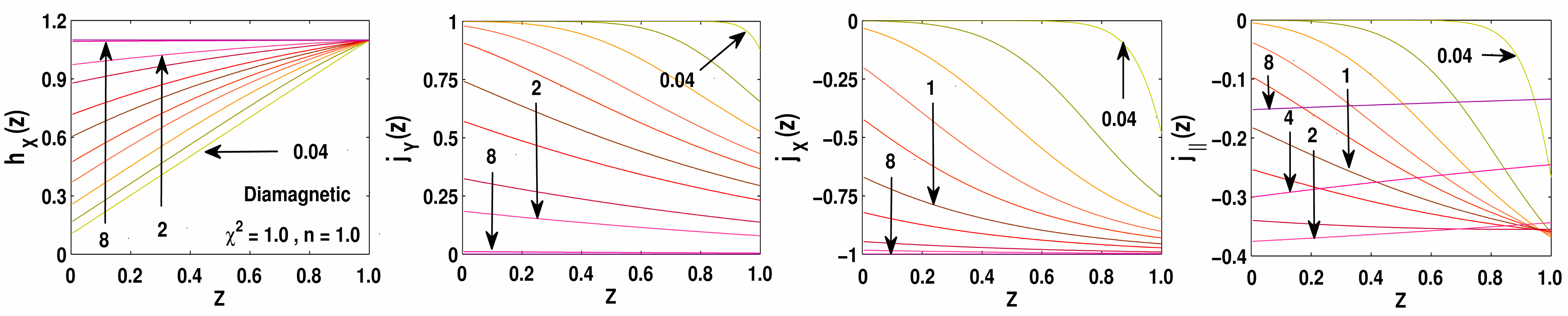}
\caption{\label{Figure_4_3} In the diamagnetic configuration and the isotropic
model ($\chi^{2}=1$,$n=1$), we show the profiles for the local magnetic field
component $h_{x}[z,h_{y}(a)]$ and their corresponding current-density profiles
$j_{y}[z,h_{y}(a)]$, starting from a \textit{first}-time step defined by
$h_{x}(a)=1.1$ and $h_{z0}=1.5$. The current component $j_{x}[z,h_{y}(a)]$ and
the cutting component $j_{\parallel}[z,h_{y}(a)]$ are also shown. The curves are
labeled according to the component of longitudinal magnetic field at
the slab surface, and correspond to the values $h_{y}(a)=0.040, 0.2, 0.4,
0.6, 0.8, 1.0, 1.2, 1.6, 2.0, 4.0,$ and $8.0$.}
\end{figure}

Below, we show the theoretical predictions for the region or material law
defined by the region $(\chi^{2}=1,n=1)$ or isotropic model (see
Fig.~\ref{Figure_4_1}), along the
magnetization processes indicated in Figure~\ref{Figure_4_2}.

Starting from a fully penetrated state with a magnetic field applied
perpendicular to the slab surfaces ($h_{z0}$), i.e., a lattice of parallel
vortices is assumed to nucleate parallel to the $z$ axis within the sample,  
 one configures either a diamagnetic or a paramagnetic critical state by
sweeping an applied parallel component $h_{x0}$ (thus inducing $j_y$). For
example, if in a first temporal branch $(t<t')$ the material is subjected to an
increasing magnetic field $h_{x}$ by means of $\Delta h_{x}(a)$, time path
steps are characterized by tilted flux lines which penetrate the specimen until
an equilibrium distribution is achieved (diamagnetic). Then, if the external
magnetic field is subsequently lowered, thereby reducing the retaining magnetic
pressure, flux lines migrate out of the sample until the equilibrium is
restored (paramagnetic). 

\begin{figure}[t]
\centering
\includegraphics[height=5cm,width=13cm]{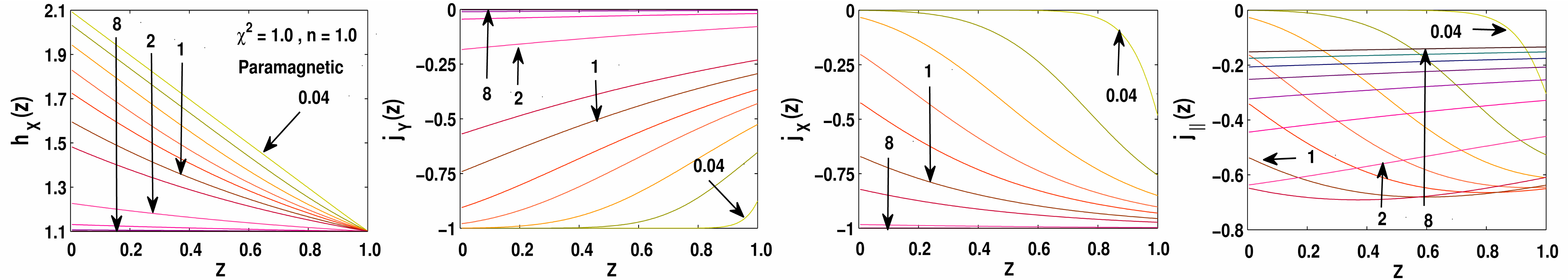}
\caption{\label{Figure_4_4} Same as figure~\ref{Figure_4_3}, but in the
paramagnetic configuration illustrated in
figure~\ref{Figure_4_2}. Here, the curves are labeled according to
the values $h_{y}(a)=0.040, 0.2, 0.4, 0.6, 0.8, 1.0, 1.2, 2.0,
3.0, 4.0, 5.0, 6.0, 7.0,$ and $8.0$.}
\end{figure}

Eventually, at $t=t'$ an increasing ramp in the other longitudinal field
component $h_{y0}$ is switched on and thus, an electric field $E_{x}$ arises at
a surface layer of the superconductor
which produces a current density $j_{x}$ that will screen the excitation. Then,
owing the restrictions on the current density vector ${\bf j}$ introduced by
the material law, the local component $j_{y}$ is affected and the corresponding
local magnetic field $h_{x}(z)$ is pushed towards the center of the sample in
the diamagnetic case (see Figure.~\ref{Figure_4_3}) or towards the external
surface in the paramagnetic one (see Figure.~\ref{Figure_4_4}). 

In detail,
figures~\ref{Figure_4_3}~\&~\ref{Figure_4_4} show how the local component
$h_{x}(0)$ increases
(diamagnetic case) or it reduces (paramagnetic case) until the specimen is fully
penetrated to satisfy the condition $h_{x}(z) = h_{x}(a)\;\forall~z$, i.e.,
$j_{y}(z)\rightarrow0$ as $h_{y}$ increases. As a consequence, no sign reversal
in the induced currents is predicted. Two important features of this model are
to be remarked.

On the one hand, it has to be recalled that by the symmetry conditions invoked
before, we may assert that, for long loops, the
contribution coming from the U turn at the far ends, exactly
equals the contribution of the long sides. This may be shown starting from the
condition $\nabla\cdot \textbf{J}=0$ (no sources) that allows us to consider the
current-density distribution as a collection of loops and ensures the equality
of the integrals over $z j_{x}$ and $z j_{y}$. Thus, the magnetic moment
components per unit area may be obtained by numerical integration of the current
density along the slab thickness, i.e.,
\begin{eqnarray}\label{Eq.4.11}
\textbf{M}=\int_{-a}^{a}\textbf{z}\times\textbf{j}dz \, .
\end{eqnarray}
Then, the saturation of the current components $j_{x}(z)$ and $j_{y}(z)$,
indicates that the isotropic hypothesis ($j_{c}=j_{c\parallel}=j_{c\perp}$
i.e., $\chi^{2}=1, n=1$) is providing a straightforward explanation of the
observed magnetization collapse~\cite{P2-ICSM_app}. Furthermore, by comparison
between the Figs.~\ref{Figure_4_3} \& \ref{Figure_4_4} with their corresponding
magnetization curves (see curves in green in Fig.~\ref{Figure_4_5}), we also
noticed that the magnetization collapse is obtained simultaneous to the
monotonic reduction of the \textit{cutting} current density or $j_{\parallel}$. 

On the other hand, another related phenomenon, the so-called {\em
paramagnetic peak effect} of the magnetic moment can not be foreseen by the
isotropic material law. As a consequence, more sophisticated models have to be
either invoked and revalidated by the study of the components of cutting and
depinning for the current density.  It has to be mentioned, that in previous
works far away of the DCSM hypothesis, this observation was explored in terms of
the so-called two velocity electrodynamic model as a crude
approximation for the real dynamics in a flux line lattice
(see details in Ref.~\cite{P2-2hydrodynamic_model})


\vspace*{1cm}
\subsection*{\label{ch-4-3}
\hspace*{0.1cm} \textsl{4.3 T-states in ``3D'' configurations}}
\vspace*{1cm}
\markboth{\hspace*{0.1cm}\textsl{4.3 T-states in ``3D'' configurations}} 
{\hspace*{0.1cm} \textsl{4.3 T-states in ``3D'' configurations}}
\addcontentsline{toc}{chapter}
{\hspace*{0.1cm} \textsl{4.3 T-states in ``3D'' configurations}}

In this section we continue the previous discussion, but now assuming a region
with an infinite band-width $(\chi\rightarrow\infty)$ or so called model for
T-states. Recalling that the magnetic field $\textbf{h}$ is measured in units of
the physically relevant
penetration field $J_{c\perp}a$ then, numerical experiments with
$h_{z0}=0.1$, $h_{z0}=1.5$, and $h_{z0}=10$ will cover the range of interest.

\begin{figure}[t]
\centering
\includegraphics[height=6cm,width=13cm]{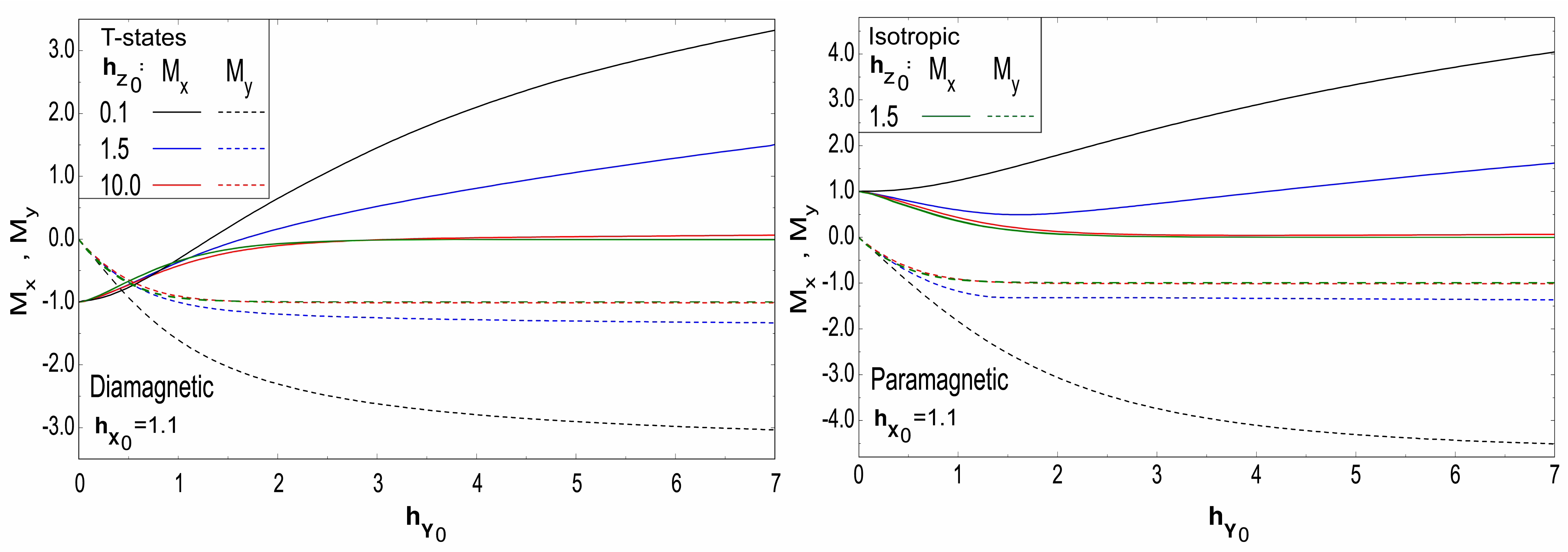}
\caption{\label{Figure_4_5} The magnetic moment $M_{x}$ (solid lines) and
$M_{y}$ (dashed lines) per unit area as a function of the applied magnetic
field component $h_{y0}$ in the diamagnetic (left pane) and paramagnetic (right
pane) initial configurations of Fig.~\ref{Figure_4_2}. Several cases are shown
accordingly to the field intensities: $h_{x0}=1.1$ together with $h_{z0}=0.1$
(black), $h_{z0}=1.5$ (blue), and $h_{z0}=10.0$ (red), for the T-states model.
Also, the corresponding curve for $h_{x0}=1.1$ and $h_{z0}=1.5$ for the
isotropic model (green) are shown. Units are $j_{c\bot}a$ for $h$ and
$j_{c\bot}a^{2}$ for $M$.}
\end{figure}

In figure~\ref{Figure_4_5} we display our results for the magnetic moment
components per unit area under the experimental conditions depicted in
figure~\ref{Figure_4_2}. The plots indicate the following features:

\begin{enumerate}

\item In general, a saturation is reached for $M_{y}(h_{y0})$, as compared to
the eventual linear increase of $M_{x}(h_{y0})$ for the highest values
of $h_{y0}$. 

\item The higher $h_{z0}$, the sooner the saturation is reached.

\item Increasing $h_{z0}$ rapidly diminishes the slope of
$M_{x}(h_{y0})$.

\item In the paramagnetic case, a minimum is observed (more evidently for $M_x$,
and more visible for $h_{z0}=1.5$), that is smoothed either for the higher or
lower values of this field component. 

\end{enumerate}

\begin{figure}[t]
\centering
\includegraphics[height=10cm,width=13cm]{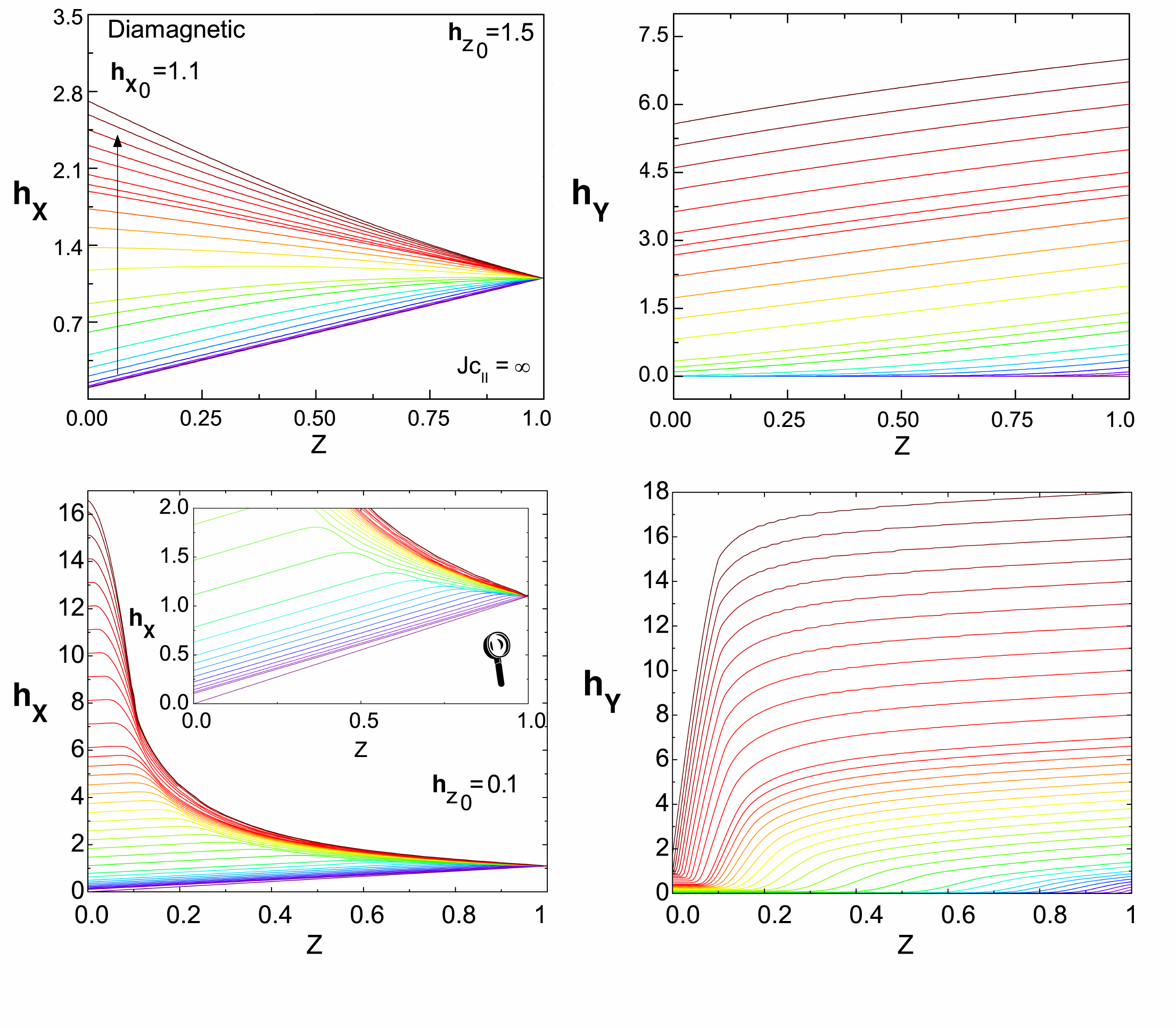}
\caption{\label{Figure_4_6} Magnetic field components $h_{x}(z)$ (left) and
$h_{y}(z)$ (right) corresponding to the current-density profiles for the
T-state limit in the diamagnetic configuration with $h_{z0}=1.5$ (top) and
$h_{z0}=0.1$ (bottom) respectively.}
\end{figure}

Here, we want to call readers' attention on the fact that our results
for moderate perpendicular
fields ($h_{z0}=1.5$ and $h_{z0}=10$) are in perfect agreement with the
differential equation approach provided by Brandt and Mikitik in
Ref.~\cite{P2-Brandt_2007} (for more details see Ref.~\cite{P2-Ruiz_PRB_2009}).
For
these cases, the underlying flux penetration profiles fully coincide with our
calculations. However, in Ref.~\cite{P2-Brandt_2007} the low field
region
was uncovered. Thus, here we will show the exotic behavior of the field and
current-density profiles for the low field regime (e.g., $h_{z0}=0.1$) in
comparison with the local electromagnetic behavior for moderate fields (e.g.,
$h_{z0}=1.5$). 

\begin{figure}[t]
\centering
\includegraphics[height=10cm,width=13cm]{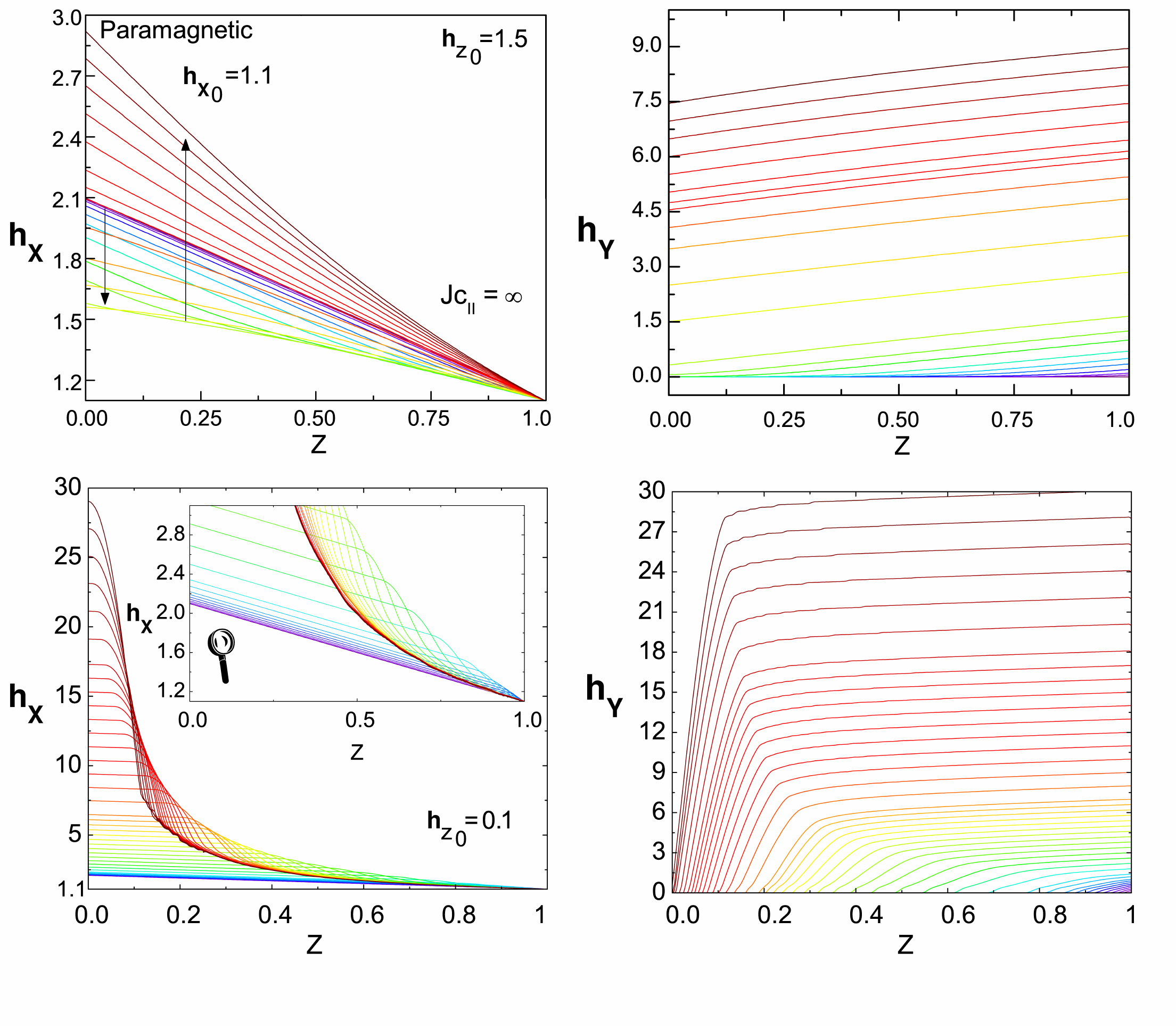}
\caption{\label{Figure_4_7} Same as Fig.~\ref{Figure_4_6} but for the
paramagnetic configuration}
\end{figure}

\begin{figure}[t]
\centering
\includegraphics[height=8cm,width=13cm]{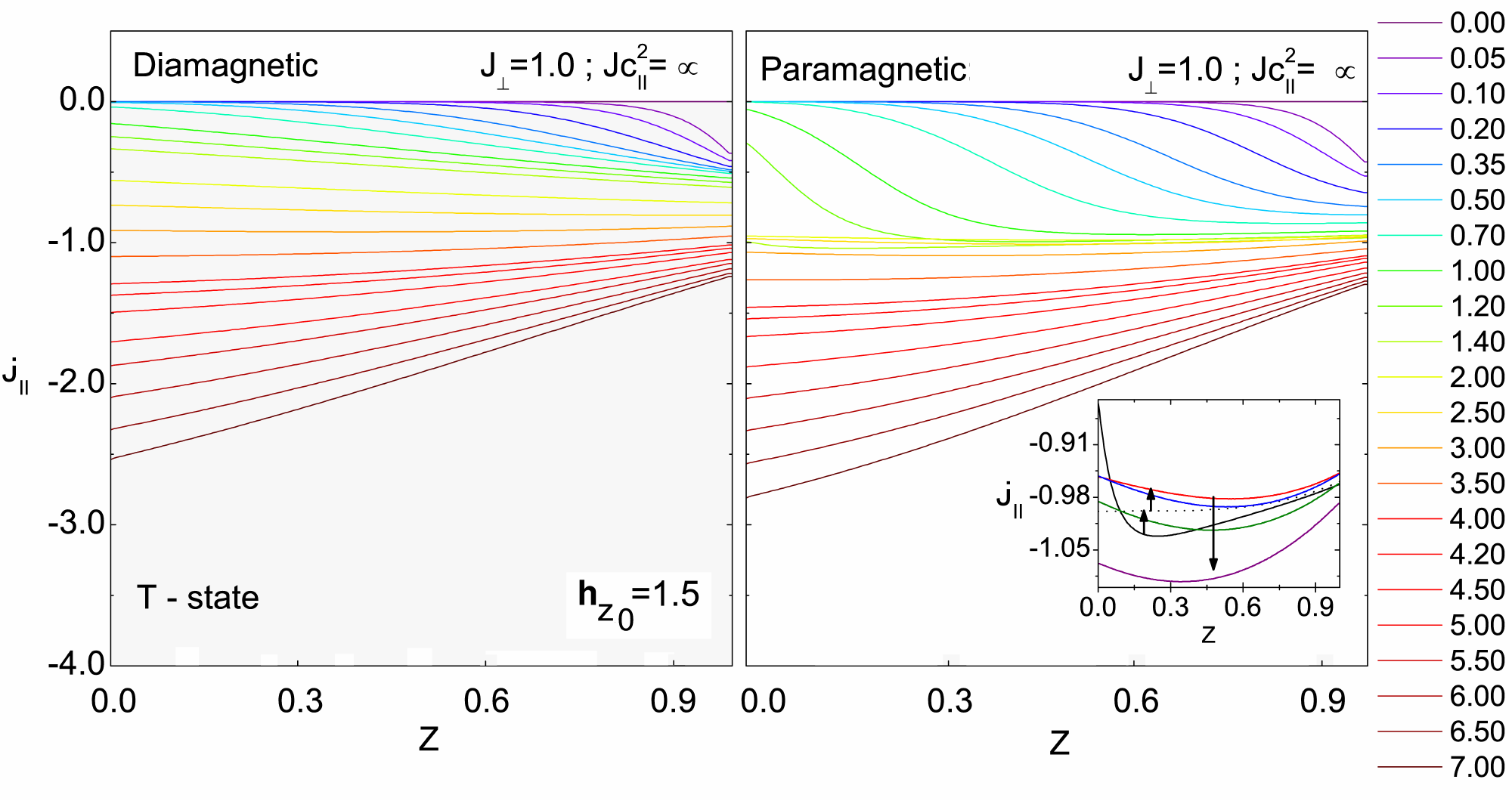}
\caption{\label{Figure_4_8} Profiles of the parallel
current component $j_{\parallel}$ for the T-state hypothesis
``$J_{c\parallel}\to\infty$ and $J_{c\perp}=1.0$''. The curves are labeled
according to the applied
field $h_{y0}$ (at right), assuming $h_{z0}=1.5$. The
diamagnetic (left pane) and paramagnetic (right pane) cases are shown. In the
paramagnetic case, the profiles of $J_{\parallel}$
for $h_{y0}=1.35,\, 1.7,\,2.0,\, 2.3,\, 2.6,\, 3.0$ are shown as an inset, and
correspond to the sign change in the slope of the magnetic moment $M_{x}$
(see figure~\ref{Figure_4_5}).}
\end{figure}

Figures~\ref{Figure_4_6}~\&~\ref{Figure_4_7} respectively display the behavior
of the in-plane magnetic field components [$h_{x}(z)$,$h_{y}(z)$] in the
diamagnetic and paramagnetic cases. On the one hand, in the diamagnetic case we
notice that for moderate fields, the local magnetic field $h_{x}(z)$ is
monotonically pushed towards the center of the sample with a nearly
homogeneous distribution of the cutting current component $j_{\parallel}$ (see
figure~\ref{Figure_4_8}). In an analogous manner, for the paramagnetic case the
dynamics of the local component $h_{x}(z)$ is also related to the dynamics of
the cutting current component.
Thus, in a first stage the array of vortices closer to the center of the
sample shows a decrease of the local component $h_{x}(z)$ until the full
penetration state for the applied magnetic field $h_{y}(z)$ is achieved. Then,
the cutting condition $j_{\parallel}\neq0$ is warranted for the whole sample.
Interestingly, once the center of the sample reaches the cutting condition a
fast change of sign in the slope of $j_{\parallel}$ is envisaged which
straightforwardly corresponds to the change of sign in the slope of the
magnetic moment $M_{x}$ or \textit{peak effect}. In turn, it leads to a second
stage which is mainly characterized by an array of vortices with a monotone
increase of the components $h_{y}(z)$ and $h_{x}(z)$ under the boundary
condition for the initial state (e.g., in our cases we have assumed
$h_{x}(a)=1.1$ for t=t', \textit{see also Fig.}~\ref{Figure_4_2}).

On the other hand, it is to be noticed that the effects induced by the
consideration of an unbounded cutting component are rather less simple for the
low field regime (e.g., $h_{z0}=0.1$). Indeed, a steep variation of
$h_{y}$ occurs for the inner region of the sample, corresponding to large values
of $j_{\parallel}$ essentially dominated by  $j_{x}$. On the contrary, $h_{x}$
displays a small slope, which relates to the condition $j_{\perp}=1$
(essentially, $j_{\perp}\approx j_{y}$ in the inner region). 

\begin{figure}[t]
\centering
\includegraphics[height=7cm,width=13cm]{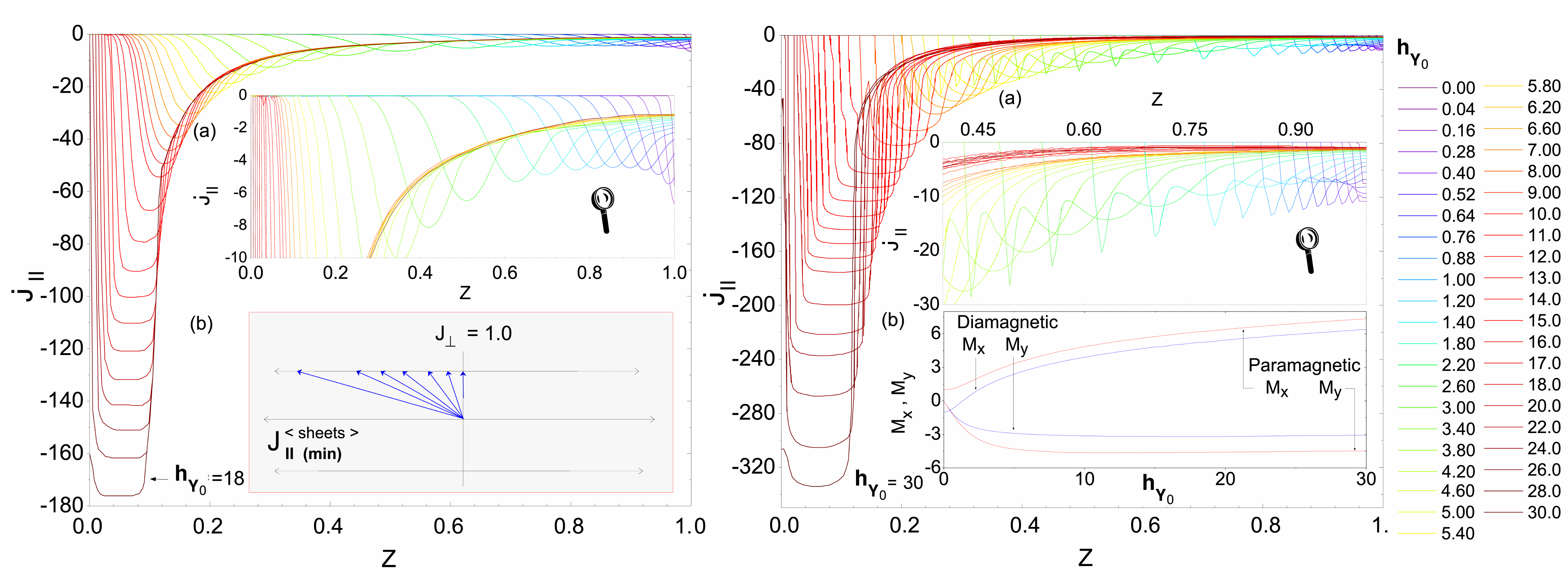}
\caption{\label{Figure_4_9} Profiles of the component
$j_{\parallel}$ for the limit $J_{c_{\parallel}}\to\infty$ (T-state) with
$h_{z0}=0.1$. In all cases the perpendicular current profiles satisfy
$J_{\perp}=J_{c\perp}=1.0$. The diamagnetic (left) and paramagnetic
(right) cases are shown. Left: inset (a) shows a zoom of $j_{{||}}$ for the
first profiles of $h_{y_{0}}$. Inset
(b) schematically shows the evolution of the vector $\bf J$ as
function of its parallel and perpendicular components. Bottom: inset (a) shows a
zoom of $j_{{||}}$
for the first profiles of increasing $h_{y0}$. Inset (b) shows the magnetic
moment components ($M_{x}$, $M_{y}$) per unit area as a function of
$h_{y0}$.}
\end{figure}

More into detail,
Fig.~\ref{Figure_4_9} displays the behavior of the projection of the current
density onto the direction of the magnetic field ($j_{\parallel}$) under the
ansatz of a T-state structure for
$h_{z0}=0.1$. It is apparent that the full penetration of the T-state
perturbation requires a high field component ($h_{y0}\approx  18$ and
$h_{y0}\approx  30$ for the diamagnetic and paramagnetic cases respectively),
and a very high ratio $J_{\parallel}/J_{c\perp}\equiv j_{\parallel}$ ($\approx
180$ for the diamagnetic case and $\approx 340$ for the paramagnetic one).
Notice that until these values are reached, one has $J_{\parallel}=0,
J_{\perp}=1$ for the inner part of the sample, and a certain distribution
$J_{\parallel}(z)$ for the outer region. We also recall a somehow complex
structure with one or two minima in between the surface of the sample and the
point reached by the perturbation. Interestingly, when $h_{y0}$ grows, the
minimum becomes very flat, corresponding to a nearly constant value of
$j_{\parallel}$. From the physical point of view, the minimum
basically represents the region where ${\bf h}$ rotates so as to accommodate the
penetration profile ${\bf h}(z)$ to the previous state of magnetization
$(h_{x},0,h_{z0})$. From the point of view of Faraday's law, this takes place as
quickly as possible so as to minimize flux variations.

Finally, we want to call readers' attention on two derived facts from this
model. On the one hand, is to be noted that within the T-states model the
magnetization
collapse does not take place at least for perpendicular fields lower than
$h_{z0}=10$. Indeed, a tiny slope on the magnetization moment curve is still
present (see Fig.~\ref{Figure_4_5}). On the other hand, as consequence of the
non constrained cutting
component, there is no restriction on the longitudinal component of the current
density that increases arbitrarily towards the center of the sample (see
Figs.~\ref{Figure_4_6}~\&~\ref{Figure_4_7}).
Thus, strictly speaking, this model can not be consider as physically
admissible although some of the experimental evidences may be reproduced.

\vspace*{3cm}
\subsection*{\label{ch-4-4}
\hspace*{0.1cm} \textsl{4.4 CT-states in ``3D'' configurations}}
\vspace*{1cm}
\markboth{\hspace*{0.1cm}\textsl{4.4 CT-states in ``3D'' configurations}} 
{\hspace*{0.1cm} \textsl{4.4 CT-states in ``3D'' configurations}}
\addcontentsline{toc}{chapter}
{\hspace*{0.1cm} \textsl{4.4 CT-states in ``3D'' configurations}}

Below, we show the theoretical predictions derived by choosing a rectangular
region for the material law $\Delta_{r}$ with a finite bandwidth $\chi$ or
so-called DCSM (i.e., $\chi\geq1$ and $n\rightarrow\infty$ within our SDCST). 
In order to keep the previous sequence of results and the underlying ideas, the
same numerical experiments depicted in Fig.~\ref{Figure_4_2} will be analyzed.

Before going into detail, let us recall that the cases analyzed in the previous
two sections directly
correspond to the lower and higher limits of the double critical state
approach, and in consequence, any material law displayed between them will be
characterized by intermediate profiles for the electromagnetic quantities (more
details in Refs.~\cite{P2-Ruiz_PRB_2009,P2-Ruiz_SUST_2010}). Thus, we can
summarize
the rich phenomenology encountered by means a thorough analysis of the magnetic
moment curves $M_{x}(h_{y0})$ and $M_{y}(h_{y0})$, and the local profiles for
the cutting current component $j_{\parallel}(z)$. 

\begin{figure}[t]
\centering
\includegraphics[height=7cm,width=13cm]{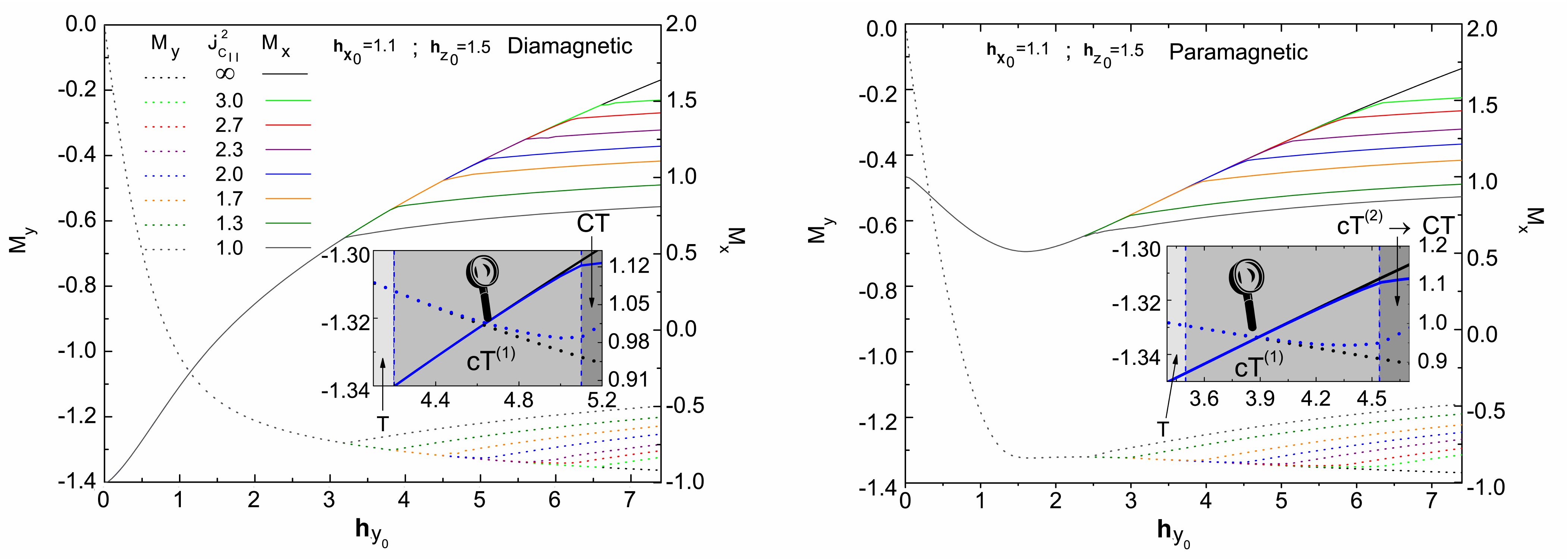}
\caption{\label{Figure_4_10} Magnetic moment curves per unit area 
($M_{x},M_{y}$) as a function of the applied field $h_{y0}$ for the
experimental configurations displayed in Fig.~\ref{Figure_4_2}. Shown are the
diamagnetic (left) and paramagnetic (right)
cases for $h_{x0}=1.1$ and the moderate perpendicular field $h_{z0}=1.5$. The
T-state curves ($j_{c\parallel}\gg 1$) are shown for comparison with
the DCSM-cases: $j_{c_{||}}^{2}=$3.0, 2.7, 2.3, 2.0, 1,7, 1.3, 1.0. The insets
show the particular case $j_{c_{||}}^{2}=2.0$ in
the region where the transition T$\rightarrow$CT is visible.}
\end{figure} 

\begin{figure}[t]
\centering
\includegraphics[height=7cm,width=13cm]{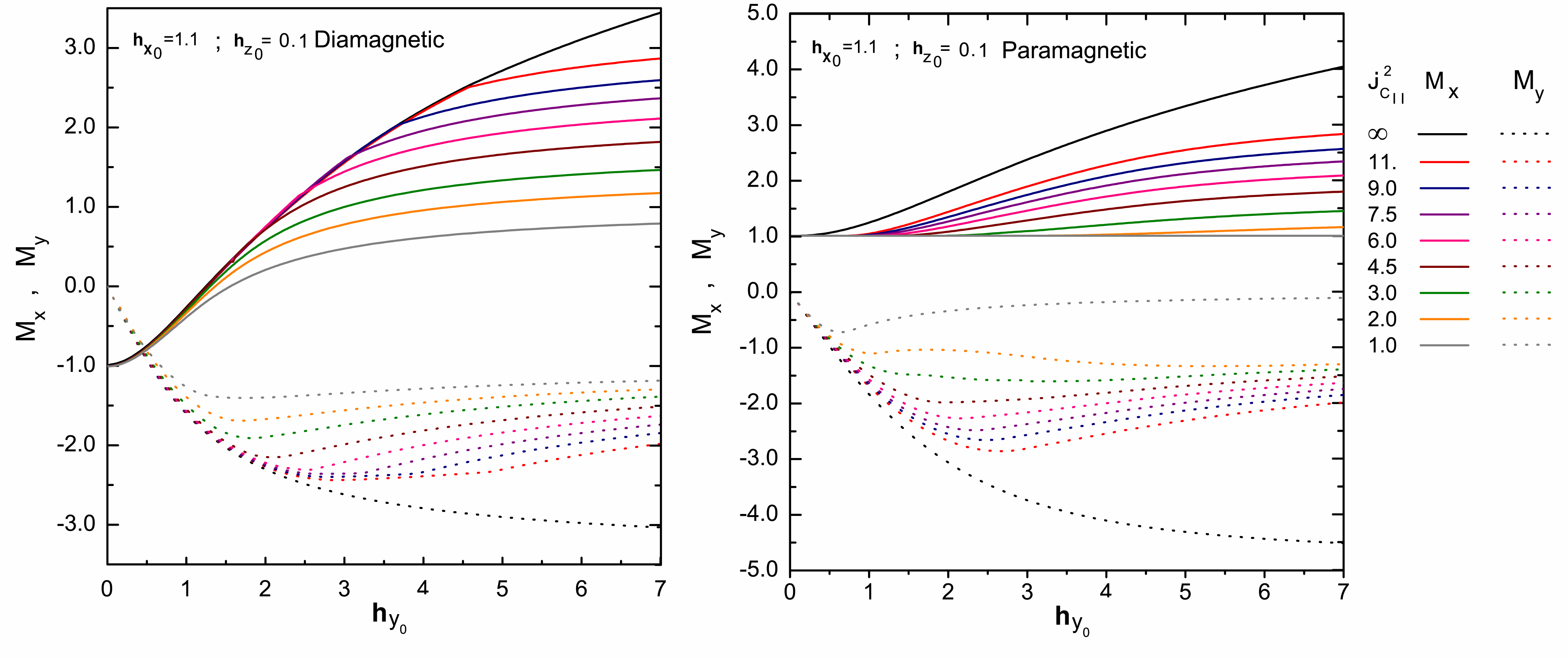}
\caption{\label{Figure_4_11} Same as figure~\ref{Figure_4_10}, but in the lower
perpendicular field regime $h_{z0}=0.1$. Here, the following DCSM-cases:
$J_{c_{\parallel}}^{2}=$ 11.0, 9.0, 7.5, 6.0, 4.5, 3.0, 2.0, and 1.0, are
shown.
}
\end{figure}

Firstly, we show the corrections to $M_x$ and $M_y$ both for the diamagnetic 
and paramagnetic cases either at a moderate perpendicular field
(Figure~\ref{Figure_4_10}) or at a low perpendicular field
(Figure~\ref{Figure_4_11}), when the
DCSM region corresponds to the aspect ratio values $\chi^{2}=1.0,
1.3, 1.7, 2.0, 2.3, 2.7$ and $3.0$. 

On the one hand, for a moderate perpendicular field (e.g.,
$h_{z0}=1.5$), it is noticeable that the limitation in $j_{c\parallel}$ produces
a {\em corner} in the magnetic moment dependencies $M_{x,y}(H_{y0})$, which
establishes the departure from the {\em master curve} defined by the T-state
model. The corner in $M_{x}$ and $M_{y}$ appears at
some characteristic field $h_{y0}^{*}$ that increases with $\chi$, eventually
disappearing within the region of interest. Thus, the higher value of
$h_{y0}$ for which the corner can not be observed ($h_{y0}\approx 3$ in the
conditions depicted into the plots of Fig.~\ref{Figure_4_10}), defines the 
acting threshold of the T-state model. The fine
structure of the corner is shown in the insets of Fig.~\ref{Figure_4_10}. Notice
that, indeed, the deviation from the master curve takes place in two steps,
being the second one that really defines the corner. 

On the other hand,  for a low perpendicular field (e.g., $h_{z0}=0.1$), the
general trends in the CT-state corrections do not very much differ from those at
moderate field values. However, some distinctive features
are worth to be mentioned for the $M_{x,y}(H_{y0})$ curves
(Figure~\ref{Figure_4_11}). To start with, we
recall that the corner structure that defines the separation of the {\em
CT-curves} from the {\em master T-state behavior} is different. Thus, as one can
notice in Fig.~\ref{Figure_4_11}, it is only for the higher values of
the parameter $\chi^{2}$ that the separations take place abruptly. In
particular, a smooth variation occurs for $\chi^{2}<6$ in all cases. Also
noticeable is the change in the behavior of the initial part of the
$M_{x}(h_{y0})$ curves for the paramagnetic
case. Recall that the minimum observed for the moderate field patterns
($h_{z0}=1.5$) has now disappeared (this can be already detected for the
T-states). Significantly, one can observe that by decreasing $\chi^{2}$, 
$M_{x}$ develops a nearly flat region at the low values of $h_{y0}$. Physically,
this means that the initial $h_{x}(z)$ profile is basically unchanged. For the
lowest values of $\chi^2$ this can take place over a noticeable range of applied
fields $h_{y0}$~\cite{P2-Ruiz_PRB_2009}. 

\begin{figure}[t]
\centering
\includegraphics[height=7cm,width=13cm]{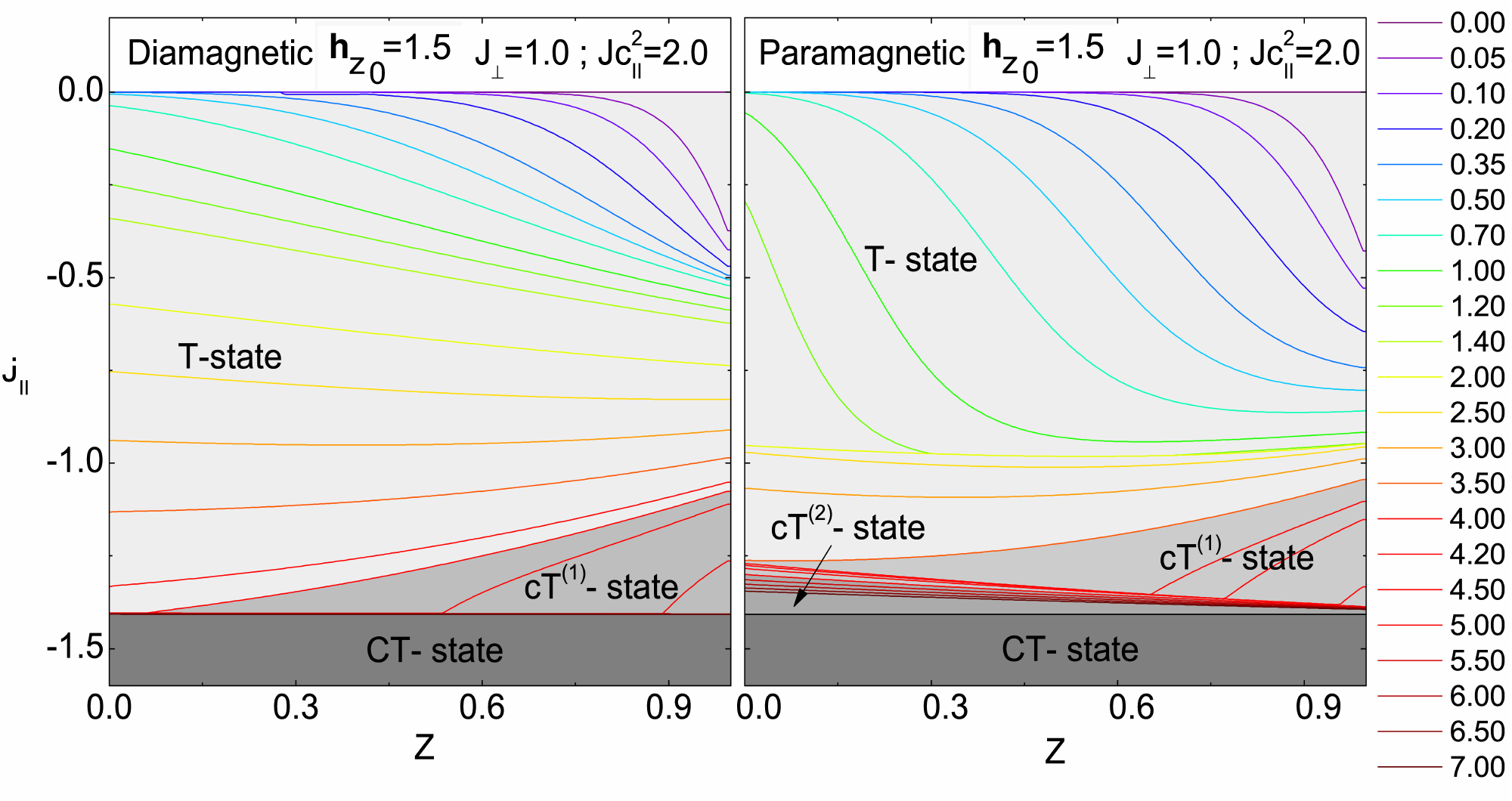}
\caption{\label{Figure_4_12} Profiles of the parallel
currents $j_{\parallel}$ for the rectangular region hypothesis
or DCSM with $J_{c\parallel}^{2}=2$ and
$J_{\perp}=J_{c\perp}=1.0$. The curves are labeled according to the applied
field $h_{y0}$ (at right), assuming $h_{z0}=1.5$. The
diamagnetic (left) and paramagnetic (right) cases are shown.}
\end{figure}

Secondly, with the aim of providing a fair understanding on how the T-states
break down for the 3D configurations studied in this chapter, in
Figure~\ref{Figure_4_12} we have plotted the local profiles of
$j_{\parallel}(z)$ for
a moderate perpendicular field $h_{z0}=1.5$ while the external magnetic
field $h_{y0}$ is increased. The left pane shows the
process of saturation in which $j_{\parallel}$
reaches the value $j_{c\parallel}$ for the diamagnetic initial condition.
Analogously, the right pane shows their corresponding behavior but for the
paramagnetic initial condition. To allow a physical interpretation of these
profiles, we have introduced the following notation: $cT$ denotes that
$j_{\parallel}$ has reached the limit $j_{c\parallel}$ only partially within the
sample, while $CT$ means that $j_{\parallel}$ equals $j_{c\parallel}$ for the
whole range $0\leq z\leq a$. For the {\em partial
penetration} cT-states, we additionally distinguish between the so-called
cT$^{(1)}$ and cT$^{(2)}$ phases. As one can see in the plot, cT$^{(1)}$
means that $j_{\parallel}$ penetrates \textit{linearly} from the surface
until the limitation is reached somewhere within the sample. For the
diamagnetic case, the profile stops at the actual value
$j_{c\parallel}$. However, for the paramagnetic case, the structure
is more complex. Thus, $j_{\parallel}$ penetrates linearly until a
{\em linear increase} (towards the center) curve is reached. This structure is
followed until the contact between both lines reaches the surface. Then, the
so-called cT$^{(2)}$ region appears. The cutting current component
$j_{\parallel}$ has reached the threshold value $j_{c\parallel}$ at the
surface, and the whole $j_{\parallel}$ curve
``pivots'' around this point until the full CT-state is reached. We
call the readers' attention that the initial separations of the
magnetic moment from the T-state master curves take place as soon as a
cT-state is obtained. Further, the corners can be clearly
assigned to the instant at which the full CT state appears.

\begin{figure}[t]
\centering
\includegraphics[height=6cm,width=13cm]{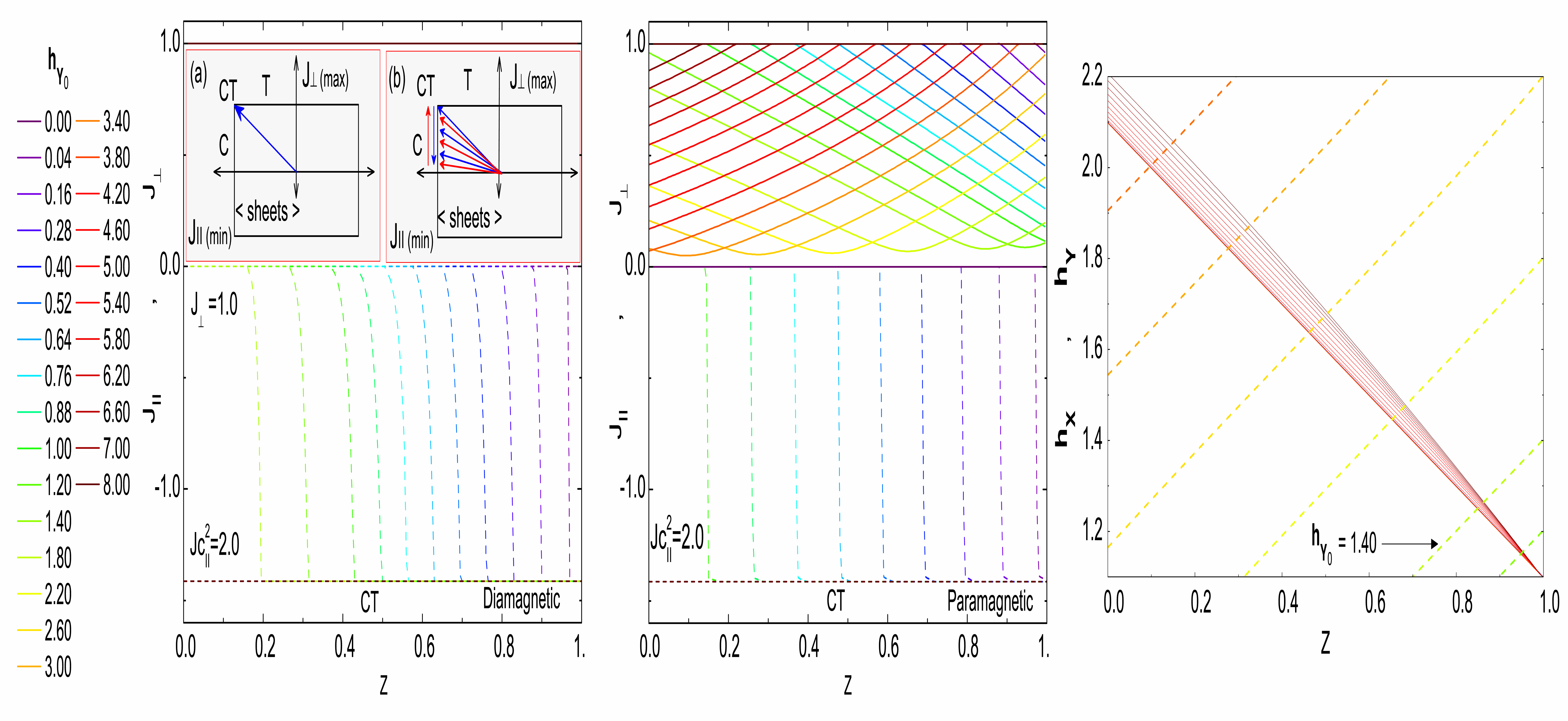}
\caption{\label{Figure_4_13} \textit{Left pane:} Profiles of the parallel
($J_{||}$) and perpendicular ($J_{\perp}$) current densities in the diamagnetic
configuration at a low perpendicular field $h_{z0}=0.1$, for the rectangular
region or DCSM law with $J_{c\parallel}^{2}=2.0$ and $J_{c_{\perp}}=1.0$.
\textit{Central pane:} Same as above, but the profiles for the paramagnetic
configuration are shown. \textit{Right pane:} Magnetic field components
$h_{x}(z)$ (solid-lines) and $h_{y}(z)$ (dashed-lines) corresponding
to the above mentioned paramagnetic configuration. For clarity, the $h_{y}(z)$
profile corresponding to $H_{y0}=1.40$ has been labeled accordingly.
All the curves follow the same color scale convention corresponding to the
values of the applied field $h_{y0}$. In left pane, inset (a) schematically
shows the CT structure of the full penetration regime in the diamagnetic case.
The CT-C structure behavior of ${\bf J}$ for the paramagnetic case is shown in
the inset (b).
 }
\end{figure}

\begin{figure}[t]
\centering
\includegraphics[height=7cm,width=13cm]{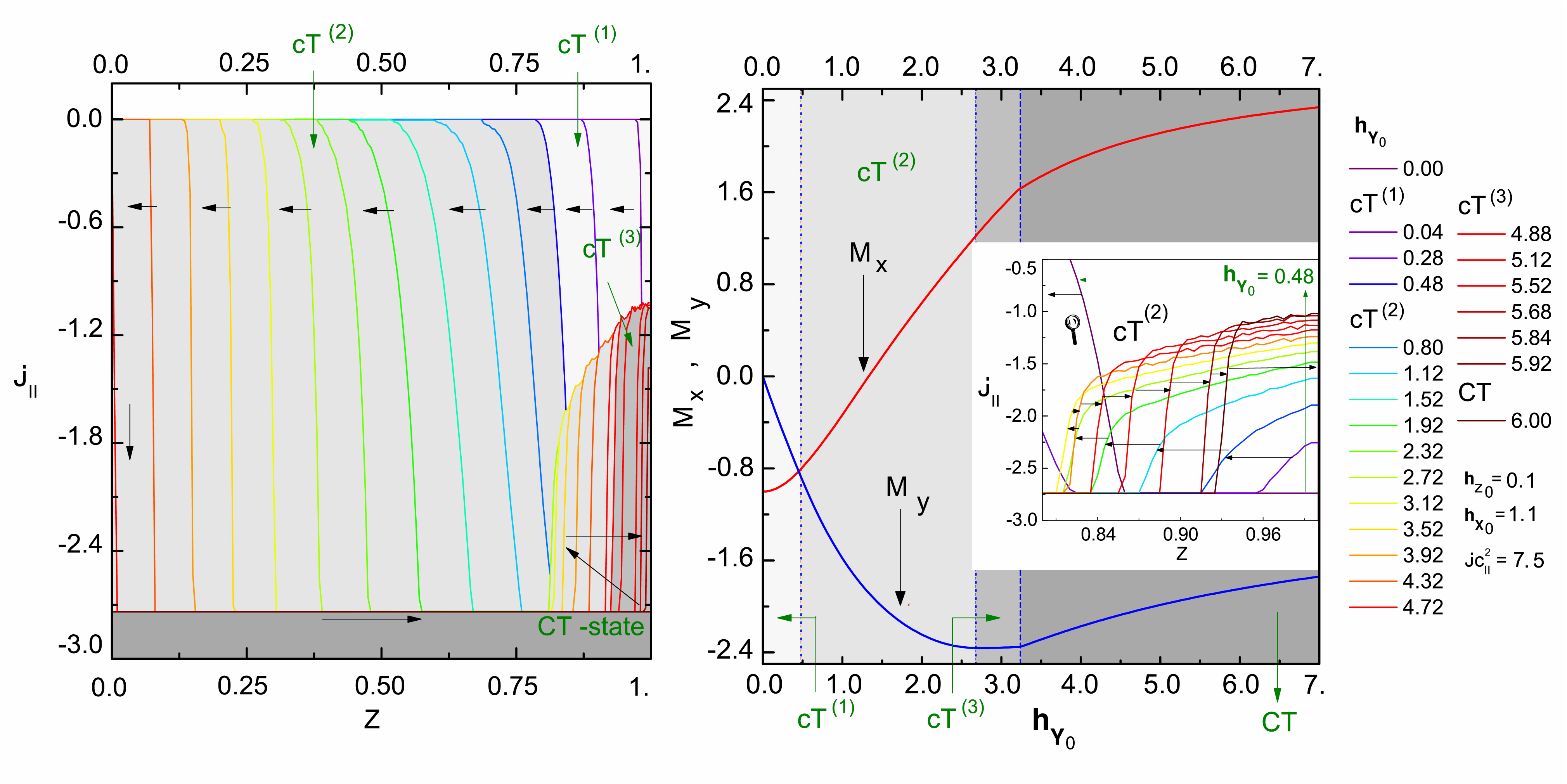}
\caption{\label{Figure_4_14} \textit{Left pane:} Profiles of $j_{||}$ for the
diamagnetic case within the rectangular DCSM with $\chi^{2}=7.5$ and
$h_{z0}=0.1$. In all cases, one gets $J_{\perp}=J_{c\perp}=1.0$.
\textit{Right pane:} The corresponding magnetic moment components
($M_{x},M_{y}$) as a function of $h_{y0}$ are shown. The evolution from the
initial full penetration T state to the final full penetration CT state takes
place in three steps that are classified according to the structure along the
sample width, by means the defined states: cT$^{(1)}\equiv$  T-CT,
cT$^{(2)}\equiv$ T-CT-T, cT$^{(3)}\equiv$ CT-T and eventually CT.
}
\end{figure}

Also interesting are the peculiarities of the cutting and depinning components
of the current density penetration profiles for low values of the
perpendicular field $h_{z0}$. They can
be observed in Figs.~\ref{Figure_4_13}~\&~\ref{Figure_4_14}, which both
reveal new physical mechanisms that do not appear for
the moderate perpendicular field values. Once more, the first observation is
that the appearance of the corner in the magnetic
moment straightforwardly relates to the current density profiles. Thus, for the
lower values of $\chi$ (no corner present), the profile $j_{\parallel}$ displays
a rather simple structure, basically jumping from $0$ to $j_{c\parallel}$ at
some point within the sample (Fig.~\ref{Figure_4_13}). On the contrary, for the
higher values of $\chi$ (those displaying a corner in $M_{xy}$) the evolution
of the cutting profiles $j_{\parallel}(z)$ is much more complex
(Fig.~\ref{Figure_4_14}). Let us analyze these plots in more detail:

Firstly, Fig.~\ref{Figure_4_13} shows the cutting
profiles
$j_{\parallel}(z)$ both for the diamagnetic and paramagnetic cases with a DCSM
region characterized by the parameter $\chi^{2}=2$.  It is to be noticed that,
in both cases, the step-like structure with $J_{\parallel}=0$ in the inner part
and $J_{\parallel}=J_{c\parallel}$ in the periphery evolves until the {\em full
penetration} state $J_{\parallel}=J_{c\parallel}\;,\;\forall\; z$ is reached.
However, an outstanding fact is that in the paramagnetic case, for the first
time along the exposition of this chapter we have met a set of conditions that
produce an excursion of $j_{\perp}$, i.e., the customary condition
$J_{\perp}=J_{c\perp}$ is violated during the
process of increasing $h_{y0}$. To be specific, $J_{\perp}$ starts from the
condition $J_{\perp}=J_{c\perp}$, given by the initial process in $h_{x0}$.
Then, a basically linear decrease from some inner point towards the surface
occurs, with an eventual reduction to a nearly null value at some regions within
the sample (\textit{C-states are basically provoked}). Further increase of
$h_{y0}$ produces a new CT-state. This behavior is shown in a pictorial form
within the insets of Fig.~\ref{Figure_4_13}. Recall that the average current
density sharply transits from a
T-state ($J_{\perp}=J_{c\perp}\,,\,J_{\parallel}=0$) to the
CT-state ($J_{\perp}=J_{c\perp}\,,\,J_{\parallel}=J_{c\parallel}$) for the
diamagnetic case, while a T $\to$ C $\to$ CT evolution happens for the initial
paramagnetic conditions. This behavior allows a physical interpretation in terms
of the evolution of the magnetic field profiles. Thus, as stated before, the
cases with small $\chi$ are characterized by a nearly frozen profile in $h_x$,
as shown in right pane of Fig.~\ref{Figure_4_13}. Then the structure of
$h_{x}(z)$ and $h_{y}(z)$ is basically a cross between two straight lines, where
the crossing point coincides with the minimum in $J_{\perp}(z)$. Thus, recalling
the interpretation of the perpendicular component of the current density
[Eqs.~(\ref{Eq.4.7})~\&~(\ref{Eq.4.8})], the minimum should be expected as
$h_{x}^{2}+h_{y}^{2}$ has a very small variation around
the crossing point of the two families of nearly parallel lines.

On the other hand, the details  about the behavior of the cutting
component $j_{\parallel}$ for the larger values of
$\chi$ are presented in Fig.~\ref{Figure_4_14}, that corresponds to the case
$\chi^{2}=7.5$. Again, owing to the complexity of the structure, we introduce
the notation cT$^{(1)}$, cT$^{(2)}$ and cT$^{(3)}$, that is explained below. Let
us first recall that the corner appears when the {\em partial penetration} 
regime cT$^{(3)}$ extinguishes and the full sample ($0<z<d/2$) satisfies the
conditions $J_{\perp}=J_{c\perp}$ and $J_{\parallel}=J_{c\parallel}$ (i.e., CT).
This property is clearly seen in the right pane of this figure. Thus, the
cT$^{(1)}$ regime is characterized by a T region in the inner part of the sample
($J_{\perp}=J_{c\perp}$ and $J_{\parallel}=0$), that abruptly becomes CT at a
point that progressively penetrates towards the center (T-CT structure). At a
certain instant, the profile becomes T-CT-T because the outermost layers develop
a {\em subcritical} $J_{\parallel}$. This is called cT$^{(2)}$. Then, the
central CT band grows towards both ends. In first instance, the inner T region
becomes CT, giving a global CT-T structure, that we call cT$^{(3)}$. In a final
step, the surface T layer shrinks again to a null width and the full profile is
a CT region. This instant establishes the appearance of the corner in the
magnetization curves.

\vspace*{1cm}
\subsection*{\label{ch-4-5}
\hspace*{0.1cm} \textsl{4.5 Smooth critical states in ``3D''
configurations}}
\vspace*{1cm}
\markboth{\hspace*{0.1cm}\textsl{4.5 Smooth critical states in ``3D''
configurations}} 
{\hspace*{0.1cm} \textsl{4.5 Smooth critical states in ``3D''
configurations}}
\addcontentsline{toc}{chapter}
{\hspace*{0.1cm} \textsl{4.5 Smooth critical states in ``3D''
configurations}}

As stated before, our smooth double critical state theory (SDCST) allows to
specify almost any critical state law by means a simple mathematical statement
that includes an index $n$ accounting for the \textit{smoothness} of the
$J_{\parallel}(J_{\perp})$  relation, and a certain \textit{bandwidth}
characterizing the magnetic anisotropy ratio $\chi=J_{c\parallel}/J_{c\perp}$
[see Eq.~(\ref{Eq.2.17})]. The systematic consideration of the influence of
these parameters is of remarkable importance as it allows a straightforward
elucidation of the relation between diverse physical processes and the actual
material law. Most of the experimental evidences reflecting accurate
observations for the influence of the cutting effects onto the macroscopic
measurements of magnetic moment for anisotropic superconducting samples,
without transport current, may be summarized along two remarkable facts:

\begin{enumerate}[(i)]

\item The occurrence of magnetization peaks which are mainly evident in
paramagnetic configurations
\cite{P2-Voloshin_1997,P2-Fisher_2000,P2-Voloshin_2001,P2-Fisher_1997,
P2-Voloshin_2010}.

\item The collapse of the magnetization curves towards an ostensible isotropic
response \cite{P2-Fisher_1997,P2-Voloshin_2010,P2-Fisher_2000_b}.

\end{enumerate}

Nowadays, it is well known that the DCSM and its precursors (the T-state and
Isotropic models) are not able to achieve a fair understanding of the above
effects in a wide number of configurations, or at least these models do not
handle environments with high magnetic fields~\cite{P2-Ruiz_PRB_2009}. This fact
has lead to consider alternative models such as the two-velocity electrodynamic
model~\cite{P2-2hydrodynamic_model}, or the helical electrodynamic model
\cite{P2-helical_model}, both lacking a solid physical basis for
the mechanisms underlying the motion of vortices. In fact, being the
threshold values for the cutting current component $(J_{c\parallel})$ and the
depinning current component $(J_{c\perp})$ the main physical observables
determining the magnetic anisotropy of a
superconductor~\cite{P2-Clem_2011_SUST,P2-Campbell_2011_SUST,P2-Karasik_1970},
within these models, other parameters have to be included ad-hoc.

Moreover, it is worth anticipating the following chapter, by mentioning that
neither of the above mentioned models allow a correct explanation for the
experimental remarks when the superconductor is also carrying a longitudinal
transport current. Thus, the purpose of this section is vindicating the physical
mechanisms of cutting and pinning, through a comprehensive study of the magnetic
anisotropy of type II superconductors, by means the modification of the {\em
conventional} DCSM which leads to the establishment of the SDCST. Such
modifications can be justified as corrections to the simplifying ideas that flux
depinning is only related to $J_{\perp}$ and the flux cutting is only related
to $J_{\parallel}$. We emphasize that in a in a general scenario, one
should consider the
dependencies $J_{c\perp}=J_{c\perp}(J_{\parallel})$ and
$J_{c\parallel}=J_{c\parallel}(J_{\perp})$
\cite{P2-Ruiz_PRB_2009,P2-Ruiz_SUST_2010,P2-Brandt_2007,P2-Clem_2011_SUST,
P2-Campbell_2011_SUST}. 

\begin{figure}[t]
\centering
\includegraphics[height=11cm,width=13cm]{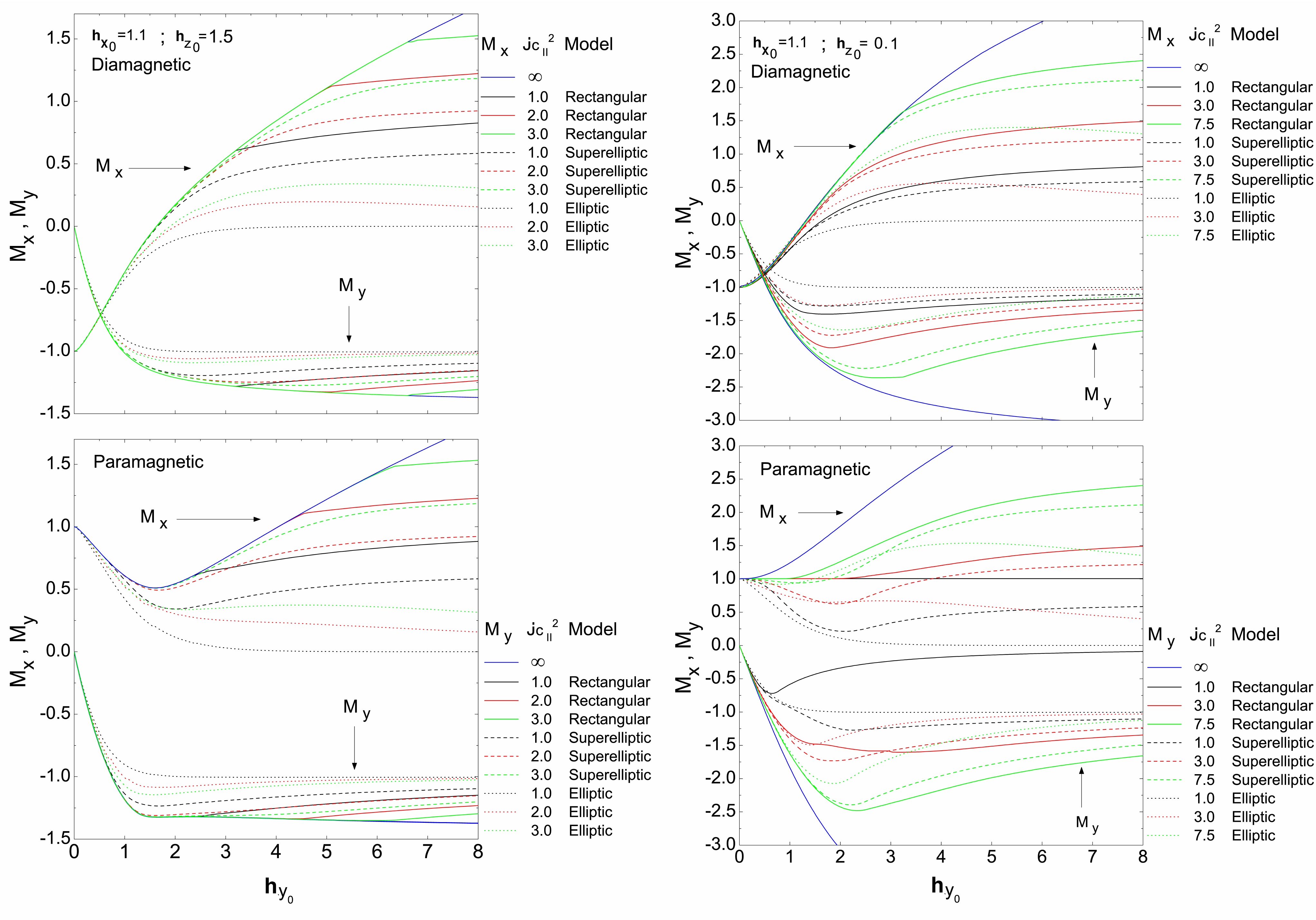}
\caption{\label{Figure_4_15} The magnetic moments $M_{x}$
and $M_{y}$ of the slab per unit area as a function of $h_{y_{0}}$
in the diamagnetic (top) and paramagnetic (bottom) cases with
$h_{x0}=1.1$ and both for, a moderate perpendicular field $h_{z0}$=1.5 (left
pane), and a low perpendicular field $h_{z0}=0.1$ (right pane). The ``infinite
band'' or T-states model (blue solid lines), the DCSM or ``rectangular regions''
(other solid lines), the SDCST's models with $n=4$ ``superelliptical regions''
(dashed-lines), and $n=1$ ``elliptical regions'' (dotted-lines), are
shown for several values of the ratio $\chi^{2}\equiv j_{c\parallel}$ for a
given $J_{c\perp}=1$. 
}
\end{figure}

Recalling that, mathematically, the effect of {\em smoothing the corners} for
the rectangular DCSM region may be represented by a {\em one-parameter} family
of superelliptical functions with the generic form given in
Eq.~(\ref{Eq.2.17}), i.e., 
\begin{eqnarray}
\left(\frac{J_{\parallel}}{J_{c\parallel}}\right)^{2n}+
\left(\frac{J_{\perp}}{J_{c\perp}}\right)^{2n}\leq 1 \, , \nonumber
\end{eqnarray}
such kind of curves cover the whole range of interest just by allowing $n$ to
take values over the positive integers. As the reader can easily verify for
$\chi^{2}>1$, the index $n=1$
corresponds to the standard ellipse and $n\geq4$ is basically a rectangle with
faintly rounded corners (see Fig.~\ref{Figure_4_1}).

\begin{figure}[t]
\centering
\includegraphics[width=1.0\textwidth]{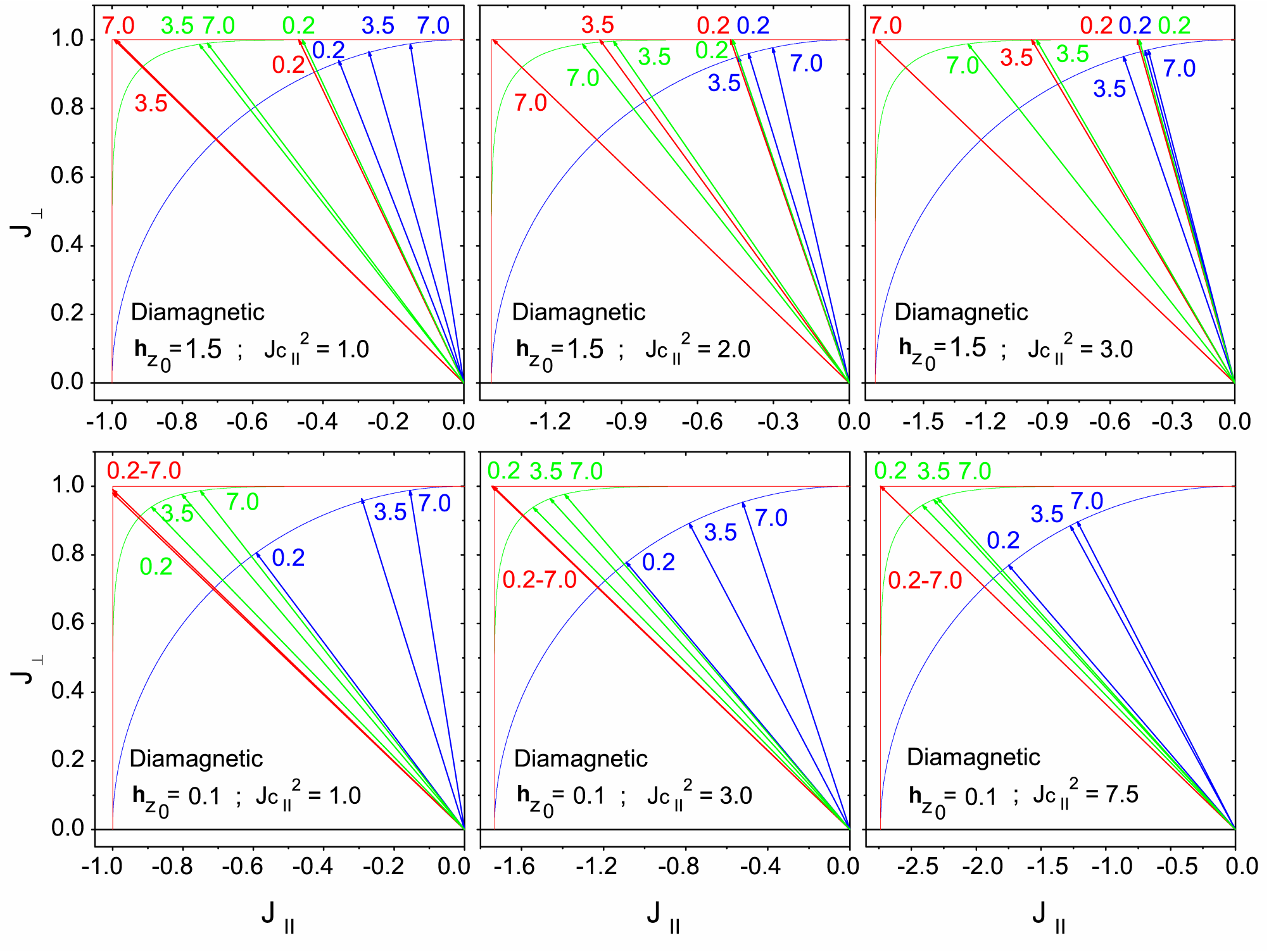}
\caption{\label{Figure_4_16} Current density vector $\textbf{J}$ in the planar
representation [$J_{\perp}$,$J_{\parallel}$] for three different material law
models, corresponding to the following $\Delta_{\textbf{r}}$ regions:
rectangular (red), superelliptical (green), and elliptical (blue). Here, the
diamagnetic case for both a moderate field $h_{z0}=1.5$
and the lower field $h_{z}=0.1$ are shown. Several vectors for several values of
the ratio $\chi=J_{c\parallel}/J_{c\perp}$ and the applied
field $h_{y0}$ are shown and labeled on each arrow. The scales on the
horizontal axes that have been re-sized for visual purposes.
}
\end{figure}

In order to illustrate the effect of smoothing the material law
$\Delta_{\textbf{r}}(\textbf{J}_{\parallel},\textbf{J}_{\perp})$ for different
bandwidths $\chi$, below we will
show the magnetization curves that are obtained for the diamagnetic and
paramagnetic configurations considered before (Fig.~\ref{Figure_4_2}). The
main results of our analysis
are depicted in Figs.~\ref{Figure_4_15}~\&~\ref{Figure_4_16}.

Figure~\ref{Figure_4_15} shows the behavior of the magnetization curves $M_x$
and $M_y$ for an external perpendicular field of either moderate intensity
$h_{z0}=1.5$ (left pane) or a lower intensity $h_{z0}=0.1$ (right pane). In
order to simplify their interpretation, at this stage we will only compare the
prediction for the \textit{smoothing} index $n=4$ (\textit{superelliptic
region}) with the limiting cases $n=1$ (\textit{elliptic region}) and
$n\rightarrow\infty$ (\textit{a DCSM or rectangular region}), under
consideration of different bandwidths $\chi$.

%
%
Firstly, for the moderate perpendicular field region (left pane of
Fig.~\ref{Figure_4_15}), we observe that the overall effect of reducing the
value of $\chi\equiv J_{c\parallel}/J_{c\perp}$ is the same for the three
material laws or $\Delta_{\textbf{r}}$ regions. The
smaller the value of $\chi$, the higher reduction respect to the T-state
($\chi\to\infty$) master curve for the magnetic moment components. On the other
hand, as regards the particular details for each model, we recall: (i) as
expected the smooth models lead to smooth variations, i.e.: the corner is not
present, (ii) the breakdown of the T-state behavior occurs before (at higher
values of $\chi$ or lower values of $h_{y0}$) for the smoother models. Strictly
speaking, the concept of T-state is only valid for the rectangular region, but
it is asymptotically generated as the superelliptic parameter $n$ grows.
Finally,
(iii) the isotropic CS limit, given by the circular region $n=1$ and $\chi=1$
produces the expected results~\cite{P2-Badia_PRL_2001}: $M_x$ collapses to zero,
and $M_y$ develops a {\em one dimensional} critical state behavior.

%
\begin{figure}[t]
\centering
\includegraphics[width=1.0\textwidth]{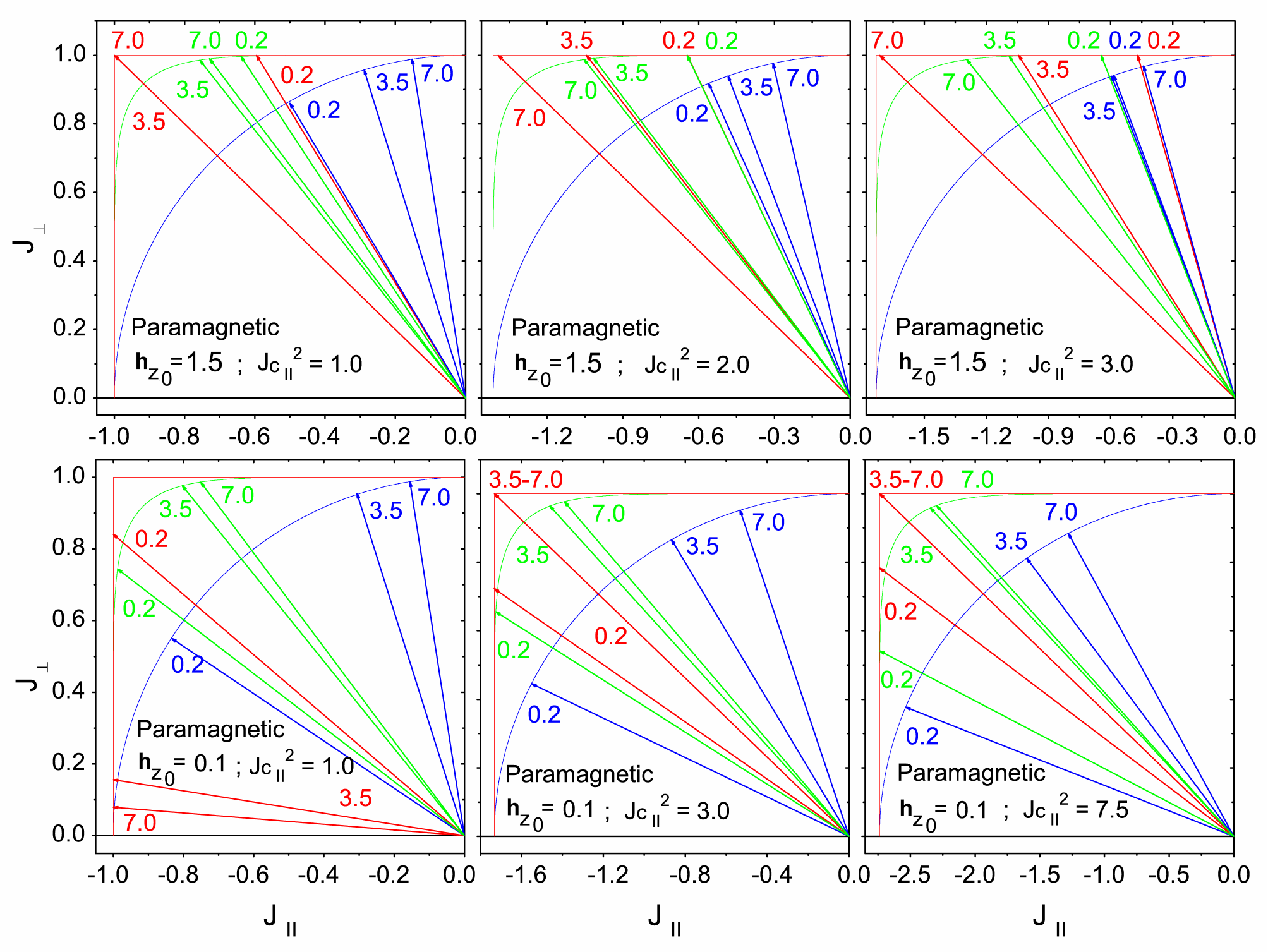}
\caption{\label{Figure_4_17} Same as Fig.~\ref{Figure_4_16}, but here the $J$
vectors corresponds to the paramagnetic case.
}
\end{figure}

Secondly, for the low perpendicular field region (right pane of
Fig.~\ref{Figure_4_15}, one can notice: (i) on the one hand, the rectangular and
superelliptical models produce very similar results for the diamagnetic case,
both for $M_x$ and for $M_y$, noticeably differing from the elliptical region
predictions, that still show a practical collapse of $M_x$ and a saturation in
$M_y$ as stated before. (ii) On the other hand, the paramagnetic case involves a
higher complexity. Thus, we recall that the already mentioned
feature of a ``\textit{flat}'' behavior in $M_{x}$ for small values of $\chi$
within the rectangular region model, is no longer observed upon smoothing of
the restriction region. On the contrary, the smooth models
involve an initial negative slope and a minimum, resembling the behavior of
$M_x$ for the rectangular model, but in moderate $h_{y0}$. As concerns $M_y$,
important differences among the three models are also to be recalled.

In order to provide a physical interpretation of the behaviors reported
in the above paragraphs for moderate and low perpendicular fields $h_{z0}$, a
comparative plot of the current density vectors for
each case is given in Figs.~\ref{Figure_4_16}~\&~\ref{Figure_4_17}
respectively. For clarity, we restrict to the
representation of the vector ${\bf J}$ at the surface of the sample ($z=a$) 
for a selected number of values of $h_{y0}$. Just at a first glance, one can
relate the  best coincidence in predicted magnetization to the more similar
critical current density structures (superelliptical and rectangular regions for
the diamagnetic case with $h_{z0}=0.1$). Recall that, in that case, the
rectangular region produces a CT-state structure
($J_{\parallel}=J_{c\parallel}$ and $J_{\perp}=J_{c\perp}$) that is represented
by a ${\bf J}$ vector, pinned in the corner. On the other hand, the vector ${\bf
J}$ related to the superelliptic model does not pin at any point, because such a
singular point does not exist. However, it is basically oriented in the same
fashion and this relates to the good agreement in ${\bf M}$. We emphasize that
the cases in which strong differences occur for the magnetic
moment are also related to important changes in the behavior of ${\bf J}$.
Thus, if one considers the paramagnetic case at small values of $h_{z0}$ and
$h_{y0}$, the significant differences in magnetization relate to an opposite
behavior in ${\bf J}$. Moreover, the rectangular model predicts a transition
towards a C-state ($J_{\parallel}=J_{c\parallel}$ and $J_{\perp}\approx 0$),
while the smooth versions produce a tendency towards the T-state (see left
bottom panel of Fig.~\ref{Figure_4_17}).


\break
\vspace*{0.5cm}
\subsection*{\label{appendix-ch-4}\textsl{Appendix I Critical angle gradient in
``3D'' configurations}}
\vspace*{1cm}
\markboth{\textsl{Appendix I Critical angle gradient in ``3D''
configurations}}{\textsl{Appendix I Critical angle gradient in ``3D''
configurations}}
\addcontentsline{toc}{chapter}{\textsl{Appendix I Critical angle gradient in
``3D'' configurations}}

On the basis of minimum complexity, in this appendix the flux cutting criterion
for 3D configurations will be revised under the assumption of a critical angle
threshold instead of a superelliptical relation. Below let me present some
results related to the concept of the critical angle gradient in 3D systems. 

First recall that the limitation on $J_{\parallel}$ appears as
related to the energy reduction by the cutting of neighboring flux lines when
they are at an angle beyond some critical
value~\cite{P2-Brandt_1979,P2-Clem_1980}.
This concept has been largely exploited in the 2D slab geometry for
fields applied parallel to the surface~\cite{P2-Clem_DCSM}, and it is introduced
by the local relation
\begin{eqnarray}\label{Eq.4.12}
\left|\frac{d\alpha}{dz}\right|=\left|\frac{J_{\parallel}}{H}\right|\leq
K_{c} \, ,
\end{eqnarray}
that establishes a critical angle gradient. Here, $\alpha$ stands
for the angle between the flux lines and a given reference within
the $XY$-plane (i.e.: an azimuthal angle). However, for the 3D cases
under consideration, the relative misorientation between flux lines
may also have a polar angle contribution, i.e.: ${\bf H}$ does not
necessarily lie within the $XY$-plane or any other given plane. 

As sketched in Fig.~\ref{Figure_2_1} (pag.~\pageref{Figure_2_1}), one has to
introduce the angle $\gamma$ within the plane defined by the pair of flux lines
under consideration. After
some mathematical manipulations, it can be shown that, for the infinite slab
geometry, with a three dimensional magnetic field one has
\begin{equation}\label{Eq.4.13}
\frac{d\gamma}{dz}=\sqrt{\frac{J_{\parallel}^{2}}{H^{2}}+\frac{H_{z}^{2}J^{2}}{
H^ { 4}}}
=\frac{1}{H}\sqrt{J_{\parallel}^{2}+\frac{H_{z}^{2}}{H^{2}}\left(J_{\parallel}^{
2}+J_{\perp}^{2}\right)}\, ,
\end{equation}
where the third component is also introduced. Actually, the above result is just
a particular case of the relation
\begin{eqnarray}\label{Eq.4.14}
\nabla \times \left( B \hat {\bf B}\right)=\left[\left(\nabla B\right) \times
\hat {\bf B}\right] +
\left[ B\;\left(\nabla\times\hat {\bf B}\right)\right]
\equiv \left[{\bf J}_{\perp ,1}\right]+\left[{\bf J}_{\perp ,2}+{\bf
J}_{\parallel}\right]\, ,
\end{eqnarray}
showing that, in general, both ${\bf J}_{\parallel}$ and ${\bf J}_{\perp}$ can
contribute to the spatial variation of the direction $\hat{\bf B}$.

Below, we display the effects of using the cutting limitation
\begin{equation}\label{Eq.4.15}
\left|\frac{d\gamma}{dz}\right|\leq \kappa_{c} \, ,
\end{equation}
instead of assuming a constant value for the parallel
critical current. Fig.~\ref{Figure_4_17} contains the main results. The
calculations have been performed for the same diamagnetic and paramagnetic
initial configurations displayed in Fig.~\ref{Figure_4_2}.

\begin{figure}[t]
\centering
\includegraphics[height=7cm,width=13cm]{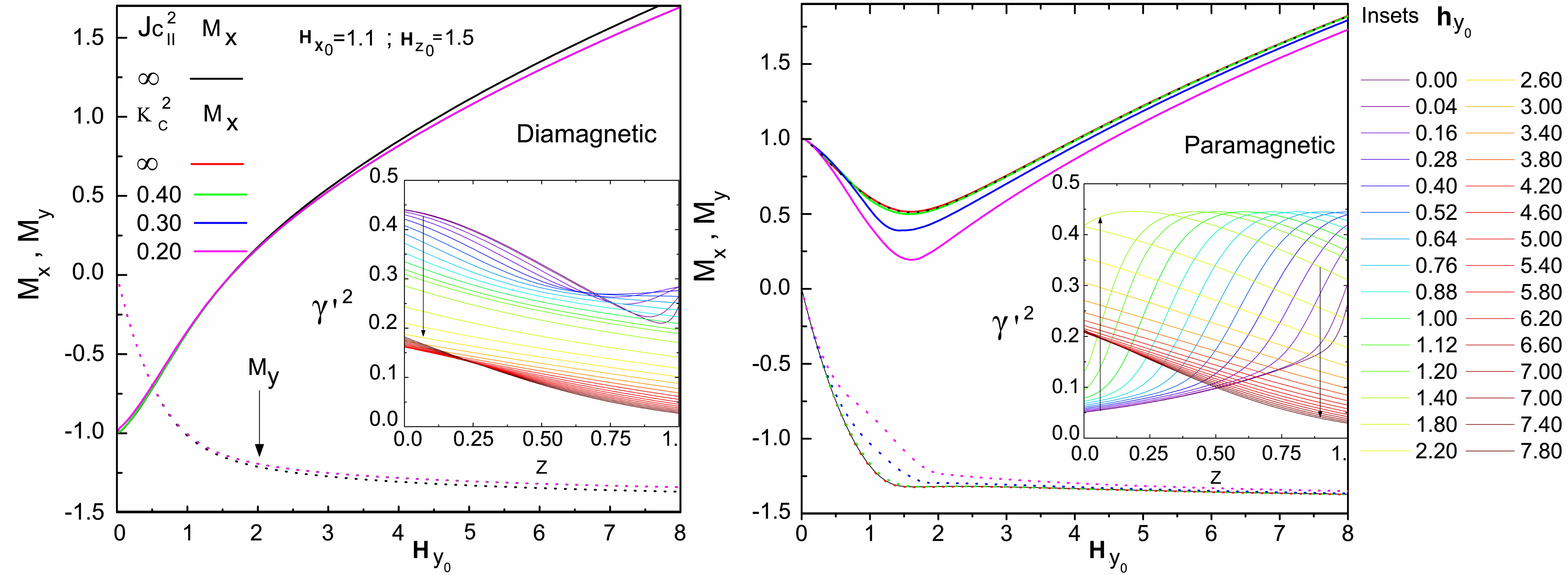}
\caption{\label{Figure_4_18} The magnetic moments per unit area $M_{x}$
(solid lines) and $M_{y}$ (dotted lines) of the slab as a function of
$h_{y_{0}}$ for the critical angle gradient model [Eq.(\ref{Eq.4.15})]. The
unrestricted case ($\kappa_{c}^{2}\to\infty$) is shown for comparison with
several cases with a restricted angle gradient: $\kappa_{c}^{2}=$0.20, 0.30 and
0.40 (dimensionless units are defined by $\kappa_{c}\equiv K_{c}a$). Shown are
the diamagnetic (left pane) an paramagnetic (right pane) cases for
$h_{x_{0}}=1.1$
and $h_{z_{0}}$=1.5. The insets detail the evolution of the angle gradient
profiles for $\kappa_{c}^{2}\to\infty$.
}
\end{figure}

In general, one can see that the smaller values for the cutting threshold in
whatever form produce the smaller magnetic moments (compare
Figs.~\ref{Figure_4_10} \& \ref{Figure_4_11} with Fig.~\ref{Figure_4_18}).
However, some important differences are to be quoted. On the one hand, the
critical angle criterion $|{\gamma}'|\leq\kappa_{c}$ produces a smooth
variation, by contrast to the corner structure induced by the critical current
one $J_{\parallel}\leq J_{c\parallel}$. On the other hand, the effect of
changing the value of $\kappa_{c}$ is much less noticeable, especially for the
diamagnetic case, in which the full range of physically meaning values of
$\kappa_{c}$ produce a negligible variation. Moreover, we call the readers'
attention that the above mentioned range for $\kappa_{c}$ is established by the
application of Eq.~(\ref{Eq.4.13}) to the initial state of the sample. Thus,
if one takes $J_{\parallel}=0,\, H_{z0}=1.5,\, H_{x0}=1.1$, the squared angle
gradient takes the value ${\gamma}'^2 = 0.19$ and one has to use $\kappa^2 >
0.19$
in order to be consistent with the initial critical state assumed.


\chapter{\label{ch-5} \sc \textbf{The Longitudinal Transport Problem}}

It is well known that various striking phenomena may occur when a type-II
superconductor with intrinsic magnetic anisotropy is under the action of a
transport current and a longitudinal magnetic
field~\cite{P2-Ruiz_PRB_2009,P2-Ruiz_SUST_2010,P2-Voloshin_2001,P2-Clem_1980,
P2-Ruiz_PRB_2011,P2-Ruiz_SUST_2011,P2-Sanchez_2010,P2-Blamire_2003,
P2-LeBlanc_2003,P2-LeBlanc_2002,P2-Matsushita_1984,P2-Matsushita_1998,
P2-Matsushita_2012,P2-LeBlanc_1993,P2-LeBlanc_1991,P2-Voloshin_1991,
P2-Cave_1978,P2-Walmsley_1977,P2-Esaki_1976,P2-Walmsley_1972,P2-London_1968,
P2-LeBlanc_1966,P2-Watanabe_1992,P2-Blamire_1986,P2-Boyer_1980,P2-Gauthier_1974,
P2-Karasik_1970,P2-Sugahara_1970,P2-Taylor_1967,P2-Sekula_1963,P2-Clem_2011_PRB,
P2-Nakayama_1972}.
In particular, a remarkable enhancement of the critical current density by
means of its \textit{compression} towards the center of the superconducting
sample has been observed in a wide number of conventional and high temperature
superconducting systems within a certain set of experimental
conditions
\cite{P2-Clem_1980,P2-Blamire_2003,P2-Sanchez_2010,P2-Watanabe_1992,
P2-Blamire_1986,P2-Boyer_1980,P2-Gauthier_1974,P2-Karasik_1970,P2-Sugahara_1970,
P2-Taylor_1967,P2-Sekula_1963}.
This property, together with other intriguing phenomena, such as the observation
of paramagnetic moments, and outstandingly, the experimental observation of a
counter intuitive phenomenon of negative resistance by the action of a parallel
magnetic field, have been reported in the course of intense experimental and
theoretical
activities~\cite{P2-Voloshin_2001,P2-LeBlanc_2003,P2-LeBlanc_2002,
P2-Matsushita_1984,P2-Matsushita_1998,P2-Matsushita_2012,P2-LeBlanc_1993,
P2-LeBlanc_1991,P2-Voloshin_1991,P2-Cave_1978,
P2-Walmsley_1977,P2-Esaki_1976,P2-Walmsley_1972,P2-London_1968,P2-LeBlanc_1966}.
Most of these works were primarily concerned with the arrangement of the
macroscopic current density ${\bf J}$ along the so-called nearly {\em force
free} trajectories~\cite{P2-Bergeron_1972}. Recall that if ${\bf J}$ is {\em
nearly parallel} to the
magnetic induction ${\bf B}$, moderate or weak pinning forces are needed for
avoiding the detrimental flux-flow losses related to the drift of flux tubes
driven by the magnetostatic force (${\bf J}\times{\bf B}$ per unit volume). More
specifically, negative voltages have been observed by different
groups
\cite{P2-LeBlanc_2003,P2-LeBlanc_2002,P2-Matsushita_1984,P2-Matsushita_1998,
P2-Matsushita_2012,P2-LeBlanc_1993,P2-LeBlanc_1991,P2-Voloshin_1991,
P2-Matsushita_1984,P2-Cave_1978,P2-Walmsley_1977} when recording the
current-voltage characteristics at specific locations on the surface of the
sample (central region). 

\begin{figure}[t]
\centering
\includegraphics[height=5.2cm,width=13cm]{Figure_5_1.pdf}
\caption{\label{Figure_5_1} (a) Schematic representation of the helical model
for the longitudinal transport problem in an infinite superconducting cylinder.
An array of parallel helical fluxoids without flux cutting is assumed. (b)
Pictorial illustration of the real situation where the flux cutting events
appear. The cylindrical and slab symmetries are shown.
}
\end{figure}

In a first approach, such resistive structure has
been intuitively understood in terms of helical domains, closely connected to
the force-free current parallel to the
flux-lines~\cite{P2-helical_model,P2-Matsushita_1984,P2-Matsushita_1998,
P2-Matsushita_2012,P2-LeBlanc_1966,P2-Gauthier_1974,P2-Esaki_1976}. The basic
idea of this model relies in the fact that the averaged direction of
the flux flow in a superconducting cylinder subjected to a longitudinal
magnetic field and transport current, is the same as the direction of the
Poynting's vector $(\textbf{E}\times\textbf{H})$ onto the external surface,
suggesting the occurrence of a continuous helical flux flow without flux
cutting [see Fig.~\ref{Figure_5_1} (a)].  Also, in order to achieve a
concordance with the experimental
evidences~\cite{P2-LeBlanc_1966,P2-Gauthier_1974,P2-Esaki_1976}, the resulting
helical flux over the cylindrical surface must be subdivided in two domains for
which the Poynting's vector is directed in two concomitant directions: inwards
(allowing the compressing of \textbf{J} towards the center of the specimen), and
outwards (allowing the occurrence of surface regions with \textbf{J} flowing in
counter direction to the flux of transport current). Nevertheless, despite the
seeming simplicity of the helical model and its intuitive explanation for the
increasing of the current density and the simultaneous occurrence of surface
negative currents, the helical symmetry shown in Fig.~\ref{Figure_5_1}(a) does
not exist in any other symmetry different to the infinite cylinder, and
furthermore, it does not include the flux cutting mechanism which causes many
derived effects [see Fig.~\ref{Figure_5_1}(b)]. 
Actually, regardless of the symmetry considered,
there are some remarkable effects which can not be explained under this
scenario. On the one hand, it has been stated that the direction of the helical
structure and that of the magnetic field at the surface are really
different~\cite{P2-Matsushita_1984,P2-Matsushita_1998}. On the other hand, by
increasing the magnetic field a continuous torsion of the helical domain should
be expected, such that the vanishing of the negative current within a finite
interval of the applied magnetic fields can not be conceived. Furthermore, it
does neither explain the bounded increase of the transport
current (i.e., the occurrence of a maximal peak on the longitudinal current
density) as one raises the magnitude of the applied magnetic field.
Finally, as a detail of fine structure, local
paramagnetic domains cannot either be predicted within the
above scenario. Therefore, a most accurate description of the diverse effects
underlying to the longitudinal transport problem, have to include the physics
behind the flux cutting mechanism. 

Relying on our theoretical approach for the superconducting critical state
problem in 3D magnetic field configurations and the aforementioned scenario,
below we present an exhaustive analysis of the electrodynamic response for the
so-called longitudinal transport problem of type-II superconductors in the slab
geometry. Remarkable numerical and conceptual difficulties related to the
implementation of the magnetic anisotropy and the relation between the
flux-line cutting (crossing and recombination) and the flux-line depinning
mechanisms, will be overcome by means simplified analytical models for extremal
cases and the further comparison with the most general solution of the smooth
double critical state theory (SDCST) for analogous material laws (subchapter
5.1). Then, supported by numerical
simulations that cover an extensive set of
experimental conditions, we put forward a much more complete physical scenario
which is based upon a set of superelliptical material laws. Thus, subchapter
5.2 is devoted to show how the striking existence of negative flow domains,
local and global paramagnetic structures, emergence of peak-like structures in
both the critical current density and the longitudinal magnetic moment, as well
as the compression of the transport current in type-II superconductors under
parallel magnetic fields, are all predicted by our general critical state
theory.
In addition, we shall introduce some ideas that could be applied for the
determination of the flux cutting threshold from local measurements of the
current density flowing along specific layers of the superconducting sample, as
correlated to the behavior of the magnetic moment components.
%
%
\vspace*{0.0cm}
\subsection*{\label{ch-5-1}
\hspace*{0.1cm} \textsl{5.1 Simplified analytical
models and beyond}}
\markboth{\hspace*{0.1cm}\textsl{5.1 Simplified analytical
models and beyond}} 
{\hspace*{0.1cm} \textsl{5.1 Simplified analytical
models and beyond}}
\addcontentsline{toc}{chapter}
{\hspace*{0.1cm} \textsl{5.1 Simplified analytical
models and beyond}}
\vspace*{0.5cm}

In this subchapter, we call the reader's attention to the fact that two
analytical approaches for the slab
geometry in extreme situations may be found in the literature. The first one
was introduced by Brandt and Mikitik in Ref.~\cite{P2-Brandt_2007} for the
regime
of strong pinning with very weak longitudinal current conditions, i.e., $h_{z0}$
must be very high as compared to the in-plane applied field $h_{xy}(a)$ (then
$J_{\parallel}\ll J_{c\perp}$).\footnotemark[1] On the
other hand, the opposite limit ($h_{z}\to 0$) was recently developed in our
group (Ref.~\cite{P2-Ruiz_PRB_2011}). Thus, in a first stage let us show how the
physical
properties of the longitudinal transport problem may be understood within our
simplified analytical model, and then we will move onto a general description of
the problem in terms of the SDCST. 

\footnotetext[1]{Recall that in order to simplify the mathematical statements,
the electrodynamic quantities are customarily normalized by defining
$\textbf{h}\equiv\textbf{H}/J_{c\perp}a$, $j\equiv\textbf{J}/J_{c\perp}$, and
$\texttt{z}=z/a$}

\break
\vspace*{0cm}
\hspace*{-0.3cm}
\textbf{\textit{5.1.1 The simplest analytical model}}
\markboth{\hspace*{0.1cm} \textsl{5.1.1 The simplest analytical
model}}{\hspace*{0.1cm}
\textsl{5.1.1 The simplest analytical model}}
\addcontentsline{toc}{chapter}{\hspace*{0.1cm} \textsl{5.1.1 The simplest
analytical model}}\label{ch-5-1-1}
\vspace*{1.0cm}

First, recall that the Amp\`ere's law takes the following form for the infinite
slab geometry considered in the previous chapter:
\begin{equation}\label{Eq.5.1}
-\frac{dh_{y}}{dz}=j_{x}
\quad ; \quad
\frac{dh_{x}}{dz}=j_{y} \, .
\end{equation}
Also notice that, in the particular case $h_{z0}=0$ [i.e., $\theta=\pi/2$
in Fig.~\ref{Figure_4_1} (pag.~\pageref{Figure_4_1})], the material law or
region $\Delta_{p}$ becomes a
rectangle with axis defined by the in-plane directions parallel and
perpendicular to $\textbf{h}$. Thus, recalling the statements issued in chapter
4.1, one can show that such expressions may be transformed into the polar form 
\begin{equation}\label{Eq.5.2}
-h\frac{d\alpha}{dz}=j_{\parallel}^{\rm p}
\quad ; \quad
\frac{dh}{dz}=j_{\perp}
\end{equation}
with $h=\sqrt{h_{x}^{2}+h_{y}^{2}}$ the modulus of the magnetic field vector,
and $\alpha={\rm atan}(h_y/h_x)$ the angle between such vector and the
$x$-axis.

Now, the thresholds of flux depinning and cutting imply the in-plane conditions
\begin{equation}\label{Eq.5.3}
|j_{\parallel}^{\rm p}|\leq j_{c\parallel}^{\rm p}(\theta =\pi
/2)=j_{c\parallel}
\quad ; \quad
|j_{\perp}^{\rm p}|\leq 1 \, .
\end{equation}

Notice that, in general, Eq.~(\ref{Eq.5.2}) and the critical constraints defined
in Eq.~(\ref{Eq.5.3}) would not straightforwardly lead to the solution of the
problem. Thus, one should also use Faraday's law, either by explicit
introduction of the related electric fields (as in
Refs.~\cite{P2-Clem_2011_SUST,P2-Campbell_2011_SUST}),
or by our variational statement. However, as in this case $\theta
=\pi/2$ and consequently $j_{c\parallel}^{\rm p}=j_{c\parallel}$,
the resolution noticeably simplifies. In fact, for the considered situation, we
will have a combination of the cases $j_{\parallel}^{\rm p}=\{0$ or $\pm
j_{c\parallel}\}$, and $j_{\perp}^{\rm p}=\{0$ or $1\}$, and then integration
of Eq.~(\ref{Eq.5.2}) is straightforward. For further mathematical ease, we will
also consider $j_{c\parallel}$ and $j_{c\perp}$ to be field independent
constants. 

Following the notation introduced in chapter~4.4 we will
refer to different zones within the sample that are basically related to 
macroscopic regions where well defined dissipation mechanisms occur. In brief,
we will speak about T-zones, where only flux depinning (transport) occurs
(${j}_{\parallel}=0\, ,{j}_{\perp}=\pm 1$), C zones, where only flux cutting
occurs (${j}_{\parallel}=\pm\chi\, ,{j}_{\perp}=0$), CT zones where both
transport and cutting occur (${j}_{\parallel}=\pm\chi\, , {j}_{\perp}=\pm 1$),
and O-zones where neither flux transport nor cutting take
place (${j}_{\parallel}=0\, , {j}_{\perp}=0$). Introducing the above set of
possibilities in Eqs.~(\ref{Eq.5.2})~\&~(\ref{Eq.5.3}) one gets the
following cases for the incremental behavior of the magnetic field in polar
components
\begin{eqnarray}\label{Eq.5.4}
dh=\left\{
\begin{array}{rr}
0\qquad \rm{(O,C)}&
\\
\pm\, dz\;\; \rm{(T,CT)}&
\end{array}
\right.
\!; \;
d\alpha=\left\{
\begin{array}{rr}
0\qquad\qquad \rm{(O,T)}&
\\
\pm\,({\chi}/{h})\,dz\;\; \rm{(C,CT)}&
\end{array}
\right. \, ,
\end{eqnarray}
and all that remains for obtaining the penetration profiles is to solve
successively (integrate) for $h$ and $\alpha$ with the corresponding boundary
conditions (evolutionary surface values $h_{0},\alpha_{0}$). The case selection
has to be made according to Lenz's law. We note, in passing, that further
specification related to the sign is usually included in the notation. Thus, a
T$_{+}$ zone will exactly mean $dh = +dz$. 

\begin{figure}[t]
\centering
\includegraphics[width=1.0\textwidth]{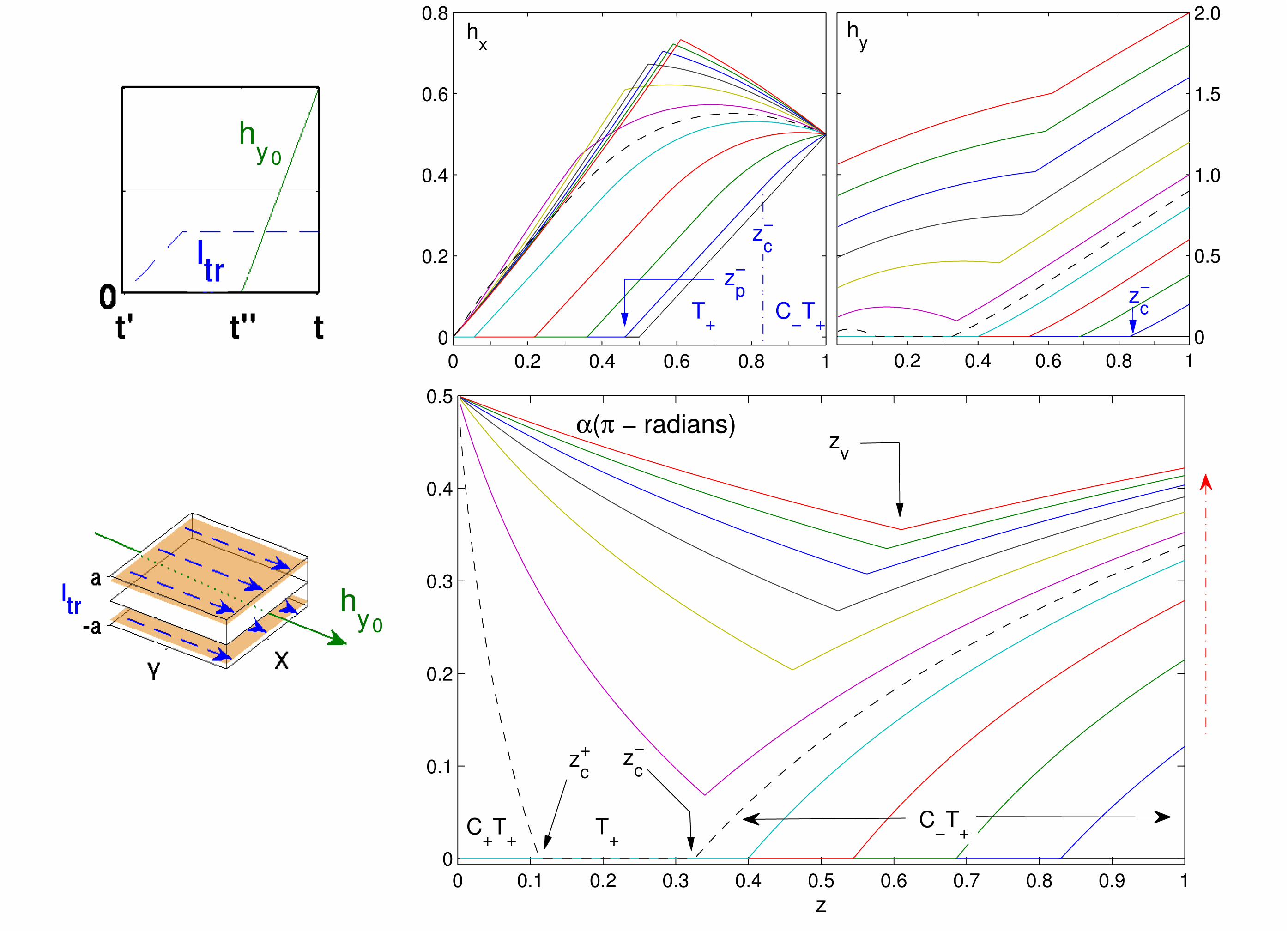}
\caption{\label{Figure_5_2} \textit{Left pane:} Schematic representation of the
simplest experimental configuration of the longitudinal transport problem which
is solved by analytical methods. \textit{Right pane} Penetration profiles of the
magnetic field components and rotation angle in the longitudinal transport
experiment ($h_{z0}=0$) for a superconducting slab of thickness $2a$, as
calculated from Eq.~(\ref{Eq.5.1}). The zone structure induced by increasing
the field
$h_{y0}$ is marked upon some of the curves. The dashed line corresponds to
the transition regime between the states $O/T_{+}/C_{-}T_{+}$ to the unstable
regime of states $C_{+}T_{+}/T_{+}/C_{-}T_{+}$ (see text).
}
\end{figure}

In detail, our experimental process starts with the application of the
transport current along the y-axis (see left pane of Fig.~\ref{Figure_5_2})
which produce a T$_{+}$ zone, i.e., the starting profile can be depicted as
follows:
\begin{eqnarray}\label{Eq.5.5}
dh= dz \quad ; \quad d\alpha = 0 
\nonumber\\
{\Downarrow}\qquad\qquad
\\
h=h_{x0}+{\tt z}-1\quad ; \quad \alpha=0 \, ,
\nonumber
\end{eqnarray}
that penetrates from the surface until the point where $h$ equals $0$, i.e.:
${\tt z}_{p0}=1-I_{tr}$. In our units, ${\tt z}_{p0}=0.5$ for
$I_{tr}=h_{x0}=0.5$. 

On the one hand, an O zone appears in the inner region
$0<{\tt z}<{\tt z}_{p0}$ as far as $I_{tr}<1$.  On the other hand, unless an
external source of magnetic field is switched on,
the above situation remains valid.  
Thus, upon increasing the external field
$h_{y0}$, a flux line rotation starts on the surface and the
perturbation propagates towards the center in the form of a C$_{-}$T$_{+}$ zone
defined by 
\begin{eqnarray}\label{Eq.5.6}
dh= dz \quad ; \quad d\alpha = -\frac{\chi}{h}\, dz\qquad\qquad
\nonumber\\
{\Downarrow}\qquad\qquad\qquad\qquad\qquad
\\
h=h_{x0}+{\tt z}-1\; ; \; \alpha=\alpha_{0}+\chi{\rm ln}\left[1+\frac{{\tt
z}-1}{h_{0}}\right] \,
,
\nonumber
\end{eqnarray}
which covers the range ${\tt z}_{c}^{-}<{\tt z}<1$, defined by $\alpha=0$
$\Rightarrow$ ${\tt z}_{c}^{-}$ $=$ $1+h_{0}$ $[{\rm exp}(-\alpha_{0}/\chi)-1]$
(see Fig.~\ref{Figure_5_2}). The former T$_{+}$ zone is pushed towards the
center and occupies the interval ${\tt z}_{p}^{-}<{\tt z}<{\tt z}_{c}^{-}$ with
${\tt z}_{p}^{-}=1-h_{0}$. Finally, an O-zone fills the core $0<{\tt z}<{\tt
z}_{p}^{-}$ until the condition ${\tt
z}_{p}^{-}=0\Leftrightarrow h_{0}=1$ is reached. Then, with further increase of
the local component $h_{y0}$, an instability towards the center of
the sample consisting of a transient structure that becomes
$C_{+}T_{+}/C_{-}T_{+}$ is induced.

In physical terms, flux vortices penetrate from the surface with some
orientation given by the components of  the vector $(h_{x},h_{y})$. Owing to the
critical condition for the penetration of the field $dh/dz = 1$, as soon as the
modulus reaches the center, flux rotation must take place there. This is needed
for accommodating the vector to the condition ${\bf h}({\tt z}=0)=(0,h_{y}({\tt
z}=0))$.
On the other hand, as the angle variation is determined by the value of
$J_{c\parallel}$, a jump is induced at the center, i.e.: $\alpha({\tt
z}=0)\to\pi /2$, and the related instability may be visualized by a critical
$C_{+}T_{+}/T_{+}/C_{-}T_{+}$ profile (dashed lines at the right pane of
Fig.~\ref{Figure_5_2}) in which the field angle decreases from its surface value
$\alpha_{0}$ to $0$ in the $C_{-}T_{+}$ region, then keeps null within the
$T_{+}$ zone, and suddenly increases to the value $\pi/2$ in the inner
$C_{+}T_{+}$ band defined by
\begin{eqnarray}\label{Eq.5.7}
dh= dz \quad ; \quad d\alpha = \frac{\chi}{h}\, dz\qquad\qquad\quad
\nonumber\\
{\Downarrow}\qquad\qquad\qquad\qquad\qquad
\\
h=h_{x0}+{\tt z}-1\; ; \;  \alpha=\pi/2-\chi{\rm ln}\left[1+\frac{{\tt
z}}{h_{0}-1}\right]\, .
\nonumber
\end{eqnarray}
In fact, the $C_{+}T_{+}/C_{-}T_{+}$ structure is stabilized with the
intersection between regions at the point [$\alpha^{{+},{+}}({\tt z}_{\rm
v})=\alpha^{{-},{+}}({\tt z}_{\rm v})$] given by 
\begin{eqnarray}\label{Eq.5.8}
{\tt z}_{{\rm v}}=1-h_{0}+\sqrt{h_{0}(h_{0}-1)}~{\rm
exp}\left[\frac{\pi/2-\alpha_{0}}{\chi}\right] \, .
\end{eqnarray}
Finally, note that upon further increasing $h_{y0}$ the point ${\tt z}_{\rm v}$
follows the rule ${\tt z}_{\rm v}(h_{y0}\to\infty)\to (1+h_{x0}/\chi)/2$

In brief, our simple analytical model allows to identify the following physical
phenomena as the longitudinal component of the magnetic field $h_{y0}$ 
(parallel to the flow direction of the transport current) is increased:
\begin{enumerate}
\item The appearance of a surface layer with negative transport current
density (mind the slope of $h_{x}$ in Fig.~\ref{Figure_5_2} in view of
Eq.~(\ref{Eq.5.1})).
\item The applied magnetic field {\em re-entry} as
related to the inner $C_{+}T_{+}$ zone, predicting the occurrence of local
paramagnetic states near of the center of the superconducting sample.
\end{enumerate}

These features will be confirmed along the forthcoming section, where the SDCST
statements are thoroughly presented and a further comparison with the
analytical model of Ref.~\cite{P2-Brandt_2007} will be displayed.

\vspace*{1.0cm}
\hspace*{-0.3cm}
\textbf{\textit{5.1.2 The SDCST statement and the BM's approach}}
\markboth{\hspace*{0.1cm} \textsl{5.1.2 The SDCST statement and the BM's
approach}}{\hspace*{0.1cm}
\textsl{5.1.2 The SDCST statement and the BM's approach}}
\addcontentsline{toc}{chapter}{\hspace*{0.1cm} \textsl{5.1.2 The SDCST statement
and the BM's approach}}\label{ch-5-1-2}
\vspace*{1.0cm}

\begin{figure}[t]
\centering
\includegraphics[height=7cm,width=13cm]{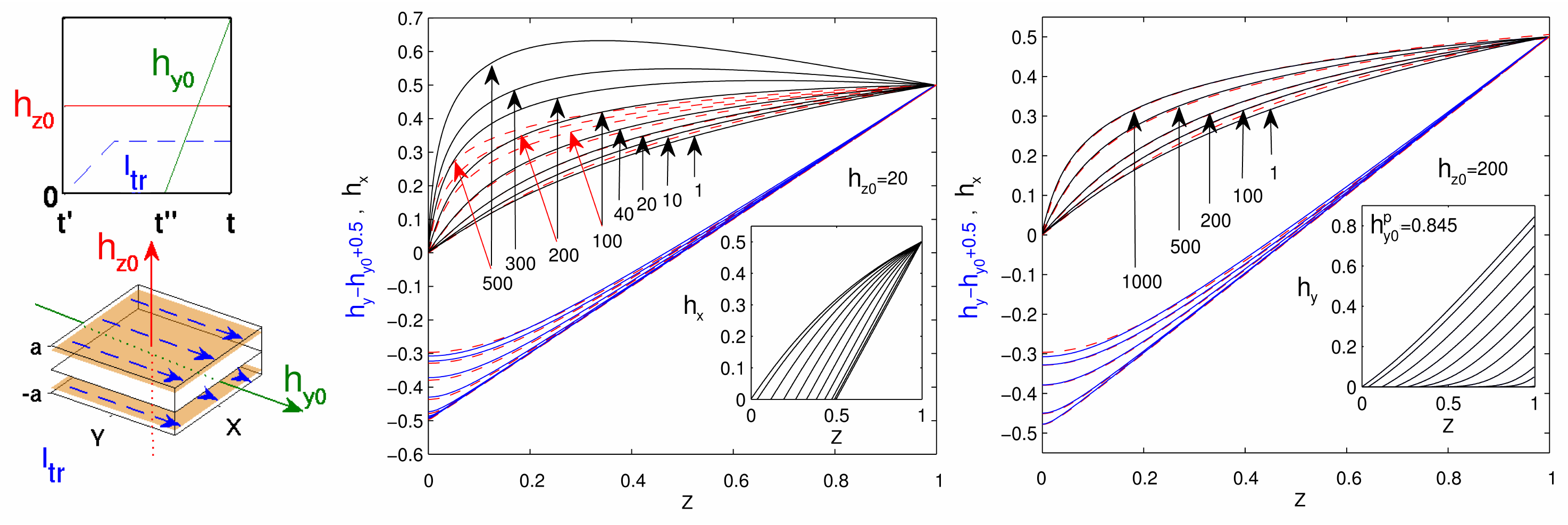}
\caption{\label{Figure_5_3} \textit{Left pane:} Schematic representation of the
3D experimental configuration of the longitudinal transport problem in the slab
geometry. \textit{Central and right panes:} Profiles of the magnetic field
components $h_x(z)$ and $h_y(z)$ for the longitudinal problem corresponding to a
transport current along the $y$ axis of value $I_{\rm
transport}\equiv J_{c\perp}a/2=0.5$
and at several increasing values of the magnetic field $h_{y0}$ as labeled in
the curves. Here, we have assumed an uniform perpendicular field $h_{z0}=20$
(central pane) and then $h_{z0}=200$ (right pane). The plot shows the comparison
of the full range numerical solution in the infinite widthband model or
so-called T-state model (continuous lines) to the analytical approximation in
Eq.~(\ref{Eq.5.15}) (dashed). The insets show the initial flux penetration
profiles for both components of the magnetic field.
}
\end{figure}

Below, let us introduce an experimental configuration rather to the opposite
side of the previous situation, and that may be used for comparison to the work
in Ref.~\cite{P2-Brandt_2007}. To be specific, based upon our well-known 3D
physical scenario for
superconductors with magnetic anisotropy, we shall consider the time evolution
of magnetic profiles $\textbf{h}_{l+1}(z)$ within an infinite superconducting
slab of thickness $2a$ (see Fig.~\ref{Figure_5_3}), cooled under the assumption
of an initial state defined by an uniform vortex lattice perpendicular to the
external surfaces, i.e., a constant magnetic field $h_{z0}$. In terms of the
geometrical interpretation for the material law, the above starting point or
initial state corresponds to choose $\theta=0$ in Fig.~\ref{Figure_4_1}
(pag.~\pageref{Figure_4_1}). Then, a transport current is
injected along the superconducting slab in the direction of the $y-$axis
inducing a rotation of the critical current region $\Delta_{\textbf{r}}$ on the
plane of currents $xy$ by means the induced magnetic field component $h_{x}$
(i.e., $\theta\neq0$ in Fig.~\ref{Figure_4_1}).
Finally, a third source of magnetic
field is switched on along the $y-$axis, which induces a new rotation of the
current density vector by means the introduced local component of magnetic field
$h_{y}$ (i.e., $\alpha\neq0$ in Fig.~\ref{Figure_4_1}). 

It is worth mentioning that, by symmetry, the current density is
confined to the $xy$-plane, such that the distribution of the current can be
displayed in a proper set of circuits naturally defined by a collection of
in-plane current layers located at the heights $z_{i}$, and each one carrying a
current density
given by $[J_{x}(z_i),J_{y}(z_i)]$ which must satisfy a certain set of
physical constraints. In particular, this means that in practice one should
impose the restriction that ${\bf J}$ belongs to the projection of the critical
current region onto the plane (${\bf J}\in\Delta_p$) and that
$\nabla\cdot\textbf{J}\equiv 0$. As a main fact it will be established
that, when building the parallel configuration, the response of the
superconductor depends on the limitations for the current density established by
the depinning threshold $J_{c\perp}$ on the orientation of the local magnetic
field, and eventually on the threshold value for the cutting component or
$J_{c\parallel}$. This is easily understood at a qualitative level
just by glancing once more the left-side of Fig.~\ref{Figure_4_1}
(pag.~\pageref{Figure_4_1}). Within this
picture, with transport current density flowing along the
$xy$-plane, notice that for moderate values of the
angle $\theta$ between the local magnetic field and the $z$-axis, the critical
current restriction or material law $\Delta_{\rm p}$ becomes an ellipse of
semi-axes $J_{c\perp}$ and $J_{c\parallel}^{\rm p}$ with
\begin{eqnarray}\label{Eq.5.9}
J_{c\parallel}^{\rm
p}={J_{c\perp}}/{\cos{\theta}}=J_{c\perp}{\sqrt{H_{x}^{2}+H_{y}^{2}+H_{z}^{2}}}/
{H_{z}} \, .
\end{eqnarray}
An increase of the in-plane magnetic field component will result in a tilt of
the cylinder by an increase of the angle $\theta$. In particular notice that,
initially the maximum value of the in-plane parallel current density,
$J_{c\parallel}^{\rm p}$, grows with the angle $\theta$,  independent of
$J_{c\parallel}$ (which is, thus, absent from the theory) until the maximum
value $\sqrt{J_{c\perp}^{2}+J_{c\parallel}^{2}}$ is reached.  Then, the ellipse
is truncated and eventually would be practically a rectangle of size
$2J_{c\parallel}\times 2J_{c\perp}$ when $\gamma \to\pi/2$. Outstandingly, for
large values of $\chi\equiv J_{c\parallel}/J_{c\perp}$ (long cylinders), the
critical current along the parallel axis $J_{c\parallel}^{\rm p}$ increases
more and more as the weight of $H_{z0}$ decreases, and furthermore, this
quantity is always beyond the individual values $J_{c\perp}$ and
$J_{c\parallel}$.

Recall that the variational statement for three dimensional configurations on
the slab geometry has been thoroughly discussed in chapter~\ref{ch-4-1}.1.
Thus, based on the numerical resolution for discretized layers\footnotemark[2],
and within the mutual inductance formulation [Eq.~(\ref{Eq.3.3})], the above
described longitudinal problem takes the following form:
\begin{eqnarray}\label{Eq.5.10} 
{\cal
F}[I_{l+1}]=&&~\frac{1}{2}\sum_{i,j}I_{i,l+1}^{x}{\rm
M}_{ij}^{x}I_{j,l+1}^{x}-\sum_{i,j}
{\rm I}_{i,l}^{x}{\rm M}_{ij}^{x}I_{j,l+1}^{x}\nonumber \\
&&+\frac{1}{2}\sum_{i,j}I_{i,l+1}^{y}{\rm M}_{ij}^{y}I_{j,l+1}^{y}
-\sum_{i,j} {\rm I}_{i,l}^{y}{\rm M}_{ij}^{y}I_{j,l+1}^{y}\nonumber \\
&&+\sum_{i}I_{i,l+1}^{y}(i-1/2)({\rm h}_{0,l+1}^{y}-{\rm h}_{0,l}^{y}) \, .
\end{eqnarray}

\footnotetext[2]{In particular, readers
have to recall that for numerical purposes, the slab has been discretized in
terms of $N_{s}$ layers of equal thickness $\delta$ ($z_{i}=\delta\, i\;,\;
\delta\equiv a/N_{s}$), each one characterized by the current intensities,
$I_{i,l+1}^{x}\equiv \delta j_{x}(z_{i},t=l+1)$ and
$I_{i,l+1}^{y}\equiv \delta j_{y}(z_{i},t=l+1)$. 
}

Moreover, we already know that the local components of the magnetic fields have
to be evaluated according to equations of the kind
(\ref{Eq.4.1})~\&~(\ref{Eq.4.2}), and the
parallel and perpendicular projections of the sheet current components are given
by 
\begin{eqnarray}\label{Eq.5.11}
I_{\perp}^{2}=\left(1-(h_{i}^{x})^{2}\right)
(I_{i}^{x})^{2}+\left(1-(h_{i}^{y})^{2}\right)(I_{i}^{y})^{2}-2h_{i}^{x}h_{i}^{y
} I_{i}^{x}I_{i}^{y}  \, ,
\end{eqnarray}
and
\begin{eqnarray}\label{Eq.5.12}
I_{\parallel}^{2}=(h_{i}^{x} I_{i}^{x})^{2} + (h_{i}^{y} I_{i}^{y})^{2}+
2h_{i}^{x}h_{i}^{y}I_{i}^{x} I_{i}^{y} \, . 
\end{eqnarray}

Technically, the main differences with the problems considered in the
above chapter are the inclusion of an additional constraint for the transport
current, and new expressions for the inductance matrices. 

On the one hand, for
our problem with transport current, one has to consider for each temporal step
the external constraint
\begin{eqnarray}\label{Eq.5.13}
\sum_{i}I_{i}^{y}(t)=I_{tr}(t) \, .
\end{eqnarray}
On the other hand, as related to the symmetry properties for the transport
configuration [$I_{i}^{y}(z)=I_{i}^{y}(-z)$ as
opposed to the antisymmetry for the case of shielding currents], here one has
to use the mutual inductance expressions
\begin{eqnarray}\label{Eq.5.14}
M_{i,j}^{x}\equiv 1+2\left[min\{i,j\} \right]\,\qquad\, , \qquad
M_{i,i}^{x}\equiv 2\left(\frac{1}{4}+i-1 \right)\nonumber \, , \\ \, \\
M_{i,j}^{y}\equiv 1+2\left[N_{s}-max\{i,j\} \right]\,\qquad\, , \qquad
M_{i,i}^{y}\equiv 2\left(\frac{1}{4}+N_{s}-i\right) \, \nonumber.
\end{eqnarray}
with $N_{s}$ the full number of layers in the discretized slab. 

Finally, within the framework of the SDCST, it is to be recalled that
the local components of the current density have to be constrained according to
one of the models contained by our generalized material law depicted in the
right pane of Fig.~\ref{Figure_4_1} (pag.~\pageref{Figure_4_1}) [Mathematically
see Eq.~(\ref{Eq.2.17}), pag.~\pageref{Eq.2.17}].
Thus, with the aim of achieving a comparison between our numerical results and
the analytical approach of Ref.~\cite{P2-Brandt_2007}, in Fig.~\ref{Figure_5_3}
we show some of the results for the T-states assumption under consideration of a
strong perpendicular field $h_{z0}=200$, and the arising discrepancies when a
perpendicular field of moderate intensity (e.g., $h_{z0}=20$) is considered.

Firstly, let us emphasize that within our SDCST there is no restriction for
the ratios $\chi^{-1}\equiv
J_{c\perp}/J_{c\parallel}$ and $\varsigma\equiv J_{c\perp}a/H_{z0}$, which
become small parameters within the analytical approach of
Ref.~\cite{P2-Brandt_2007}. Notice that the smallness of
$\chi^{-1}$ means that the arising critical state is approximated by the
unbounded band region $|j_{\perp}|=1, 0 < |j_{\parallel}|
<\infty$ described in the chapter 4.6 (T-states). The smallness of $\varsigma$
was meant to indicate a small deviation of the full magnetic field respect to
the $z-axis$. Then, moderate values of $j_{\parallel}$ are expected.
Notoriously, the above hypotheses of Ref.~\cite{P2-Brandt_2007} allow to state
the problem by a set of approximate analytic formulas for the electromagnetic
quantities which in turn allow to bypass the numerical solution of the
differential equations. However, as it will be shown below, the range of
application is narrower than expected.

Figure~\ref{Figure_5_3} shows the comparison between the penetration profiles
for the local magnetic field components $h_{x}$ and $h_{y}$ obtained from our
SDCST with $\chi\rightarrow\infty$ and $j_{c\perp}=1$, and from the analytic
expressions in Ref.~\cite{P2-Brandt_2007}, i.e., 
\begin{eqnarray}\label{Eq.5.15}
h_{x}&=&\frac{\alpha}{{\rm cos}\,\theta}{\rm
arcsinh}\left(\frac{z}{\alpha}\right)
\nonumber\\
h_{y}&=&h_{y0}-\alpha\left(\sqrt{1+\frac{1}{\alpha^{2}}}-\sqrt{1+\frac{z^2}{
\alpha^2}}\right)
\nonumber\\
{\rm cos}\,\theta &=& 2\alpha\,{\rm arcsinh}\left(\frac{1}{\alpha}\right) \, .
\end{eqnarray}
Here, $\alpha$ has to be obtained for each value of the applied field from the
condition $cos\theta=h_{z0}/\sqrt{h_{z0}^{2}+h_{y0}^{2}}$. One can notice that
the agreement is rather good for the higher value of the perpendicular field
$h_{z0}=200$ (right pane of Fig.~\ref{Figure_5_3}), whereas remarkable
differences appear for a moderate field $h_{z0}=20$ as $h_{y0}$ increases
(central pane of Fig.~\ref{Figure_5_3}). Our
interpretation of the above facts is as follows. 

On the one hand, as regards to the establishment of the full penetration
profile, we have straightforwardly obtained this condition through the
step-by-step integration starting from the state $h_{y0}=0$ (the evolution is
shown in the insets of the Fig.~\ref{Figure_5_3}). Whereas the value 0.796 is
estimated for the penetration field $h_{y0}^{p}$ within the analytical approach
of Ref.~\cite{P2-Brandt_2007}, by the straightforward method described above we
get $h_{y0}^{p}=0.845$. Remarkably, in spite of some small differences for the
low field profiles $h_{x}$ and $h_{y}$, at moderate values of the transverse
field ($h_{y0}<h_{z0}$) the curves always coincide. On the other
hand, the failure of the analytical approximation for the higher
values of $h_{y0}$ is readily explained by the observation of the
plot. Thus, increasing $h_{y0}$ compresses the transport current
towards the center of the sample (as indicated by the slope of
$h_{x}(z)$). For the case of $h_{z0}=20$ (central pane of
Fig.~\ref{Figure_5_3}), one gets $j_{y,max}\approx 5$
when $h_{y}\approx 100$ and $j_{y,max}\approx 50$
when $h_{y}\approx 1000$, then a considerable value of $j_{\parallel}$ is
obtained. This leads to a not so good approximation from the analytic condition
in the approximation of Ref.~\cite{P2-Brandt_2007}, which one is only
valid for small values of this quantity. However, when comparison is made
for $h_{z0}=200$, one gets $j_{y,max}\approx 1$
when $h_{y}\approx 100$ and $j_{y,max}\approx 5$
when $h_{y}\approx 1000$. Then, a much better performance is obtained for the
analytical limit even for very high applied fields $h_{y0}$.

In brief, from the above discussion we may conclude that our SDCST overcomes
previous limitations related to the {\em weak longitudinal current}
conditions. By contrast, in the
following section we will show a wide set of numerical calculations for material
laws even most complicated than the simplest T-state model here considered,
allowing to display the corrections needed in the general critical states.

\break
\vspace*{0.0cm}
\subsection*{\label{ch-5-2}
\hspace*{0.1cm} \textsl{5.2 Magnetic anisotropy and the uncommon
effects}}
\vspace*{1cm}
\markboth{\hspace*{0.1cm}\textsl{5.2 Magnetic anisotropy and the uncommon
effects}} 
{\hspace*{0.1cm} \textsl{5.2 Magnetic anisotropy and the uncommon
effects}}
\addcontentsline{toc}{chapter}
{\hspace*{0.1cm} \textsl{5.2 Magnetic anisotropy and the uncommon
effects}}

This section will be devoted to unveil the features of longitudinal
transport problems under general critical state conditions, and to identify the
influence of a number of physical parameters along the different stages of the
magnetization process. Here, for the experimental configuration depicted in the
left pane of Fig.~\ref{Figure_5_3} and based on the numerical resolution of the
variational statement, a complete tour along the whole set of values for the
perpendicular field will be presented. We shall concentrate on
the effect of the flux cutting boundary ($j_{c ||}$) considering several
conditions for the material law introduced by our SDCST (see left pane at
Fig.~\ref{Figure_4_1}, pag.~\pageref{Figure_4_1}).  Firstly, the extreme case
$\chi\to\infty$ or infinite bandwidth model (T-states model) will be considered
(Subchapter~5.2.1). Secondly, several anisotropic models characterized by
the \textit{superelliptic} relation [Eq.~(\ref{Eq.2.17}),
pag.~\pageref{Eq.2.17}]
\begin{eqnarray}\label{Eq.5.16}
j_{\perp}^{2n}+(j_{\parallel}/\chi)^{2n}\leq1
\end{eqnarray}
will be thoroughly analyzed (Subchapter~5.2.2). In particular, we will set of
material laws defined by the parameters $\chi=1$, 2, 3, and 4 with the
\textit{smoothing}
index n=4.

Remarkably, our procedure will reveal
the fingerprints of the cutting and depinning mechanisms, thus being a
theoretical pathway for the reconstruction of the material law, represented by
the proper region $\Delta_{\textbf{r}}$.

\vspace*{0.5cm}
\hspace*{-0.3cm}
\textbf{\textit{5.2.1 Extremal case: The T-states model}}
\markboth{\hspace*{0.1cm} \textsl{5.2.1 Extremal case: The T-states
model}}{\hspace*{0.1cm}
\textsl{5.2.1 Extremal case: The T-states model}}
\addcontentsline{toc}{chapter}{\hspace*{0.1cm} \textsl{5.2.1 Extremal case: The
T-states model}}\label{ch-5-2-1}
\vspace*{0.5cm}

The T-state model for three dimensional configurations of the applied magnetic
field has been exhaustively studied in chapter 4.3. Here, although the depicted
scenario is similar, the assumption of a longitudinal transport current allow a
straightforward understanding of the physical scopes of this model beyond the
analytical approximations. First, we will analyze the properties of the
local field [$\textbf{h}(z)$] and the current density profiles
[$\textbf{j}(z)$] for a longitudinal configuration built in the fashion
described in the left pane of Fig.~\ref{Figure_5_3} (a third component of the
magnetic field $h_{z0}$ is incorporated).

\begin{figure}[t]
\centering
\includegraphics[height=8.59cm,width=13cm]{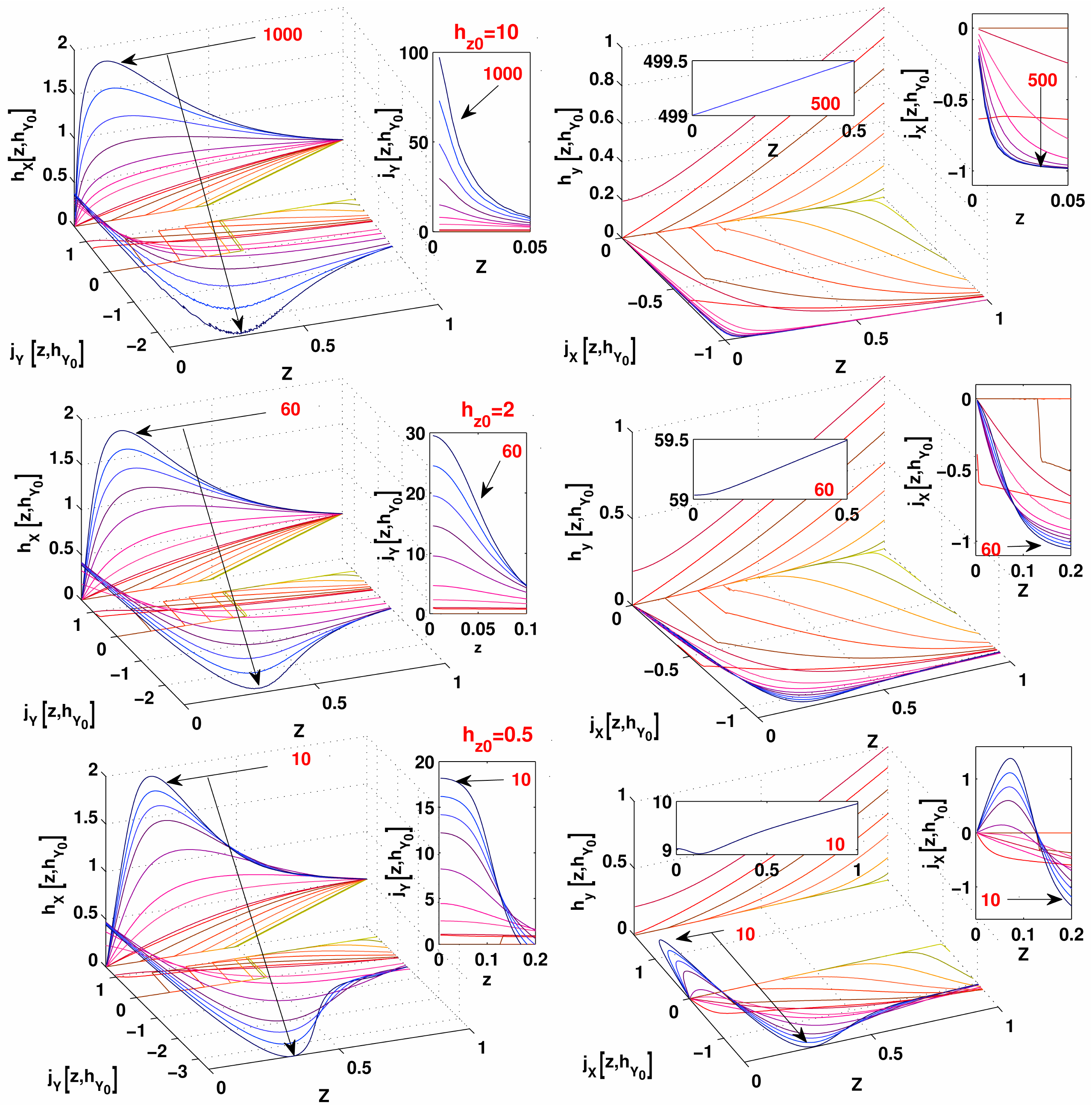}
\caption{\label{Figure_5_4} Profiles of the local magnetic field components
$h_{x}[z,h_{y0}]$ and $h_{y}[z,h_{y0}]$, and the corresponding
current-density profiles $j_{y}[z,h_{y0}]$ and  $j_{x}[z,h_{y0}]$ for the
\textit{T}-state model and perpendicular magnetic field components $h_{z0}=10$
(top), $h_{z0}=2$ (middle) and $h_{z0}=0.5$ (bottom). The different curves
correspond to the following sets of values for the applied longitudinal field at
the surface: (i) top row: $h_{y0}=0.005, 0.050, 0.170, 0.340, 0.500, 0.680,
0.845,1.0, 40.0,80.0, 150.0, 300.0, 500.0, 750.0$, $1000$, (ii) middle row:
$h_{y}(a)=0.0050.050,0.170,0.340,0.500,0.680,0.845,1.0,5.0,10.0$,
$20.0,30.0,40.0,50.0,60.0$, (iii) bottom row: $h_{y}(a)=0.001,0.050,
0.170,0.340,0.500$, $0.680, 0.845, 1.0, 2.0, 3.0,5.0,7.0, 8.0, 9.0, 10.0$.
Insets to the middle pane correspond to the $j_{y}[z,h_{y0}]$ profiles
closer to the center of the sample. Analogously, insets to the right correspond
to the $j_{x}[z,h_{y0}]$ profiles. Finally, inner insets at right pane
correspond to a specific profile of $h_{y}[z,h_{y0}]$ so as to highlight the
occurrence of magnetic field reentry at low values of $h_{z0}$.
}
\end{figure}

Fig.~\ref{Figure_5_4} shows the behavior of the magnetic field profiles and the
induced currents subsequent to the application of the transport current for
three different initial conditions: $h_{z0}=10$, $h_{z0}=2$, and
$h_{z0}=0.5$, all of them under assumption of the T-state model. The initial
state for the transport current condition ($I_{tr}=J_{c \perp}a/2$) establishes
the initial transport profile $j_{y}\{0\leq z<a/2\}=0$ and $j_{y}\{a/2\leq z\leq
a\}=1$. As the transport current is no longer modified, the condition
$h_{x}(a)=0.5$ can be applied in what follows. On the other hand, by symmetry,
one has the condition $h_{x}(0)=0$ at the center of the slab.

\begin{figure}[t]
\centering
\includegraphics[height=10.5cm,width=12cm]{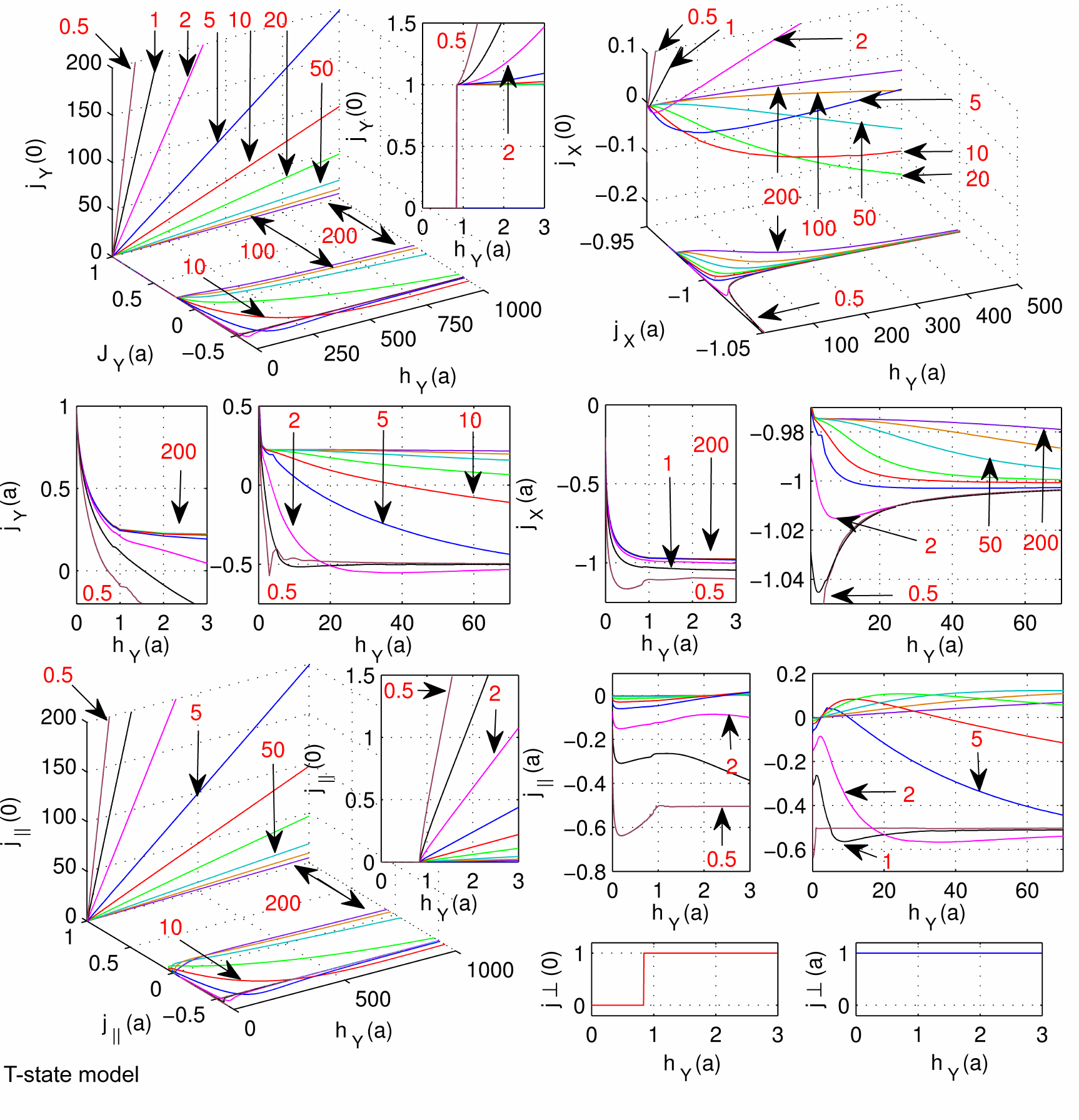}
\caption{\label{Figure_5_5} Dynamics of the local current density as a function
of the applied longitudinal magnetic field $h_{y0}=h_{y}(a)$ along the central
and external sheets of the slab. The results are shown for the T-state model
($J_{c||}\to\infty ~\&~ J_{c\perp}=1.0$). Top: the components $j_{y}$ and
$j_{x}$ at $(z=0)$ and $(z=a)$. Middle: details of the above behavior. Bottom:
dynamics of the parallel and perpendicular components of $\bf j$ in the same
conditions as above. The different curves correspond to the values of the
perpendicular magnetic field given by $h_{z0}= 0.5, 1, 2, 5, 10, 20, 50, 100,
200$. Several scales have been displayed to avoid information loss with all
plots
running over the same color scale.
}
\end{figure}

In detail, when the external magnetic field $[h_{y0}=h_{y}(a)]$ (the applied
parallel field) is linearly increased from
$h_{y}(a)=0$, a current density $j_{x}$ is induced from the superconducting
surface as an effect of Faraday's law.  Simultaneously, the local component
of the magnetic field $h_{x}(z)$ increases monotonically following two
continuous stages fulfilling the aforementioned boundary conditions. First, the
superconducting sample is fully penetrated by the transport current when
$h_{y}^{\star}(a)=0.845\pm0.003$ and eventually, the condition $j_{y}(0)=1$ is
reached as soon as $h_{y}(a)\rightarrow 0.860$. 
We notice that the value of $h_{y}^{\star}(a)$ for the full penetration profile
is basically independent of $h_{z0}$ (at least to the numerical precision of our
numerical calculations\footnotemark[3]), in agreement
with our analytical solution~\cite{P2-Ruiz_PRB_2011} introduced in the previous
section (Chapter~5.1.1). Second, a remarkable enhancement of the transport
current density occurs around the center of the slab as $h_{y}(a)$ increases
beyond $h_{y}^{\star}(a)$. Furthermore, an eventual negative current density
appears shielding the positive transport current around the center of the slab.
It is to be noticed that the appearance of negative current flow is enhanced
when the magnetic component $h_{z0}$ is decreased (Fig.~\ref{Figure_5_5}). 

\footnotetext[3]{The optimization ends only when the following conditions are
achieved: (i) The norm of the constraint functions, if any, is smaller than
1E-6. (ii) The norm of the projected gradient of the merit function (the
objective function in the bound-constrained case and the augmented Lagrangian
function when there are more general constraints present) is smaller that 1E-7.}

It is noteworthy that for the range of values
$h_{z}(0)<h_{y}^{\star}(a)$ negative surface current appears even for the
partial penetration regime, e.g., for $h_{z0}=0.5$ one has $j_{y}(a)<0$ for
$h_{y}(a)>0.722$. Further, another outstanding property is the occurrence of
profiles with magnetic field reentry (paramagnetism in the
component $h_{y}$ around the center of the slab) for $h_{z0}\lesssim 1$ and 
under relatively low applied magnetic fields $h_{y}(a)$ (see
Fig.~\ref{Figure_5_4}). In fact, we call the readers' attention that
the above mentioned effects, local paramagnetism and negative current zones,
have both been shown analytically in the limiting case $h_{z0}=0$ (Chapter
5.1.1). Along this line, as a general rule,
we can conclude that the smaller the value of $h_{z0}$, the sooner the surface
of negative transport current and even paramagnetic local effects appear. 

On the other hand, figure~\ref{Figure_5_5} displays the evolution of the
current density vector focusing us on the specific values at the
superconducting surface ($z=a$) and at the center of
the superconducting slab ($z=0$) as the longitudinal magnetic field $h_{y}(a)$
is increased. Outstandingly, the involved physical features are
straightforwardly explained by the polar decomposition of the current density
introduced in chapter~4.1 [See Eqs.~(\ref{Eq.4.6})~-~(\ref{Eq.4.8}),
pag.~\pageref{Eq.4.6}]. In brief, notice that the unbounded parallel current
density allows unconstrained rotations for the flux lines as the applied
magnetic field increases. In particular, this leads to negative values of
$j_{y}(a)$ (slope of $h_{x}(a)$), simultaneous to high $j_{y}(0)$ (slope of
$h_{x}(0)$). Moreover, it should be noticed that negative values of the
transport current are favored by smaller and
smaller values of the field component perpendicular to the surface of the sample
$h_{z0}$. Also, notice that at the center of slab the
flux line dynamics is mainly governed by the longitudinal transport current
density $j_{y}(0)$. The basic idea is that for moderate values of $h_z$, when
$h_{y}$ increases $j_{y}$ practically becomes $j_{\parallel}$. As this component
is unconstrained, it grows indefinitely at the center.

For a closer connection with real experiments, below let's concentrate on
the magnetostatic properties by means of the
\textit{global} sample's magnetization curve \textbf{M}(\textbf{H}). Thus, we
have calculated \textbf{M} as a function of the longitudinal magnetic field
$h_{y}(a)$. 

\begin{figure}[t]
\centering
\includegraphics[height=8cm,width=10cm]{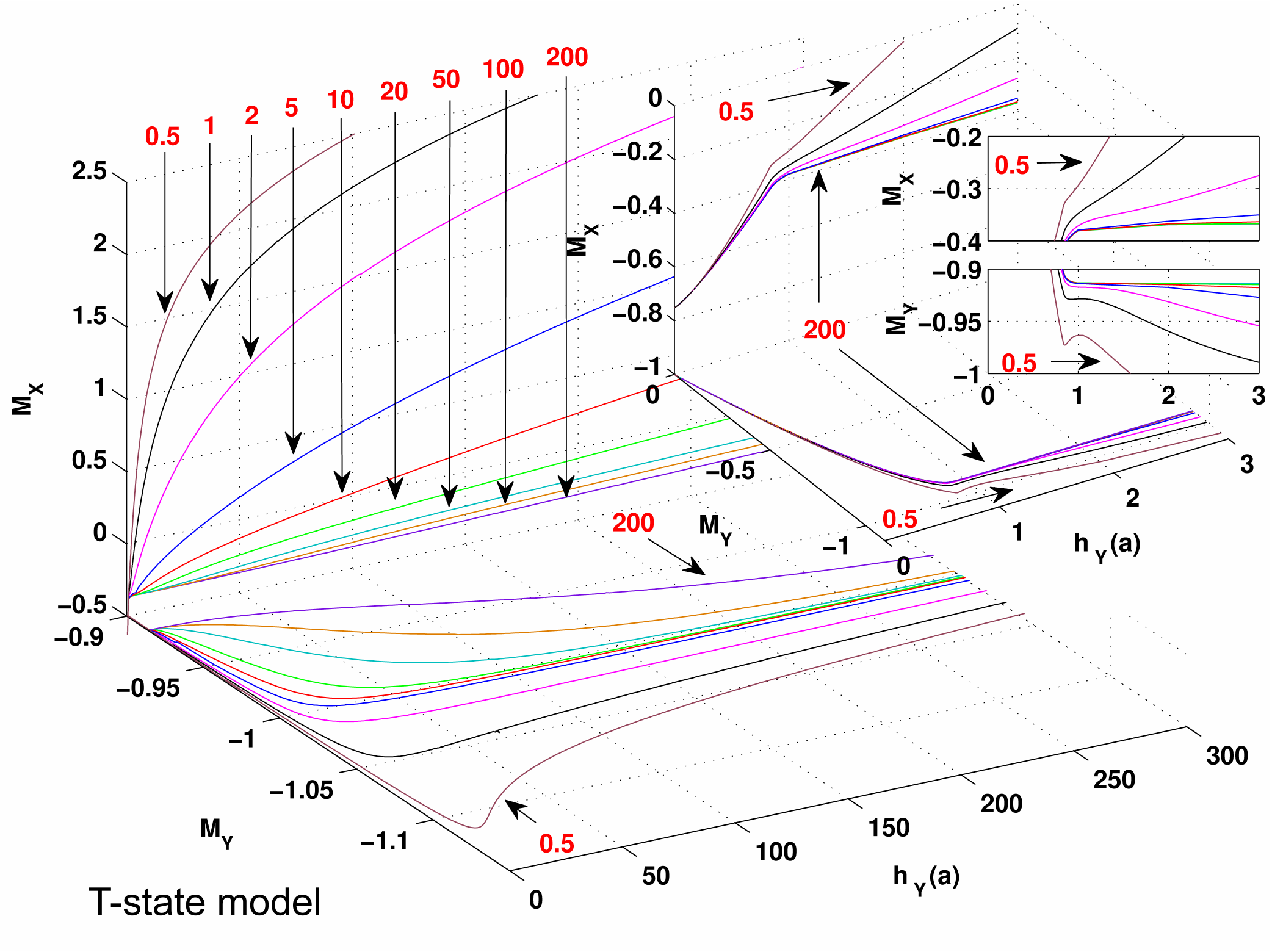}
\caption{\label{Figure_5_6} The magnetic moment components $(M_{x},M_{y})$ of
the slab as a function of the applied magnetic field component $h_{y}(a)$ for
the T-state model. The curves are labeled according to the perpendicular
magnetic field $h_{z0}= 0.5, 1, 2, 5, 10, 20, 50, 100, 200$. The insets
show different zooms of the magnetic moment components for low values of the
applied longitudinal field. The same color scale applies to all the plots.
}
\end{figure}

Fig.~\ref{Figure_5_6} displays the magnetic moment components
$M_{x}(h_{y0})$ and $M_{y}(h_{y0})$ in units of $J_{c\perp}a^{2}$. First, notice
that within the partial penetration regime ($h_{y}(a)\leqslant
h_{y}^{\star}(a)$) the magnetic moment components are almost independent of the
perpendicular magnetic field $h_{z0}$ (at least for non small values of this
quantity). On the contrary, when $h_{z0}<1$ and the patterns of negative current
even occur before the fully penetrated state, the magnetization slightly
increases. This is accompanied by faint field {\em reentry} effects that are
also shown within the figure. Furthermore, as the threshold cutting current
density $j_{c\parallel}$ is unbounded for the T-state model, the magnetic moment
$M_{x}$ always increases as related to the diverging behavior of $j_{y}(0)$.

Let us emphasize that, as the unbounded behavior for the parallel current
density assumed above leads to the prediction of arbitrarily high values of the
transport current density, consequently the T-state model must be physically
reconsidered. For example, on the one hand, the trend of
the magnetic moment $M_{x}$ and also the unbounded longitudinal current density
$j_{y}$ disagree with the experimental evidences recollected in
Refs.~\cite{P2-Voloshin_2001,P2-LeBlanc_2003,P2-LeBlanc_2002,P2-Esaki_1976,
P2-LeBlanc_1966,P2-Gauthier_1974,P2-Karasik_1970,P2-Sugahara_1970,
P2-Taylor_1967,P2-Sekula_1963}. On the
other hand, in Fig.~\ref{Figure_5_5} one can observe that, as soon as the flow
of
negative current along the superconducting surface is reached, it never
disappears notwithstanding the longitudinal magnetic field remains increasing.
By contrast, the disappearance of the patterns of superficial negative current
were detected in
Refs.~\cite{P2-Matsushita_1984,P2-Matsushita_1998,P2-Matsushita_2012,
P2-Voloshin_1991,P2-Esaki_1976,P2-LeBlanc_1966}. These observations have lead to
consider
$j_{c\parallel}$~{\em bounded} descriptions as satisfactory solutions of the
peculiar phenomena involved on the longitudinal transport current
problem~\cite{P2-LeBlanc_2003,P2-LeBlanc_2002,P2-Matsushita_1984,
P2-Matsushita_1998,P2-Matsushita_2012}. Rather
recent experimental data on high temperature
superconductors~\cite{P2-Clem_2011_SUST,P2-Campbell_2011_SUST}
also indicate that physical bounds are to be considered for both components of
the critical current.

Thus, as will be described below in section 5.2.2, our generalized SDCST suits
the necessity of dealing with a physically acceptable description of both local
and global issues about the electromagnetic quantities involved in the
longitudinal transport current
problem~\cite{P2-Ruiz_PRB_2009,P2-Ruiz_PRB_2011,P2-Ruiz_SUST_2011}.

\vspace*{0.5cm}
\hspace*{-0.3cm}
\textbf{\textit{5.2.2 Material laws with magnetic
anisotropy: CT$\chi$~--~models}}
\markboth{\hspace*{0.1cm} \textsl{5.2.2 Material laws with magnetic
anisotropy: CT$\chi$~--~models}}{\hspace*{0.1cm}
\textsl{5.2.2 Material laws with magnetic
anisotropy: CT$\chi$~--~models}}
\addcontentsline{toc}{chapter}{\hspace*{0.1cm} \textsl{5.2.2 Material laws with
magnetic
anisotropy: CT$\chi$~--~models}}\label{ch-5-2-2}
\vspace*{0.5cm}

More realistic models for the material law are presented below. Henceforth, we
shall use the simplified notation T or CT$\chi$ as regards to the
infinite bandwidth model (T by transport) or the \textit{superelliptical}
critical state models with anisotropy $\chi=|J_{c ||}/J_{c \perp}|$ (CT by
cutting and transport, and $\chi$ the index controlling the magnetic anisotropy
of the model). Recall that neither the isotropic model (n=1) nor the
rectangular models ($n\rightarrow\infty$) are able to achieve a full
explanation of the involved physical events. Thus, hereafter a value $n=4$ will
be chosen for the smoothing index, corresponding to an intermediate regime
between the above mentioned material laws. As an additional advantage, we want
to mention that, technically, the use of smooth models produces stable and
faster numerical convergence.

\begin{figure}[t]
\centering
\includegraphics[height=9.4cm,width=13cm]{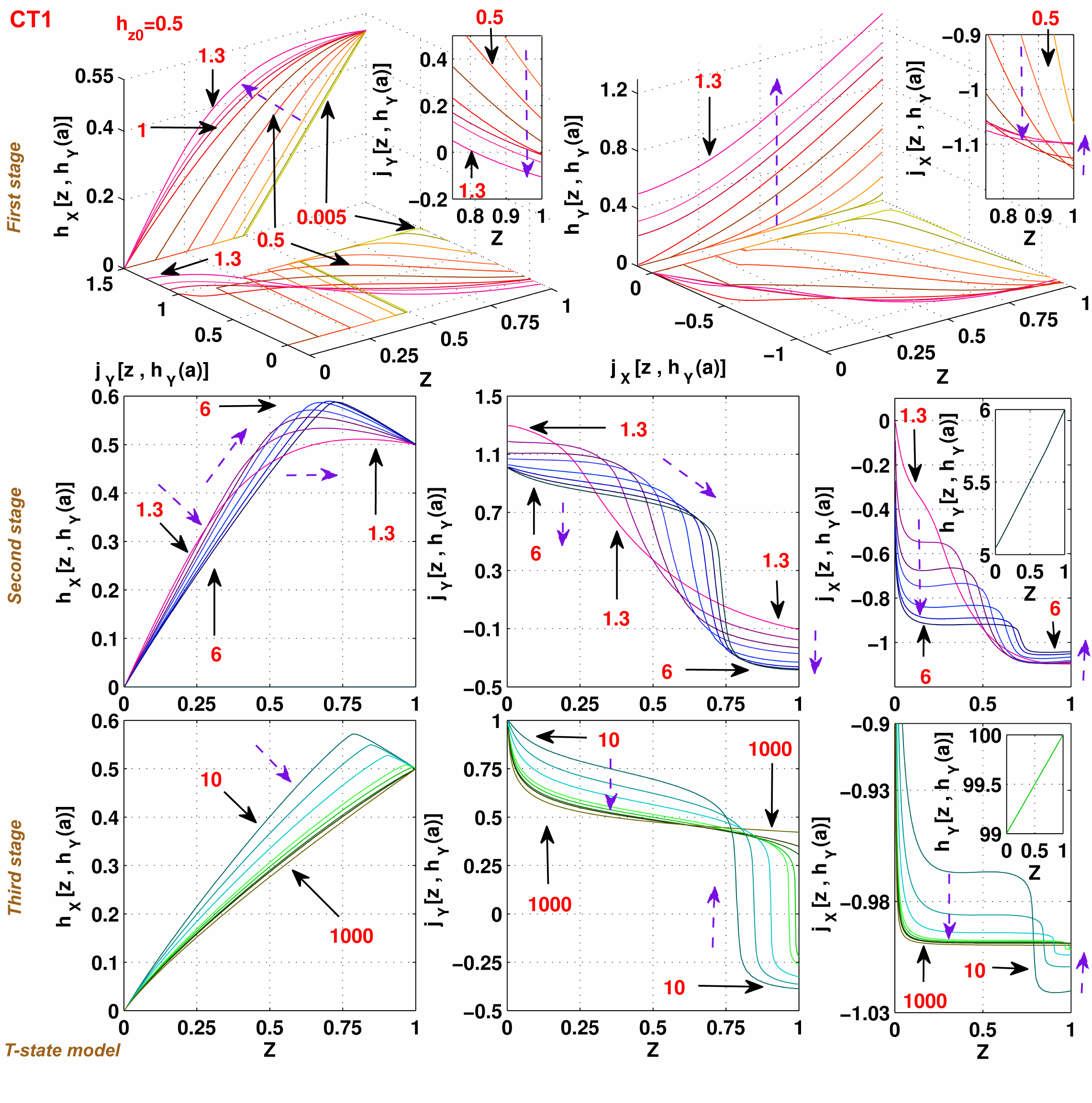}
\caption{\label{Figure_5_7} Profiles of the magnetic field components
$h_{x}[z,h_{y0}]$ and $h_{y}[z,h_{y0}]$ for a perpendicular field component
$h_{z0}=0.5$. Also included are the corresponding current-density profiles
$j_{y}[z,h_{y0}]$ and  $j_{x}[z,h_{y0}]$ for the CT1 model. For clarity,
the longitudinal magnetic field is applied in three stages: Top,
$h_{y}(a)=$0.005, 0.050, 0.170, 0.340, 0.500, 0.680, 0.845, 1.0, 1.1, 1.3; 
Middle,
$h_{y}(a)=$1.3, 1.6, 1.9, 2.2, 3.0, 4.0, 5.0, 6.0, and Bottom $h_{y}(a)=10,
20, 40, 80, 100, 125, 150, 1000$. Insets at top pane 
show a zoom of the current density profiles close the external sheet of the
superconducting slab. Insets at either middle or bottom panes show the shape
of one of the profiles $h_{y}[z,h_{y}(a)]$ for the corresponding stage. Dashed
line arrows  are included to supply a trace method from the initial profile
until the last one, in correspondence to the aforementioned stages.
}
\end{figure}

\begin{figure}[t]
\centering
\includegraphics[height=8.1cm,width=13cm]{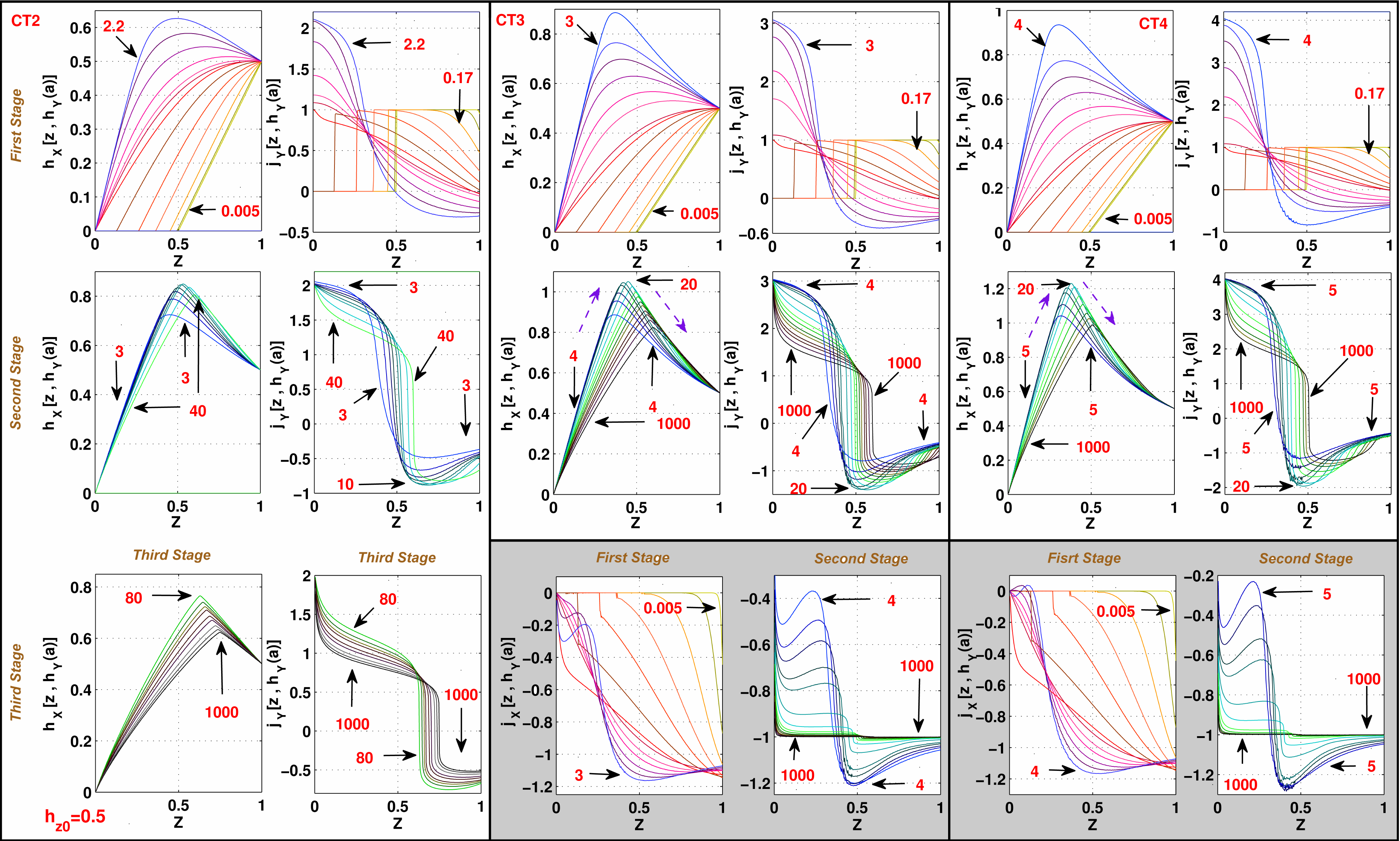}
\caption{\label{Figure_5_8} Profiles of $h_{x}[z,h_{y0}]$ and
$j_{y}[z,h_{y0}]$ for the 1st (top) and 2nd (middle) stages of the magnetic
dynamics described in the cases CT2 (left pane), CT3 (central pane) and CT4
(right pane), all under a field $h_{z0}=0.5$. The 3rd stage
is only defined for the CT2 case (left pane - bottom), as we have not
noticed remarkable difference between the 2nd and 3rd stages for
cases CT3 and CT4, at least for values of the longitudinal applied field not
beyond of $h_{y}(a)=1000$. So, in return, the lower panes in cases CT3
and CT4 show the profiles for $j_x[z,h_{y}(a)]$ for the 1st and
2nd stages. $h_{y}(a)$ is as follows.
CT2: (\textit{1st stage})  $h_{y}(a)=$0.005, 0.050, 0.170, 0.340, 0.500,
0.680, 0.845, 1.0, 1.1, 1.3, 1.6, 1.9, 2.2; (\textit{2nd stage}) 
$h_{y}(a)=$3, 4, 5, 6, 8, 10, 20, 40; and (\textit{3rd stage}) $h_{y}(a)=$ 80,
120, 160, 200, 300, 400, 600, 800, 1000. CT3: (\textit{1st stage})
$h_{y}(a)=$0.005, 0.050, 0.170, 0.340, 0.500, 0.680, 0.845, 1.0, 1.5, 1.8, 2.2,
2.6, 3.0; and (\textit{2nd stage}) $h_{y}(a)=$4, 5, 6, 8, 15, 20, 40, 70, 100,
150, 200, 300, 400, 600, 1000. CT4: (\textit{1st stage}) $h_{y}(a)=$0.005,
0.050, 0.170, 0.340, 0.500, 0.680, 0.845, 1.0, 1.5, 1.8, 2.2, 2.6, 3.0, 4.0; and
(\textit{2nd stage}) $h_{y}(a)=$5, 6, 8, 15, 20, 40, 100, 200, 400, 600, 1000. 
}
\end{figure}

\begin{figure}[t]
\centering
\includegraphics[height=9.4cm,width=13cm]{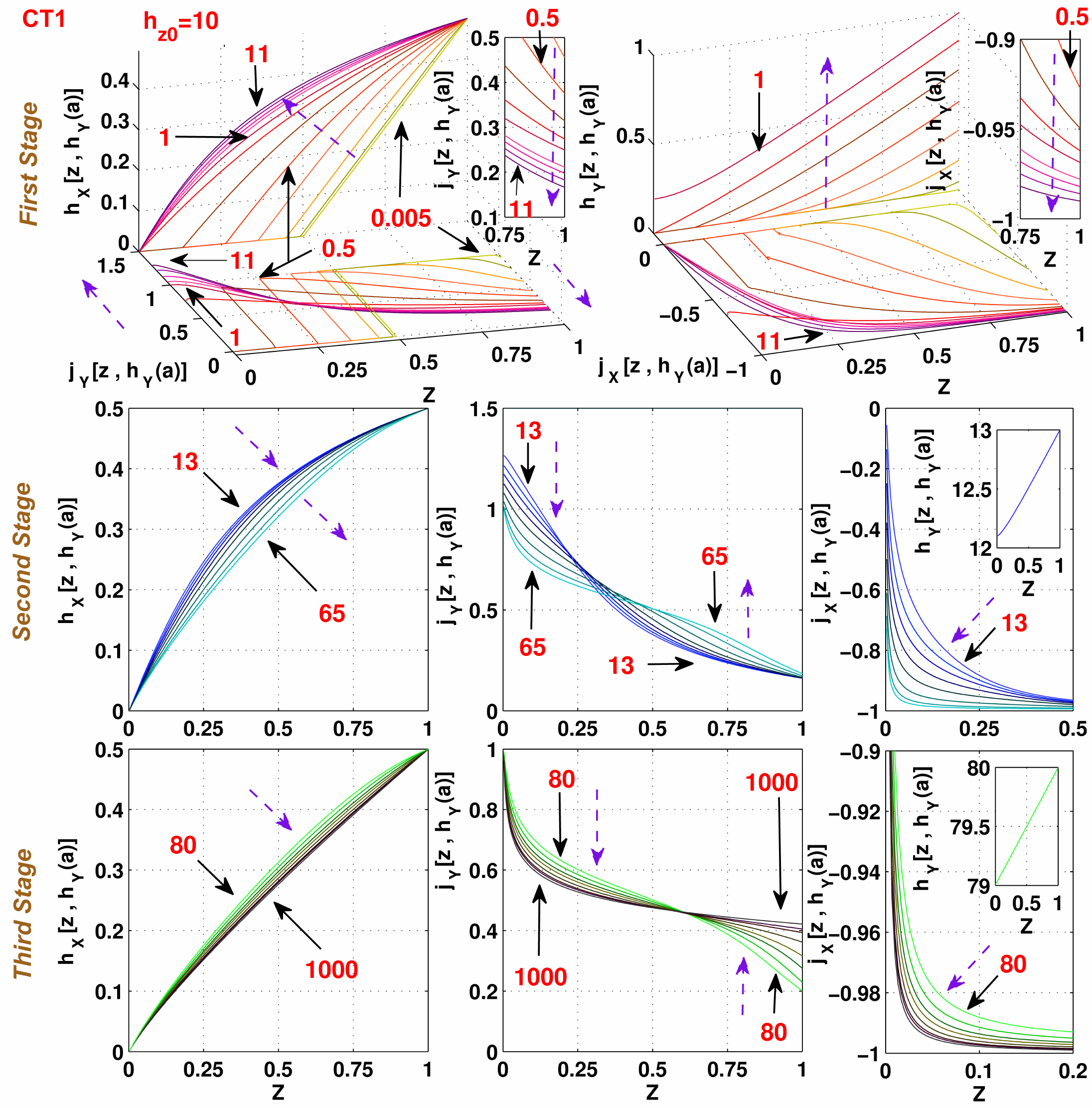}
\caption{\label{Figure_5_9} Same as Fig.\ref{Figure_5_7}, but for $h_{z0}=10$
and the values of the longitudinal field: (Top) $h_{y}(a)=$ 0.005, 0.050, 0.170,
0.340, 0.500, 0.680, 0.845, 1.0, 4.0, 6.0, 8.0, 11, (Middle) $h_{y}(a)=$ 13,
15, 17, 20, 25, 35, 50, 65, (Bottom) $h_{y}(a)=$ 80, 100, 125, 150, 200, 300,
400, 1000.
}
\end{figure}

\begin{figure}[t]
\centering
\includegraphics[height=9.4cm,width=13cm]{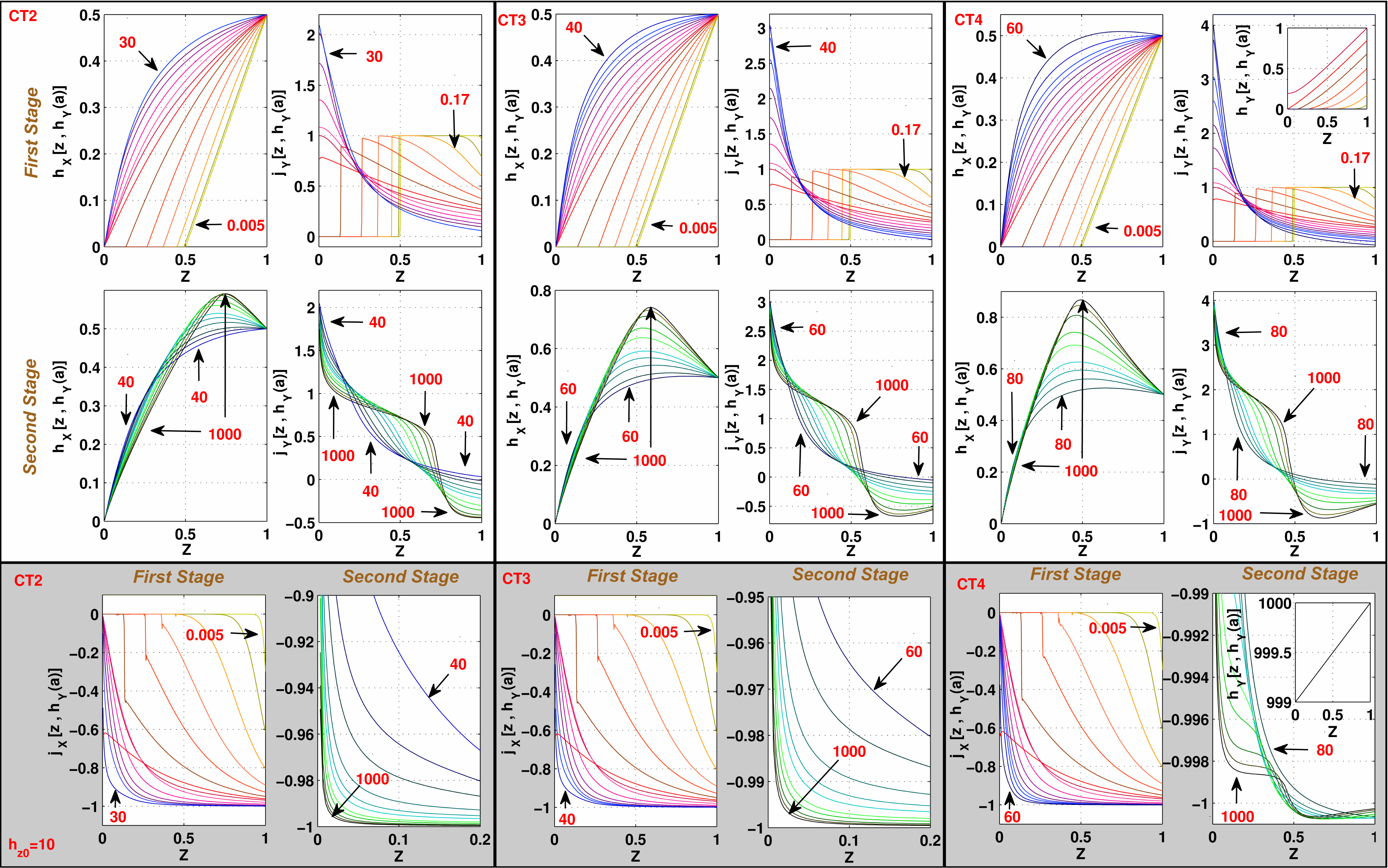}
\caption{\label{Figure_5_10} Similar to Fig.~\ref{Figure_5_8}, but for
$h_{z0}=10$. The curves are labeled as follows. CT2
(left pane): 1st stage (top) $h_{y}(a)=$ 0.005, 0.050, 0.170, 0.340, 0.500,
0.680, 0.845, 1.0, 5.0$, $10, 15, 20, 30, and 2nd stage (middle) $h_{y}(a)=$ 40,
60, 80, 120, 160, 200, 300, 400, 600, 800, 1000. CT3 (middle pane): 1st stage
(top) $h_{y}(a)=$ 0.005, 0.050, 0.170, 0.340, 0.500, 0.680, 0.845, 1.0, 5.0,
10, 15, 20, 25, 30, 40, and 2nd stage (middle) $h_{y}(a)=$60, 80, 120, 160, 200,
300, 400, 600, 800, 1000. CT4 (right pane): 1st stage (top) $h_{y}(a)=$0.005,
0.050, 0.170, 0.340, 0.500, 0.680, 0.845, 1.0, 5.0, 10, 15, 20, 35, 30, 40,
60, and 2nd stage (middle) $h_{y}(a)=$ 80, 120, 160, 200, 300, 400, 600, 800,
1000. In the bottom of panes CT2, CT3 and CT4, the corresponding profiles of
$j_x[z,h_{y}(a)]$ are shown.
}
\end{figure}

\begin{figure}[t]
\centering
\includegraphics[height=11.5cm,width=13cm]{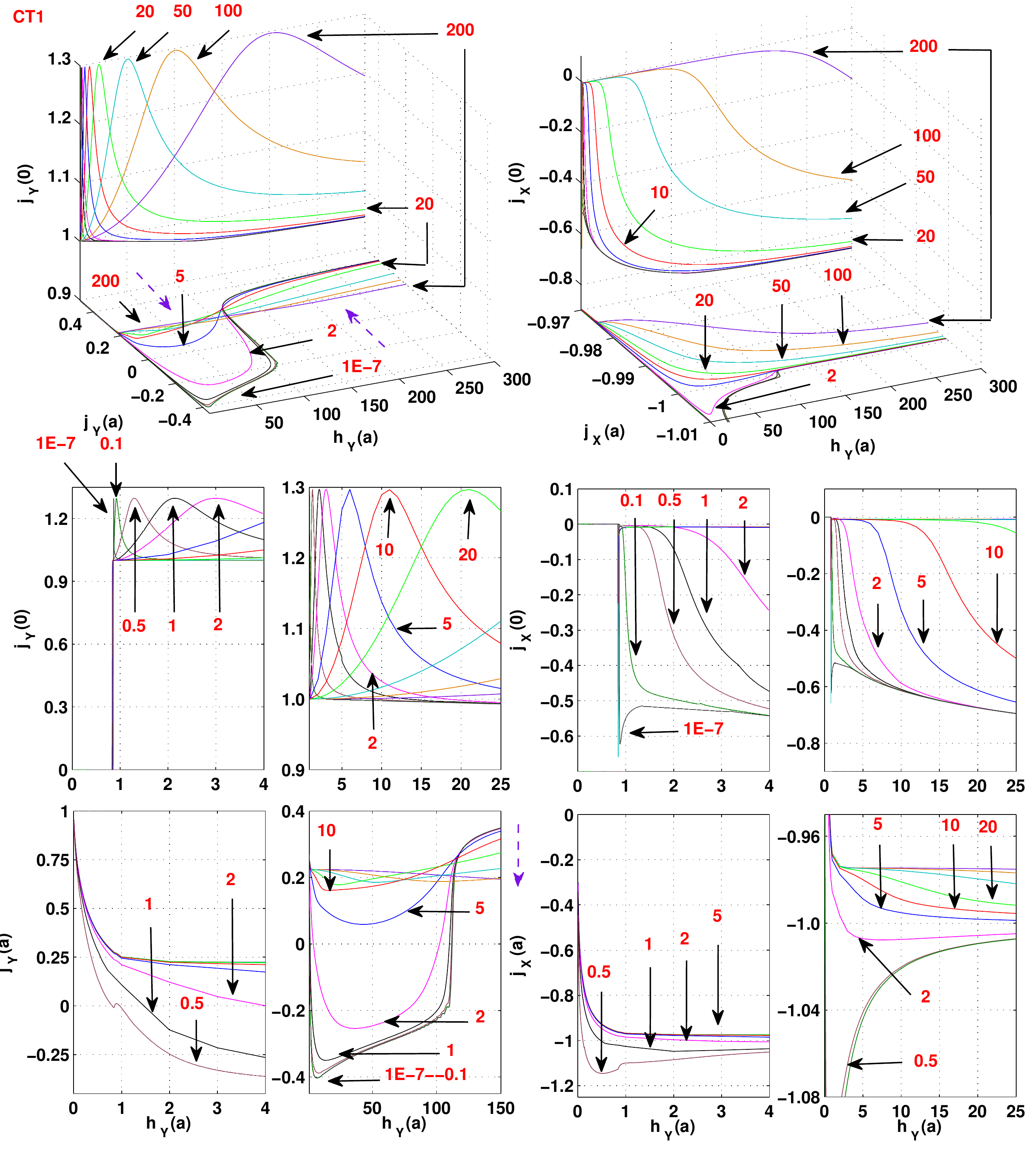}
\caption{\label{Figure_5_11} Evolution of the current density vector for the CT1
model. To the left, we show the patterns of transport current over the central
layer ($j_{y}(z=0)$) and external layer ($j_{y}(z=a)$) of the slab. To the
right, we show the variation of the perpendicular component $j_{x}(z=0)$ and
$j_{x}(z=a)$.  The curves are labeled according to the perpendicular magnetic
field component $h_{z0}=$ 0.1, 0.5, 1, 2, 5, 10, 20, 50, 100, 200.
}
\end{figure}

\begin{figure}[t]
\centering
\includegraphics[height=10cm,width=13cm]{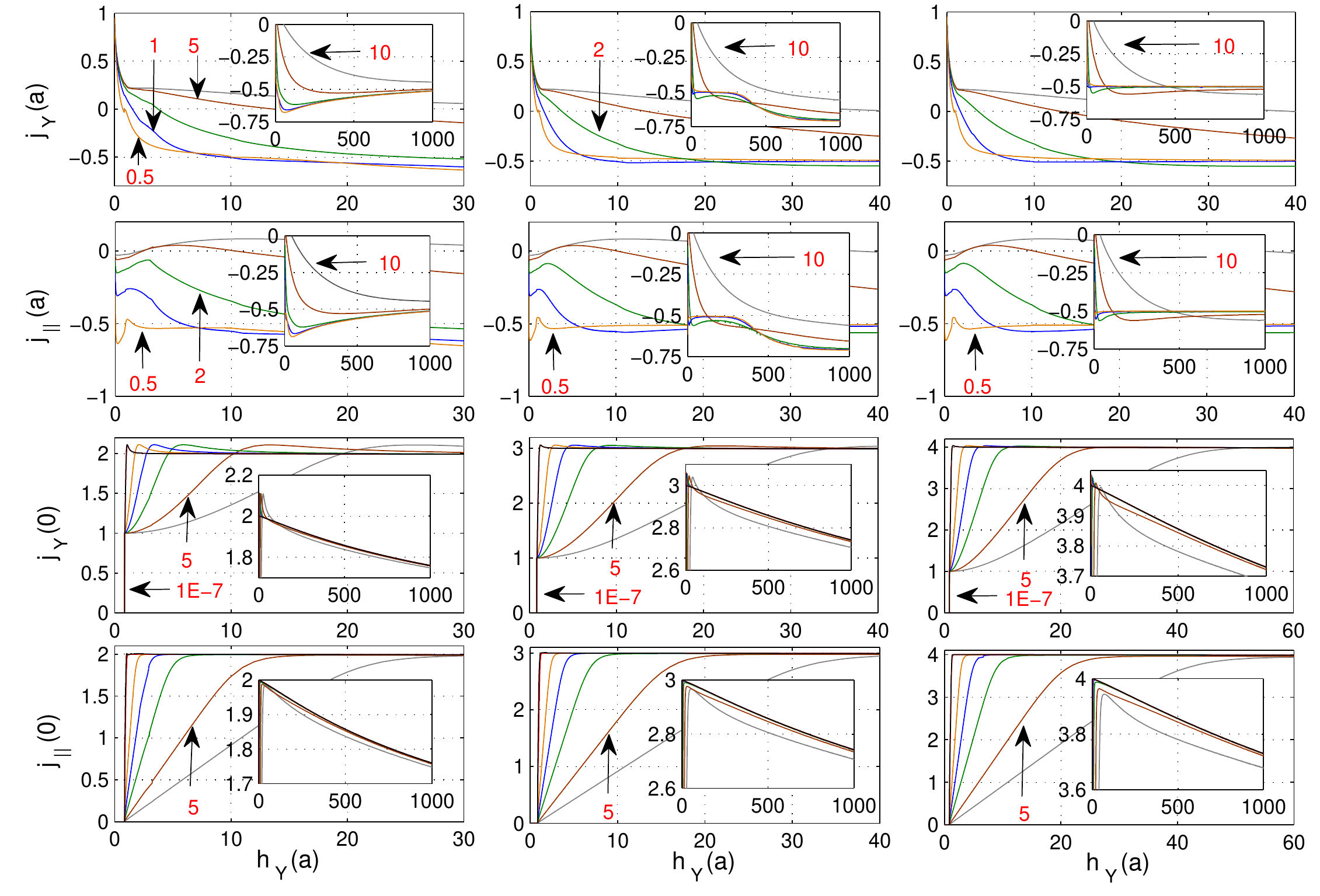}
\caption{\label{Figure_5_12} Dynamics of the current density vector for the
models CT2 (left pane), CT3 (middle pane) and CT4 (right pane). The first row of
plots corresponds to the patterns of transport current along the surface
layers of the slab [$j_{y}(z=a)$]. The second one, corresponds to the respective
values for the parallel component of the current density [$j_{||}(a)$]. The
dynamics of
the transport current $j_{y}$ and the component $j_{||}$ at the central layer of
the superconducting slab $(z=0)$ is shown at the third and fourth rows,
respectively. The curves are labeled according to the perpendicular magnetic
field component $h_{z0}=$1E-7, 0.5, 1, 2, 5, 10 with all plots having the same
color scheme.}
\end{figure}

\begin{figure}[t]
\centering
\includegraphics[height=12cm,width=13cm]{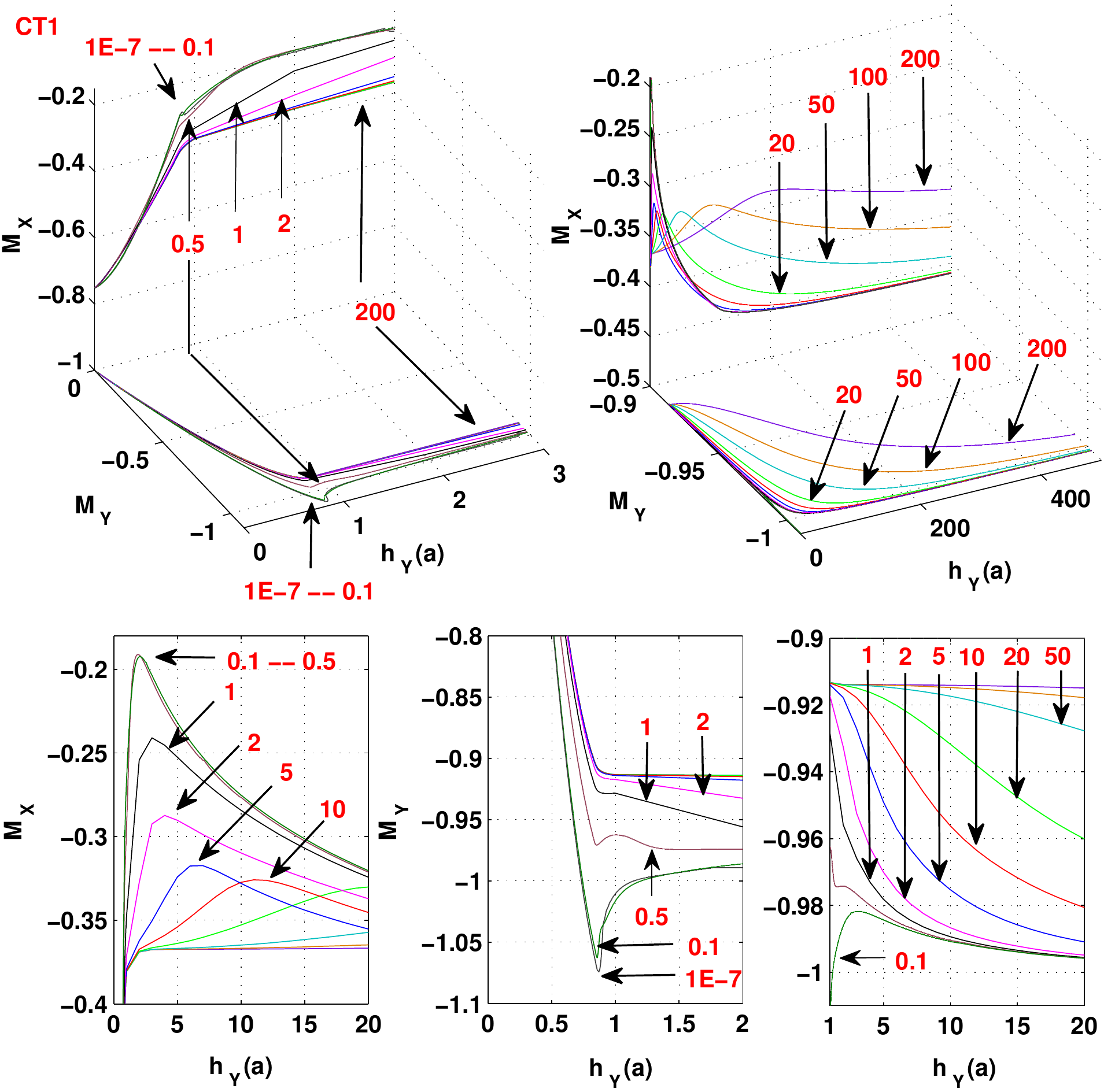}
\caption{\label{Figure_5_13} Magnetic moment components $M_{x}$ and $M_{y}$ of
the slab as a function of the applied magnetic field $h_{y}(a)$ for
the CT1 model. The curves are labeled according to the perpendicular magnetic
field component $h_{z0}$ for each.}
\end{figure}

\begin{figure}[t]
\centering
\includegraphics[height=12cm,width=13cm]{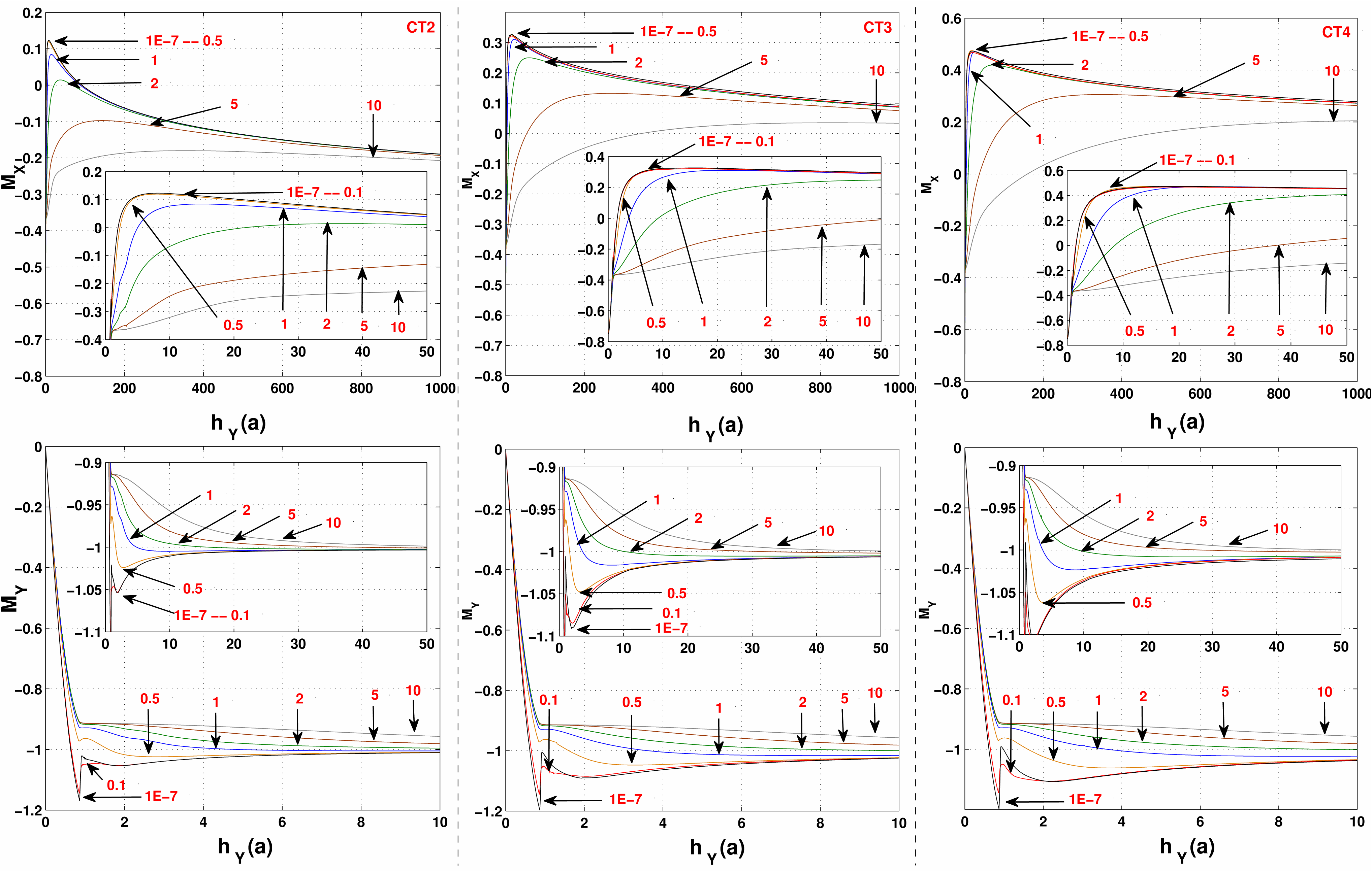}
\caption{\label{Figure_5_14} Same as Fig.~\ref{Figure_5_13}, but here the
curves are corresponding to the CT2 (left pane), CT3 (central pane), and CT4
(right pane) critical state models.
}
\end{figure}

In order to obtain continuity with the T-state results obtained above,
the electrodynamic quantities of interest have been obtained under the same
experimental configuration shown at the left pane of Fig.~\ref{Figure_5_3}, and
a similar analysis scheme to that developed in the previous section will be
tackled here. Thus, with the aim of getting a
detailed physical interpretation on how the longitudinal and transverse magnetic
fields affect the dynamics of the transport current problem, we will show the
magnetic penetration profiles for low and high perpendicular fields, i.e.,
$h_{z0}=0.5$ (Figs.~\ref{Figure_5_7}~\&~\ref{Figure_5_8}) and $h_{z0}=10$
(Figs.~\ref{Figure_5_9}~\&~\ref{Figure_5_10}), for different $CT\chi$
conditions. Then, for completeness, the set of initial conditions $h_{z0}$ is
extended in Figs.~\ref{Figure_5_11}~\&~\ref{Figure_5_12}, by means a thorough
analysis of the dynamics of the current density over the central layer
$[j_{y}(z=0)]$ and external layer $[j_{y}(z=a)]$ of the slab. Finally, it will
be shown that the fingerprint of the different CT$\chi$ models is identified as
a peak effect in the curves for the magnetic moment components
(Fig.~\ref{Figure_5_13}~\&~\ref{Figure_5_14}) caused by the maximal enhancement
of the transport current density along the central layer.

\paragraph{(~\textit{A.}~) \textit{Penetration Profiles and
current density behavior}\vspace*{0.5cm}\\}

On the one hand, let us recall that under the SDCST the square condition is
given by $\chi=1$ (CT1) assuming the customary condition
$J_{c\perp}=1$ (i.e., $j_{c\parallel}=1$ as $\chi\equiv
J_{c\parallel}/J_{c\perp}$). This can be considered as a lower bound for such
quantity because the experimental values reported in the literature are
typically above unity. On the other hand, as we have observed an intricate
behavior for the local dynamics of the electromagnetic quantities under the
CT$\chi$ conditions, henceforward, we will split the experimental process in
three successive stages as the longitudinal magnetic field component $h_{y}(a)$
is increased. We mean:\footnotemark[1]

\begin{itemize}
\item[(\textit{i})] The current density at the center $j_{y}(0)$ increases until
a maximum value is obtained. The set of profiles for the physical quantities
[$h_{x}(z)$,$j_{y}(z)$] and [$h_{y}(z)$,$j_{x}(z)$] in the partial penetration
regime are also included within this stage (see e.g. top of
Figs.~\ref{Figure_5_7}~-~\ref{Figure_5_10}). The occurrence of possible negative
values for the longitudinal current density along the superconducting surface
$j_{y}(a)$ is also focused for low values of the perpendicular magnetic field
$h_{z0}$.
\item[(\textit{ii})] The longitudinal current density profiles $j_{y}(z)$ show a
\textit{bow tie pattern}, whose evolution is shown until the minimum value for
the longitudinal current density along the superconducting surface, $j_{y}(a)$,
is reached (see e.g., the middle row of
Figs.~\ref{Figure_5_7}~-~\ref{Figure_5_10}).
\item[(\textit{iii})] Eventually, the longitudinal current density
$j_{y}(a)$ grows up by increasing the longitudinal applied field $h_{y}(a)$, and
further it stabilizes around a certain value (e.g.,
$j_{y}(a)\approx0.5$ for CT1 case as it is shown
in the bottom of Figs.~\ref{Figure_5_7}~-~\ref{Figure_5_9}).
\end{itemize}

\footnotetext[1]{Mind the color scheme displayed for
the profiles evolution in Figs.~\ref{Figure_5_7}~-\ref{Figure_5_10}}

As far as concerns to the field and current density penetration profiles, it
is to be noticed that the trend of the profiles for the partial penetration
regime is fairly independent of the perpendicular magnetic field $h_{z0}$
(Figs.~\ref{Figure_5_7}~-~\ref{Figure_5_10}). Moreover, the partial penetration
regime in which the transport current zone progressively penetrates the sample
is nearly independent of the magnetic anisotropy of the critical state
(compare the above profiles to their respective ones for the T-state condition
in Fig.~\ref{Figure_5_4}). 

We call readers' attention to the fact that, in the
aforementioned first stage, the negative current patterns
are also found under the low applied magnetic fields $h_{z0}<0.5$. However, by
contrast to the results within the previous section, it is to be recalled that
for the T-state model the condition $j_{c ||}\rightarrow\infty$ allows unbounded
values for the longitudinal current $j_{y}$ at the center of the sample. By
contrast, for the
bounded cases CT$\chi$, the magnetic anisotropy of the material law
$\Delta_{r}$ defines
the maximal current density for the critical state regime. In other words, the
maximal length of the vector $\textbf{j}$ corresponds to the optimal
orientation of the region $\Delta_{r}(\chi,n)$ in which the biggest distance
into the superelliptical condition is reached, i.e., such situation corresponds
to the maximal transport current allowed in the superconductor, and can be
calculated by the following analytical expression
\begin{eqnarray}\label{Eq.5.17}
{\rm Max}\{j_{c\parallel}(\Delta_{\textbf{r}})\}=j_{y}^{max}=
\left( 1+\chi^{2n/(n-1)} \right)^{(n-1)/2n} \, . 
\end{eqnarray}
%

\paragraph{(~\textit{B.}~) \textit{Central and superficial patterns of the
current
density}\vspace*{0.5cm}\\}

Note that Eq.~\ref{Eq.5.17} allows us to obtain the maximum value expected for
$j_{y}$ in terms of the actual critical state model in use. Thus, as we have
assumed n=4 for the different CT$\chi$ models considered here, one gets
$j_{y}^{max}=1.2968$ for the CT1 case, $j_{y}^{max}=2.1127$ for the CT2 case, 
$j_{y}^{max}=3.0591$ for the CT3 case, and $j_{y}^{max}=4.0369$ for the CT4
case. These values may be checked by means our numerical results in
Figs.~\ref{Figure_5_11}~\&~\ref{Figure_5_12} for the intensity of the transport
current density in the central layer of the slab, i.e., the obtained patterns
$j_{y}(0)$. 

In addition, note that by a thorough analysis of the current density components
along the central ($z=0$) and external layers ($z=|a|$) of the superconducting
slab, several remarkable physical properties can be straightforward depicted
(Figs.~\ref{Figure_5_11}~\&~\ref{Figure_5_12}). On the one hand, notice that the
full penetration regime can be clearly distinguished from the
partial penetration regime, and remarkably the emergence of negative
states for the transport current density close to the external surface of the
superconducting sample it is more evident either when $h_{z0}$ is reduced
and/or the current density anisotropy factor $\chi$ is increased. On the other
hand, outstandingly the maximal value of the current density along the original
direction of the transport current ($y-$axis) can be depicted in terms of
$j_{y}(0)$, where its maximal value defined by Eq.~(\ref{Eq.5.17}) turns out to
be
independent of the perpendicular applied magnetic field at least as regards the
existence of the \textit{peak effect} in the transport current density. Thus,
the enhancement of the transport current density can be either accelerated or
decelerated with the tilt of the applied magnetic field, but in general terms,
its maximum directly relates to the limitation introduced by the cutting
mechanism. Physically, this means that the role played by the magnetic
anisotropy of the material law may be characterized by the influence of the
threshold cutting value on the enhancement of the critical current density.

In order to confirm the above physical interpretation, in
Fig.~\ref{Figure_5_12} we show the magnetic dynamics of
the longitudinal current density $j_{y}$, and the cutting current component
$j_{||}$ for the conditions CT2, CT3, and CT4. We have taken a wide
set of values for the perpendicular field component ($h_{z0}$). 
On the one hand, as far as concerns the sample's surface, we have observed that
the longitudinal current density $j_{y}(a)$ does not display significant
differences when one has $\chi\geq2$ (see also Fig.~\ref{Figure_5_5} for the
T-state model with $\chi\rightarrow\infty$). Hence, the disappearance of
the negative current flow along the external superconducting surface does not
occur despite a very high applied magnetic field has been considered
($h_{y}(a)=1000$). On the other hand, it is important
to notice that the patterns of the parallel current density along the
superconducting surface ($j_{||}(a)$) are almost indistinguishable as soon as
the condition $\chi\geq 2$ (CT2) is reached (upper half of
Fig.~\ref{Figure_5_12}). This implies that for an accurate picture of the
parallel critical current, surface properties do not provide a useful
information.

However, Fig.~\ref{Figure_5_12} shows that the threshold value for the
cutting current density can be estimated from the experimental measurement of
the transport current density along the central sheet of the superconducting
sample. Moreover, regardless to the experimental conditions
($h_{z0}$, $h_{y}(a)$) and also for different bandwidths $\chi$ no significant
change occurs in the parallel current density around the central sheet of the
sample (lower half of Fig.~\ref{Figure_5_12}). 

\paragraph{(~\textit{C.}~) \textit{Features on the magnetic
moment}\vspace*{0.5cm}\\}

Outstandingly, it is to be noticed that the limitation introduced by the
flux cutting mechanism imposes a maximal compression of the current density
within the sample. Thus, the peak effects both for the transport current density
$j_{y}$ (Figs.~\ref{Figure_5_11} \& \ref{Figure_5_12}) and for the magnetic
moment component $M_{x}$ (Figs.~\ref{Figure_5_13} \& \ref{Figure_5_14}) are
defined by the instant at which the maximal transport
current density occurs. Additionally, upon further increasing the longitudinal
applied magnetic field component $h_{y}(a)$, the profile $h_{x}(z)$ will be
forced to decrease from the central sheet ($z=0$) towards the external surface
($z=a$). This reversal generates a local distortion of the longitudinal current
density $j_{y}$ in a bow tie shape (see the middle row of
Figs.~\ref{Figure_5_7}~\&~\ref{Figure_5_10}). Likewise, as soon as the profile
$j_{||}(0)=j_{c||}$ is reached,
the magnetic moment $M_{x}$ starts decreasing as one can see
by comparison of Figs.~\ref{Figure_5_11}~\&~\ref{Figure_5_13} and
Figs~\ref{Figure_5_12}~\&~\ref{Figure_5_14}.

Finally,  one additional feature is to be noted: the interval between the
instant at which the maximal transport current condition is reached
($j_{y}(0)=j_{y}^{max}$), and the instant at which the slope of the magnetic
moment $M_{x}$ changes sign, could be assumed as transient or stabilization
period required for an accurate determination of the value $j_{c ||}$ when
measurements are performed in terms of the applied longitudinal field
$h_{y0}=h_{y}(a)$.
Apparently, this transient increases with the value of the perpendicular field
$h_{z0}$. From this point on, $j_{y}(0)$ may be basically identified with
$j_{c\parallel}$.


\chapter{\label{ch-6} \sc \textbf{Electromagnetism For Superconducting
Wires}}

Type-II superconducting wires are deemed promising elements for large-scale
technological applications such as power transmission cables, magnet systems for
large particle accelerators, and magnetic-based medical techniques such as
MRI. The usefulness of these kind of technology is straightforwardly linked to
the local electromagnetic response of the superconductor under variations of
the ambient magnetic field and the customary condition of transport-current.
Special interest relies on determining the value of the maximum
dissipation-free current and characterizing the mechanisms of reduction of
electric power dissipation due to alternating fields and/or alternating current
flow (commonly called AC losses).

Despite the electric power hysteretic losses associated to a
superconducting material are somewhat smaller than the range of power
dissipated in normal metals, from the practical point of view, the
superconducting technology is still not so attractive to replace the power
technology based in copper wires, because in order to keep
the temperature of the superconductor below $T_{c}$, heat removal requires a
sophisticated and costly cryogenic system. Thus, in order to make the
superconducting devices more attractive and competitive with respect to other
technologies, it is of utter importance to understand, predict, and eventually,
reduce the AC losses of superconducting wires under practical configurations.

Major features of the macroscopic electromagnetic behavior of type-II
superconducting wires have been captured in Bean's model of the critical
state~\cite{P2-Bean_1964}. In this framework, magnetization currents of density
$\bf J$ are induced within the superconductor during variations of the magnetic
flux which accordingly redistribute themselves to screen the penetrating flux
within the sample. Their magnitude adopts the critical value $J_c$ at a
given temperature and specified field.
Although simple for idealized configurations, the electrodynamics underlying
Bean's model becomes cumbersome when realistic configurations are addressed.
Thus, penetration of magnetic flux must be typically obtained by sophisticated
numerical methods. 

For example, in order to comprehend the distributions of the field and current
in two- and three-dimensional samples subject to time-varying electromagnetic
fields a sort of free-boundary problem should be solved. In a superconducting
slab or a large cylinder (radius much smaller than their length) exposed either
to a parallel magnetic field or a longitudinal transport current, the magnetic
field has only one component, and analysis of the field and current distribution
is simple~\cite{P2-Bean_1964}. Nevertheless, if the situation is such that the
local magnetic field has two components that are functions of two spatial
variables, then solving the free-boundary problem for arbitrary relations of the
external excitations becomes more and more complicated. In fact, for real
applications of superconducting wires, the scenario is such that a simultaneous
field and transport current condition must
be satisfied. Then, in addition to the ambient field one has the magnetic field
generated by the transport current itself~\cite{P2-Ruiz_APL_2012}. This is the
situation, for example, in superconductor windings where each wire is subjected
to the magnetic field of its neighboring wires. Thus, as this is the
configuration very often met in practice, this chapter is devoted to study
round superconducting wires under different configurations of transverse
magnetic field and transport current flow. 

Subchapter 6.1 is devoted to introduce the theoretical framework of the
problem with special attention to the numerical procedure of our variational
approach. It is to be noticed that the theoretical statements developed along
this section are not only valid for round wires but rather for long
superconductors with any topology of the transverse section, e.g., it may
include strips of rectangular cross section, wires of elliptical or circular
cross section, also any intermediate shape between them, and even multifilament
structures as the observed in practical superconducting wires.

Then, in order to introduce our numerical results and to get a fairly clear
understanding of the phenomena involved in the dynamics of the electromagnetic
quantities for superconducting wires with simultaneous transport current and
applied magnetic field, subchapter 6.2 deals with the simpler cases where
the superconducting wire is only subjected to a transverse magnetic field in
absence of transport current, or on the contrary, assume that the
superconducting wire is only subjected to a transport current condition. For a
closer connection with the experimental quantities, we have calculated
the wire's magnetic moment ($\textbf{M}$) and the hysteretic losses ($L$) as a
function of AC external excitations, and their comparison with classical
analytical approaches are featured. On the other hand, we present a thorough
discussion about the observed patterns for the local dynamics of the
electromagnetic quantities such as the inner distribution of current density
$\textbf{J}$, the components of the magnetic flux density $\textbf{B}$ and the
isolevels of the vector potential $\textbf{A}$ (i.e., the lines of magnetic
field), as well as the local distribution of the density of power dissipation
$\textbf{E}\cdot\textbf{J}$, whose profiles are displayed at the end of the
chapter in the corresponding section of supplementary material.

In subchapter 6.3, numerical simulations of filamentary type II
superconducting wires under simultaneous AC transport current and oscillating
transverse magnetic field are performed. A wide number of configurations have
been considered, with special attention to the aforementioned experimental
quantities, $\textbf{M}$ and $L$. On the one hand, according to the temporal
evolution of the AC sources, exotic magnetization loops are
envisaged. On the other hand, remarkable numerical corrections to several widely
used approximate formulae for the hysteretic losses are
identified. Also, in view of the differences found between the different cases
studied,  we present a comprehensive study of the local
dynamics of the electromagnetic quantities in analogy to the previous section,
which allows to reveal some of the main ``\textit{control}'' factors affecting
the losses into the superconducting wires. The full set of profiles describing
the temporal dynamics for the electromagnetic quantities $\textbf{J}$,
$\textbf{B}$, and $\textbf{E}\cdot\textbf{J}$ within the superconducting wire
are shown as function of the external excitations at the end of this chapter.
In more detail, subchapter 6.3.1 addresses the study of cases with synchronous
oscillating excitations, where significant differences between the obtained AC
losses and those predicted by regular approximation formulas are reported.
Furthermore, noticeable non-homogeneous dissipation and field distortions are
displayed, as well as an outstanding {\em low pass filtering} effect intrinsic
to the magnetic response of the system is described. Then, premagnetized
superconducting wires are also considered, i.e.: a magnetic moment is induced
previous to switching on the synchronous oscillating excitation. Thus, the above
envisioned results are straightforwardly generalized. Finally, subchapter 6.3.2
deals with asynchronous oscillating sources, focusing on the calculations of
the double frequency effects provided by one or the other source. This closes
our discussion about the main parameters controlling the hysteretic losses in
superconducting wires, and how this knowledge can help in the design of new
kinds of applications.

\break

\vspace*{0.0cm}
\subsection*{\label{ch-6-1}
\hspace*{0.1cm} \textsl{6.1 Theoretical framework and general
considerations}}
\vspace*{1cm}
\markboth{\hspace*{0.1cm}\textsl{6.1 Theoretical framework and general
considerations}} 
{\hspace*{0.1cm} \textsl{6.1 Theoretical framework and general
considerations}}
\addcontentsline{toc}{chapter}
{\hspace*{0.1cm} \textsl{6.1 Theoretical framework and general
considerations}}

From the theoretical point of view and as has been stated
in the previous chapters, the major features of the macroscopic
electromagnetic behavior of type-II superconductors have been captured by Bean
in the phenomenological model of the critical state~\cite{P2-Bean_1964}, and
its ideas have been extended in our generalized critical state
theory or also called smooth double critical state theory
(SDCST)~\cite{P2-Ruiz_PRB_2009,P2-Ruiz_SUST_2010}. It is worth mentioning
that despite our SDCST allows to include any experimental dependence of the
critical current density on the local properties of the superconducting
specimen (e.g., $J_{c}(H,T)$), the simple Bean's statement $(J\leq
J_{c})$ allows to achieve a clearest interpretation of the
physical phenomena involved in the electromagnetic dynamics of type II
superconductors. Moreover, it allows to establish the limiting values expected
for macroscopic quantities such as magnetization and
energy losses. Thus, in what follows we will assume that
the critical current density does not depend on the applied magnetic field (at
least for the intensities here considered) neither on the temperature. 

Being more specific, the technical problem in the critical state theory consists
of solving a free boundary problem for the distribution of penetrating current
(or magnetic flux) for a given time of the external electromagnetic excitation.
From the analytical point of view, in simplified configurations such as infinite
slabs and cylinders, either considering transport current or parallel
magnetic field, the inner flux-free region can be straightforwardly depicted
by a planar front of flux or radial flux fronts centered in the symmetry axis of
the superconductor respectively \cite{P2-London_1963,P2-Hancox_1966}. Thus,
once a method is found to obtain the actual size of the flux free region, all
the magnetic properties such as the magnetic field lines, magnetization, and AC
losses can be deduced from the knowledge of the penetrating current profiles,
also called flux fronts. However, for a superconducting strip or cylinder
exposed to a transverse
magnetic field the flux fronts are not radially symmetric, and the free boundary
problem is not so easy to be tackled. Nevertheless, already in early
calculations as those performed in 1970s and 1980s various analytical
simplifications have often been used, as for example sinusoidal or elliptical
ansatz for the flux-free region
\cite{P2-Carr_1975,P2-Zenkevitch_1980,P2-Minervini_1989}, these studies being
summarized in more detail in the books by Carr \cite{P2-Carr_2001} and Gurevich
et
al. \cite{P2-Gurevich_1997}. Even for extremal cases such as infinitely thin
strips, exact analytical solutions can be achieved
\cite{P2-Swan_1968,P2-Halse_1970,P2-Norris-1970,P2-Brandt_1993,P2-Zeldov_1994}.
Most recently, analytical expressions for the magnetic field and current
distributions within the CS model for hollow superconducting tubes of thickness
much smaller than their external radius were reported by
Mawatari~\cite{P2-Mawatari_2011}. On the other hand, either for the case of a
strip with finite thickness or a bulk superconducting cylinder exposed to a
transverse magnetic field, exact analytical solutions have not been obtained.
Moreover, when the problem is such that the superconductor is simultaneously
subjected to a transport current, asymmetric deformations of the flux free
region can appear, and analytical approaches are even less conceivable. Thus,
implementation of numerical procedures becomes mandatory when handling intricate
configurations where simultaneous alternating transport
current and transverse applied field occur. 

First, a consistent implementation of a variational approach allowed Ashkin
\cite{P2-Ashkin_1979} to trace out the true structure of the
partly flux-penetrated state of a superconducting wire subject to a transverse
magnetic field, whose results were confirmed in the meantime by various
numerical and analytical calculations
\cite{P2-Pang_1981,P2-Telschow_1994,P2-Kuzovlev_1995,P2-Bhagwat_2001,
P2-Gomory_2002,P2-Haken_2002}. For example, Telschow and Koo
\cite{P2-Telschow_1994} suggested an integral-equation approach for
determining the flux-front profile, thus reducing the problem (for the case of a
constant critical current density $J_{c}$) to solving a single Fredholm integral
equation of the first kind which may be performed by several algorithms; the
method applying generally to a sphere or long cylinders exposed to axisymmetric
external fields. In a similar approach, Kuzovlev has
established the integral equation for the flux-free zone and an exact value of
the full penetration field ($B_{p}$) for a three-dimensional superconductor of
an arbitrary
axisymmetric form \cite{P2-Kuzovlev_1995}. Bhagwat and Karmakar developed a
method allowing for determination of the flux-front form in cylinder
superconductors of different cross-sections and field orientations by solving a
(formally infinite, in fact: large) system of nonlinear ordinary differential
equations for coefficients determining the front \cite{P2-Bhagwat_2001}. On the
other hand, in a similar fashion to the numerical solution for the ellipsoid
geometry developed in Ref.~\cite{P2-Navarro_1991}, minimization procedures were
used in Refs.~\cite{P2-Gomory_2002} \& \cite{P2-Haken_2002} to optimize the
trial boundary of the flux-free region avoiding assuming an \textit{a priori}
shape for the flux fronts as was customarily done in the precedent works.

At this point, it is worth of mention that currently the most popular trend in
the analysis of magnetic flux dynamics in superconductors are the
numerical simulations implementing finite-element methods in conjunction with
nonlinear power-law voltage-current characteristics 
\cite{P2-Amemiya_2001,P2-Nibbio_2001,P2-Stavrev_2002,P2-Hong_2008}. Also of
mention is the development of new formulations of the critical state
model \cite{P2-Prigozhin_2004,P2-Barrett_2010,P2-Campbell_2007,P2-Ruiz_PRB_2009,
P2-Ruiz_SUST_2010}, or new algorithms to approach the CS in commercial
finite element codes~\cite{P2-Farinon_2010}. For example, on the one hand,
Campbell et al.~\cite{P2-Campbell_2007} have suggested that the critical state
model could be made amenable in COMSOL-multiphysics by modifying the so called
material law by an explicit function of the vector potential, and Farinon et
al.~\cite{P2-Farinon_2010} have proposed a special algorithm to be implemented
in commercial ANSYS-code approaching the CS by an iterative adjustment of the
material resistivity. On the other hand, a dual formulation approach to the free
boundary problem was developed by Prigozhin \cite{P2-Prigozhin_2004} which
allows consideration of a wide class of variational problems, particularly the
treatment of the critical states in superconductors of complicated shapes
without assuming a priory specific shape of the flux free zone. Likewise, in a
way similar to the spirit of Prigozhin's work, our above
introduced variational statement for the most general
SDCST~\cite{P2-Ruiz_PRB_2009,P2-Ruiz_SUST_2010} allowing for arbitrary mutual
orientation of the external field and transport
current and implementing finite-element methods is able to tackle in a very
efficient way the so called \textit{front tracking} problem for superconducting
wires \cite{P2-Ruiz_APL_2012}. In the present case of interest (a
superconductor of thickness much smaller than its length, subjected to time
dependent transverse magnetic field and a simultaneous transport current
condition), the cumbersome analysis of the intrinsic anisotropy effects may be
straightforwardly avoided, as the streamlines preserve only one
direction (perpendicular to the applied magnetic field), which means a
significant reduction of the computational time.

\paragraph{(~\textit{A.}~) \textit{Statements For The
Variational Approach}\vspace*{0.5cm}\\}

Following the same methodology introduced in the previous chapters, here
the whole superconducting region is involved in the calculation and the
free boundary is obtained as a part of the solution of the minimization
procedure. In this sense, the shape of the superconducting sample may be
arbitrary and it is related to the mesh design (see Fig.~\ref{Figure_3_1}).
Going into detail, in our case of interest we are enabled to discretize the
samples according to their cross-section area ($\Omega$) through a
collection of points ($\textbf{r}_{i}$) depicting the straight
infinite elementary filaments fulfilling the condition $r_{i}\in\Omega$. Thus,
for a
sufficiently large mesh, a uniform current density can be assumed within each
elementary wire such that $I_{i}=J_{i}s_{i}$ with $s_{i}$ the cross
sectional area of the filament. Then, the problem can be straightforwardly
written in terms of local contributions of the vector potential $A_{i}(r_{i})$
accordingly to cylindrical filaments of section $s_{i}=\pi a^{2}$ with
$a\ll R$, being $R$ the maximal distance from
the geometrical center of the superconducting sample to its external surface
(in our case, $R$ defines the radius of the circular section $\Omega$).
Therefore, the vector potential of each filament ($A_{i}$) splits up into
two expressions, one within the filaments of radius $a$:
\begin{eqnarray}\label{Eq.6.1}
 A_{i}(r_{i}\in s_{i})=\frac{\mu_{0}}{4\pi} \left[2\pi
a^{2}J_{i}\ln{\left(a\right)}-\pi (a^{2}-{r}_{i}^{2})J_{i}\right]+
C_{1} \,~\,~,~\,
\forall ~\, ~\, {r}_{i}<a \, ,
\end{eqnarray}
and one outside the filament:
\begin{eqnarray}\label{Eq.6.2}
 A_{i}(r_{j}\notin s_{i})=-\frac{\mu_{0}}{4\pi} \left[ J_{i}
\ln(r_{ij}^{2}/a^{2})\right] + C_{2} \,~\,~,~\, \forall ~\,
~\, r_{ij} \, .
\end{eqnarray}
Here, $r_{ij}$ denotes the distance between the centers of filaments $i$ and
$j$, and $C_{1}$ and $C_{2}$ are arbitrary integration constants, one of them
determined by continuity at $r_{i}=a$ and the other one can be absorbed in a
global constant for the whole section $\Omega$ (\textit{C in what follows}). In
fact, as it was established in chapter 2.1 any arbitrary constant may be added
to the vector potential $A_{i}$ without altering the magnetic field produced by
the wires, and therefore one can choose $C\equiv0$ to solve the critical state
problem according to the minimization functional in Eq.~(\ref{Eq.2.11}).
However, some care must be taken when electric fields related to flux
variations are calculated. Let us be more specific. In general, an electrostatic
term $\nabla\varPhi$ enters the definition of the electric field
($\textbf{E}=-\partial_{t}\textbf{A}-\nabla\varPhi$). In the long wire
geometry,
$\nabla\phi$ may be argued to be spatially constant ($C_{t}$) by symmetry
reasons. Then,
as related to gauge invariance, the vector potential $\textbf{A}$ may be
recalibrated in the form $A\rightarrow \tilde{A}+\nabla\varPhi$, and therefore,
arbitrary constants may be induced so as to fit the physical condition $E=0$ for
those regions with absence of magnetic flux variations. In fact, if the
minimization functional has not to be constrained by a transport current
condition, i.e. $I_{tr}(t)=0$, $C_{t}$ disappears in the optimization process.

We recall that in quasi-steady regime (excellent approximation
for the large scale application frequencies) the discrete form of Faraday's law
($\delta\textbf{B}_{i} =- \nabla\times\textbf{E}_{i}(\textbf{J}_{i})\,\delta t$)
in a mesh of circuits that carry the macroscopic electric current density
$J_{i}$, is obtained in general terms by minimizing the action of an
averaged field Lagrangian (~of density ${\cal L}=[{\bf
B}({t+\delta t})-{\bf B}({t})]^{2}/2$~) coupling successive time layers. Thus,
by
using this procedure, and introducing the magnetic vector potential, the
quantity to be minimized transforms to the so called objective function
[viz., Eqs. (\ref{Eq.3.2}) \& (\ref{Eq.3.3})]
\begin{eqnarray}\label{Eq.6.3}
\frac{1}{2}\sum_{i,j}I_{i,l+1}{\rm M}_{ij}I_{j,l+1}-\sum_{i,j}I_{i,l}{\rm M}_{
ij } I_ { j , l+1 }+\sum_{i}I_{i,l+1}\Delta A_{0} \, ,
\end{eqnarray}
with $\{I_{i,l+1}\}$ the set of filaments with unknown current for the time
steps $l+1$, $A_{0}$ the vector potential related to \textit{non-local}
sources, and $M_{ij}$ the mutual inductance matrix between filaments $i$
and $j$, which accordingly to Eqs.~(\ref{Eq.6.1}) \& (\ref{Eq.6.2}), for
filaments of cylindrical cross section $s_{i}$ centered at the positions
$r_{i}\in\Omega$ and subject to uniform distributions of current density
$J_{i}\in s_{i}$ may be defined as:
\begin{eqnarray}\label{Eq.6.4} 
{\rm M}_{ij}=
\left\{
\begin{array}{ll}
\dfrac{\mu_{0}}{8\pi}  \,~\,~ & ,~\, \forall ~\,~\, r_{i}=r_{j} \in
\Omega \, 
\\ \, \\
-\dfrac{\mu_{0}}{4\pi}\ln(r_{ij}/a)\,~\,~ & ,~\, \forall ~\, ~\,
r_{i}\neq r_{j}
\in \Omega \, ~ \, ~ \, .
\end{array}
\right. 
\end{eqnarray}

For our cases of interest, $A_{0}$ corresponds to the magnitude
of the vector potential produced by a uniform transverse magnetic field
$\textbf{B}_{0}$, which can be calculated from the components of the vectorial
expression
\begin{eqnarray}\label{Eq.6.5}
\textbf{A}_{0}(\textbf{r}_{i})=\textbf{B}_{0}\times \textbf{r}_{i} \, .
\end{eqnarray}

Furthermore, when required, optimization must to be performed under the
restriction of applied transport current, i.e., 
\begin{eqnarray}\label{Eq.6.6}
\sum_{i\in\Omega}I_{i}=I_{\rm tr} \, .
\end{eqnarray}
and the physically admissible solutions have to be constrained by the 
CS material law for the current density, that in this case reads
$|J_{i}|\leq J_{c}$. 

Minimization is done under prescribed sources $(B_{0},I_{tr})$ for the time
step $l$ and the above CS material law, and as result of the optimization
procedure one gets the distribution of current filaments along the cross
section of the superconducting sample at the time step $l+1$. Eventually, the
vector potential can be evaluated in the whole space by superposition of
Eqs.~(\ref{Eq.6.1}), (\ref{Eq.6.2}), and (\ref{Eq.6.5}). Then, one may plot
the magnetic flux lines as the isolevels of the total vector potential
$\textbf{A}$, and the components of the magnetic flux density can be evaluated
according to its definition $\textbf{B}=\nabla\times\textbf{A}$. 

Furthermore, in order to achieve a closer connection with experiments the
sample's magnetic moment per unit length ($l$) has been calculated by means the
vectorial expression
\begin{eqnarray}\label{Eq.6.7}
\textbf{M}=\frac{l}{2}\int_{\Omega}\textbf{r}\times\textbf{J} d\Omega\, ,
\end{eqnarray}
and the hysteretic AC losses per unit time and volume ($\Phi$) for cyclic
excitations of frequency $\omega$ can be calculated by integration of the local
density of power dissipation ($\textbf{E}\cdot\textbf{J}$) as follows
\begin{eqnarray}\label{Eq.6.8}
L=\omega\oint_{f.c.}dt\int_{\Phi}\textbf{E}\cdot\textbf{J} d\Phi \, .
\end{eqnarray}
Here, $f.c.$ denotes a full cycle of the time-varying electromagnetic sources.

Applications of the above statements are developed along the following
subchapters,  in a systematic study of infinite cylindrical wires under a
wide variety of different experimental conditions.

\paragraph{(~\textit{B.}~) \textit{Numerical procedure}\vspace*{0.5cm}\\}

Some technical details are worth of mentioning as far as concerns the
numerical procedure. 

On the one hand, the mesh utilized for the whole set of
calculations presented along this chapter is defined in terms of a rectangular
grid with filaments equally distanced under the prescribed condition $r_{ij}\geq
2a$ to satisfy Eq.~(\ref{Eq.6.4}). The number of filaments which
have been considered to fill out the cross section of the superconducting
cylinder is 3908. However, owing to the planar symmetry of the problem one is
allowed to reduce the number of variables to 1954 (i.e., 977 filaments
per quadrant), which is still a large number because the
objective function to be minimized is highly nonlinear. To be more specific, the
number of quadratic terms in Eq.~(\ref{Eq.6.3}) involve minimizing the
action of 1910035 elements, i.e., the sum of elements produced by the
mutual inductance terms between filaments (977*977*2), and the
self inductance terms (977).

On the other hand, contrary to the common choice of a sinusoidal oscillation
process we have run the simulations for triangular oscillating processes, which
indeed do not change the electromagnetic response of the superconductor if
resistive currents are neglected. In fact, an instantaneous response takes place
in the absence of resistance. Also, under the critical state framework, Joule
heat release may be calculated by $\dot{L}=J_{c}E$ as overcritical flow
($J>J_{c}$) is neglected because instantaneous response is assumed
[Fig.~\ref{Figure_1_1} 
(pag. \pageref{Figure_1_1})].


\vspace*{0.5cm}
\subsection*{\label{ch-6-2}
\hspace*{0.1cm} \textsl{6.2 SC wires subjected to isolated external
sources}}
\vspace*{0.5cm}
\markboth{\hspace*{0.1cm}\textsl{6.2 SC wires subjected to isolated external
sources}} 
{\hspace*{0.1cm} \textsl{6.2 SC wires subjected to isolated external
sources}}
\addcontentsline{toc}{chapter}
{\hspace*{0.1cm} \textsl{6.2 SC wires subjected to isolated external
sources}}

\vspace*{0.5cm}
\hspace*{0.3cm}
\textbf{\textit{6.2.1 ~Wires with an injected AC transport
current}}
\markboth{\hspace*{0.1cm} \textsl{6.2.1 ~Wires with an injected AC transport
current}}{\hspace*{0.1cm}
\textsl{6.2.1 ~Wires with an injected AC transport
current}}
\addcontentsline{toc}{chapter}{\hspace*{0.1cm} \textsl{6.2.1 ~Wires with an
injected AC transport current}}\label{ch-6-2-1}
\vspace*{0.5cm}

\begin{figure}[t]
\centering
\includegraphics[height=7cm,width=13cm]{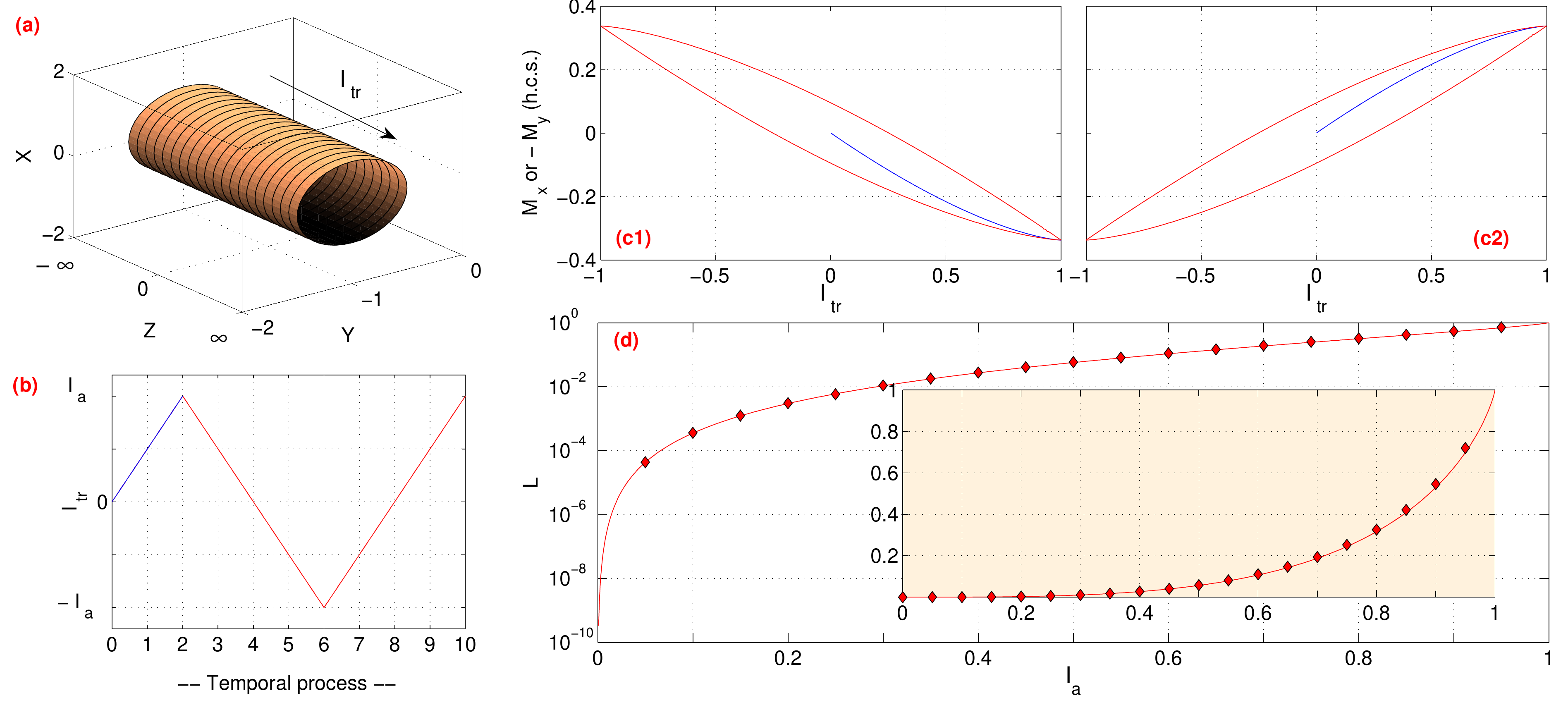}
\caption{\label{Figure_6_1} For the experimental configuration displayed in
subplot (a), and for an injected AC transport current of amplitude $I_{\tt{a}}$
according to the temporal process depicted in subplot (b), we show the 
calculated components of the magnetic moment for half cross section of the
superconducting wire, where (c1) corresponds to $\Omega^{+}$, i.e.:
$(x,y)\in\Omega~|~x>0$, and (c2) to $\Omega^{-}$, i.e.:
$(x,y)\in\Omega~|~x<0$. Notice that both sections cancel each other to
satisfy $\textbf{M}(\Omega)=0$. In subplot (d) we show the calculated
hysteretic AC losses per excitation cycle (since time-step 2 to 10). Red solid
line corresponds to the exact analytical expression of Eq.~(\ref{Eq.6.11}), and
solid diamonds corresponds to our numerical results. Inset shows the same
results in linear scale. Here and henceforth, units are $\pi
R^{2}J_{c}\equiv{I_c}$ for $I_{\rm tr}$, $(\mu_{0}/4\pi)J_{c}R$ for B,
$J_{c}R^{3}$ for M, and $(\mu_{0}/4\pi)\omega R^{2}J_{c}^{2}$ for the
hysteretic losses per cycles of frequency $\omega$.
}
\end{figure}

As it is well known, the magnetic flux penetrates a superconducting material
first entering from the surface towards the center while is shielded by 
screening currents flowing at the critical value $J_{c}$. Penetration occurs to
a depth known as the flux front boundary, where the
magnetic flux density drops to zero. Trivially, for long cylindrical wires
(length much higher than its radius $R$) with transport current [see
Fig.~\ref{Figure_6_1}(a)], the flux front profile may be defined in terms of a
set of circular front boundaries tracking the time evolution of the injected
transport current. For example, and mainly to illustrate how the
patterns of the main electromagnetic
quantities evolve along the cyclic process depicted in the
Fig.~\ref{Figure_6_1}(b), our numerical results for the local profiles of
current $I_{i}$, the components of the magnetic flux density $\textbf{B}$ and
the corresponding isolevels of the vector potential $\textbf{A}$ (i.e., the
flux lines of $\textbf{B}$), as well as the local distribution of the density of
power dissipation $\textbf{E}\cdot\textbf{J}$ are shown in the section of
supplementary material [Figs. S1 \&
S2 (pags.~\pageref{Figure_S_1},~\pageref{Figure_S_2})]. These figures will be of
much help for eventual discussion about the AC hysteretic losses when
simultaneous oscillating excitations must to be
considered. However, already for this simple case, several aspects have to be
noticed. 

First of all, let us to focus on the circulating transport currents so as to
visualize their contribution to the total magnetic moment of the superconducting
sample. Recall that, according to Eq.~(\ref{Eq.6.7}), those cases where only
transport current is applied, the total magnetization of the superconducting
sample becomes zero and the screening currents can be simply called
``\textit{injected current lines}''.  They distribute accomplishing the critical
state condition
$j_{i}=j_{c}$ in the regions where $B(r_{i})\neq0$. In fact, although these
currents produce local patterns of magnetization as it is shown in
Fig.~\ref{Figure_6_1}(c), by symmetry, the sum over the whole sample is fully
compensated. Nevertheless, for this case, \textbf{M}$(\Omega)=0 $ does not
imply the absence of hysteretic losses (see Fig.~\ref{Figure_6_1}(d)) which are
actually produced by the Joule heat release in those regions where flux
transport occurs. On the other hand, as one can see in Fig.~S1
(pag.~\pageref{Figure_S_1}), the local density of power
dissipation ($\textbf{E}\cdot \textbf{J}$) is not uniform despite the material
law assumes the critical state condition $j_{i}=j_{c}$. This means that,
locally one cannot assume a unique value for the electric field or $E=E_{c}$. In
fact, for this case the maximal power dissipation always occurs over the
superconducting surface decreasing to zero beyond the flux front.

Eventually, it is worth of mention that for this simple case the current
distribution can be reliably calculated by following the boundary flux front
defined through axisymmetric circumferences of radius
$\tilde{r}=R\sqrt{1-I_{tr}/I_{c}}$,
whose section $\tilde{r}<r<R$ produces the intensity of magnetic flux (in polar
coordinates):
\begin{eqnarray}\label{Eq.6.9}
B_{\phi}=\frac{\mu_{0}I_{c}}{2\pi r}
\left[\frac{I_{tr}}{I_{c}}+\left(\frac{r^{2}}{R^{2}}-1\right)\right] \, ~ \, ~
\,
\forall \, ~ \, ~ \, \tilde{r}<r<R \, .
\end{eqnarray}
Notice that here, $I_{tr}$ stands for the applied transport current at a given
time, and $I_{c}\equiv J_{c}\pi R^{2}$. Thus, the electric field can be
analytically calculated by the one dimensional Maxwell equation
$\partial_{r}E_{z}=\partial_{t}B_{\phi}$ satisfying
the condition $E_{z}(r<\tilde{r})=0$, i.e., 
\begin{eqnarray}\label{Eq.6.10}
E_{z}=\frac{\mu_{0}}{2\pi}\ln{\left[\frac{r}{R}\left(1-\frac{I_{tr}}{I_{c}}
\right)^{
-1/2 }\right]} \dot{I}_{tr}
 \, ~ \, ~ \,
\forall \, ~ \, ~ \, \tilde{r}<r<R \, .
\end{eqnarray}

In fact, if the temporal evolution of the injected transport current is
monotonic the average specific hysteretic loss rate per unit length can be
calculated by integration of the local density of power dissipation
($\textbf{E}\cdot \textbf{J}$),
\begin{eqnarray}
\dot{L}_{m}(I_{tr}(t))=
\frac{1}{\pi R^{2}}\int_{0}^{2\pi}\int_{\hat{r}}^{R}E_{z}J_{c}~r d\phi dr=
-\frac{\mu_{0}}{4\pi^{2}R^{2}}
\left[I_{tr}+I_{c}\ln\left(1-\frac{I_{tr}}{I_{c}}\right)\right]\dot{I}_{tr} \,
,\nonumber
\end{eqnarray}
and integrating it with respect to time, the monotonic hysteretic losses is
then 
\begin{eqnarray}\label{Eq.6.11}
{L}_{m}(I_{tr}(t))=
\frac{\mu_{0}}{4\pi^{2}R^{2}}
\left[I_{tr} I_{c} \left(1-\frac{I_{tr}}{2I_{c}}\right)
+I_{c}^{2}\left(1-\frac{I_{tr}}{I_{c}}\right)
\ln\left(1-\frac{I_{tr}}{I_{c}}\right)\right]
\end{eqnarray}

Moreover, if $I_{tr}(t)$ is periodic, the dependence of the hysteretic
loss density per period on $I_{tr}=I_{\tt{a}}$ ($\tt{a}$ for the amplitude of
the oscillating source) remains the same, and therefore the magnitude of the
loss per cycle may be straightforwardly obtained from the monotonic first
branch as, $L_{f.c.}=4 L_{m}$.

\vspace*{0.5cm}
\hspace*{-0.3cm}
\textbf{\textit{6.2.2 ~Wires under an external AC
magnetic flux}}
\markboth{\hspace*{0.1cm} \textsl{6.2.2 ~Wires under an external AC
magnetic flux}}{\hspace*{0.1cm}
\textsl{6.2.2 ~Wires under an external AC
magnetic flux}}
\addcontentsline{toc}{chapter}{\hspace*{0.1cm} \textsl{6.2.2 ~Wires under an
external AC
magnetic flux}}\label{ch-6-2-2}
\vspace*{0.5cm}

With the aim of providing a clear picture of the effects related to the
occurrence of local \textit{magnetization currents} (screening currents produced
by external magnetic fields), the second case under consideration will
correspond to a superconducting wire under zero transport current condition
($I_{tr}(t)=0$) and subjected to an external magnetic field perpendicular to its
surface. Here, we must call readers' attention to the fact that for this
seemingly simple case an exact analytical solution for the dynamics of the
flux front profiles has not been reported, although remarkable efforts have
been done along the last five decades to implement diverse analytical and
numerical approaches~\cite{P2-Ashkin_1979,P2-Carr_1975,P2-Zenkevitch_1980,
P2-Minervini_1989,P2-Carr_2001,P2-Gurevich_1997,P2-Gomory_2002}. In fact,
for an arbitrary relation between the amplitude of the applied field ($H_{a}$)
and the full penetration value $H_{p}$, the cyclic hysteretic losses can only be
found numerically. Nevertheless,
for cylindrical superconducting wires subjected to a monotonic source
$H_{0}(t)$, the losses may be approached by the so-called Gurevich's
relation \cite{P2-Gurevich_1997}:
\begin{eqnarray}\label{Eq.6.12} 
L_{m}(B_{0}(t))=\dfrac{2B_{p}^{2}}{3\mu_{0}}
\left\{
\begin{array}{ll}
\left(\dfrac{B_{0}}{B_{p}}\right)^{3} \left(1-\dfrac{B_{0}}{2B_{p}}\right)
\,~\,~ & ,~\, \forall ~\, ~\, B_{0}<B_{p} \, 
\\ \, \\
\dfrac{B_{0}}{B_{p}}-\dfrac{1}{2}\,~\,~ & ,~\, \forall ~\, ~\, B_{0}\geq B_{p}
\, ~ \, ~ \, .
\end{array}
\right. 
\end{eqnarray}
Here, the customary relation for superconducting materials
$\textbf{B}=\mu_{0}\textbf{H}$ has been
assumed, and $B_{p}$ is given by:
\begin{eqnarray}\label{Eq.6.13}
B_{p}=\frac{2}{\pi}\mu_{0}J_{c}R \, .
\end{eqnarray}

Just as in the previous case (only transport current), if $B_{0}(t)$ is
periodic, the dependence of the hysteretic loss density per period remains the
same, and the magnitude of the cyclic losses is higher by a factor of four
regarding the monotonic branch.

\begin{figure}[t]
\centering
\includegraphics[height=8cm,width=13cm]{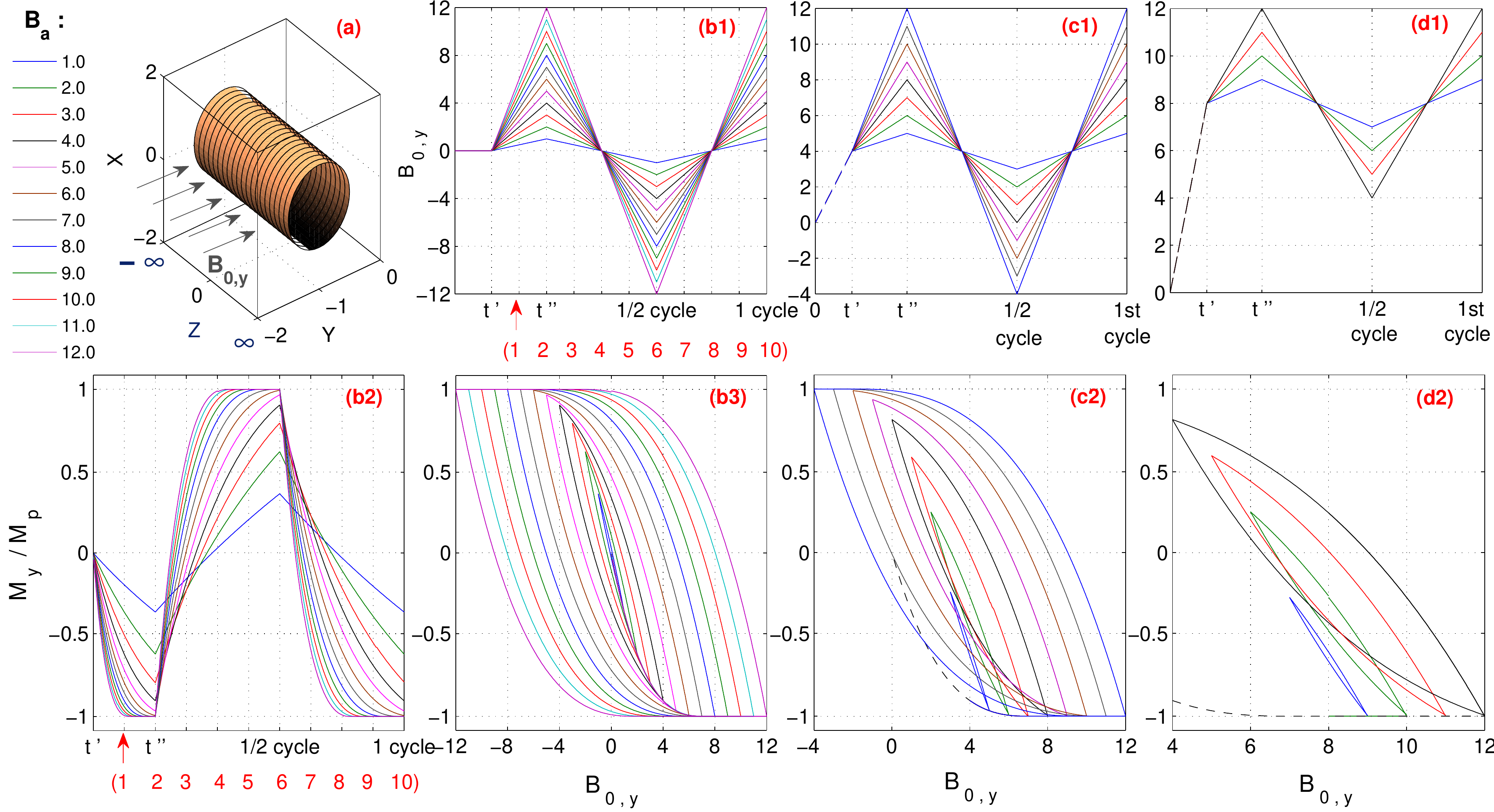}
\caption{\label{Figure_6_2} For the experimental configuration represented in
subplot (a), and before applying an oscillating magnetic flux with amplitude
$B_{\tt{a}}=B_{0,y}(t'')-B_{0,y}(t')$, three different initial
states $(B_{0,y}(t'))$ have been considered. Firstly, a non magnetized wire has
been assumed accordingly to the magnetic temporal process displayed in subplot
(b1). Subplots (b2) and (b3) show the evolution of the  dimensionless magnetic
moment $M_{y}/M_{p}$ as a function of the temporal process and the applied
magnetic field $B_{0,y}$ respectively. Secondly, a premagnetized sample with
$B(t')<B_{p}$ has been considered (subplot c1), and the corresponding 
$M_{y}/M_{p}$ curves are displayed (subplot c2). Thirdly, in (d1) the initial
state has been assumed to be $B(t')=B_{p}$, and the corresponding 
$M_{y}/M_{p}$ curves are displayed in (d2). For all plots, curves are labeled
according to $B_{\tt{a}}$. In subplots (b1) and (b2) an additional temporal
scale in red has been incorporated to allow an straightforward interpretation
for the sequence of profiles displayed in the section of
supplementary material, Figs. S3 (pag.~\pageref{Figure_S_3}) \& S4
(pag.~\pageref{Figure_S_4}). Recall that units for B are
$(\mu_{0}/4\pi)J_{c}R$, and $J_{c}R^{3}$ for M.
}
\end{figure}

\paragraph{(~\textit{A.}~) \textit{Features on the magnetic
response}\vspace*{0.5cm}\\}

For a major understanding of the electromagnetic quantities
involved, we have considered the
experimental configuration depicted in Fig.~\ref{Figure_6_2}(a) under different
amplitudes for the applied magnetic flux $B_{0,y}$. Just for completeness,
three different magnetization pre-conditions have been considered as relates
the first branch of the temporal processes $t<t'$ (see top pane in
Fig.~\ref{Figure_6_2}). Thus, subplots (b1) to (b3) correspond to a non
magnetized wire for which one gets magnetization loops centered at the
coordinates (0,0) of the plane ($M_{y},B_{0,y}$). However, in those cases where
$B_{0,y}(t')\neq0$ [Figs.~\ref{Figure_6_2} (c1) and (d1)] a different behavior
reveals. An initial magnetization branch is present when the external AC
magnetic flux is  switched on ($t=t'$) [see
dashed lines in Figs.~\ref{Figure_6_2} (c2) and (d2)]. Then, for $t>t'$ the SC
wire is subjected to an oscillating magnetic flux of
amplitude $B_{\tt{a}}$, such that the first quarter of the subsequent periodic
excitation occurs along the interval $t'<t<t''$, and consequently the AC losses
are calculated relative to cycles departing from $t''$. For the oscillating
process in the above figure, the alternating magnetic field is applied along the
same direction of the previous magnetization under the flux conditions
$B_{0,y}(t')<B_{p}$ and $B_{0,y}\geq B_{p}$ respectively. The following facts
have to be noticed:\footnotemark[2]

\footnotetext[2]{In what follows, the reader must recall that dimensionless
units have been defined according to Fig.~\ref{Figure_6_1} as follows: $\pi
R^{2}J_{c}\equiv{I_c}$ for $I_{\rm tr}$, $(\mu_{0}/4\pi)J_{c}R$ for B,
$J_{c}R^{3}$ for M, and $(\mu_{0}/4\pi)\omega R^{2}J_{c}^{2}$ for the
hysteretic losses per cycles of frequency $\omega$.}

(\textit{i}) Concerning to the numerical accuracy, it has to be noticed that the
mutual inductance matrices used in this fore, corresponds to the exact
analytical solution for filaments of cylindrical cross section. Thus, some
discrepancies between the analytical and numerical quantities can be expected.
In fact, in the classical results of Ref.~\cite{P2-Ashkin_1979} it may be
noticed that the magnetization curve saturates to a flat value ($M_{p}$) before
reaching the analytical limit $B_{p}=8$, which is in agreement with our
numerical predictions in Fig.~\ref{Figure_6_2} ($B_{p}\simeq
7\pm5\times10^{-3}$). Accordingly to our
numerical method, discrepancies with
the analytical solution are explained in geometrical terms. Notice that even for
a highly refined mesh, by assuming cylindrical filaments, the entire
superconducting area cannot be filled, and therefore some deviation should be
expected. We emphasize that other choices for the mesh elements filling
the superconducting area could be done, but the complication seems unnecessary
if one considers that the accuracy in the obtained physical quantities is
already very high, and furthermore this mesh has been recognized to fit well to
experimental evidences in the same geometry~\cite{P2-Ashkin_1986}.

(\textit{ii}) Recalling Eq.~(\ref{Eq.6.7}), the sum of the local
magnetic moments associated to each of the filaments over the entire
superconducting sample increases as $B_{0,y}$ grows monotonically until the
condition $B_{0,y}=B_{p}$ is met. Then, the magnetization of the sample
saturates to
the analytical value $M_{p}=2J_{c}R^{3}/3$ (in our dimensionless units
$M_{p}=2/3\approx0.6667$). Regarding to our numerical results we have
obtained $M_{p}\approx0.655\pm5\times10^{-3}$.

(\textit{iii}) Once the cyclic process starts ($t>t''$), and regarding the 
magnetization loops characterized by AC cycles of external magnetic flux with
amplitudes higher than $B_{p}$, the magnetic moment saturates at different
values given by the dimensionless relation
\begin{eqnarray}\label{Eq.6.14}
B_{p^{\dag}}=\mp \left( 2 B_{p}- B_{\tt{a}} \mp B_{0}(t')-1/2\right) \, ,
\end{eqnarray}
where the signs choice is made simultaneously, it for consider the time
derivative of the cyclic excitation.

For example, in Fig.~\ref{Figure_6_2}(b3) if $B_{0}(t')=0$ and the amplitude of
the external magnetic flux is $B_{\tt{a}}=8$, into the cyclic process $(t>t'')$
the magnetization of the superconducting wire saturates at $B_{0,y}=\mp5.5$. On
the other hand, if the sample has been previously premagnetized, the center of
the magnetization loop is displaced in the axis $B_{0,y}$ of the plane
$(M_{y},B_{0,y})$ by the amount $B_{0,y}(t')$ [see Fig.~\ref{Figure_6_2}
(c2,d2)], and therefore the sample saturates at two
different values of the applied magnetic field (e.g., for $B_{\tt{a}}=8$,
$M_{y}$ is equals to $M_{p}$ at $B_{0,y}=-1.5$, and $M_{y}$ is equals to
$-M_{p}$ at $B_{0,y}=9.5$). 

(\textit{iv}) Remarkably, the set of magnetization loops displayed in
Fig.~\ref{Figure_6_2} serves as a map for any magnetization loop and any
arbitrary relation between the experimental parameters $B_{0,y}(t')$ and
$B_{\tt{a}}$. In
fact, it can be done by the simple interpolation of the known shape
of the magnetization loops where the first corner of $M_{y}$ (corresponding to
the higher excitation peak for the first half of the excitation cycle) always
falls over the magnetization branch for $t<t''$. Moreover, if
$B_{\tt{a}}+B_{t'}>B_{p}$ the position of the corners is straightforwardly
given by the saturation values $B_{p^{\dag}}$.

\paragraph{(~\textit{B.}~) \textit{Flux penetration and local
power density}\vspace*{0.5cm}\\}

For a clear understanding of the different terms
affecting the calculation of the hysteretic losses in superconducting wires, it
is advisable to get familiar with the local dynamics of the electromagnetic
quantities in the same way as it was done in the previous section. Therefore, in
the section of supplementary material, readers will find out some of our
numerical results for one of the experimental processes depicted in
Fig.~\ref{Figure_6_2}. In particular, in Fig.~S3 (pag.~\pageref{Figure_S_3}) we
show the flux penetration profiles for an external AC magnetic flux of amplitude
$B_{\tt{a}}=6$ at intervals of $\Delta B_{0,y}=3$, assuming an initially non
magnetized wire [see by reference
the temporal process depicted in Fig.~\ref{Figure_6_2}(b)]. Also shown are
corresponding patterns for the local density of power dissipation across the
section of the superconducting wire. The direction of the magnetic
field can be tracked from the dynamics of the Cartesian components $B_{x}$ and
$B_{y}$ both displayed in Fig.~S4 (pag.~\pageref{Figure_S_4}). 

It becomes clear that the distribution of magnetization currents across
the section of the superconducting wire preserves some symmetry respect to
both Cartesian axes, although rotational invariance characteristic for
transport problems [Fig.~S1 (pag.~\pageref{Figure_S_1})] is not
fulfilled. Actually, we can argue that for cases with no rotating
transverse magnetic field the numerical problem may be reduced to considering
only two of the four Cartesian quadrants according to the following
symmetry condition: (\textit{i}) $I_{i}(r_{i}(y^{+})=I_{i}(r_{i}(y^{-})$, being
``$y$'' the axis of the applied magnetic field $B_{0}$. Moreover, for our
current case also the symmetry condition  (\textit{ii})
$I_{i}(r_{i}(x^{+})=-I_{i}(r_{i}(x^{-})$, being $x$ the axis perpendicular to 
the direction of $B_{0}$ may be called. However, the latter can only
be fulfilled as long as the transport current condition $I_{tr}=0$ occurs.

As it may be observed in Fig.~S3 (pag.~\pageref{Figure_S_3}), once the external
magnetic flux $B_{0,y}$ is
switched on, a set of screening currents symmetric along the $y$-axis but
antisymmetric by sign along the $x-$axis appears so as to expell the magnetic
field
from the inner sample [see also Fig.~S4 (pag.~\pageref{Figure_S_4})].
Actually, for transverse magnetic
fields, it is the Faraday's law which produces the
simultaneous occurrence of positive and negative screening currents
distributed along the positive and negative semi-axis, but both orthonormal to
the direction of the applied magnetic field. Thus, if the rate $\Delta
B_{0,y}(t)$ is monotonic, the shape of the flux free region approaches to
ellipses with their foci along the $y-$axis, but with acute nodes in the
boundaries of their major
axis. Moreover, the associated lengths to the major and minor semi-axis of
the ellipse also change responding to the variation of the external magnetic
flux in the different points where the screening currents are allocated, which
is in concordance with
Refs.~\cite{P2-Ashkin_1979,P2-Telschow_1994,P2-Kuzovlev_1995,P2-Bhagwat_2001,
P2-Navarro_1991,P2-Gomory_2002,P2-Brandt_1996}.

As regards the local power loss density [right pane in
Fig.~S3 (pag.~\pageref{Figure_S_3})], we have observed some fine structure
details which are worth of
mention. First, if the rate of magnetic flux $\Delta B_{0,y}$ is monotonic
(without change of sign), it is to be noticed that the specific local power
density preserves the same kind of pattern observed for the dynamics of the
local profiles of the magnetic flux density component $B_{y,i}$. However, the
role played by the orthonormal component $B_{x,i}$ is rather different and
geometry dependent [see subplots (1) and (2) in Figs.~S3
(pag.~\pageref{Figure_S_3}) \& S4 (pag.~\pageref{Figure_S_4})]. Then, as soon
as $\Delta B_{0,y}$ changes sign, the local distribution of power
density is much more complex as the core enclosed by the flux fronts satisfying
the condition $E=0$ must be shielded [see the sequence of subplots (3) to (10)
in Figs.~S3 (pag.~\pageref{Figure_S_3}) \& S4 (pag.~\pageref{Figure_S_4})].

\begin{figure}[t]
\centering
\includegraphics[height=8cm,width=10cm]{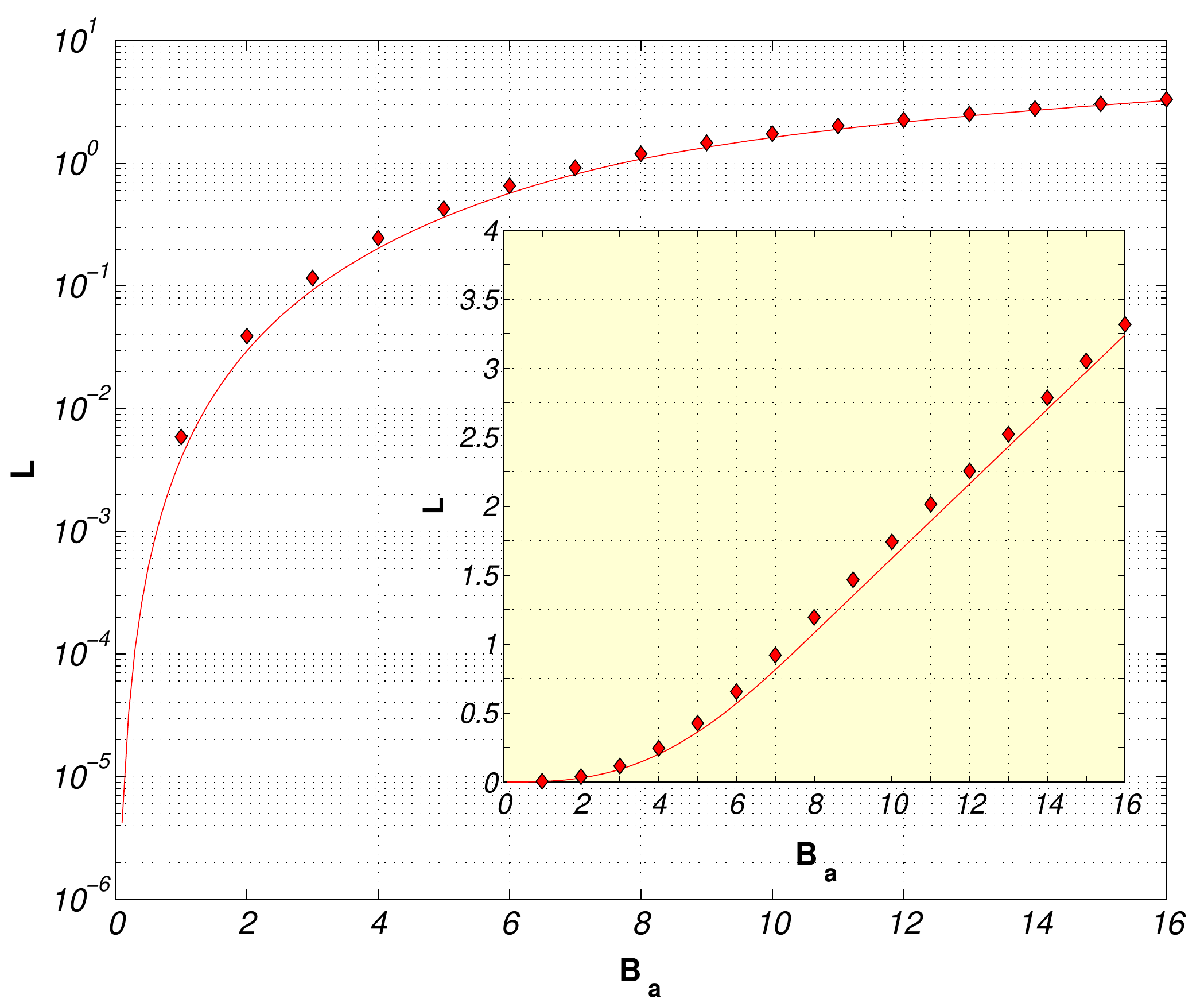}
\caption{\label{Figure_6_3} Hysteretic AC losses per cycle 
for an oscillating magnetic field of amplitude $B_{\tt{a}}$. Red solid
line corresponds to the analytical approach of Eq.~(\ref{Eq.6.12}), and
diamonds correspond to our numerical calculations.
Inset shows the same results in linear scale. Units are $(\mu_{0}/4\pi)J_{c}R$
for $B_{0,y}$, and $(\mu_{0}/4\pi)\omega R^{2}J_{c}^{2}$ for the hysteretic
losses per cycles of frequency $\omega$.
}
\end{figure}

In general, power losses appear for the whole section where the
magnetization currents are displayed, but their maximum contribution always
occurs over the external surface of the superconducting sample. On the other
hand, for the hysteretic losses per cycle the total amount of heat release can
be straightforwardly calculated based on energy transport over the external
surface of the superconducting sample as explained below.

\vspace*{1.0cm}
\hspace*{-0.3cm}
\textbf{\textit{6.2.3 ~Ultimate considerations on the AC losses}}
\markboth{\hspace*{0.1cm} \textsl{6.2.3 ~Ultimate considerations on the AC
losses}}{\hspace*{0.1cm}
\textsl{6.2.3 ~Ultimate considerations on the AC losses}}
\addcontentsline{toc}{chapter}{\hspace*{0.1cm} \textsl{6.2.3 ~Ultimate
considerations on the AC losses}}\label{ch-6-2-3}
\vspace*{1.0cm}

At this point and after gaining some experience by the analysis of wires under
transport or applied magnetic field, we are in the position of
casting Eq.~(\ref{Eq.6.8}) in different forms for
convenient interpretation of the AC losses. Thus, taking
advantage of the knowledge generated in the above sections, below we reformulate
the argument $\dot{L}$ of Eq.~(\ref{Eq.6.8}) in terms of the definitions for
$\textbf{E}$ or $\textbf{J}$. This will allow a physical interpretation of the
mechanism responsible for AC losses in complex configurations.

\paragraph{\label{B-oriented}(~\textit{A.}~) \textit{Conventional approach
[B-oriented]}\vspace*{0.5cm}\\}

First of all, recalling that the electric field may be defined for discretized
time layers as $\textbf{E}=-\Delta \textbf{A} / \Delta t$, and furthermore, that
it can be computed from the sum of the external vector potential
$\textbf{A}_{0}(\textbf{B}_{0})$ [see Eq.~(\ref{Eq.6.5})] and the vector
potential induced by the screening currents
$\textbf{A}_{ind}(\textbf{J}(r_{i}))$
[see Eqs.~(\ref{Eq.6.1}) \& (\ref{Eq.6.2})], the losses of power density
$\dot{L}\equiv \Delta L/\Delta t$ may be rewritten as
\begin{eqnarray}\label{Eq.6.15}
\Delta L=-\int_{\Omega} \Delta\textbf{A}_{0}\cdot \textbf{J} ~ dr
-\int_{\Omega} \Delta\textbf{A}_{ind}\cdot \textbf{J} ~ dr \, .
\end{eqnarray}
Thus, recalling Eq.~(\ref{Eq.6.7}), the first term on the
right-hand side of Eq.~(\ref{Eq.6.15}) becomes
\begin{eqnarray}\label{Eq.6.16}
\Delta L_{0}(\Delta B_{0,y})=-\int_{\Omega} \Delta \textbf{A}_{0}\cdot
J ~ dr \equiv \int_{\Omega}M_{y} ~ dB_{0,y} \, ,
\end{eqnarray}
which corresponds to the losses produced by an external excitation of magnetic
flux density $B_{0}(t)$. 

In fact, since the work by Ashkin (Ref.~\cite{P2-Ashkin_1979}), it is well
known that for screening currents produced by external variations of an applied
magnetic field of flux density $B_{0,y}$, the so called magnetization currents,
the average AC losses can be straightforwardly calculated as the enclosed area
by the magnetization loop $M_{y}$ as a function of the excitation $B_{0,y}$
(Fig.~\ref{Figure_6_2}). Thus, as long as the experimental configuration
is such that $I_{tr}(t)=0$, physically the AC hysteretic losses may
be understood as surface losses due to remagnetization.

Nevertheless, as in most applications of superconducting wires, the system is
subjected to a simultaneous oscillating transport current, a most careful
analysis of the second term at the right-hand side of Eq.~(\ref{Eq.6.15}) is
needed.

First recall that in the CS theory the MQS approach allows us to define
Ampere's law as the spatial condition
$\mu_{0}\nabla\times\textbf{B}=\textbf{J}$, where $\textbf{B}_{i}$ is
the local
density of magnetic flux produced by the induced screening currents of density
$\textbf{J}(\textbf{r}_{i})$ plus the density of magnetic flux
produced by the external excitation $\textbf{B}_{0}(\textbf{r}_{i})$, i.e.,
$\textbf{B}_{i}=\textbf{B}_{0}(\textbf{r}_{i})+\textbf{B}_{ind}(\textbf{r}_{ij}
)$, so that the second term at the right-hand side of
Eq.~(\ref{Eq.6.15}) may be rewritten as
\begin{eqnarray}\label{Eq.6.17}
\Delta L_{ind}=-\mu_{0}\int_{\Omega} \Delta\textbf{A}_{ind}\cdot
\nabla\times\left(\textbf{B}_{0}+\textbf{B}_{ind}\right) ~ dr \, .
\end{eqnarray}

On the other hand, accordingly to the CS statement, $\Delta{J_{i}}=\pm
J_{c}\neq 0$ accomplishes for the flux front profile, ergo $\Delta
\textbf{A}_{ind}\equiv {\rm M}_{ij}\Delta J_{i} \neq0$. Then, for systems only
subjected to external magnetic fields, the distribution of screening currents
over the cross section of the superconducting sample is such that the induced
magnetic field over flux front is rotationally invariant with respect to
the direction of the applied magnetic field. For example, considering the first
monotonic branch in cases of Fig.~\ref{Figure_6_2}, the flux front profile is
defined by $B_{ind}(t)=-B_{0,y}(t)$. Said in other words, the induced magnetic
field flows only in opposite direction to $B_{0}$ so that $\Delta L_{ind}=0$ as
long as $I_{tr}(t)\equiv 0$, and therefore the hysteretic losses may be
straightforwardly computed  as $\Delta L\equiv \Delta L_{0}$ in the fashion of
Eq.~(\ref{Eq.6.16}). Nevertheless, if $I_{tr}(t)\neq 0$ the local density of
magnetic flux is not rotationally invariant with respect to the components of
the induced magnetic field. For example, for the cases described in the
previous subchapter (Fig.~\ref{Figure_6_1}), where $B_{0,y}(t)\equiv0$ (absence
of magnetization currents), the magnetic flux density produced by the
screening currents $J_{i}$, so called there \textit{injected current lines},
shows two components $B_{x,i}$ and $B_{y,i}$ so that
$\nabla\times\textbf{B}_{i}\neq0$,
and thence that $\Delta L\equiv\Delta L_{ind} \neq 0$ despite
$\int_{\Omega}M(\textbf{r}_{i})\equiv0$.

In order to understand the underlying physics behind the the concept
of injected current lines into the critical state theory, at least for the 2D
configurations studied along this chapter, it becomes
useful to further analyze Eq.~(\ref{Eq.6.17}) by the vectorial definition
$\textbf{A}\cdot\nabla\times\textbf{B}=$
$\nabla\cdot\textbf{B}\times\textbf{A}+$
$\textbf{B}\cdot\nabla\times\textbf{A}$, such that we can write
\begin{eqnarray}\label{Eq.6.18}
\Delta L_{ind}=-\mu_{0}\int_{\Omega}
\nabla\cdot(\textbf{B}\times \Delta \textbf{A}_{ind}) ~ dr -
\mu_{0}\int_{\Omega}\frac{\textbf{B}\cdot\Delta \textbf{B}}{2} ~ dr
 \, . 
\end{eqnarray}
Into this framework the first integral turns to a
surface integral over the curved walls of the cylinder, which does not
contribute to the average losses because the contribution to the surface
integral from the end planes vanishes as
$\textbf{B}\times\textbf{A}_{ind}\cdot \hat{\textbf{z}}=0$, while the integral
over the lateral surface turns zero in a closed cycle. Therefore, the
contribution of the injected current lines may be understood by the simple
relation 
\begin{eqnarray}\label{Eq.6.19}
\Delta L_{ind}=-\mu_{0}\int_{\Omega}\frac{\textbf{B}\cdot\Delta
\textbf{B}}{2} ~ dr 
 \, ,
\end{eqnarray}
where it is pointed out that significant reductions of the hysteretic losses
may be achieved by reducing the magnitude of the local inductive magnetic field.

\paragraph{\label{S-oriented}(~\textit{B.}~) \textit{Alternative approach
[S-oriented]}\vspace*{0.5cm}\\}

In order to justify that the hysteretic losses may be calculated by the
knowledge of the
electromagnetic quantities over the external surface of a superconducting
material, it is interesting to transform Eq.~(\ref{Eq.6.8}) in terms of a
second formulation based upon the definition of $\textbf{J}$ instead of
$\textbf{E}$. Indeed, assuredly the specific losses can be evaluated by the
conservation energy principle defining
$\textbf{J}=\nabla\times\textbf{H}-\partial_{t}\textbf{D}$ and by using
the divergence theorem as follows:

\begin{eqnarray}\label{Eq.6.20}
\dot{L}=\int_{\Phi}\textbf{E}\cdot\left(\nabla\times\textbf{H}-\partial_{t}
\textbf{D}\right) dr =
-\int_{\Phi}(\textbf{E}\cdot\partial_{t}\textbf{D}+\textbf{H}
\cdot\partial_{t}\textbf{B}) ~ dr -
\oint_{s}\textbf{S}\cdot\hat{\textbf{n}} ~ ds \, , \nonumber \\
\end{eqnarray}

\noindent where $\Phi$ is introduced to distinguish the volume integral over the
entire superconducting sample from the surface integral over the flux fronts
defined by the Poynting's vector $\textbf{S}=\textbf{E}\times\textbf{H}$, and 
$\hat{\textbf{n}}$ the unit vector normal to its surface
element $(d\hat{\textbf{s}}_{i}=ds\hat{\textbf{n}})$.

The first term on the right-hand side of Eq.~(\ref{Eq.6.20}) represents the
total electromagnetic energy stored within the superconductor volume, and the
second one corresponds to the energy flow produced by the local variations
of magnetic field as a consequence of the occurrence of screening currents.
Then, if one is only interested on the hysteretic losses per closed cycles, one
is entitled to evaluate the AC losses between two well-defined stationary
regimes,
\begin{eqnarray}\label{Eq.6.21}
L= - \omega \int_{peak~a}^{peak~b} dt
\oint_{s}\textbf{S}\cdot\hat{\textbf{n}} ~ ds
= -2 \omega \int_{h.f} dt
\oint_{s}\textbf{S}\cdot\hat{\textbf{n}} ~ ds
 \, , 
\end{eqnarray}
because the total electromagnetic energy is a conserved quantity between two
consecutive peaks defined by the wavelength of the
oscillating source\footnotemark[2]. In fact, the calculation can be simplified
following the same argument to consider only half cycle ($h.f$) of the
excitation process. Remarkably, this fact can be straightforwardly observed by
comparison between the local profiles $\textbf{E}\cdot \textbf{J}$ for the
excitations of Figs.~\ref{Figure_6_1}(a) or \ref{Figure_6_2}(b1), accordingly
to the steps (6) and (10) in  Figs.~S1 (pag.~\pageref{Figure_S_1}) and S3
(pag.~\pageref{Figure_S_3}), respectively.

\footnotetext[2]{For the temporal processes displayed in
Figs.~\ref{Figure_6_1}(a) or \ref{Figure_6_2}(b1), the aforesaid peaks
in Eq.~(\ref{Eq.6.21}) refers to the pointed steps 2 (\textit{peak a}) and
10 (\textit{peak b}).}

Finally, taking advantage of the two dimensional symmetry of our problem (as
we have assumed wires of infinite length), Eq.~(\ref{Eq.6.18}) may be
transformed to a path integral for the flux
of energy over the external surface of the superconducting wire, which for
cylindrical wires in polar coordinates is equivalent to say that the hysteretic
losses per unit length can be reliably calculated by the following expression:

\begin{eqnarray}\label{Eq.6.22}
L = -2 \omega \int_{h.f} dt \oint_{l}\textbf{S}\cdot
d\hat{\textbf{\textit{l}}} = \left. -2 \omega R \int_{h.f} dt 
\oint_{l}\textbf{S}\cdot \hat{\textbf{r}} ~ d\phi ~ \right|_{r=R}
\, .
\end{eqnarray}
%


\vspace*{1.0cm}
\subsection*{\label{ch-6-3}
\hspace*{0.1cm} \textsl{6.3 SC wires under simultaneous
AC excitations $(\textbf{B}_{0}$ , $I_{tr})$}}
\vspace*{1.0cm}
\markboth{\hspace*{0.1cm}\textsl{6.3 SC wires under simultaneous
AC excitations $(\textbf{B}_{0}$ , $I_{tr})$}} 
{\hspace*{0.1cm} \textsl{6.3 SC wires under simultaneous
AC excitations $(\textbf{B}_{0}$ , $I_{tr})$}}
\addcontentsline{toc}{chapter}
{\hspace*{0.1cm} \textsl{6.3 SC wires under simultaneous
AC excitations ($\textbf{B}_{0}$ , $I_{tr}$)}}

As we have mentioned before, the implementation of superconducting wire
technology straightforwardly depends on the demonstration of their reliability,
competitive advantages in terms of improved efficiency and reduced operating
costs, with capital costs comparable to those of conventional devices. Thus, for
the development of competitive devices for the industry, it is important to
precisely understand the AC loss properties when realistic non-trivial AC
excitations have to be considered. In fact, almost in all the conceived
applications for superconducting wires, is well known that each one of the wires
holds an AC transport current and experiences an additional AC magnetic field
due to the neighboring wires. This situation is found, for example, in
superconductor windings for AC magnets, generators, transformers and motors,
where each turn feels the magnetic field of all the
others~\cite{P2-Larbalestier_2001,P2-Oomen_2003,P2-Hull_2003,P2-Paul_2005,
P2-Yunis_1995,P2-Takeuchi_1998}. 

The first conceptualization of the problem of
superconducting wires under configurations of simultaneous alternating current
and applied magnetic field was provided in 1966 by
Hancox~\cite{P2-Hancox_1966}, who studied the AC losses through simplified
analytical methods for determining the flux front profile in an infinite slab
subjected to a field applied parallel to the direction of the injected transport
current. This work went almost unnoticed for over a decade, until a similar
approach was proposed in 1979 by Carr~\cite{P2-Carr_1979}. Then, the same kind
of experimental configurations but for monotonic
rates of the experimental sources $(B_{0},I_{tr})$ has been studied since the
1990's, under the assumptions of very thin superconducting strips to allow
different analytical
considerations~\cite{P2-Brandt_1993,P2-Zeldov_1994,P2-Schonborg_2001}. However,
in more realistic situations, where the cross section of the superconducting
sample cannot be reduced one dimension, the use of exact analytical methods
is not feasible. Thus, the use of numerical methods as the described in previous
sections becomes in the more attainable procedure for the forecast and
understanding of the electromagnetic observables such as the magnetization
curves and the AC power density losses. 

It is worth mentioning that despite the fact that there is a significant
number of works assuming isolated superconducting wires of diverse geometries,
mainly strips subjected to synchronous
excitations~\cite{P2-Carr_1979,P2-Amemiya_1998,
P2-Yazawa_1998,P2-Yazawa_1999,P2-Amemiya_2001,P2-Schonborg_2001,
P2-Zannella_2001,P2-Stavrev_2002,P2-Tebano_2003,P2-Tonsho_2003,P2-Ogawa_2003,
P2-Enomoto_2004,P2-Stavrev_2005,P2-Nguyen_2005a,P2-Nguyen_2005b,
P2-Vojenciak_2006,P2-Pardo_2007,P2-Pardo_2005,P2-Pi_2010,P2-Thakur-2011a,
P2-Thakur-2011b}, a thoroughly study
of cylindrical superconducting wires was
still absent, and therefore some outstanding predictions had not been reported
before~\cite{P2-Ruiz_APL_2012}. 

\begin{figure}[t]
\centering
\includegraphics[height=8cm,width=13cm]{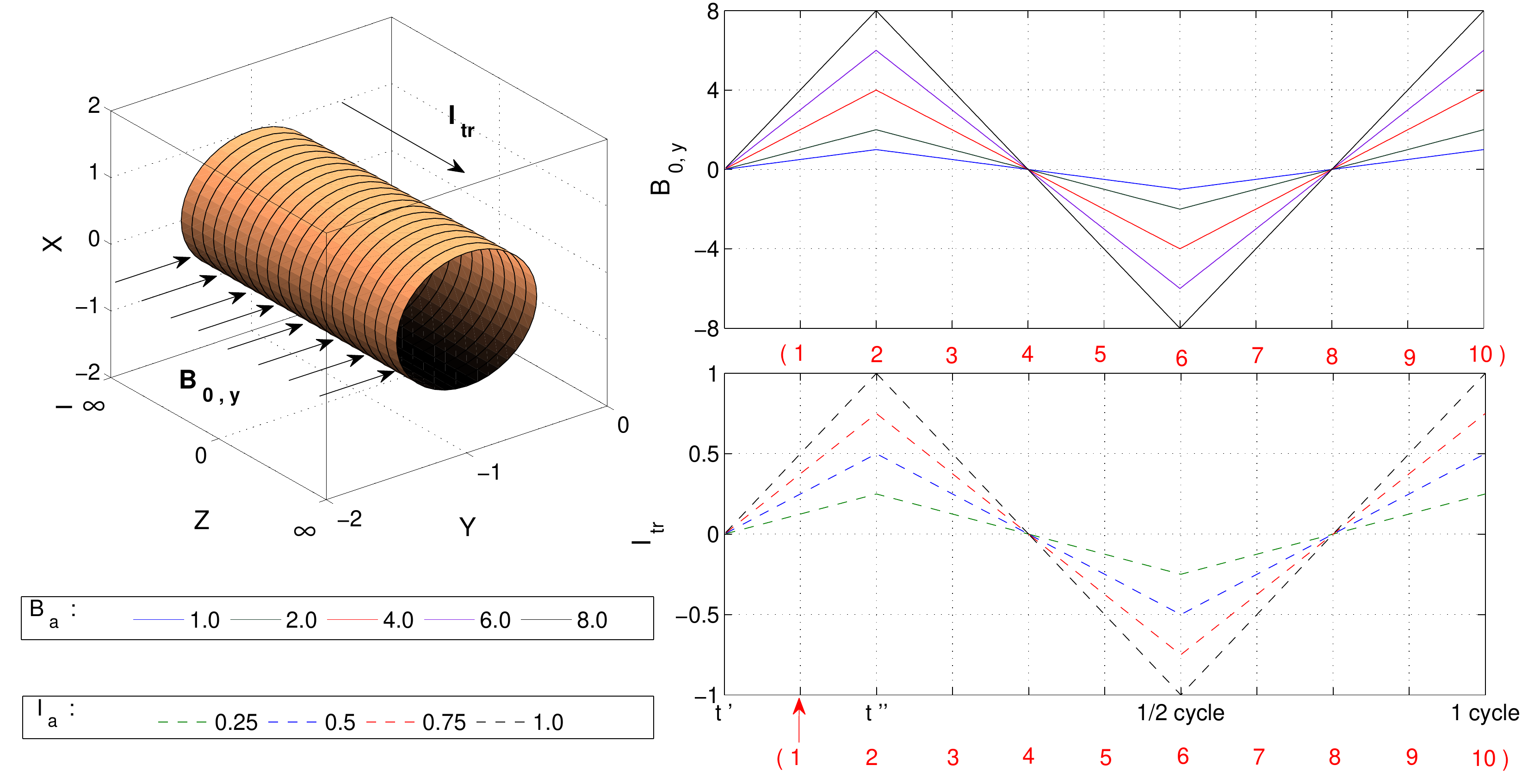}
\caption{\label{Figure_6_4} Sketch of some of the experimental processes
analyzed along this chapter. Here, cylindrical SC wires subjected to
synchronous oscillating excitations $B_{0,y}$ and $I_{tr}$, of amplitudes 
$B_{\tt{a}}$ and $I_{\tt{a}}$ have been considered, 
}
\end{figure}

\begin{figure}[t]
\centering
\includegraphics[height=10cm,width=12cm]{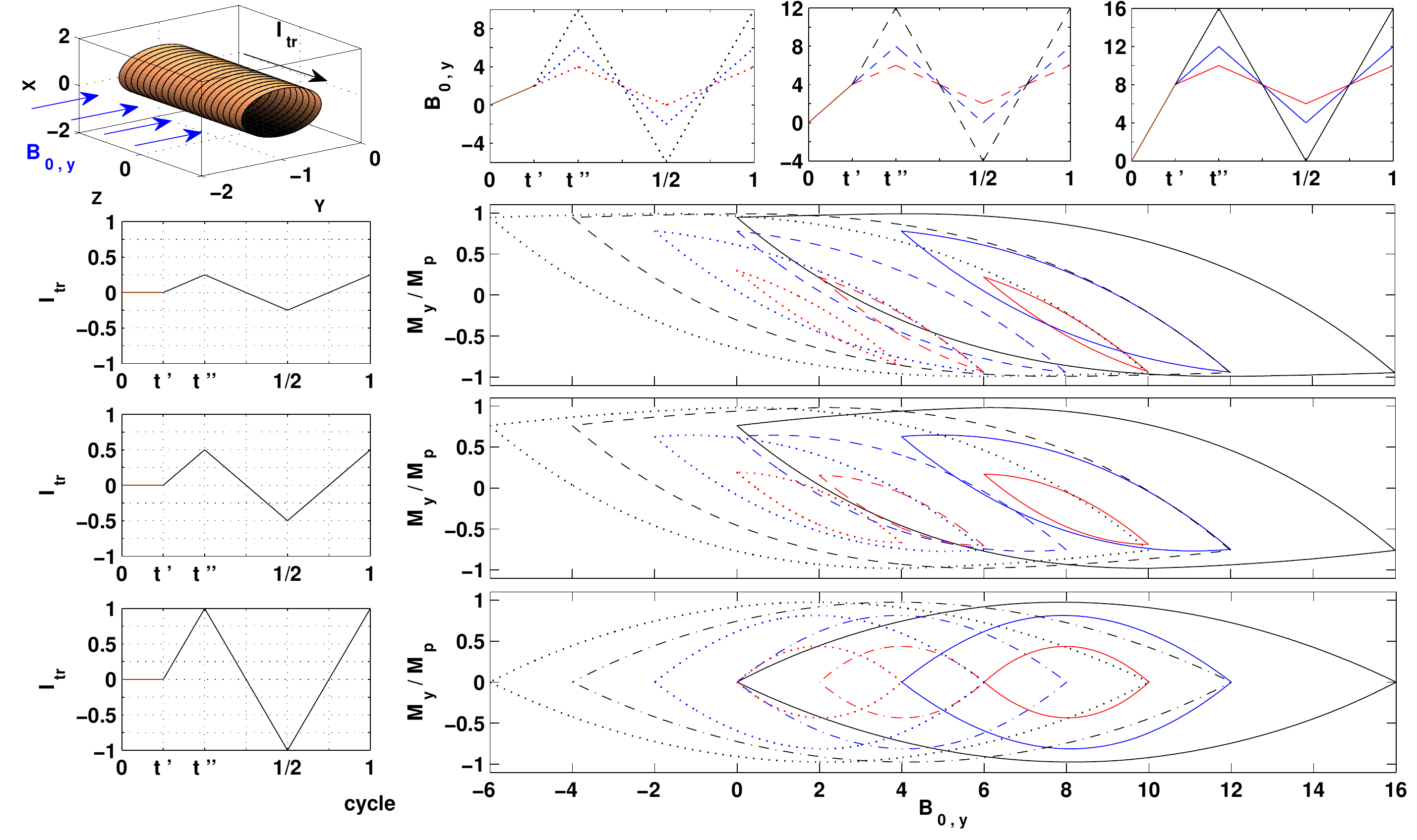}
\caption{\label{Figure_6_9} Dimensionless magnetic moment
$M_{y}/M_{p}$ as a function of the applied magnetic field $B_{0,y}$ for cycles
of simultaneous ac excitations $B_{0,y}$ and $I_{tr}$ of amplitudes
$(B_{\tt{a}},I_{\tt{a}})$. Although the excitation peak to peak of both sources
assumes
synchronous and with equal frequency, several premagnetized samples have been
considered according to: $B(t')=2$, $B(t')=4$ and
$B(t')=8$ (see 1st row), for the time instant when the ac current
$I_{tr}(t')$ is switched on. Regarding to the cyclic process (i.e, from $t''$
to 1), several cases are shown accordingly to the amplitudes $I_{tr}(t'')=0.25$
(see 2nd row), $I_{tr}(t'')=0.5$ (see 3rd row), and $I_{tr}(t'')=1$ (see 4th
row), as well as to $B_{\tt{a}}=$2 (dotted lines), $B_{\tt{a}}=$4 (dashed
lines) and $B_{\tt{a}}=$8 (straight lines) respectively.}
\end{figure}

\begin{figure}[t]
\centering
\includegraphics[height=8cm,width=13cm]{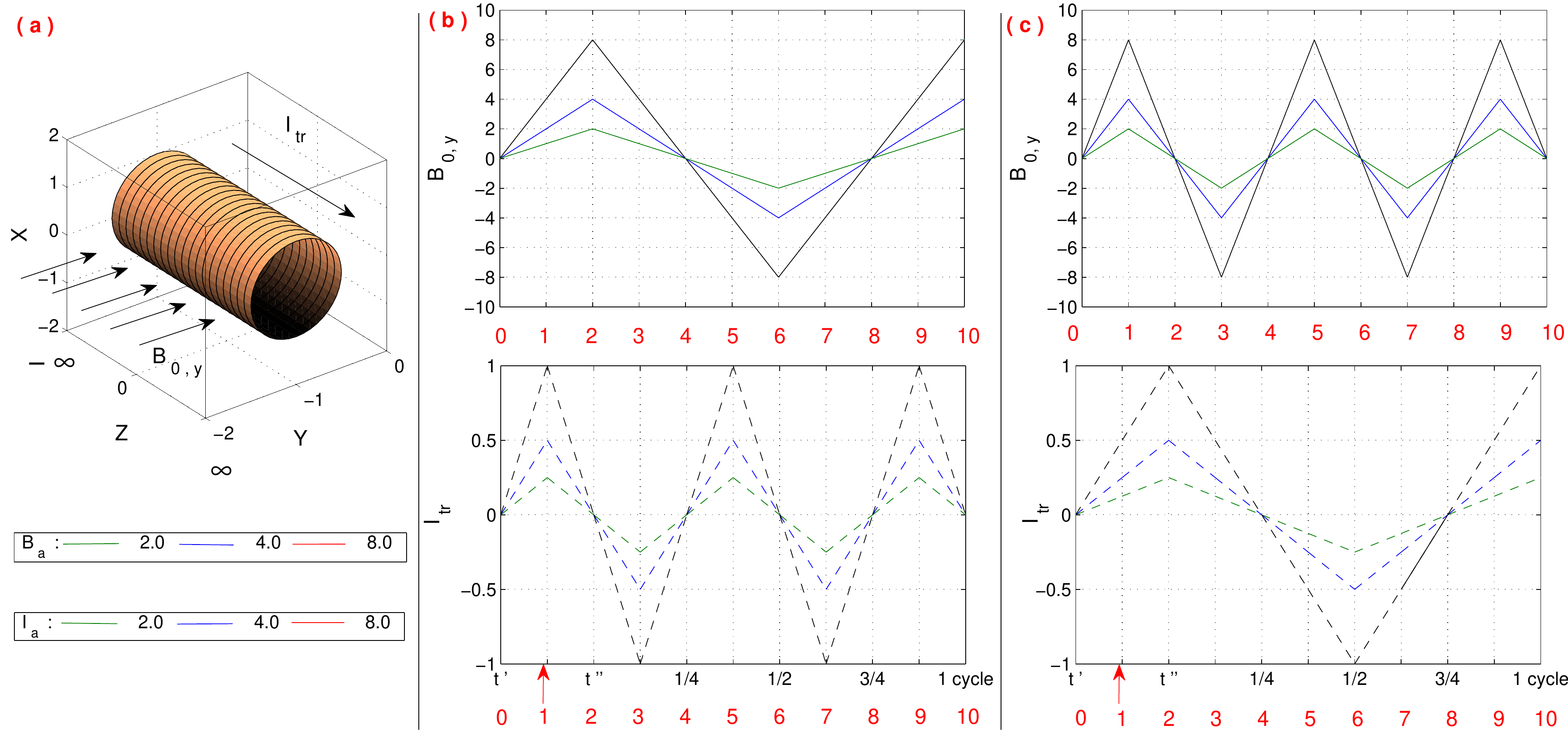}
\caption{\label{Figure_6_12} Sketch of some of the experimental processes
analyzed along this chapter. Here, a cylindrical SC wire subjected to
asynchronous oscillating excitations in the configuration
shown in pane (a) are considered according to the temporal processes
depicted in panes (b) and (c).
}
\end{figure}

In the present subchapter, we show a comprehensive study of the physical
features associated to the local electrodynamics of superconducting wires under
the combined action of AC current and AC magnetic field, which continues our
previous discussion and constitutes a step forward in the understanding of the
electromagnetic observables and the local effects associated to the AC losses.
Section~6.3.1 is restricted to the situation of a \textit{synchronous} AC
excitation $(B_{0},I_{tr})$,
corresponding to uniform AC magnetic field $B_{0,y}$ in phase with the
injected transport current $I_{tr}$, both with the same oscillating
frequency (see Fig.~\ref{Figure_6_4}). Also, synchronous
excitations are considered in situations where the superconducting wire has been
premagnetized (see Fig.~\ref{Figure_6_9}, pag.~\pageref{Figure_6_9}). On the
other hand, section~6.3.2 addresses the effects related to the consideration
of \textit{asynchronous} excitations, in which, both sources may be out
of phase and apply at different frequencies (see Fig.~\ref{Figure_6_12},
pag.~\pageref{Figure_6_12}). Premagnetized wires subjected to synchronous or
asynchronous sources, may be found in superconducting multicoils for the
production of high magnetic fields~\cite{P2-Iwasa_2009}, accelerator magnet
technologies~\cite{P2-Iwasa_2009,P2-Wanderer_2006,P2-Kashikhin_2006}, and
superconducting magnetic energy storage systems~\cite{P2-He_2010}.

\vspace*{0.5cm}
\hspace*{-0.3cm}
\textbf{\textit{6.3.1 ~Synchronous excitations}}
\markboth{\hspace*{0.1cm} \textsl{6.3.1 ~Synchronous
excitations}}{\hspace*{0.1cm}
\textsl{6.3.1 ~Synchronous excitations}}
\addcontentsline{toc}{chapter}{\hspace*{0.1cm} \textsl{6.3.1 ~Synchronous
excitations}}\label{ch-6-3-1}
\vspace*{0.5cm}

Accordingly to the cases explored in previous sections and holding the aim
of achieving a clearest understanding of the electromagnetic quantities involved
in the actual use of superconducting wires, is continued by our discussion
considering the experimental scenario displayed in Fig.~\ref{Figure_6_4}. 

\paragraph{(~\textit{A.}~) \textit{Flux penetration
profiles}\vspace*{0.5cm}\\}

\begin{figure}[t]
\begin{center}
\includegraphics[height=5.0cm,width=3cm]{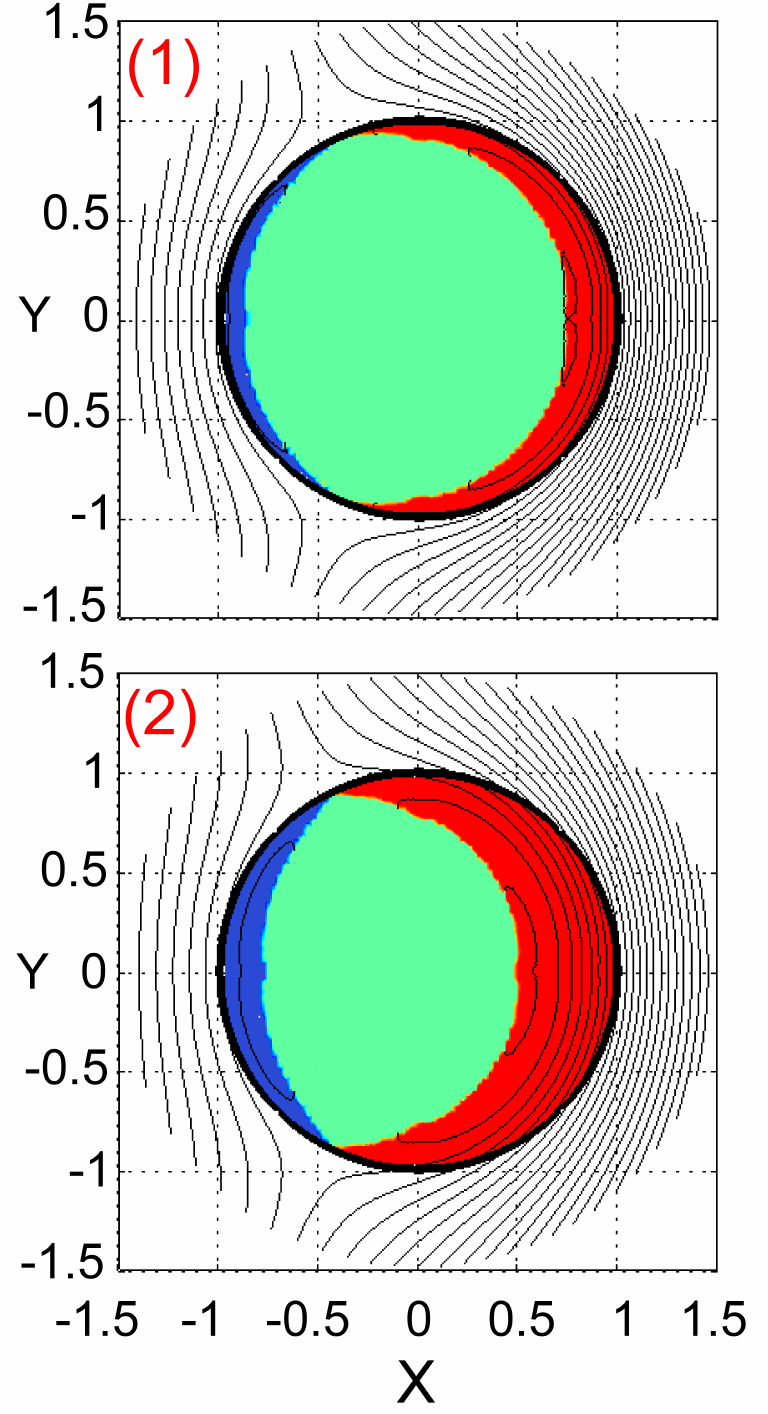} \hspace*{-0.1cm}
\includegraphics[height=5.0cm,width=3cm]{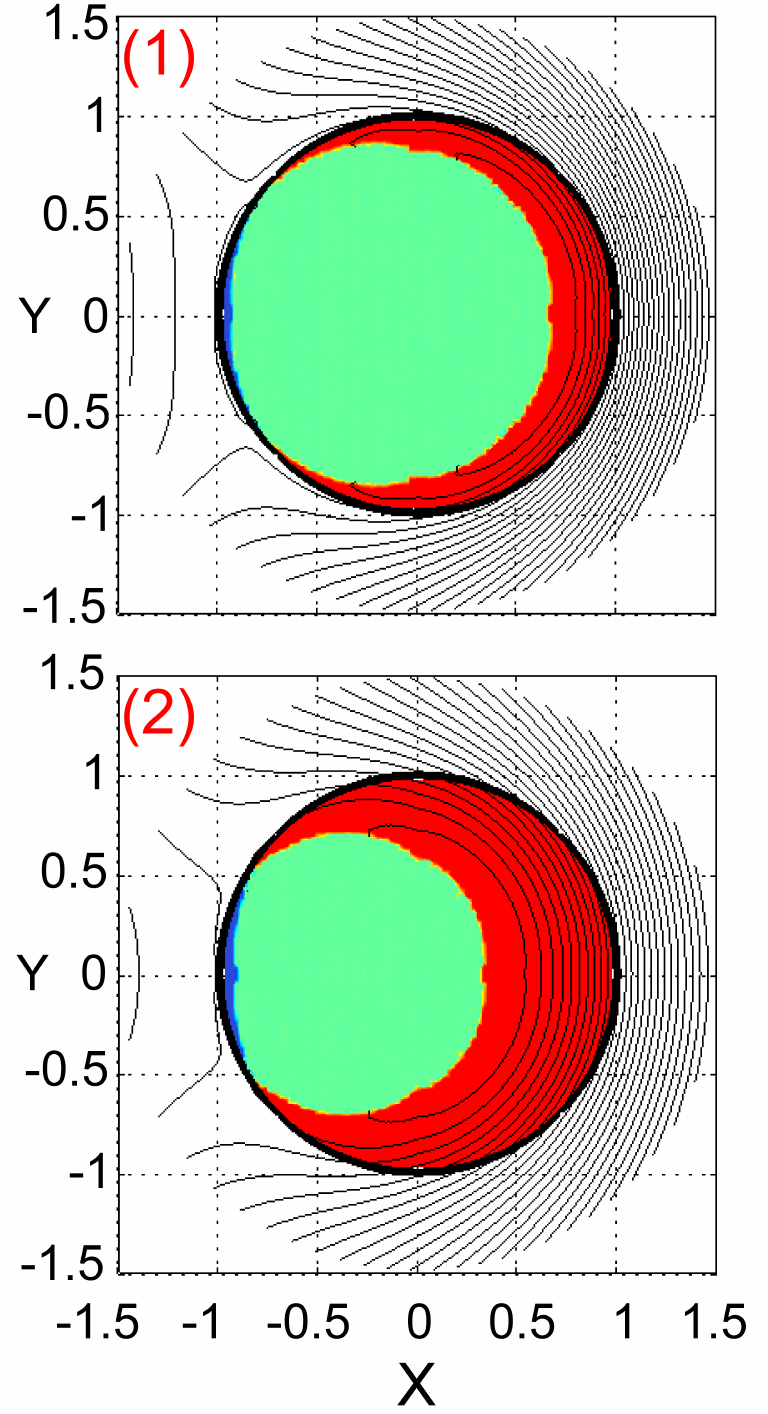} \hspace*{-0.1cm}
\includegraphics[height=5.0cm,width=3cm]{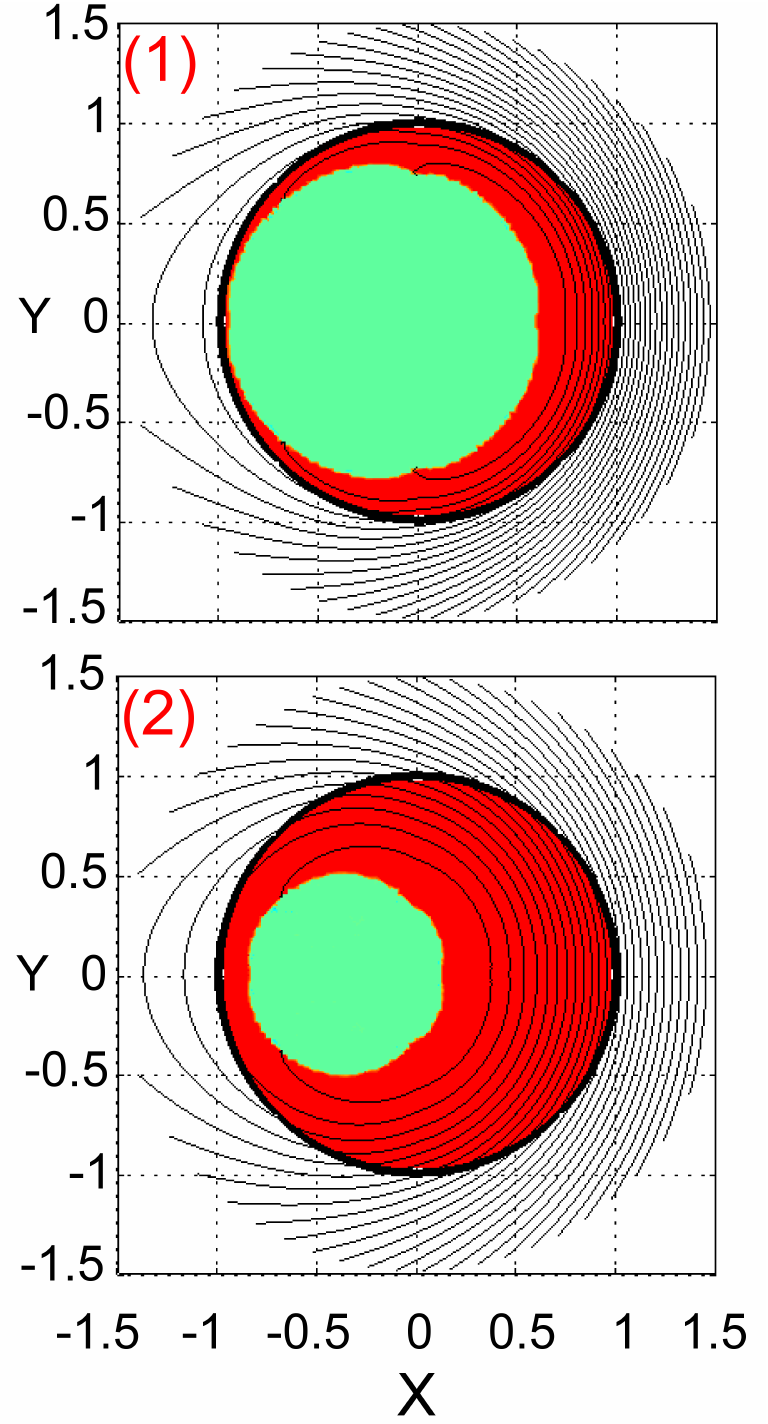} \hspace*{-0.1cm}
\includegraphics[height=5.0cm,width=3cm]{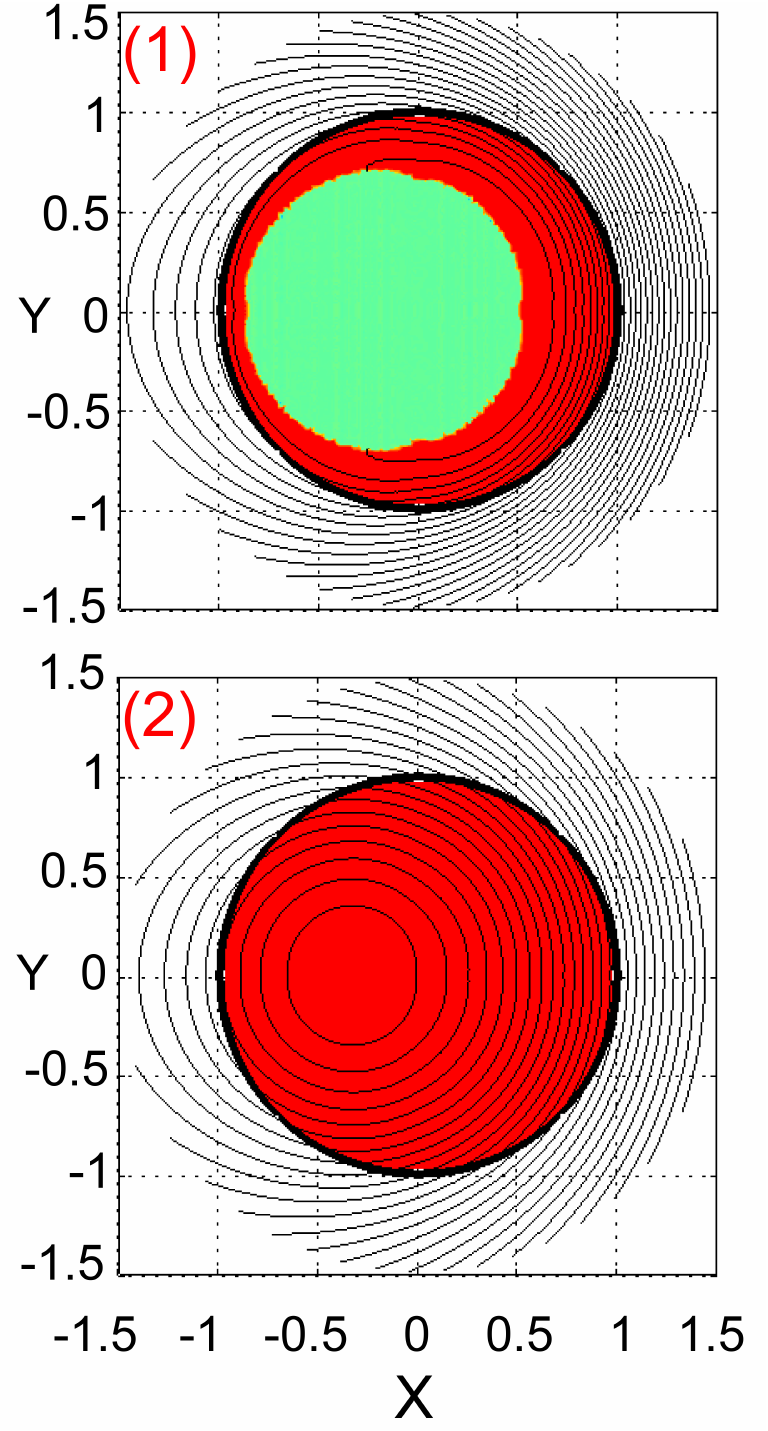} \hspace*{-0.1cm}
\end{center}
\caption{\label{Figure_6_5} Evolution of the magnetic flux lines and their
corresponding profiles of current
with simultaneous oscillating sources $(B_{0,y},I_{tr})$
of amplitudes $B_{\tt{a}}=2$ (\textit{low field}) and, \textit{1st column}:
$I_{\tt{a}}=0.25$,
\textit{2nd column}: $I_{\tt{a}}=0.5$, \textit{3rd column}: $I_{\tt{a}}=0.75$,
and \textit{4th column}: $I_{\tt{a}}=1$. Subplots are labeled according to
the monotonic branch of the experimental processes depicted in
Fig.~\ref{Figure_6_4}. For the branches corresponding to the synchronous cyclic
excitation see Fig.~S5 (pag.~\pageref{Figure_S_5}).
}
\end{figure}

\begin{figure}[t]
\begin{center}
\includegraphics[height=5.0cm,width=3cm]{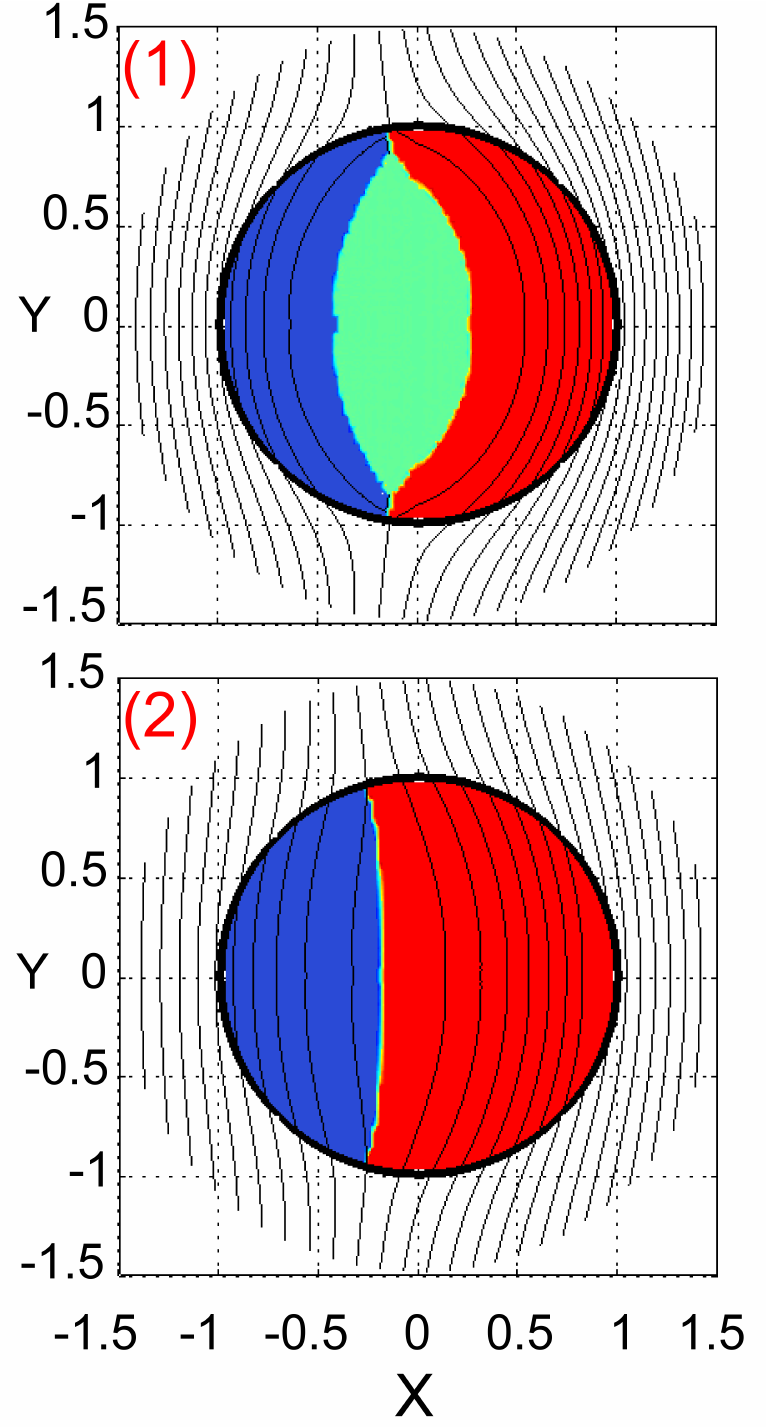} \hspace*{-0.1cm}
\includegraphics[height=5.0cm,width=3cm]{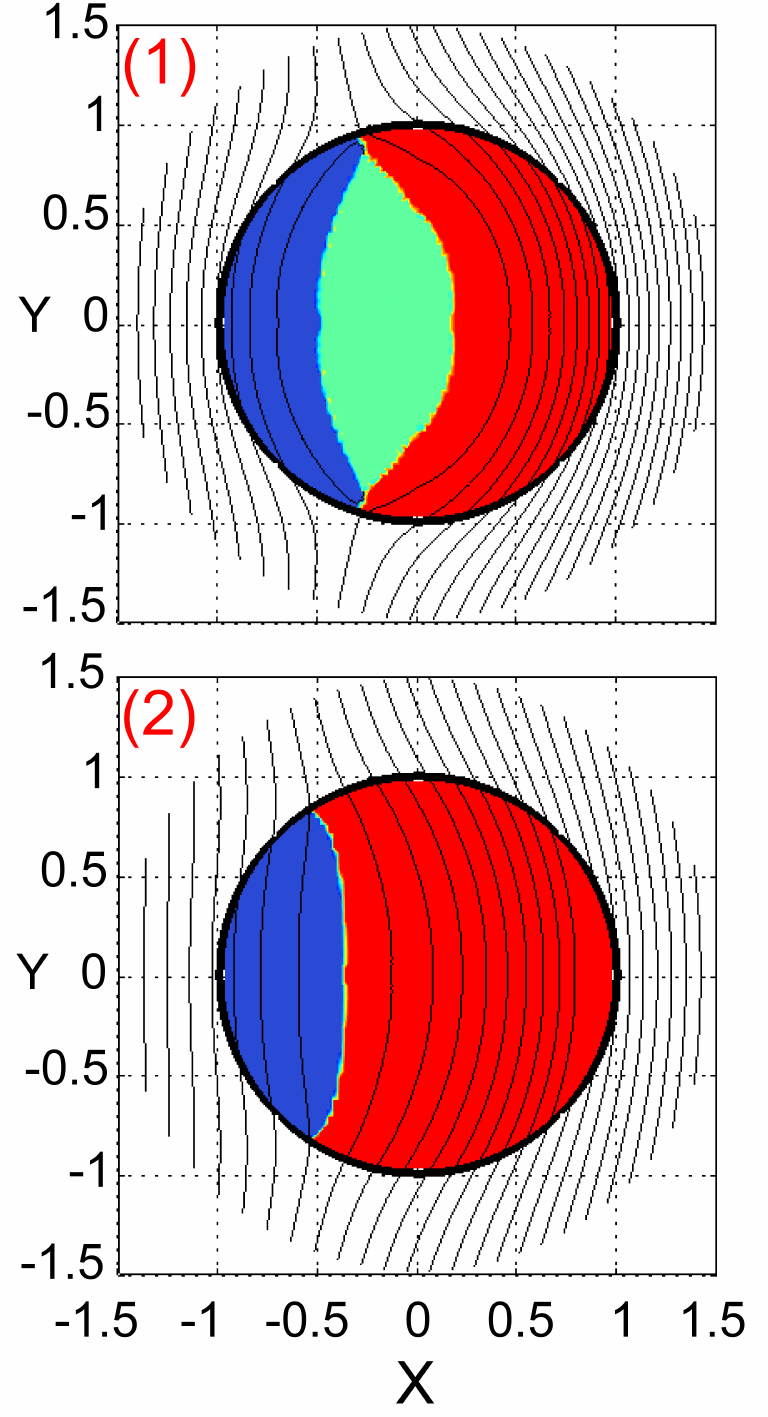} \hspace*{-0.1cm}
\includegraphics[height=5.0cm,width=3cm]{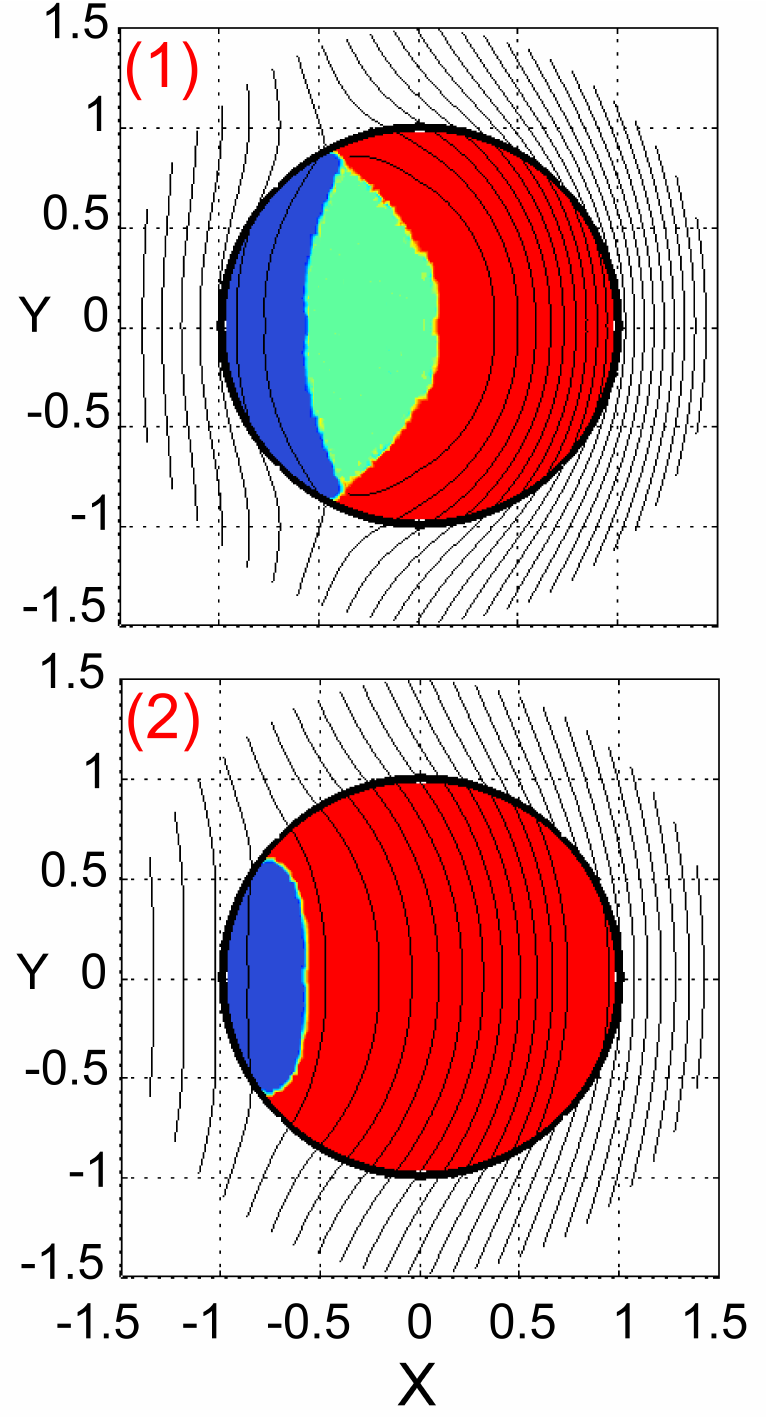} \hspace*{-0.1cm}
\includegraphics[height=5.0cm,width=3cm]{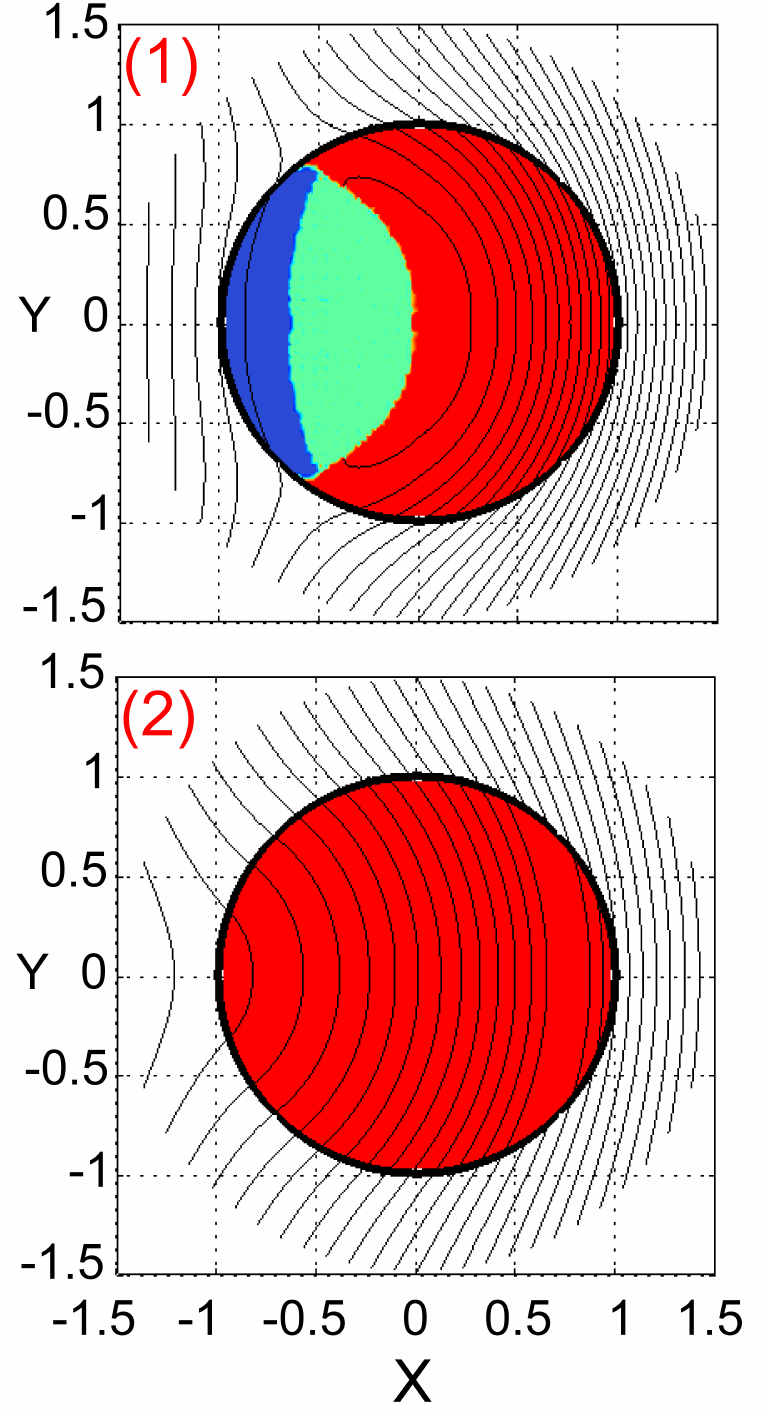} \hspace*{-0.1cm}
\end{center}
\caption{\label{Figure_6_6} Evolution of the magnetic flux lines (projected
isolevels of the vector potential) and their corresponding profiles of current
with simultaneous oscillating sources $(B_{0,y},I_{tr})$
of amplitudes $B_{\tt{a}}=8$ (\textit{high field}) and, \textit{1st column}:
$I_{\tt{a}}=0.25$,
\textit{2nd column}: $I_{\tt{a}}=0.5$, \textit{3rd column}: $I_{\tt{a}}=0.75$,
and \textit{4th column}: $I_{\tt{a}}=1$. Subplots are labeled according to
the monotonic branch of the experimental processes depicted in
Fig.~\ref{Figure_6_4}, i.e., label (1) identifies the time-step corresponding
to half of the first branch, and (2) the first excitation peak. For visualizing
the electromagnetic response in the following branches (cyclic response) reader
is advised to see Fig.~S8 (pag.~\pageref{Figure_S_8}).
}
\end{figure}

Certainly, the flux front profile in the initial stage penetrates from the
surface as the intensity of the external excitations ($B_{0,y}$,$I_{tr}$)
increases (see Figs.~\ref{Figure_6_5} \& \ref{Figure_6_6}). Recall that
screening currents produced by the external excitations have
been conveniently introduced in terms of two different groups depending on the
nature of the source. On the one hand, we speak about \textit{magnetization
currents} produced by the excitation magnetic flux density $B_{0,y}$, and on
the other hand, we refer to the \textit{injected current lines} which must
accomplish the additional global constraint
$\sum_{i}{I}_{i}(t)\equiv{I}_{tr}(t)$ [Eq.~(\ref{Eq.6.6})]. When the action of
isolated sources is conceived, the distribution of screening currents preserves
a well defined symmetry. However, for simultaneous application of both sources 
(Fig.~\ref{Figure_6_4}), the consumption of the magnetization
currents by the injected current lines distorts the axisymmetric orientation of
the flux-front, by displacing the current free core to the left during the
monotonic branch (Figs.~\ref{Figure_6_5} \& \ref{Figure_6_6}).

For low magnetic fields (Fig.~\ref{Figure_6_5}), the profiles of current
density are rather similar to those obtained for $B_{0,y}=0$. The basic
difference is that the center of the current free core moves towards the
semiaxis $x$-negative. In fact, if the intensity of the transport current is
high enough, the flux front becomes nearly circular (\textit{current-like}).
Then, the distribution of screening currents may be understood as the
straightforward overlapping of the profiles of current density for isolated
sources $I_{tr}$ [Fig.~S1 (pag.~\pageref{Figure_S_1})] and $B_{0,y}$ [Fig.~S3
(pag.~\pageref{Figure_S_3})], and additionally displacing the center of the
core devoid of electric current and magnetic flux (green zone) by the respective
difference between the known flux fronts. It should be mentioned that such
assumption has been made in Refs.~\cite{P2-Brandt_1993} \&
\cite{P2-Zeldov_1994} for calculating the current profiles for thin strips with
synchronous excitations
$(B_{0},I_{tr})$. Nevertheless, recently it has been proved that even in
this simple configuration, the overlapping principle for the flux front
tracking may be only fulfilled for high current and low applied
field~\cite{P2-Pardo_2007}. 

Likewise, if the applied magnetic field is intense enough as compared to the
transport current (see e.g., left side in Fig.~\ref{Figure_6_6}) the
distribution of screening currents is \textit{field-like}. Nevertheless, the
inherent existence of injected current lines makes it impossible to discern
which filaments correspond to the so called magnetization currents, and which
are the injected current lines. Certainly, for the monotonic branch of the
cyclic excitation and before attaining a full penetration state by the screening
currents, the ``active'' zone (blue) where $I_{i}$ takes the value 
$-I_{c}$ straightforwardly corresponds to the so called magnetization currents.
However, the remaining ``active'' zone (red) defined by screening currents
$I_{i}$ taking the value $I_{c}$ is not spatially symmetric
as regards the direction of the applied magnetic field, which means that
a certain amount of the magnetization currents are contributing in the same
direction as the transport current, whilst another part has been consumed by the
injected current lines. As it will be shown below, a parallel effect is that 
the density of magnetic flux increases in the
``active'' zone, where the patterns of injected current lines dominate.

For the cyclic processes displayed in Fig.~\ref{Figure_6_4} tracking the flux
front for synchronous excitation with low magnetic field is intuitive
[Fig.~S5 (pag.~\pageref{Figure_S_5})], although following up the components of
the magnetic flux density $B_{x}$ [Fig.~S6 (pag.~\pageref{Figure_S_6})] and
$B_{y}$ [Fig.~S7 (pag.~\pageref{Figure_S_7})] is not. For high magnetic
fields, ascertaining the distribution of screening currents in the cyclic stage
is much more elaborated, as long as the electromagnetic history is not erased by
the maximal condition for the amplitude of the AC transport current
$I_{\tt{a}}=I_{c}$. Actually, if $I_{\tt{a}}<I_{c}$ the flux fronts do not
overlap to a unique contour
line defined by the filaments with current alternating between $I_{c}$ and
$-I_{c}$ [Fig.~S8 (pag.~\pageref{Figure_S_8})]. Likewise, describing the
evolution of the magnetic flux density [Figs.~S9-S10
(pags.~\pageref{Figure_S_9}-\pageref{Figure_S_10})] is also complicated if one
compares them with the simplest cases in which isolated sources are assumed
[Fig.~S2 (pag.~\pageref{Figure_S_2}) \& Fig.~S4 (pag.~\pageref{Figure_S_4})].

Although the analysis of the magnetic flux density $B$ is complicated,
one of the most outstanding observations for considering synchronous excitations
as
shown in Fig.~\ref{Figure_6_4}, is that the local distribution of magnetic field
preserves the same kind of pattern along the cyclic stage, independently of the
intensity of the external sources. Thus, one can notice that the maximal
density of magnetic flux occurs always to the right side of the superconducting
wire, which corresponds to the ``active'' zone where the injected current lines
are dominating the system. Concomitantly, substantial distortions of the
magnetic flux density outside the wire appear. These are particularly marked
when $B_{0,y}$ and $I_{tr}$ tend to zero during the excitation.

\begin{figure}[t]
\centering
\includegraphics[height=4.1cm,width=13cm]{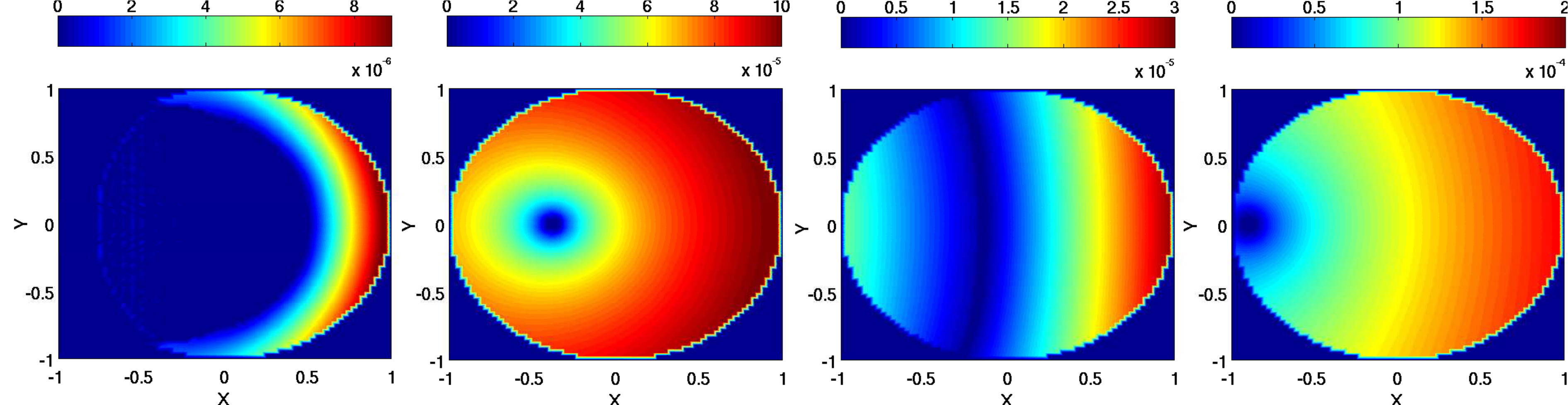}
\caption{\label{Figure_6_7} Local density of power dissipation E$\cdotp$J at
the time frame of full cycle (step number 10 in Fig.~\ref{Figure_6_4}) for
synchronous oscillating sources of amplitudes: (\textit{1st pane})
$B_{\tt{a}}=8$ and $I_{\tt{a}}=0.25$,
(\textit{2nd pane}) $B_{\tt{a}}=8$ and $I_{\tt{a}}=1.0$, (\textit{3rd pane})
$B_{\tt{a}}=2$ and $I_{\tt{a}}=0.25$, (\textit{4th pane}) $B_{\tt{a}}=2$ and
$I_{\tt{a}}=1.0$. The local dynamics for the aforementioned quantities in the
full cyclic process including the initial monotonic branch, can be inferred from
the supplementary figures for the low-field regime [Fig.~S14
(pag.~\pageref{Figure_S_14})], and the
high-field regime [Fig.~S15 (pag.~\pageref{Figure_S_15})].
}
\end{figure}

Remarkably, the strong localization of the inner density of magnetic flux
density produces a significant change in the local distribution of
density of power dissipation $\textbf{E}\cdot \textbf{J}$, which rises from
low-value parts (blue) to high-value parts (red) [see, Fig.~\ref{Figure_6_7}],
in such manner that the heat release from the superconducting wire is highly
localized too. In fact, this asymmetric distribution of power losses regarding
the perpendicular direction to the orientation of $\textbf{B}$, remains along
the entire cyclic process as long as both excitations evolve synchronous
[see e.g., Figs. S14-S16 (pags.~\pageref{Figure_S_14}-\pageref{Figure_S_16})].
Therefore, its pronounced bias unfolding across the wire could increase the
probability of quench. 

\begin{figure}[t]
\centering
\includegraphics[height=10cm,width=12cm]{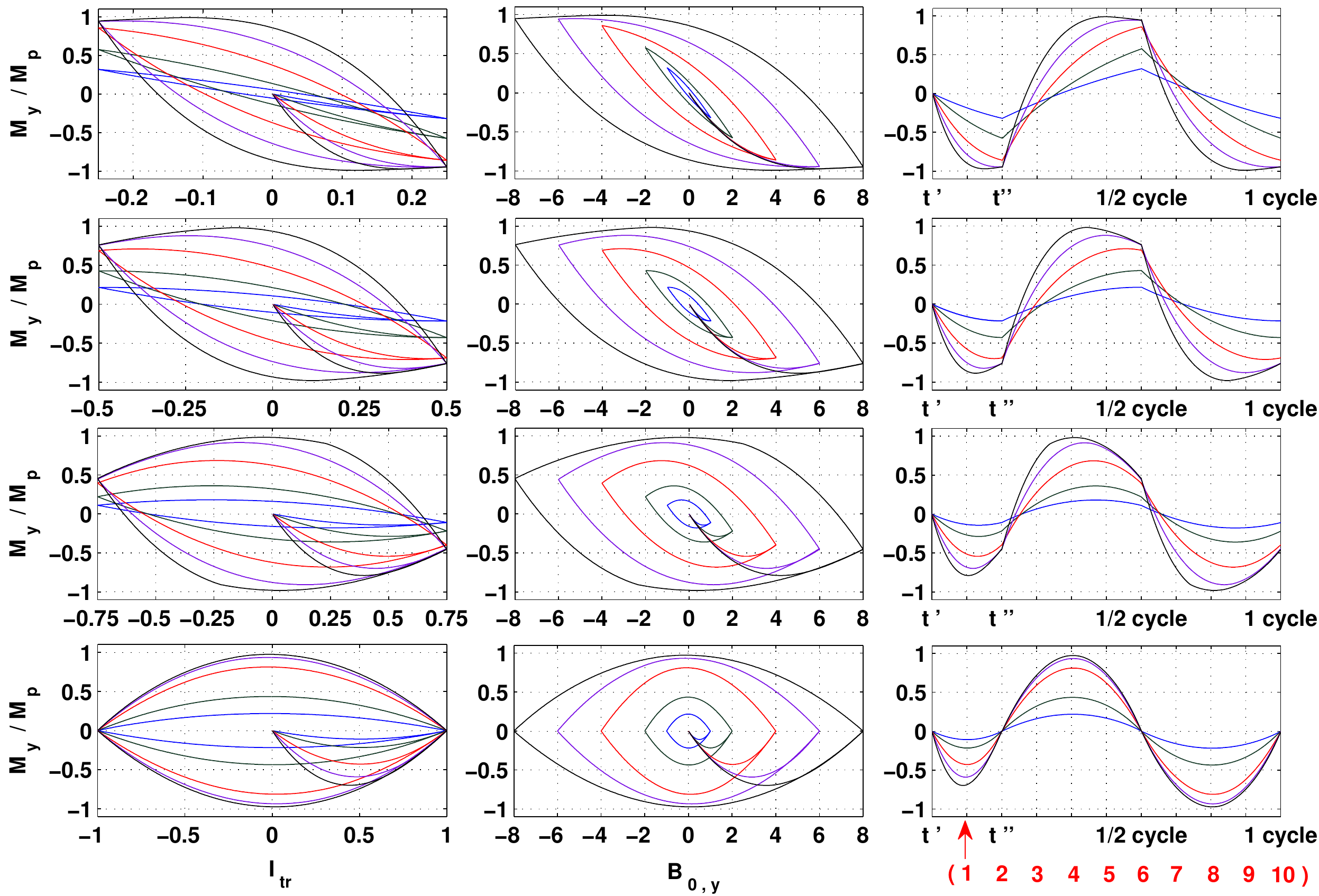}
\caption{\label{Figure_6_8} The dimensionless magnetic moment
$M_{y}/M_{p}$ for the synchronous AC excitations
displayed in Fig.~\ref{Figure_6_4}. Curves are shown as function of the
injected transport current $I_{tr}$ (\textit{left pane}), the applied magnetic
field $B_{0,y}$ \textit{(central pane)}, or either
by its temporal evolution \textit{(right pane)}. In this figure, the amplitudes
for the electromagnetic AC sources can be extracted either from color comparison
with curves in Fig.~\ref{Figure_6_4}, or from the respective limits along the
abscissas in left and right panes.
}
\end{figure}

\paragraph{(~\textit{B.}~) \textit{Magnetic response}\vspace*{0.5cm}\\}

The above described behavior for the local flux distributions gives way to the
following features on the magnetic moment response.

For the set of cases displayed in Fig.~\ref{Figure_6_4}, we have analyzed the
dynamics of the magnetic moment component $M_{y}$ as a function of the
amplitudes of the electromagnetic sources $B_{\tt{a}}$ and $I_{\tt{a}}$
(Fig.~\ref{Figure_6_8}).
Results are shown accordingly to the temporal dependence with the synchronous
AC excitations (right pane), and also by their dependence with each one of the
electromagnetic sources, say $I_{tr}$ (left pane) and $B_{0,y}$ (middle
pane). We realize that only for small values of the amplitude of the ac
transport current, almost Bean-like loops of $M_{y}$ obtain. However, as
$I_{\tt{a}}$ grows we notice a progressive disappearance of the flat saturation
behavior for values of $B_{\tt{a}}$ higher than $B_{p}$. Actually, the notorious
change of sign for the slope of the magnetic moment curve along a monotonic
branch of the synchronous AC excitation, allows an unambiguous glimpse of the
consumption of magnetization lines by effect of injected current lines.
Remarkably, this phenomenon ends up with a symmetrization of the loops, both as
functions of $B_{y}$ and $I_{tr}$, into characteristic lenticular shapes. As a
consequence of this process, a distinct low-pass filtering effect comes to the
fore which, in the case of the triangular input excitations considered here,
yields a nearly perfect sinusoidal (first-harmonic) output signal $M_{y}(t)$.

Interestingly, from the cycles of $M_{y}$, it furthermore appears that a proper
determination of the ``active'' zones depends on the history of the virgin
branch, thus bearing witness to the system's memory. 
For example, a positive slope in the synchronous excitation $B_{y}$ and
$I_{tr}$ produces a higher power dissipation in the positive 
$x$-direction perpendicular to the wire. On the other side, if the
superconducting wire has been premagnetized before switching on the synchronous
AC excitation [$t=t'$ at Fig.~\ref{Figure_6_9}], the magnetic moment curve
is displaced in such a manner that the center of the magnetization loop lies
over the master curve for
the isolated excitation $B_{0,y}$ at $M_{y}(t')$ (see Fig.~\ref{Figure_6_2}),
and the nodes move towards the boundaries $B_{0,y}(t')\pm B_{\tt{a}}$.

\paragraph{(~\textit{C.}~) \textit{AC Losses}\vspace*{0.5cm}\\}

\begin{figure}[t]
\centering
\includegraphics[height=7cm,width=13cm]{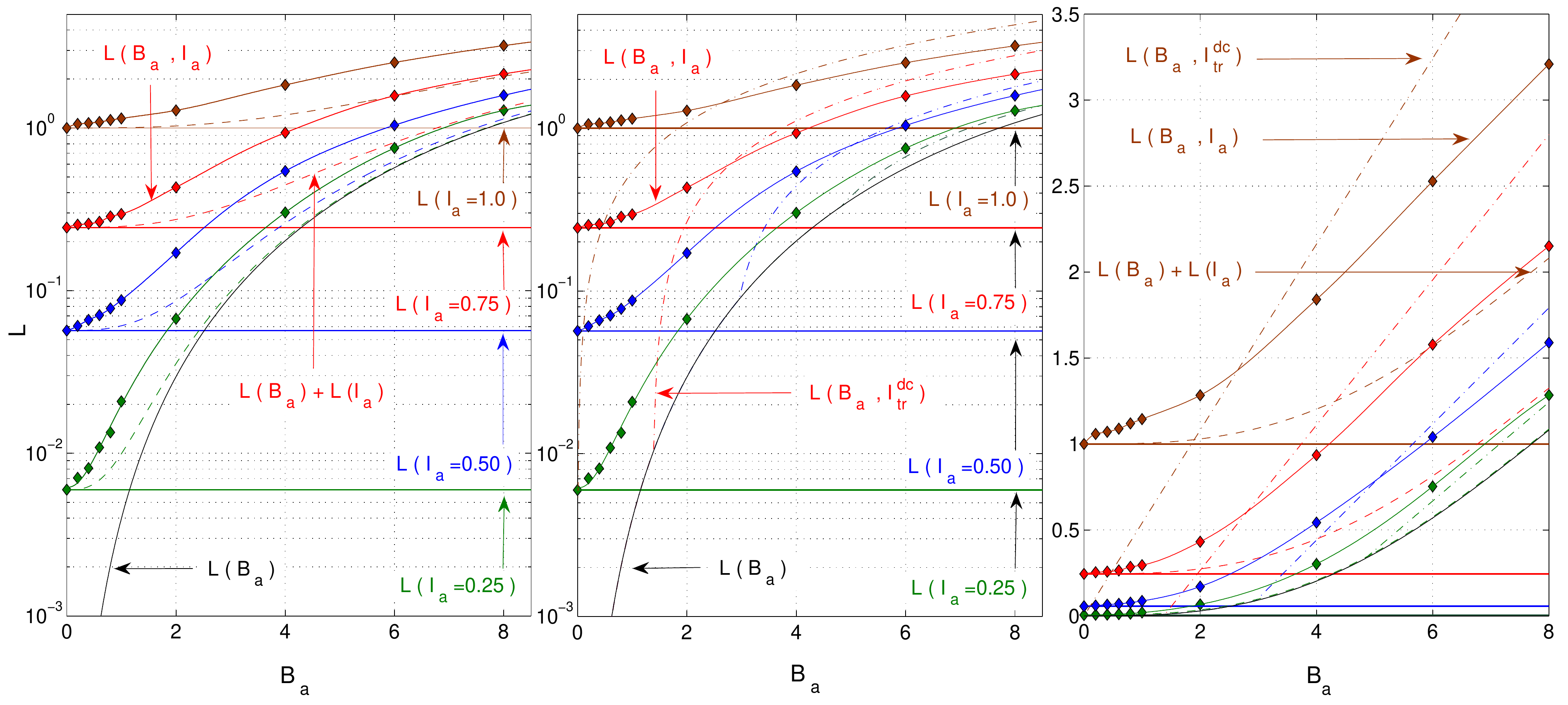}
\caption{\label{Figure_6_10} Hysteretic ac losses per cycle
for synchronous AC magnetic flux density and oscillating transport current of
amplitudes $(B_{\tt{a}},I_{\tt{a}})$ accordingly to the
Figs.~\ref{Figure_6_4} \& \ref{Figure_6_9}. Results of this work are shown as
color solid lines with markers. Comparisons with
results from conventional approaches are shown for,
\textit{(i) Left pane}: separate excitations $L(B_{\tt{a}})$ (black solid
line) and $L(I_{\tt{a}})$ (straight color lines), as well as their linear
superposition $L(B_{\tt{a}})+L(I_{\tt{a}})$ (color dashed lines);
\textit{(ii) Central pane}: an ac magnetic field together with
a dc transport current of intensity $I_{\rm tr}^{\rm dc}=I_{\tt{a}}$
, $L(B_{\tt{a}},I_{\rm tr}^{dc})$;
\textit{(ii) Right pane}: the whole set of results
is also plotted in linear scale. Units for losses are $(\mu_{0}/4\pi)\omega
R^{2}
J_{c}^{2}$.}
\end{figure}

Regarding the power density losses attained along the premagnetization process,
it does not seem to play any role in the calculation of the AC losses
(Fig.~\ref{Figure_6_10}). At least, this was observed to the precision of our
calculations. However, the
definition of the flux front profile becomes much more tangled, because multiple
domains enclosed by contour lines defined by the screening currents alternating
from $I_{c}$ to $-I_{c}$ arise [see Fig.~S11 (pag.~\pageref{Figure_S_11})].

Notwithstanding, the Bean-like magnetic moment curves as the described above,
and the ostensible explanation for the distribution of screening currents in
terms of the consumption of magnetization lines by the injected current
lines, is actually insufficient for the proper interpretation of the actual
AC losses produced by synchronous excitations. In fact, despite the collection
of reliable experimental data is quite laborious (because there are many
pitfalls in the measurement procedures), there is an extended consensus that
the heat release by the superconducting wire may come from the 
electromagnetic sources $B_{0}$ and $I_{tr}$ in independent
manners~\cite{P2-Vojenciak_2006,P2-Zannella_2001,P2-Tonsho_2003,P2-Ogawa_2003}.
Moreover, many works dealing with this issue argue that the transport
loss and the magnetization loss can be separately determined by electromagnetic
measurements at least for low values of the magnetic field and high currents or
vice versa \cite{P2-Ogawa_2011,P2-Inada_2005,P2-Jiang_2004,P2-Rabbers_2001,
P2-Reuver_1985}. However, as we have discussed before, for general cases, the
competence between the magnetization
currents and the injected current lines involving axisymmetric distributions
of the screening currents, generates a strong localization of the local density
of magnetic flux, as well as of the local density of power losses, which
makes it difficult discriminating the role of the AC losses
introduced by the inductive terms [Eq.~(\ref{Eq.6.19})]. Therefore, from the
theoretical point of view, it is advisable using experimental methods based upon
the S-oriented approach (pag.~\pageref{S-oriented}), such as calorimetric
methods which directly make a measurement of the release flux of energy over the
superconducting
surface~\cite{P2-Nguyen_2005a,P2-Nguyen_2005b,P2-Ramos_2010,P2-Ashworth_2000,
P2-Magnusson_2000}. 

As we will show below, the feasibility of approaching the total AC loss by
overlapping the isolated contributions, strongly depends on the relative
magnitudes of the AC field $B_{0}$, and the AC transport current $I_{tr}$.
Accordingly to the numerical experiments shown in Figs.~\ref{Figure_6_4} \&
\ref{Figure_6_9}, eventually, we will present a percentage analysis of the
actual AC loss for synchronous excitations, $L(B_{\tt{a}},I_{\tt{a}})$, in
comparison with the most celebrated approaches.
Our numerical results for 
$L(B_{\tt{a}},I_{\tt{a}})$ are displayed in
Fig.~\ref{Figure_6_10} both in logarithmic and linear -scales with the aim of
remarking the actual differences at low and high magnetic fields.

First, let us recall that, according to Eq.~(\ref{Eq.6.11}) for isolated
sources, the AC transport loss $L(I_{\tt{a}})$ may be calculated according to
\begin{eqnarray}\label{Eq.6.23}
L(i_{\tt{a}})\equiv
\frac{\mu_{0} I_{c}^{2}}{\pi^{2}R^{2}}
\left[
i_{\tt{a}}\left(1-\frac{i_{\tt{a}}}{2}\right)
+(1-i_{\tt{a}})\ln(1-i_{\tt{a}})
\right]
\,~\,~\, \forall ~\, ~\, 0< i_{\tt{a}} \leq 1\, \, ,
\end{eqnarray}
\noindent where the dimensionless parameter $i_{\tt{a}}=I_{\tt{a}}/I_{c}$ has
been introduced.

On the other hand, the hysteretic loss produced by magnetization effects may
be estimated from Eq.~(\ref{Eq.6.12}), in such manner that $L(B_{\tt{a}})$ is
calculated from
\begin{eqnarray}\label{Eq.6.24} 
L(b_{\tt{a}})\equiv 
\dfrac{8B_{p}^{2}}{3\mu_{0}}
\left\{
\begin{array}{ll}
b_{\tt{a}}^{3} ~ \left(1-\dfrac{1}{2}b_{\tt{a}}\right)
\,~\,~ & ,~\, \forall ~\, ~\, 0 < b_{\tt{a}} \leq 1 \, 
\\ \, \\
b_{\tt{a}}-\dfrac{1}{2}
\,~\,~ & ,~\, \forall ~\, ~\, b_{\tt{a}} \geq 1
\, ~ \, ~ \, .
\end{array}
\right. 
\end{eqnarray}

\noindent with the dimensionless parameter 
$b_{\tt{a}}=B_{\tt{a}}/B_{p}$. Recall that full penetration is given by
$b_{\tt{a}}\geq1$ (or $i_{\tt{a}}=1$).

The simplest approach for determining the AC losses of
cylindrical superconducting wires subjected to synchronous AC excitations relies
in the linear superposition of the separate contributions,
$L(B_{\tt{a}})+L(I_{\tt{a}})$. Another possibility is to assume
that the actual AC losses for synchronous excitations do not strongly differ of
the hysteretic losses for superconducting samples carrying a constant transport
current $I_{tr}^{dc}$ and a simultaneous oscillating magnetic field of
amplitude $B_{\tt{a}}$ \cite{P2-Gurevich_1997}. This idea was applied in the
analytical approach by Zenkevitch et al. in Ref.~\cite{P2_Zenkevitch_1983}.
In such a framework, the hysteretic loss for a period is
approximated by:
\begin{eqnarray}\label{Eq.6.25} 
L(B_{\tt{a}},I_{tr}^{dc})\equiv 
\dfrac{8 B_{p}^{2}}{3\mu_{0}}
\left\{
\begin{array}{ll}
b_{\tt{a}}^{3} \left(1-\dfrac{1}{2}b_{\tt{a}}\right) 
\,~\,~ & ,~\, \forall ~\, ~\, b_{\tt{a}} < i_{\tt{a}}^{\dag} \, 
\\ \, \\
i_{\tt{a}}^{\dag^{3}} \left(1-\dfrac{1}{2}i_{\tt{a}}^{\dag}\right)+
\left(1+i_{\tt{a}}^{2}\right) 
\left(b_{\tt{a}}-i_{\tt{a}}^{\dag}\right)
\,~\,~ & ,~\, \forall ~\, ~\,  b_{\tt{a}} \geq i_{\tt{a}}^{\dag}
\, ~ ,
\end{array}
\right. 
\end{eqnarray}

\noindent where we have introduced the dimensionless parameter
$i_{\tt{a}}^{\dag}\equiv1-i_{\tt{a}}^{2/3}$ and the condition $b_{\tt{a}} \geq
i_{\tt{a}}^{\dag}$. Here, $I_{\tt{a}}\equiv I_{tr}^{dc}$. Thus,
$b_{\tt{a}}\geq1$ or $i_{\tt{a}}^{\dag}\equiv0$, both define a full penetrated
sample.

Fig.~\ref{Figure_6_10} shows our numerical results for the variation of
the actual AC losses of cylindrical SC wires in terms of the amplitude of the
synchronous oscillating sources, $L(B_{\tt{a}},I_{\tt{a}})$
(\textit{solid-diamond lines}), compared to the customary approaches
$L(B_{\tt{a}})+L(I_{\tt{a}})$ (\textit{dashed lines at the left pane}) and
$L(B_{\tt{a}},I_{tr}^{dc})$ (\textit{dash-dotted lines at the central pane}),
for four different amplitudes of the AC/DC transport current $I_{\tt{a}}$, and a
set of amplitudes of the AC density of magnetic flux $B_{\tt{a}}$. The
whole set of results is also plotted in linear scale at the right pane of this
figure. 

Comparison reveals the important fact that a linear superposition of
contributions due to either type of excitation may be only appropriate for 
high amplitudes of the magnetic field ($B_{\tt{a}}\geq B_{p}$) and low currents
($I_{\tt{a}}<I_{c}/4$), or the converse limit, very low magnetic fields
($B_{\tt{a}}\leq 1$) and extremely high currents $I_{\tt{a}}\lessapprox I_{c}$;
a finding which adds to previous work dealing with a rectangular
geometry~\cite{P2-Pardo_2007} and sheds new light on the validity of approximate
formulae at the same time. Actually, notice that the actual AC loss
$L(B_{\tt{a}},I_{\tt{a}})$ is always higher than the instinctive approach
$L(B_{\tt{a}})+L(I_{\tt{a}})$, whilst the deviation respect to
$L(B_{\tt{a}},I_{tr}^{dc})$ strongly depends on the intensity of the
electromagnetic excitations. Consequently, approximations such as $
L\left(B_{\tt{a}} \right) + L\left( I_{\tt{a}} \right)$ and $ L\left(
B_{\tt{a}}, I_{tr}^{dc} \right)$  can drastically under-
or overestimate the true losses.

\begin{figure}[t]
\begin{center}
\includegraphics[height=6cm,width=6.6cm]{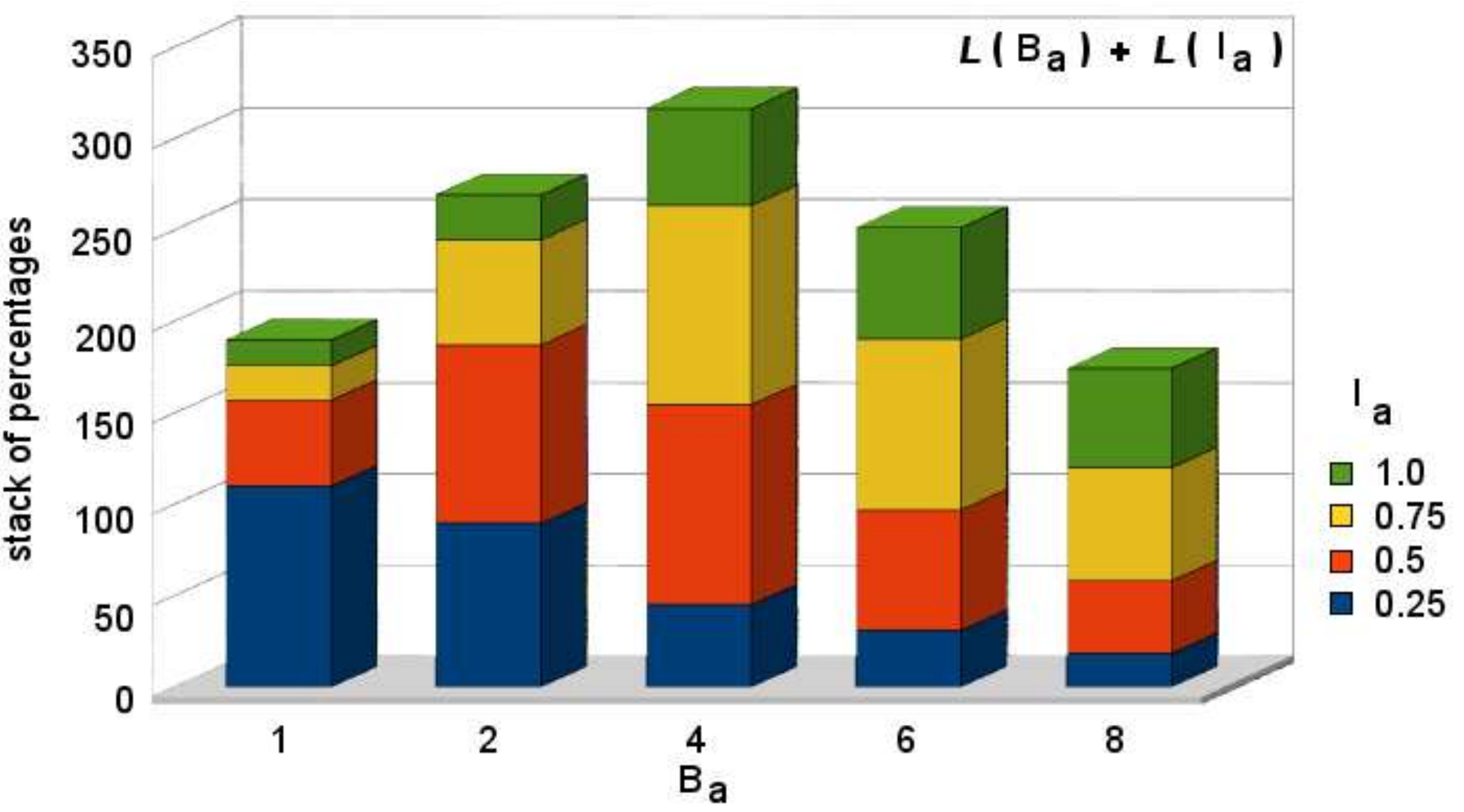}
\hspace*{-0.1cm}\vline
\includegraphics[height=5.8cm,width=6.2cm]{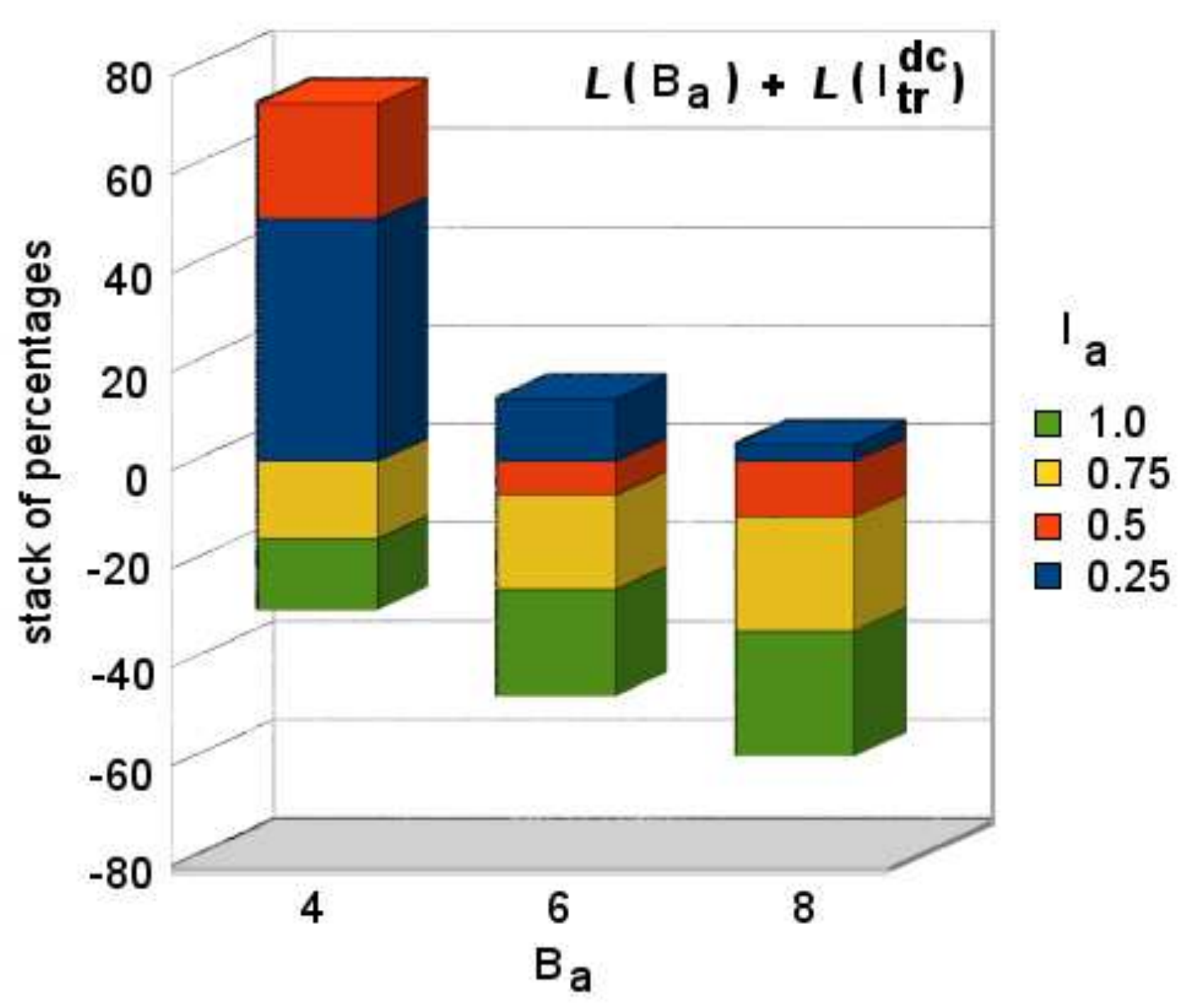}
\end{center}
\caption{\label{Figure_6_11} 
Percent change between the actual AC loss $L(B_{\tt{a}},I_{\tt{a}})$
numerically calculated and the intuitive approaches
$L(B_{\tt{a}})+L(I_{\tt{a}})$ (left pane) and $L(B_{\tt{a}})+L(I_{tr}^{dc})$
(right pane). 
}
\end{figure}

For further understanding of the above behavior, Fig.~\ref{Figure_6_11} shows
the percentage relation between the actual AC loss, $L(B_{\tt{a}},I_{\tt{a}})$, 
and the intuitive approaches, $ L\left(B_{\tt{a}}\right) + L\left(I_{\tt{a}}
\right)$ (\textit{left pane}) and $ L\left(B_{\tt{a}},I_{tr}^{dc} \right)$
(\textit{right pane}), stacked according to the values of $I_{\tt{a}}$. On
the one side, we note that for the approach $L\left(B_{\tt{a}}\right) +
L\left(I_{\tt{a}}\right)$, and for small values of $I_{\tt{a}}$ (e.g.
$I_{\tt{a}}=0.25$), the deviation is gradually reduced as one increases the
amplitude of the magnetic field $B_{\tt{a}}$. However, as $I_{\tt{a}}$
increases deviations may either reduce (for low values of $B_{\tt{a}}$, e.g.,
$B_{\tt{a}}=1$) or increase (for high values of $B_{\tt{a}}$, e.g.,
$B_{\tt{a}}=8$). Moreover, for moderate fields (e.g., $2\leq B_{\tt{a}}\leq6$)
percentage deviations first grow as a function of $I_{a}$ until
$I_{\tt{a}}=0.5I_{c}$, and then decrease as $I_{\tt{a}}$ approaches the
current limit $I_{c}$. On the other side, the approach
$L\left(B_{\tt{a}},I_{tr}^{dc}\right)$ is not even comparable with the actual
AC losses $L(B_{\tt{a}},I_{\tt{a}})$ for the regime of low magnetic fields
(particularly for $b_{\tt{a}}<i_{\tt{a}}^{\dag}$), a range in which the
approximation conceals the effect of $I_{a}$. However, for moderate and high
magnetic fields, the percentage ratio between the actual AC loss and the AC loss
predicted by the
latter approach decrease as $I_{\tt{a}}$ increases even reaching negative
values. Hence, $L\left(B_{\tt{a}},I_{tr}^{dc}\right)$ may either overestimate or
underestimate the actual AC loss for wires subjected to synchronous
oscillating sources. Remarkably, for high amplitudes
of the oscillating magnetic flux density $B_{\tt{a}}$, we note that a
synchronous oscillating transport current of amplitude $I_{\tt{a}}$ produces a
lower amount of hysteretic losses per period, than those predicted when the
superconducting sample is carrying a constant transport current $I_{tr}^{dc}$.

\vspace*{0.5cm}
\hspace*{-0.3cm}
\textbf{\textit{6.3.2 ~Asynchronous excitations}}
\markboth{\hspace*{0.1cm} \textsl{6.3.2 ~Asynchronous
excitations}}{\hspace*{0.1cm}
\textsl{6.3.2 ~Asynchronous excitations}}
\addcontentsline{toc}{chapter}{\hspace*{0.1cm} \textsl{6.3.2 ~Asynchronous
excitations}}\label{ch-6-3-2}
\vspace*{0.5cm}

In many of the large-scale power applications for superconducting wires, such
as windings of motors, transformers, generators and power three-phase
transmission lines, the SC wire is subjected to diverse configurations of
electromagnetic excitations, where the AC transport current flowing
through and the magnetic field to which the wire is exposed could not fulfill
the synchronous conditions referred above (same phase and frequency). Moreover,
remarkable experimental differences between the AC loss measured for
superconducting wires or tapes with synchronous and asynchronous oscillating
transport current and perpendicular magnetic field have been already reported by
several authors~\cite{P2-Nguyen_2005a,P2-Nguyen_2005b,P2-Vojenciak_2006}.

\paragraph{(~\textit{A.}~) \textit{General considerations on
the ``asynchronous'' AC losses}\vspace*{0.5cm}\\}

Our analysis of the AC loss formulae for cylindrical superconducting
wires subjected to simultaneous oscillating transport current $(I_{tr})$ and
perpendicular magnetic flux density ($B_{0,y}$), has revealed that the total AC
loss may be controlled by reducing the inductive magnetic flux density produced
by the superposition between the external magnetic field and the contribution by
the whole set of screening currents [recalling Eq.~(\ref{Eq.6.9})]. This can be
achieved just by a time shift respect to one of the AC electromagnetic sources
(either $B$ or $I$) respect to the other, so that
the occurrence of the peaks of excitation for each one of the electromagnetic
sources is no longer synchronous with the other. Thus, some eccentric branches
with opposite temporal derivatives appear between two consecutive peaks of the
dominant excitation (i.e., the excitation with lower frequency this is the
case), which may counterbalance the local variation of the magnetic flux density
produced by the other one in the zone of maximum heat release. 

Evidently, by competition between the magnetization currents and the so-called
injected current lines, the magnitude of the local density of magnetic
flux ($B_{i}$) may be reduced in half section of the superconducting cylinder
shifting the relative phase between the electromagnetic excitations. Thus, as
long as the electromagnetic excitations have the same oscillating
period, reductions of the actual AC loss could be announced for shifts
in the relative phase measured between the synchronous case and a temporal
displacement of half period [i.e., $\Delta\phi=\pi$ if both sources accomplish
the generic relation $f=f_{0}cos(\omega t+\phi)$], as it has been observed in
Refs.~\cite{P2-Nguyen_2005a,P2-Nguyen_2005b,P2-Vojenciak_2006}. 
Recall that, we have shown that the total AC loss decreases ensuring minimal
variation of $\Delta B_{i}$ along the whole sample. In fact, if the
relative phase shift equals half a period, it means that the occurrence
of the excitation peaks for the electromagnetic sources are likewise synchronous
but are pointing in opposite directions. Thus, as the total AC loss may be
calculated by integration of the excitations peak to peak, for this case
$\Delta B_{i}$ is maximum and therefore also the actual AC loss. On the other
hand, the forecast of the minimal variation of the set of integrands in
Eq.~(\ref{Eq.6.9}) and consequently the total AC loss, may be done by
considering a temporal displacement of a quarter of period [i.e., $\phi=\pi/2$
for circular excitation functions], so that the local competition between the
magnetic flux densities $\mid B_{0,i} \mid$ and $\mid B_{ind,i} \mid$ minimizes
as $B_{0,y}(\Omega)=0$ when $\mid I_{tr} \mid =I_{a}$. Then, under
this simplified scenario, and at least for cases where the local distribution
of screening currents is current-like (very low magnetic field,
$b_{a}\approx 0$, and high transport current, $i_{a}\approx1$), or
field-like (high magnetic field, $b_{a}\gtrsim1$, and very low transport
current, $i_{a}\approx0$), the minimal AC loss is envisioned to appears for a
relative phase difference of a quarter of period. Likewise, the
maximal AC loss may be predicted when both sources are fully synchronous or
when there is a relative phase shift of half period. Latter facts agree with
the analytical approaches for the slab~\cite{P2_Takacs_2007} and strip
\cite{P2_Mawatari_2006} geometries. Also, in further agreement with the
experimental evidences
of Refs.~\cite{P2-Nguyen_2005a,P2-Nguyen_2005b,P2-Vojenciak_2006}, within our
statement we predict that as long as a phase shifting occurs, at least a
minimal reduction of the AC loss should be observed.

When the distribution of screening currents is ``\textit{nothing-like}'',
i.e., it shows a strong deformation when compared to the obtained profiles for
the isolated excitations, it is not obvious to deduce a general rule for the
position of the maximum and minimum of the total AC loss, as the nonhomogeneous
interplay between the injected current lines and the magnetization currents
affects the total AC loss. In fact, the situation may be very much complicated
for the actual applications of superconducting transformers and three-phase
transmission lines
\cite{P2_Hamajima_2006,P2_Hamajima_2007,P2_Hamajima_2008,P2_Gouge_2005}, by the
fact that the self induced magnetic field and
the external magnetic field may differ considerably in phase. Especially, one
can foretell complicated behaviors when effects of phase transposition
and, frequency shifts appear. Here, we will show how the effects of double
frequency which may be occasionally found in the power supply networks, can
drastically alter the efficiency of the superconducting machines.

Fig.~\ref{Figure_6_12}, shows the configuration analyzed below. Notice that, 
in what follows, we consider the effect of introducing one of the excitations
with an oscillating frequency twice as big as the other. Thus, calculation of
the AC loss, i.e., integration of the local density
of power dissipation $\textbf{E}\cdot\textbf{J}$ is made along the smaller
frequency excitation. In detail, we have considered the following cases:

\begin{enumerate}[I.]

\item The injected transport current is the source within the
double frequency regime [see Fig.~\ref{Figure_6_12} (b)], and

\item The temporal dynamics of the magnetic flux density
associated to the external source of magnetic field shows a double frequency
behavior [see Fig.~\ref{Figure_6_12} (c)].

\end{enumerate}

\noindent Most remarkable features for the flux dynamics, magnetic response, and
AC losses for the above mentioned configurations are detailed below.

\begin{figure}[t]
\begin{center}
\includegraphics[height=7.5cm,width=6.4cm]{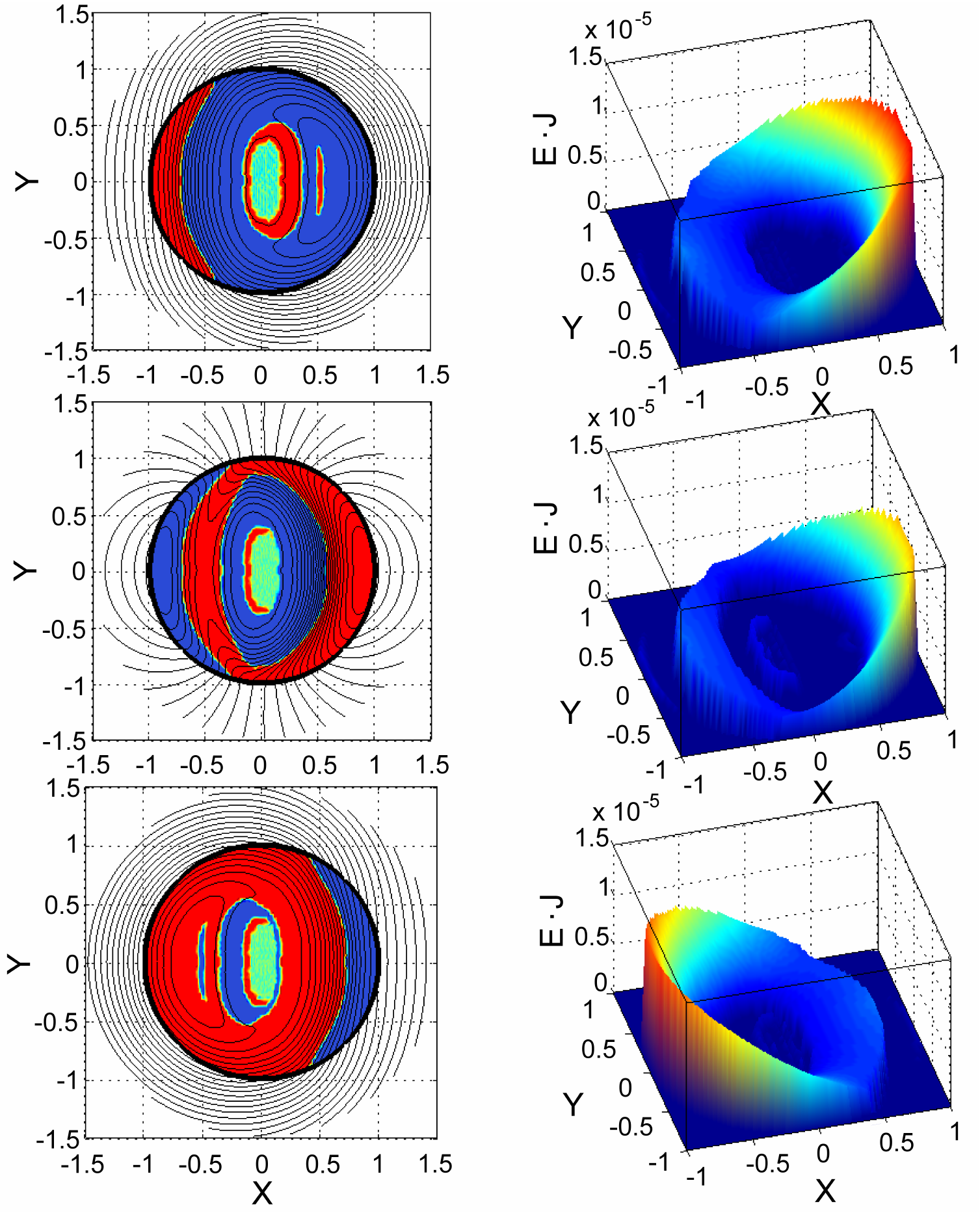}
\hspace*{-0.1cm}\vline
\includegraphics[height=7.5cm,width=6.4cm]{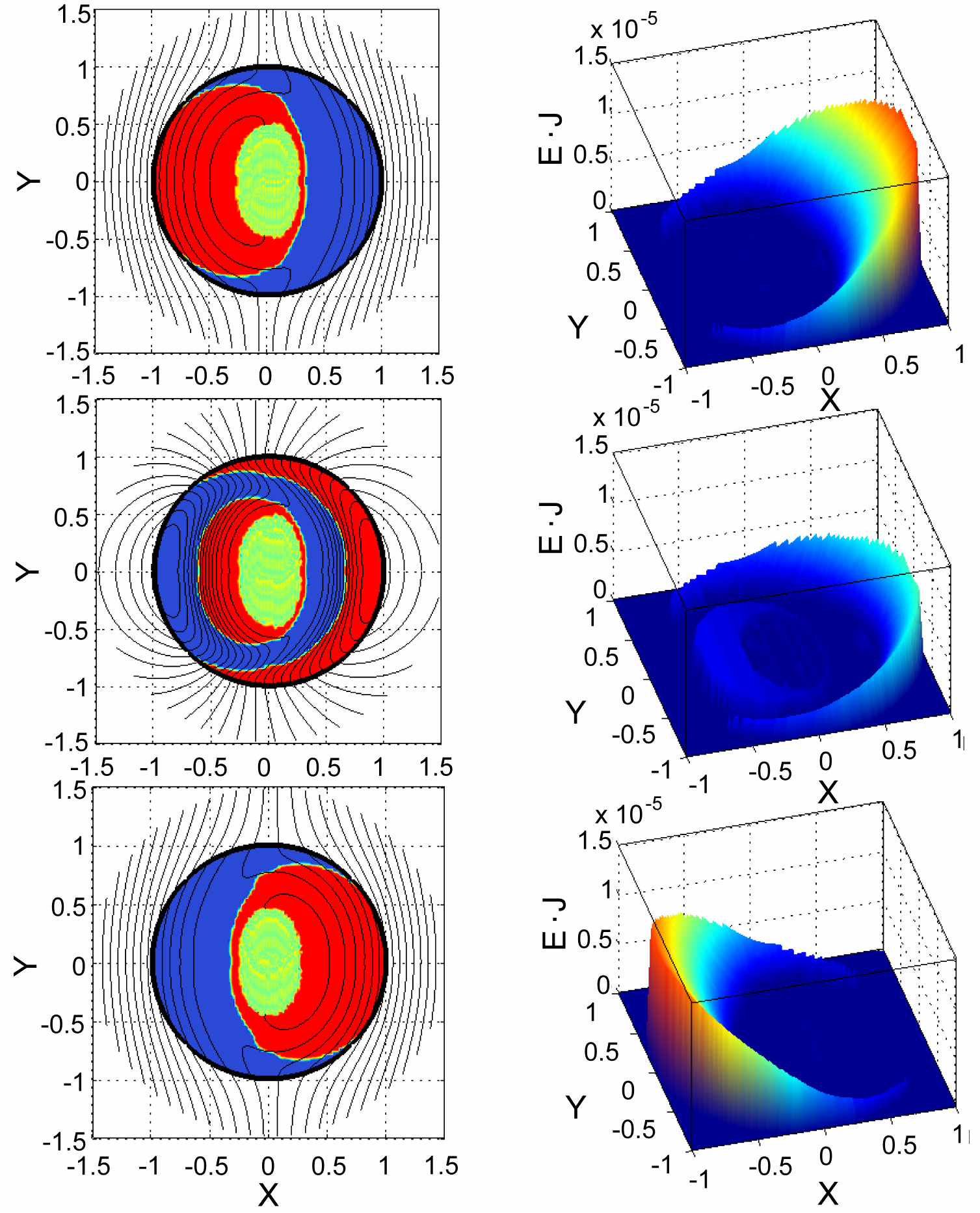}
\end{center}
\caption{\label{Figure_6_13} 
Evolution of the magnetic flux lines and their corresponding profiles
of current with asynchronous oscillating sources $B_{0,y}$ and $I_{tr}$ of
amplitudes $B_{\tt{a}}=4$ and $I_{\tt{a}}=0.5$ (left side into each pane),
accordingly to the temporal processes displayed into Fig.~\ref{Figure_6_12}(b)
``left pane herein'' and Fig.~\ref{Figure_6_12}(c) ``right pane herein''. Also
the corresponding profiles for the local density of power dissipation
\textbf{E}$\cdotp$\textbf{J} are shown (right side into each pane). In
particular, in this figure we show the set of results for the last branch of the
dominant excitation according to the time-steps marked with the labels (6), (8),
and (10) in Fig.~\ref{Figure_6_12}. More details to follow up the
electromagnetic quantities along the cyclic process are found in the section of
supplementary material, pages \pageref{Figure_S_17}-\pageref{Figure_S_20}.
}
\end{figure}

\paragraph{(~\textit{B.}~) \textit{Flux dynamics}\vspace*{0.5cm}\\}

Before discussing the results obtained for the total AC loss in the set of
experiments displayed in Fig~\ref{Figure_6_12}, some outstanding facts related
to the rich phenomenology found for the local dynamics of the electromagnetic
quantities are worth of mention. For example, when the applied magnetic
field and the transport current are synchronous, and their associated amplitudes
$(B_{\tt{a}},I_{\tt{a}})$ are weak enough such that a flux-free core
remains along the AC cycles (i.e., as long as $i_{\tt{a}}^{*}\neq1$
and $b_{\tt{a}}^{*}<1$), it is more or less simple to identify the active
zones in the AC cycles via the previous knowledge of the virgin branch (see
Figs.~\ref{Figure_6_5} \& \ref{Figure_6_6}, pags.
\pageref{Figure_6_5}-\pageref{Figure_6_6}), and therefore, explaining and
obtaining the AC loss may be achieved if the distribution of screening currents
is well known for the first half of the AC period. However, when the temporal
dynamics of the isolated excitations shows an asynchronous response, this is not
longer valid. The reason is, that the hysteretic losses produced along the
virgin branch are not monotonic
concerning the temporal evolution of both electromagnetic excitations, such that
the accruing hysteretic losses for the lower limit in the first integral of
Eq.~(\ref{Eq.6.21}) are different for the first
and second half of the cyclic period. In other words, the distribution of
screening currents in the first peak of the excitation with smaller frequency
or below so called dominant excitation [time step 2 in
Fig.~\ref{Figure_6_12}] may drastically differ from those conceived in the
second and third peaks [time steps 6 and 10].\footnotemark[4] Hence, the
distribution of screening currents for the first half period of
the AC cycle cannot be fetched through their distribution in the virgin branch.
Therefore, we want to call reader's attention to the fact that for making use of
Eq.~(\ref{Eq.6.21}) for the calculation of AC losses, the steady regimes
for the limits of the time-integral have to be defined for the excitation peaks
defining the second half period of the dominant excitation. Thus, a proper
description of the profiles of current density in an asynchronous AC regime must
be done at least for this temporal branch (see Fig.~\ref{Figure_6_13}).

\footnotetext[4]{A thorough follow-up of the local distribution of
current density and lines of magnetic field for asynchronous sources with double
frequency effects, including the virgin branch has been drawn in
Fig.~S17, supplementary material, pag.~\pageref{Figure_S_17}.
Likewise, their corresponding profiles for the components of magnetic flux
density are shown in Figs. S18  (pag.~\pageref{Figure_S_18}) and S19
(pag.~\pageref{Figure_S_19}), as well as the evolution of the local density of
power dissipation $\textbf{E}\cdot\textbf{J}$ in Fig.~S20
(pag.~\pageref{Figure_S_20}), for the time steps labeled in
Fig.~\ref{Figure_6_12}.}

\begin{figure}[t]
\centering
\includegraphics[height=10cm,width=13cm]{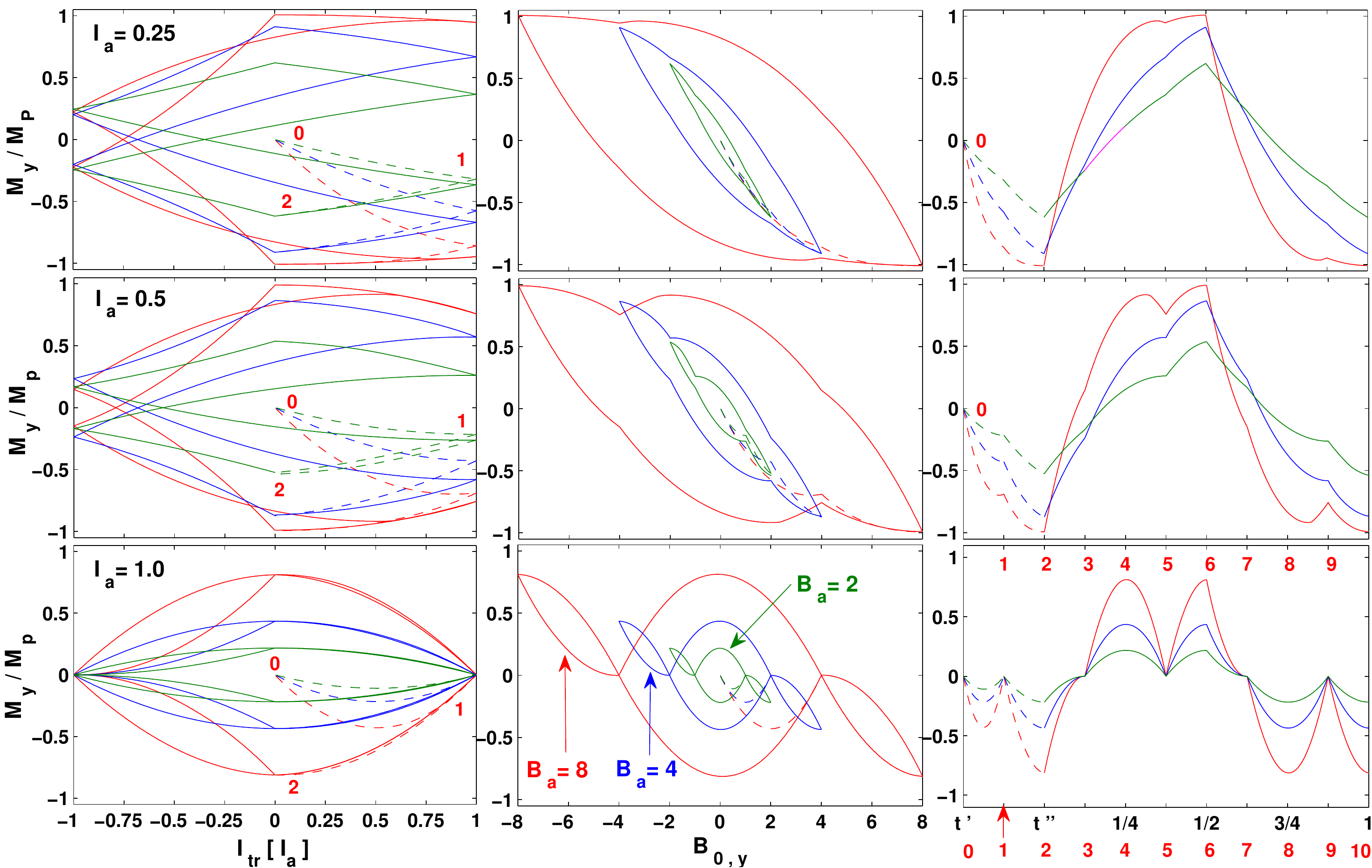}
\caption{\label{Figure_6_14} The dimensionless magnetic moment
$M_{y}/M_{p}$ for the AC asynchronous excitations
displayed in Fig.~\ref{Figure_6_12}(b) where the applied magnetic field have
the role of dominant excitation. Curves are shown as function of the
injected transport current $I_{tr}$ in units of their amplitude $I_{\tt{a}}$
(\textit{left column}), the applied magnetic
field $B_{0,y}$ \textit{(central column)}, or either
by its temporal evolution \textit{(right column)}. Same color scheme 
to point out the amplitude of the AC magnetic field ($B_{\tt{a}}$) has been used
in all subplots.
}
\end{figure}

Analyzing the distribution of current density profiles in
Fig.~\ref{Figure_6_13} and their corresponding profiles for the local density of
power dissipation $\textbf{E}\cdot\textbf{J}$, we found at least two interesting
facts which are worth of mention. On one side, whether it is the magnetic field
or the transport current, the excitation which leads the role of  dominant,
multiple domains or active zones connected between them may appear, which makes
it impossible to find out a feasible analytical solution for
the flux front boundary in the infinite spectra of combinations between the
amplitudes $B_{\tt{a}}$ and $I_{\tt{a}}$, specially if the pattern of current
density is far away of the approaches for profiles of the kind
\textit{current-like} or
\textit{field-like}. On the other side, a most striking fact revealed in
Fig.~\ref{Figure_6_13} is that contrary to the behavior displayed for the
local profiles of density of power dissipation $\textbf{E}\cdot\textbf{J}$, when
both electromagnetic excitations are synchronous (see e.g.
Fig.~\ref{Figure_6_7}, pag.~\pageref{Figure_6_7}), in asynchronous cases the
zone of heat release is no longer localized in one side of the
superconducting sample. Thus, the idea of focusing heat release in some part of
the superconductor requires a special attention in the synchronization of
sources.

\paragraph{(~\textit{C.}~) \textit{Magnetic response}\vspace*{0.5cm}\\}

\begin{figure}[t]
\centering
\includegraphics[height=10cm,width=13cm]{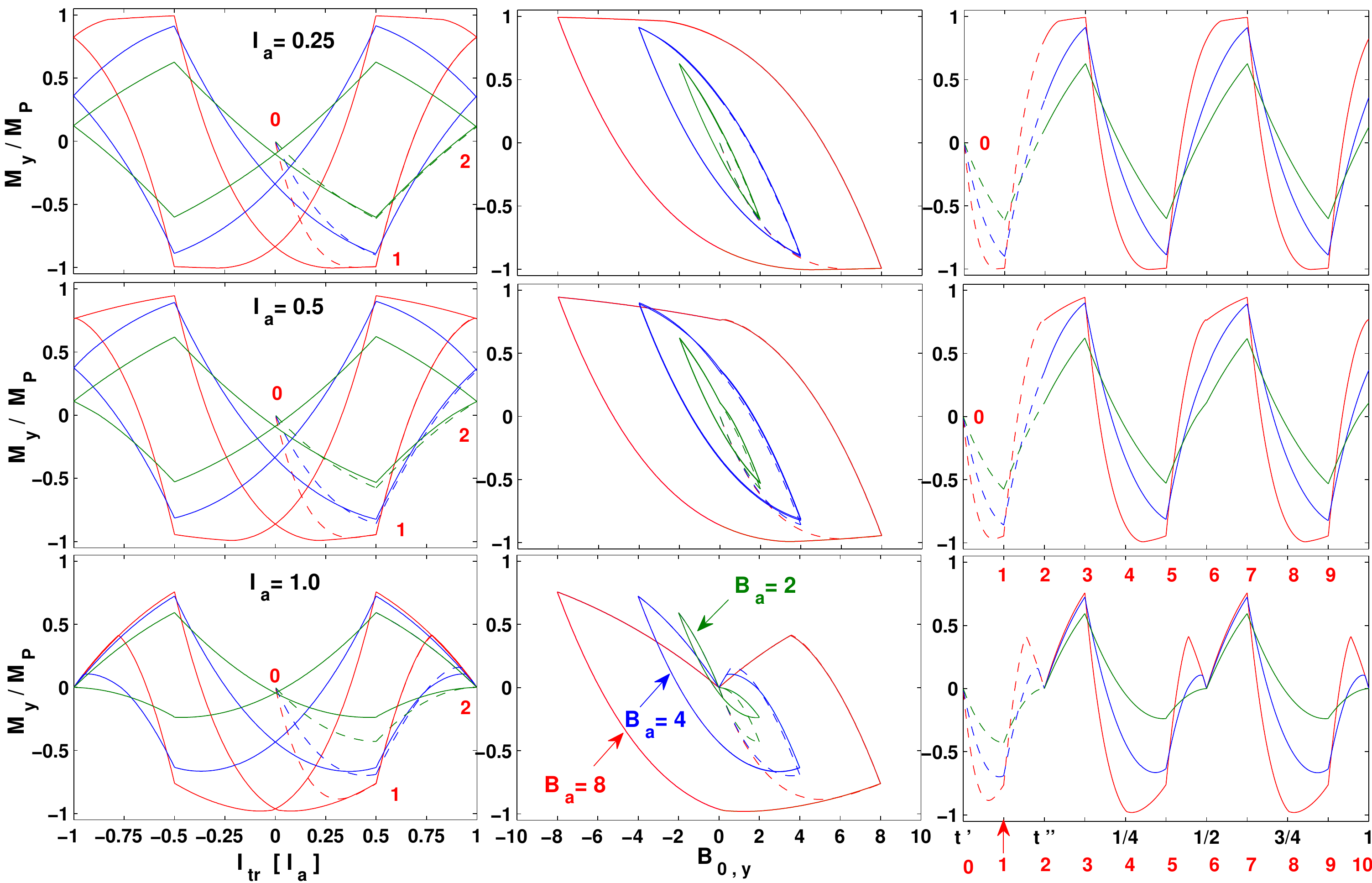}
\caption{\label{Figure_6_15} The dimensionless magnetic moment
$M_{y}/M_{p}$ for the AC asynchronous excitations
displayed in Fig.~\ref{Figure_6_12}(c) where the transport current has the
role of dominant excitation. Curves are shown as function of the
injected transport current $I_{tr}$ in units of their amplitude $I_{\tt{a}}$
(\textit{left column}), the applied magnetic
field $B_{0,y}$ \textit{(central column)}, or either
by its temporal evolution \textit{(right column)}. Same color scheme 
to point out the amplitude of the AC magnetic field ($B_{\tt{a}}$) has been used
in all subplots.
}
\end{figure}

Another interesting feature which derives from the study of asynchronous
excitations is the actual possibility of finding ``exotic'' magnetization loops
as a function of the AC sources (see Figs.~\ref{Figure_6_14} \&
\ref{Figure_6_15}), where the straightforward competition between the
magnetization currents (by consumption) and the injected current lines, may be
visualized in terms of a non local macroscopic measurement. 

In Fig.~\ref{Figure_6_14}, the component of magnetic moment $M_{y}$ is displayed
for the AC process in Fig.~\ref{Figure_6_12}(b), where the
AC magnetic field dominates the cyclic period of excitation. The whole set of
results for $M_{y}$ have been renormalized according
to the maximal expected value for the magnetic moment when only applied
magnetic field is considered ($M_{p}=2/3$). Thus, curves are shown as function
of the isolated electromagnetic excitations, $I_{tr}$ (in units of their
associated AC amplitude $I_{\tt{a}}$) at the left column, $B_{0,y}$ at the
central column, as well as by the time defining the first
steady-period (right column). Notice that,  the virgin branch which does not
play any role for the integration of the AC losses per cyclic periods is
shown through dashed lines. Furthermore, results have been organized
accordingly to the associated amplitudes for the applied density of magnetic
flux $B_{\tt{a}}$, such that $B_{\tt{a}}=2$ corresponds to the green curves,
$B_{\tt{a}}=4$ to the blue curves, and $B_{\tt{a}}=8$ to the red curves.
Likewise, the set of curves shown for each row can be straightforwardly
associated to a single value for the amplitude of the AC transport current,
$I_{\tt{a}}=0.25$ (first row), $I_{\tt{a}}=0.5$ (second row),  and
$I_{\tt{a}}=1$ (third row). Analogously, the corresponding set of curves
obtained for the component of magnetic moment $M_{y}$ in those cases where the
AC transport current dominates the cyclic period of excitation, are shown in the
same fashion above described in Fig.~\ref{Figure_6_15}.

Outstandingly, whether $B_{0,y}$ (Fig.~\ref{Figure_6_14}) or $I_{tr}$
(Fig.~\ref{Figure_6_15}) is the dominant excitation, and for low values of
$I_{\tt{a}}$, the magnetization loops as function the magnetic flux density
$M_{y}(B_{0,y})$ show a Bean-like behavior. As the value of
$I_{\tt{a}}$ increases, notorious deformations of the Bean-like structures for
the magnetic moment appear. Nevertheless, the behavior is radically
different comparing the double frequency effects provided by one or another
excitation, as it is explained below.

\paragraph{\indent(~\textit{i.}~) \textit{Transport current with double
frequency}\vspace*{0.5cm}\\}

On the one hand, when it is the AC transport current, $I_{tr}$, that  shows a
double frequency, the magnetization curves $M_{y}(B_{0,y})$ display a symmetric
behavior in the regions (left-right) of the periods
$[B_{\tt{a}}\Rightarrow-B_{\tt{a}}]$ and $[-B_{\tt{a}}\Rightarrow B_{\tt{a}}]$
(see Fig.~\ref{Figure_6_14}). On the contrary, for $M_{y}(I_{tr})$, one can
notice the existence of a symmetrization of the curves of magnetic moment
regarding their positive and negative values (\textit{up/down}). Evidently, the
steady-states where the maximum consumption of the magnetization currents
occurs, always arise when the asynchronous AC excitation $[B_{0,y},I_{tr}]$
reaches the values for the current's peaks [see Fig.~\ref{Figure_6_12}(b)],
i.e., for the time-steps (3) $[B_{\tt{a}}/2,-I_{\tt{a}}]$, (5)
$[-B_{\tt{a}}/2,I_{\tt{a}}]$, (7) $[-B_{\tt{a}}/2,-I_{\tt{a}}]$, and (9)
$[B_{\tt{a}}/2,I_{\tt{a}}]$. Then, by increasing (decreasing) the value of
$I_{\tt{a}}$, a progressive decreasing (increasing) of the magnetic moment at
these points ends up in the simultaneous collapsing of the magnetization curves
$(M_{y}\equiv0)$ for the half periods of the dominant excitation $B_{0,y}$, as
long as $I_{\tt{a}}=\pm I_{c}$. Latter fact is followed by the symmetrization of
the loops, either as
functions of $B_{0,y}$ and $I_{tr}$, into characteristic lenticular shapes
bounded by two non-connected magnetization curves, both defined by the elapsed
periods in which the time derivative of $I_{tr}(t)$ is positive, i.e., for the
temporal branches $(3\Rightarrow5)$ and $(7\Rightarrow9)$. Then, the connecting
curves for the abovementioned magnetization branches shows characteristic
lashing shapes when the time derivative of $I_{tr}(t)$ is negative. Remarkably,
as a consequence of this process, the output signal $M_{y}(t)$ does not show the
low-pass filtering effects conceived for synchronous excitations. In fact, given
our study, we prove that the low-pass filtering effect for superconducting
wires, may only be envisaged when the temporal evolution of the injected AC
transport current and the perpendicular magnetic field is fully synchronous (in
phase and frequency).

\paragraph{\indent(~\textit{ii.}~) \textit{Applied magnetic field with double
frequency}\vspace*{0.5cm}\\}

On the other hand, when it is the AC density of magnetic flux, $B_{0,y}$,
the electromagnetic source disclosing the double frequency effect
[Fig.~\ref{Figure_6_12} (c)], the calculated curves of magnetization
$M_{y}(B_{0,y})$, are outstandingly different. Thus, there are no symmetry
conditions which may be observed in this representation (2nd column in
Fig.~\ref{Figure_6_15}).
However, for the set of magnetization curves as a function of the transport
current, $M_{y}(I_{tr})$, we have noticed a well-defined symmetrization of 
the magnetization loops regarding to the positive and negative
values of $I_{tr}(t)$ (\textit{left/right}). Notwithstanding, in terms of the
magnetization curves $M_{y}(B_{y})$, there is also a further fact to be
mentioned. Strikingly, by using this representation it is possible to note that
the steady-states where the consumption of the magnetization currents becomes
evident, are mainly present along the period in which the time
derivative of $B_{0,y}(t)$ is negative, whilst the AC transport current evolves
through the current's peaks, i.e., for the temporal branches defined by the
time-steps (1) $[B_{\tt{a}},I_{\tt{a}}/2]$ to (3) $[-B_{\tt{a}},I_{\tt{a}}/2]$,
and (5) $[B_{\tt{a}},-I_{\tt{a}}/2]$ to (7) $[-B_{\tt{a}},-I_{\tt{a}}/2]$. 
Moreover, the magnetic response of the superconductor is not monotonic along
these branches. For example, from the time step (5) until the time step (6)
$[0,-I_{\tt{a}}]$, the electromagnetic excitations, $B_{0,y}$ and $I_{tr}$,
have the same tendency (both decreasing). However, if $I_{\tt{a}}$ tends to
the limiting value $I_{c}$, $M_{y}$ may increase and
decrease within the same period. Then, from the time step (6) to the
time step (7) both evolve in opposite directions, but however the magnetic
moment always increases along this period. On the other hand, for the following
analogous branch in the AC period of magnetic field, say the time-steps (9) to
(11), the magnetic moment curve is the same, but the competition between the
electromagnetic sources $B_{0,y}$ and $I_{tr}$ is opposite to the
aforementioned evolution. Thus, depending on the intensities of the
electromagnetic sources, both reduction or enhancement of the hysteretic AC
losses may envisaged when there is a difference between the
oscillating excitation frequencies. Recall that, in contrast, by assuming only
a relative difference in phase, only reductions of the actual AC loss may be
foretell.

\begin{figure}[t]
\centering
\includegraphics[height=6.5cm,width=13cm]{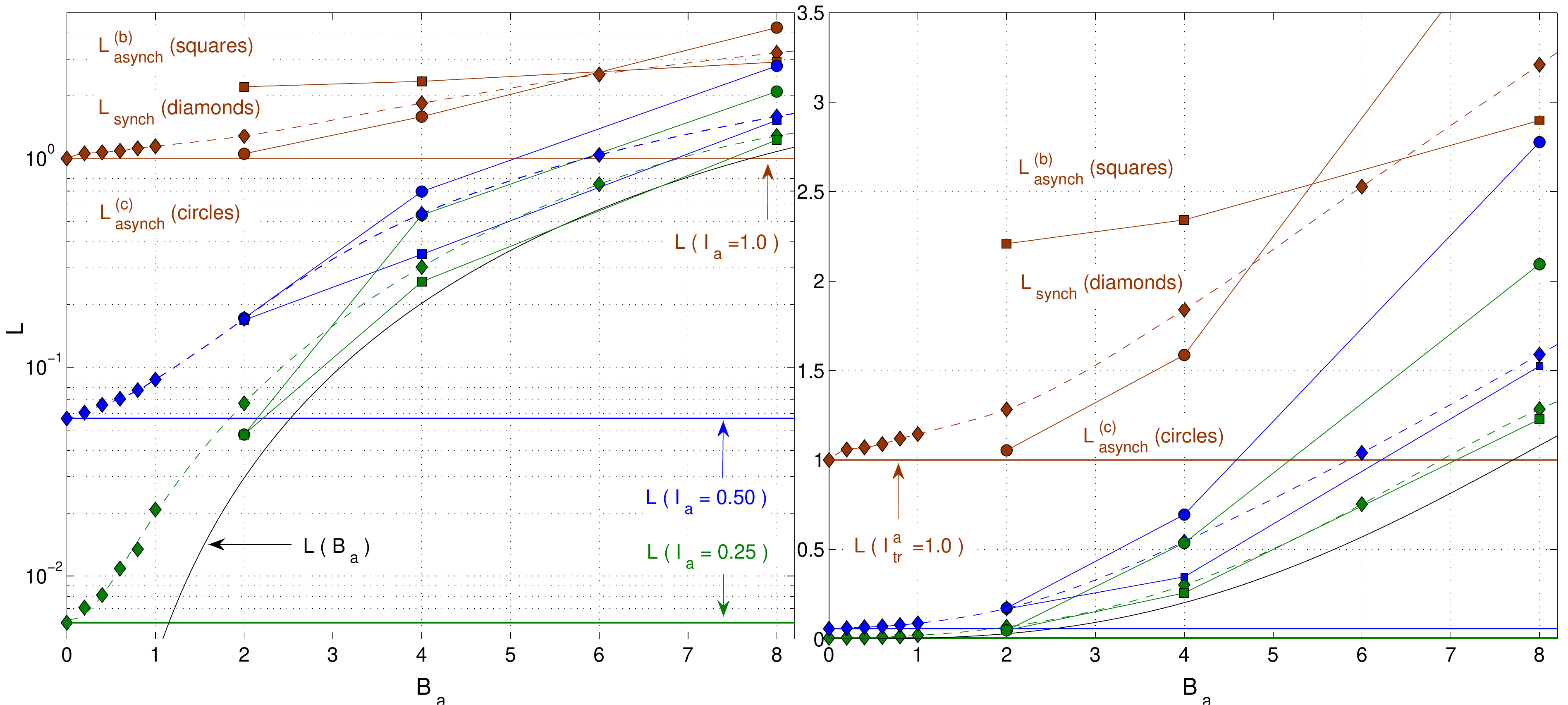}
\caption{\label{Figure_6_16} Hysteretic AC losses per cycle
for asynchronous sources accordingly to the excitations shown
in Fig.~\ref{Figure_6_12}(b) ``Herein, $L_{asynch}^{(b)}$ :
square-solid-lines'', and  Fig.~\ref{Figure_6_12}(c) ``Herein,
$L_{asynch}^{(c)}$ : circle-solid-lines''. The results are compared with the
curve of losses for synchronous sources, $L_{synch}\equiv
L(B_{\tt{a}},I_{\tt{a}})$ predicted above (Fig.~\ref{Figure_6_10}), and the
curves for isolated excitations $L(B_{\tt{a}})$ and $L(I_{\tt{a}})$. The whole
set of results
is also plotted in linear scale. Units for losses are $(\mu_{0}/4\pi)\omega
R^{2} J_{c}^{2}$.}
\end{figure}

\paragraph{(~\textit{D.}~) \textit{AC Losses in asynchronous
systems}\vspace*{0.5cm}\\}

\begin{figure}[t]
\begin{center}
\includegraphics[width=0.7\textwidth]{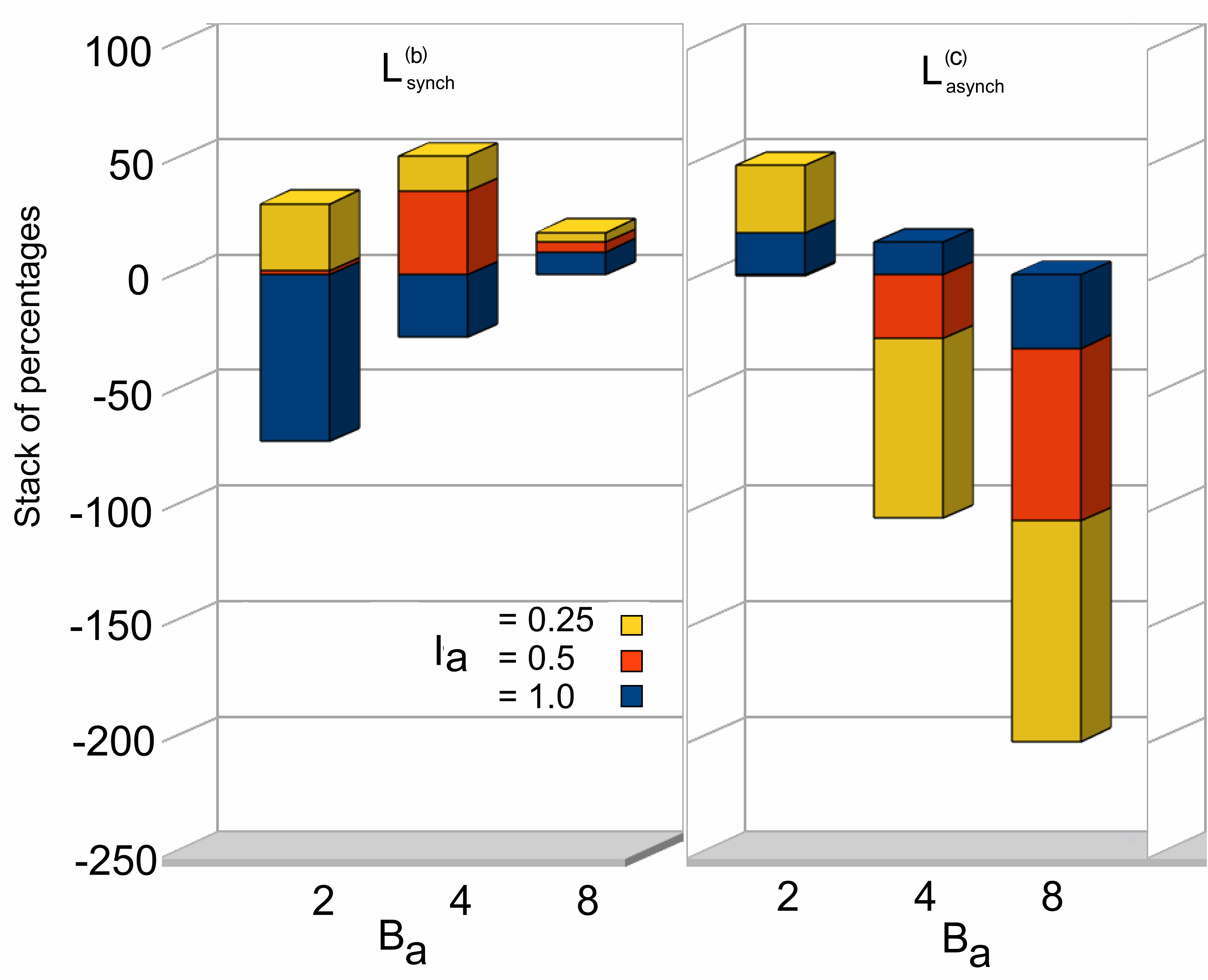}
\end{center}
\caption{\label{Figure_6_17} 
Percent change between the AC loss for synchronous excitations,
$L_{synch}$, and the losses $L_{asynch}^{(b)}$ (at the left-side) and
$L_{asynch}^{(c)}$ (at the right-side), for combinations of three different
amplitudes $B_{\tt{a}}$ and $I_{\tt{a}}$.
}
\end{figure}

In Fig.~\ref{Figure_6_16}, the hysteretic AC loss calculated for the
experimental configurations conceived in Fig.~\ref{Figure_6_12} [panes (b) and
(c)] are shown. To be specific, $L_{asynch}^{(b)}$ and $L_{asynch}^{(c)}$, are
shown in terms of the amplitude of the applied density of magnetic flux
$B_{\tt{a}}$, whilst the different values for $I_{\tt{a}}$ are pointed in terms
of the sequence of colors for the DC loss
curves depicted in Fig.~\ref{Figure_6_10} (pag.~\pageref{Figure_6_10}).
Likewise, results are compared with the corresponding curves for the actual AC
loss when the synchronous electromagnetic excitations were considered, i.e.,
$L_{synch}\equiv L(B_{\tt{a}},I_{\tt{a}})$. Outstandingly, for both cases,
remarkable variations of the AC loss occur. Thus, with the aim of providing
a clearest understanding of the range of variations in the AC loss curve for the
configurations abovementioned, and further help the readers in the visualization
of the numerical data, in Fig.~\ref{Figure_6_17} we show the percentage
relation between the calculated losses for synchronous excitations,
$L_{synch}$, and the calculated losses for the asynchronous
cases, $L_{asynch}^{(b)}$ and $L_{asynch}^{(c)}$.

Notice that, on the one hand, when the applied magnetic field provides the
dominant oscillating period (the impressed AC transport current shows
a relative double frequency, Fig.~\ref{Figure_6_12}(b)), the resulting
comparison between the calculated losses $L_{asynch}^{(b)}$ and $L_{synch}$ for
high values of $B_{\tt{a}}$ [Fig.~\ref{Figure_6_17}], shows a small but
sizable increase of the hysteretic loss as a consequence of the double frequency
effect ($\sim4-10\%$ for $B_{\tt{a}}\equiv B_{p}=8$). Then, assuming that
$I_{\tt{a}}\equiv I_{c}=1$, we have found that the AC loss reduces as
$B_{\tt{a}}$ decreases. However, an outstanding fact is that for the lowest
values of $B_{\tt{a}}$ and $I_{\tt{a}}$, a notorious increase of the
actual AC losses appears. For example, for $B_{\tt{a}}=B_{p}/2=4$  and
$I_{\tt{a}}=I_{c}/2=0.5$, deviation is about $36\%$. Likewise, for
$B_{\tt{a}}=B_{p}/4=2$  and $I_{\tt{a}}=I_{c}/4=0.25$, deviation is about $29\%$

On the other hand, when the electromagnetic source with the double frequency is
the applied magnetic field [Fig.~\ref{Figure_6_12}(c)], the resulting AC loss
($L_{asynch}^{(c)}$) for high values of $B_{\tt{a}}$ shows a significant
reduction as compared to the predicted losses for synchronous configurations
($\sim95\%$ for $B_{\tt{a}}\equiv B_{p}=8$ and $I_{\tt{a}}\equiv I_{c}/4=0.25$).
However, by reducing $B_{\tt{a}}$ a notorious increase of the AC loss may be
revealed depending on the value of $I_{\tt{a}}$. In fact, we call readers'
attention about the relative increase of the AC losses as compared to those of
the synchronous cases: $\sim14\%$ for ($B_{\tt{a}}=4$ , $I_{\tt{a}}=1$), and
$\sim18\%$ for ($B_{\tt{a}}=2$ , $I_{\tt{a}}=1$). Moreover, for those cases
with the lower
values of $B_{\tt{a}}$, i.e., $B_{\tt{a}}=2$,  increases of the AC loss are also
found for $I_{\tt{a}}=0.25$ ($\sim29\%$),
whilst for the intermediate case [$B_{\tt{a}}=2$,$I_{\tt{a}}=0.5$], it shows a
reduction of the AC loss of less than 1\%. The remarkable point here, is that
for asynchronous excitations of $B_{0}$ and $I_{tr}$, reductions of the AC
losses can be only asserted if both sources evolve with the same frequency,
i.e.: if one is restricted to shifts in phase.


 \chapter*{}
 \vspace*{-3cm}
 \section*{\Huge{Conclusions II}}
 \markboth{\sc \textbf{Conclusions II}}{\sc \textbf{Conclusions II}}
 \addcontentsline{toc}{chapter}{\sc \textbf{Conclusions II}}
 \vspace*{2cm}

In this part, we have shown that our general critical state theory for the
magnetic response of type-II superconductors in the framework of optimal
control variational theory and computational methods for large scale
applications may be applied in an extensive number of configurations. 

In order to summarize the main physical features extracted from our numerical
experiments, the conclusions are presented according to the
previous sequence of chapters as follows:

\vspace*{0.5cm}
\paragraph{\textbf{\textit{Chapter 4 \vspace{1.0cm}\\}}}

General critical state problems have been solved for a wide number of examples 
within the infinite slab geometry. All of them share a three dimensional 
configuration for the magnetic field, i.e., ${\bf H}=(H_{x},H_{y},H_{z})$, under
various magnetic processes, and different models for the critical current
restriction or
material law. Thus, we have considered several physical scenarios classified by
the ansatz for the flux depinning and cutting processes (basically affecting the
critical current thresholds $J_{c\perp}$ and $J_{c\parallel}$), their
relative importance (given by $\chi\equiv J_{c\parallel}/J_{c\perp}$), and a
\textit{coupling} index $n$ which controls the smoothness of the
material law. In summary, the following scenarios have been analyzed:

\begin{enumerate}

\item Isotropic solutions, in which the limiting case with $\chi^{2}=1$ and
$n=1$ produces states under the 1D constraint $J=J_{c}$.

\item T-state solutions, in which the approximation $~\chi\gg 1$ produces the
result $J_{\perp}=J_{c\perp}$, and $J_{\parallel}$ may  be arbitrarily high. Our
predictions show an excellent agreement with previous analytical results in the
literature, and extend the theory to the full range of applied magnetic fields.

\item CT-state solutions in which $\chi\geq 1$ for several cases within the
{\em rectangular region} given by the threshold conditions $J_{\perp}\leq
J_{c\perp}$ and
$J_{\parallel}\leq J_{c\parallel}$ are analyzed. Outstandingly,
the appearance of the flux cutting limitation takes place as a sudden corner in
the magnetic moment curves in many cases. The corner establishes a criterion for
the range of application of T-state models.

\item SDCST solutions, in which the possible coupling between the flux depinning
and cutting limitations has been studied through the solution of {\em smoothed}
DCSM cases. In particular, we consider the effect of rounding the corners of the
rectangular region $J_{\perp}\leq J_{c\perp}$ and $J_{\parallel}\leq
J_{c\parallel}$, by the {\em superelliptic} region criterion $(J_{\parallel}/
J_{c\parallel})^{2n}+(J_{\perp}/ J_{c\perp})^{2n}\leq 1$ with $1\leq n <
\infty$. It is shown that, under specific conditions (paramagnetic initial state
and low perpendicular fields), important differences in the predictions of the
magnetic moment behavior are to be expected. The differences in ${\bf M}$ have
been related to the behavior of the critical current vector ${\bf J}_c$ around
the corner of the rectangular region.

\end{enumerate}

\begin{figure}[t]
\centering
\includegraphics[width=1.0\textwidth]{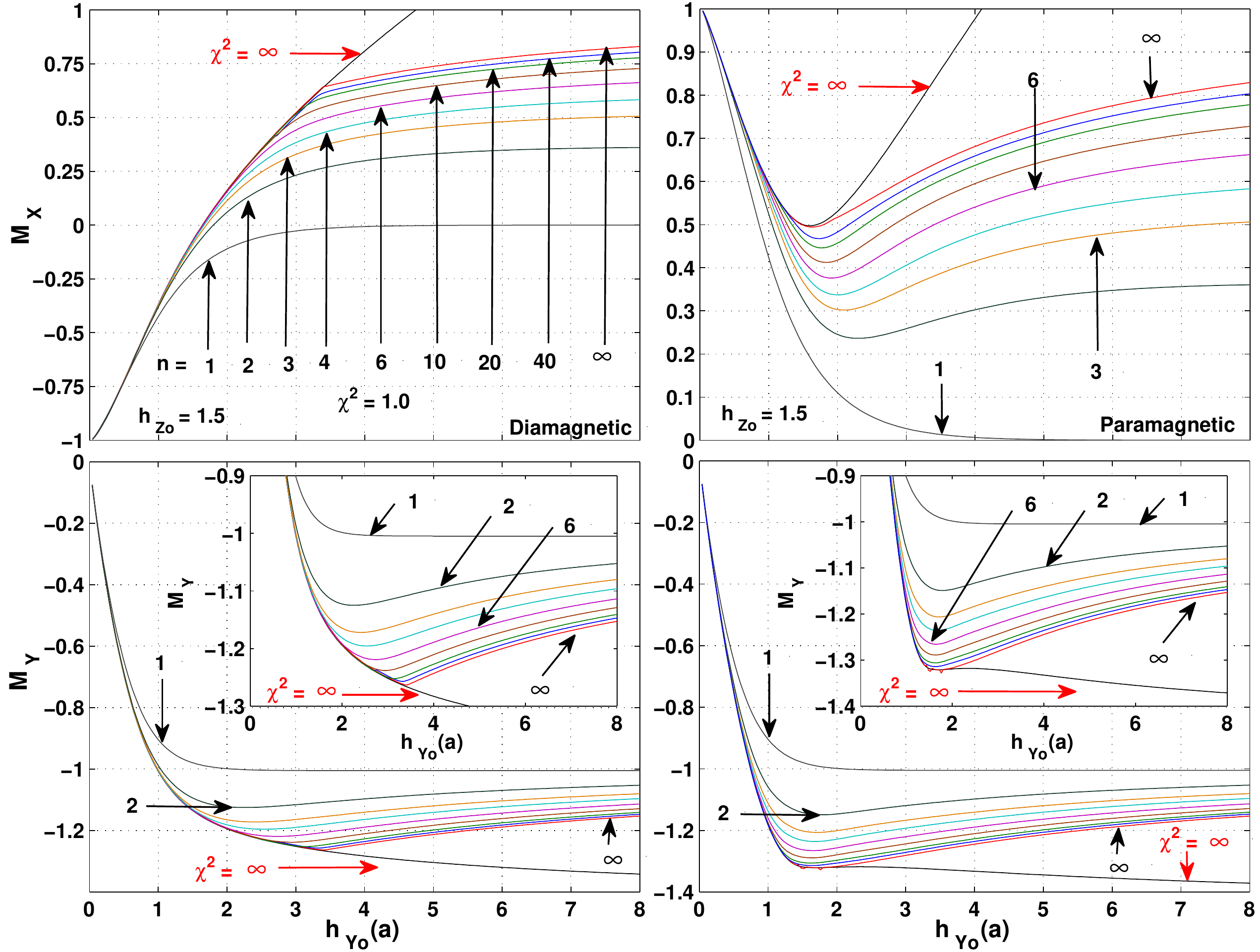}
\caption*{\label{Figure_Conc_4_1} Figure II-1: The magnetic moment
components of the slab $M_{x}$ (top) and $M_{y}$ (bottom) per unit area  as a
function of the applied magnetic field $h_{y_{0}}$ for the experimental
configurations depicted in Fig.~\ref{Figure_4_2}. By comparison, results for
several models in the diamagnetic (left panes) and paramagnetic (right panes)
configurations are shown. In terms of our SDCST we display the magnetization
curves for: the infinite bandwidth model or so called model of T-states
($\chi^{2}\rightarrow\infty$, $J_{c\perp}\neq0$), the conventional DCSM
($\chi^{2}=1$, $n\rightarrow\infty$), several material laws defined by the
superelliptical regions $\chi^{2}=1$ and $n=2$, 3, 4, 5, 6 and 10, and finally
the isotropic model or superelliptical region with $\chi^{2}=1$ and n=1.
}
\end{figure}

Remarkably, the whole set of physical features linked to the different material
laws may be depicted in terms of the magnetization curves for the
aforementioned experimental configurations [see Fig.~\ref{Figure_4_2},
pag.~\pageref{Figure_4_2}]. The main findings are synthesized in
figure~\hyperref[Figure_Conc_4_1]{II-1}. 

Firstly, we have noticed a pronounced peak effect in
both components of the magnetic moment. We emphasize that whatever region is
considered [excepting the limiting cases ``$\chi^{2}=1$, $n=1$'' (isotropic
model), and ``$\chi^{2}\to\infty$'' (T- or infinite bandwidth- model)], the peak
effect in the paramagnetic case is predicted for both components of the
magnetization. Thus, we argue that the peak effect cannot be
interpreted as a direct evidence of an elliptical material law. Instead of this,
it is a universal signal of the anisotropy effects involved in a general
description of the material law. The evolution of the peak effect as a function
of $\chi^{2}$ has been shown in Fig.~\ref{Figure_4_15}
(pag.~\pageref{Figure_4_15}). There, we note that an increase of the bandwidth
$\chi^2$ produces a stretched magnetic peak. Consequently, paramagnetic effects
are visible over a wider range as the cutting threshold value $J_{c \parallel}$
increases. We also emphasize that the overall effect of increasing the value
$\chi^{2}=(J_{c \parallel}/J_{c \perp})^{2}$ is that the components of ${\bf M}$
get closer to the {\em master} curves defined by $\chi\to\infty$. 

Secondly, some additional and distinctive signals for the
different models have been also observed. On the one hand, for the isotropic
model, the collapse of the magnetization is achieved while $J_{\parallel}$ is
monotonically reduced [Figs.~\ref{Figure_4_3} -- \ref{Figure_4_5},
pags. \pageref{Figure_4_3} -- \pageref{Figure_4_5}]. When the
material law is the infinite bandwidth model ($\chi^{2}=\infty$, or so called
T-state model) the magnetization collapse does not occur, and there
is no restriction on the longitudinal component of the current that increases
arbitrarily towards the center of the sample [Figs.~\ref{Figure_4_6} --
\ref{Figure_4_10}, pags. \pageref{Figure_4_6} -- \pageref{Figure_4_10}]. This
corresponds to the absence of flux cutting, i.e.: $J_{\parallel}$ does not
saturate by reaching a threshold value $J_{c \parallel}$. For rectangular or
{\em smooth rectangular} regions [Figs.~\ref{Figure_4_11} -- \ref{Figure_4_15},
pags. \pageref{Figure_4_11} -- \pageref{Figure_4_15}], together with the absence
of collapse, one also observes that $J_{\parallel}$ basically saturates to a
value that depends on the smoothing parameter $n$ (exactly to $J_{c\parallel}$
for the very rectangular case $n\to\infty$). Remarkably, when a rectangular
section is assumed, the sample globally reaches the CT state (corner of the
rectangle). As a consequence of the sharp limitation for $J_{\parallel}$, a
well-defined corner in the magnetic moment components $M_{x}$ and $M_{y}$
appears, both for the diamagnetic and paramagnetic cases (see e.g.,
Fig.~\hyperref[Figure_Conc_4_1]{II-1}). This clear trace of the DCSM establishes
the departure from the master curves defined by the T-state, and has been
assigned to the instant at which the sample reaches the CT state. 

Let us call the readers' attention about a noticeable gap in
Fig.~\hyperref[Figure_Conc_4_1]{II-1}, separating the isotropic
model ($\chi^{2}=1,~n=1$) and the square model ($\chi^{2}=1,~n\to\infty$).
In fact, if one compares Fig.~\hyperref[Figure_Conc_4_1]{II-1} and
Fig.~\ref{Figure_4_15} (pag.~\pageref{Figure_4_15})
one can realize that smooth models for a given ratio
$\chi\equiv J_{c\parallel}/J_{c\perp}$ will fill the gap between the master
limiting curves defined by the rectangular ($\chi,n\to\infty$) and elliptic
($\chi,n=1$) models, and their corresponding curves for different values of
$\chi$ will intersect in a complicated fashion. In this sense, we argue that
the magnetization curves by themselves do not provide unambiguous information
on the material law which defines the critical state dynamics in type II
superconductors. Moreover, notice that in the regime of low fields
$H_{z}\sim h_{y}(a)$ the material law is
indistinguishable and the magnetic moment may be reproduced even by the
isotropic model. However, we have noticed that, although the dynamics of the
profiles $H_{x}$, $J_{y}$, and $J_{x}$  is almost indistinguishable between the
smooth and rectangular models a clear distinction arises by analyzing
$J_{c\parallel}$. On the one hand, when the rectangular model is assumed
$J_{\parallel}$ reaches the threshold value $J_{c\parallel}$, and the entire
specimen verifies a CT-state as the applied
magnetic field increases. On the other hand, when the rectangular region is
smoothed by the index $n$, the parallel component of the current density
eventually decreases to a value that depends on the values of $n$ and $\chi$.
Thus, further research along this line is suggested~\cite{P2-Clem_2011_PRB},
i.e.: the design of some experimental routine that defines a well
posed inverse problem for the determination of $\Delta_{\textbf{r}}$. 

Finally, in Appendix-I the critical angle (between vortices) criterion that
establishes the limitation on $J_{\parallel}$ has been modified for 3D
problems. It is shown that, in general, the concept may involve both
$J_{\parallel}$ and $J_{\perp}$ as one can see in
Eqs.~(\ref{Eq.4.12})~\&~(\ref{Eq.4.15}). Nevertheless, the influence of the
local magnetic anisotropy and the underlying effects at the
flux cutting mechanism are much less noticeable, especially for the
diamagnetic case, in which the full range of physically meaning values of
$\kappa_{c}$ produce a negligible variation.

\vspace*{0.5cm}
\paragraph{\textbf{\textit{Chapter 5 \vspace{1cm}\\}}}

Despite of extensive experimental and theoretical studies about the
electrodynamic response of type-II superconductors in longitudinal geometries,
much uncertainties remain about the interaction between flux depinning and
cutting mechanisms, and their influence in such striking observations as the
appearance of negative transport current flow, the enhancement of the critical
transport current density, and the observation of peak effects on the
magnetization curves. In this chapter, and based on the application of our
SDCST, we have reproduced theoretically the existence of
negative flow domains, local and global paramagnetic structures, emergence of
peak-like structures in the longitudinal magnetic moment, as well as the
compression of the transport current density for a wide number of experimental
conditions.

Here, the longitudinal transport problem in superconducting slab geometry
has been studied as follows: on the one hand, we have considered a
superconducting slab lying at the $xy$ plane and subjected to a transport
current density along the $y$ direction as it is shown in the left pane of
Fig.~\ref{Figure_5_3} (pag.~\pageref{Figure_5_3}). The slab is assumed to be
penetrated by a uniform vortex array along the $z$ direction, so that the local
current density along the thickness of the sample is entirely governed by the
depinning component ($J_{\perp}$) perpendicular to the local magnetic field.
Subsequently, a magnetic field source parallel to the transport current
direction is switched on. Then, the experimental conditions have been changed
through the value of the
external magnetic field $H_{y0}$.

The dynamical behavior of the transport current density is shown to rely on the
interaction between the cutting and depinning mechanisms. Moreover, the
intensity of the inherent effects has been shown to depend on the perpendicular
component $H_{z0}$, being more prominent as this quantity is reduced. In fact,
for restricted situations (infinite slab geometry and only fields parallel to
the surface, $H_{z0}=0$), we have shown that the prediction of
the counterintuitive effect of negative current flow in type-II superconductors
may be even predicted within a simplified analytical model (section 5.1.1).
Then, the three-dimensional effects are straightforwardly incorporated by
numerical methods when the third component of the local magnetic field ($H_{z}$)
is considered.

By means of our SDCST that allows to modulate the influence of
the different physical events, by using a \textit{superelliptical} material
law that depends on two parameters ($\chi\equiv J_{c\|}/J_{c\perp}$ and $n$)
accounting respectively for the intrinsic material anisotropy and for the
smoothness of the $J_{\parallel}(J_{\perp})$ law, we have quantitatively
investigated the influence of the flux cutting mechanism and shown that the peak
structures observed in the magnetization curves and the patterns of the
transport current along the central section of a superconducting sample are both
directly associated with the local structure of the vortex lattice.
Such dependence may become more pronounced as the extrinsic pinning of the
material is reduced, in favor of the flux cutting interactions. 
The same conclusion was pointed out from the experimental measurements of
Blamire et.al. (Ref.~\cite{P2-Blamire_2003,P2-Blamire_1986}) for high critical
temperature and conventional superconductors. It has been done by comparing the 
T-state model ($\chi\to\infty$), and the
smooth double critical state conditions CT$\chi$ with $\chi=1$, 2, 3 and 4, all
of them with the {\em smoothing} index $n=4$ and $J_{c\perp}=1$. Going into
detail, when the cutting threshold is high ($J_{c\|}\gg J_{c\perp}$
or $\chi \gg 1$) the emergence of negative current patterns is ensured because
unbounded parallel current density allows unconstrained rotations for the flux
lines as the longitudinal magnetic field increases. Thus, under a range of
conditions, the peak effects in the magnetic moment and a modulation of the
negative surface currents have been predicted. 

Only for completeness, it is to be mentioned that from our theoretical
framework we have obtained that the isotropic model (circular
region: $\chi=1$, $n=1$) does neither predict the appearance of negative current
patterns nor the peak effects in the magnetic moment curves. However, as long as
a clear distinction between the depinning and the cutting components of $\bf J$
is allowed (by letting $n > 1$), several remarkable facts can be explained.

In order to understand the different consequences and physical phenomena derived
from the SDCST for the longitudinal transport problem, our numerical results may
be summarized as follows: 

\begin{enumerate}
\item For the magnetic process under consideration [Fig.~\ref{Figure_5_3}, pag.
\pageref{Figure_5_3}], and concentrating on the local properties within the
sample [see e.g., Fig.~\ref{Figure_5_4}, pag.~\pageref{Figure_5_4}], a clear
independence of the field and current density profiles relative to the
anisotropy level of the material law has been obtained for the {\em partial
penetration regime}, in which the flux free core progressively shrinks to zero
[top pane of Figs. \ref{Figure_5_7} -- \ref{Figure_5_10}, pags.
\pageref{Figure_5_7} -- \pageref{Figure_5_10}].

 \item Negative values for the transport current density $j_{y}$
are neither obtained for the T or CT$\chi$ states when $h_{z0}$ is high
($h_{z0}\gtrsim 50$) until extreme values of the longitudinal field
($h_{y0}\gtrsim 500$) are
reached [see e.g., $j_{y}(a)$ profiles in
Figs.~\ref{Figure_5_5} (pag. \pageref{Figure_5_5}), \ref{Figure_5_11} (pag.
\pageref{Figure_5_11}) \& \ref{Figure_5_12} (pag. \pageref{Figure_5_12})]. On
the
contrary, one can early find negative current flow
for both cases when $h_{z0}\leq2$. Notice by Eq.~(\ref{Eq.5.1}), pag.
\pageref{Eq.5.1}, that the reduction of
the perpendicular component of the magnetic field may be understood as an
enhancement of the cutting current component. Thus, the negative values of
${j}_{y}(z)$ are obtained for smaller and smaller
$h_{y0}$ as $h_{z0}$ also decreases. In fact, negative values can happen even
for the partial penetration regime ($h_{y0}\lesssim 0.845$) when $h_{z0}$ tends
to $0$, in accordance with the analytical model presented before
(Section 5.1.1).

\item If $j_{\parallel}$ is unbounded (T states) the
${j}_{y}(z)$ structure becomes rather inhomogeneous as $h_{y0}$ increases and
takes the form of a highly positive layer in the center {\em shielded}  by a
prominent negative region, i.e., the transport current is essentially
\textit{compressed} toward the center of the sample by the effect of the
shielding currents. Also, it is worth of mention that, when simulating
experiments in which the transport current is applied subsequent to the field,
the SDCST does not predict negative flow values at all. On the contrary, in such
cases, what one gets is a {\em compression} of the original field penetration
profile, until the increasing transport current leads to dissipation. 

\item When $j_{\parallel}$ is bounded (CT states) one
observes a negative layer at the surface that eventually disappears when
$h_{y0}$ increases more and more [e.g., $h_{y0}>100$ for CT1 case - see bottom
pane of Fig.~\ref{Figure_5_11} (pag.~\pageref{Figure_5_11})]. Thus, as a
general rule, the smaller the value of $h_{z0}$, the sooner the negative
transport current is found. In the CT cases, this also increases the range of
longitudinal field for which negative values are observed.

\item  The peaked structure of $j_{y}(z)$
for the T-states at $h_{z0}=0.5$ is
accompanied by a similar behavior in $j_{x}(z)$ that relates to a subtle
magnetic field reentry phenomenon in $h_{y}(z)$ [see curves labeled
$h_{y0}=10$ in the bottom pane of Fig.~\ref{Figure_5_4}
(pag.~\pageref{Figure_5_4})].  For the
corresponding CT$\chi$ cases, the occurrence of this phenomena is linked to the
choice of widthbands larger than $\chi=1$ [by comparison, see profiles for
the labeled first stage at the bottom of
Fig.~\ref{Figure_5_8} (pag.~\pageref{Figure_5_8})].

\item Unlimited growth of the global magnetic moment component $M_{x}$ as a
function of the longitudinal magnetic field $h_{y0}$ occurs for the T-states in
which $j_{\parallel}$ is unbounded. From the local point of view, this relates
to an unlimited growth (\textit{compression}) of the current density
at the center of the slab [$j_{y}(0)$]. Remarkably, the appearance of a
peak structure in $M_{x}(h_{y0})$ correlates with the maximum value of the
transport current density at the center of the slab for the bounded CT states.
For example, for the CT1 case (square with $\chi=1$ and a smoothed corner
by $n=4$) the obtained maximum value $j_{y}^{max}(0)= 1.2968$ corresponds to the
optimal orientation of the region $\Delta_{\textbf{r}}$ in which the biggest
distance within the superelliptic hypothesis is reached. Such situation is
sketched in Fig.~\ref{Figure_4_1} (pag. \pageref{Figure_4_1}) and one may check
the numeric result from the expression
\begin{eqnarray}
{\rm Max}\{j_{c\parallel}(\Delta_{\textbf{r}})\}=j_{y}^{max}=
\left( 1+\chi^{2n/(n-1)} \right)^{(n-1)/2n} \, .\nonumber
\end{eqnarray}
Notice that, as a limiting case, it produces the expected value $2^{1/2}$  for
the diagonal of a perfect square [i.e., $n\to\infty$
in Eq.~(\ref{Eq.2.17}), pag.~\pageref{Eq.2.17}]. 

\end{enumerate}

\begin{SCfigure}
\centering
\includegraphics[height=7.0cm,width=8.5cm]{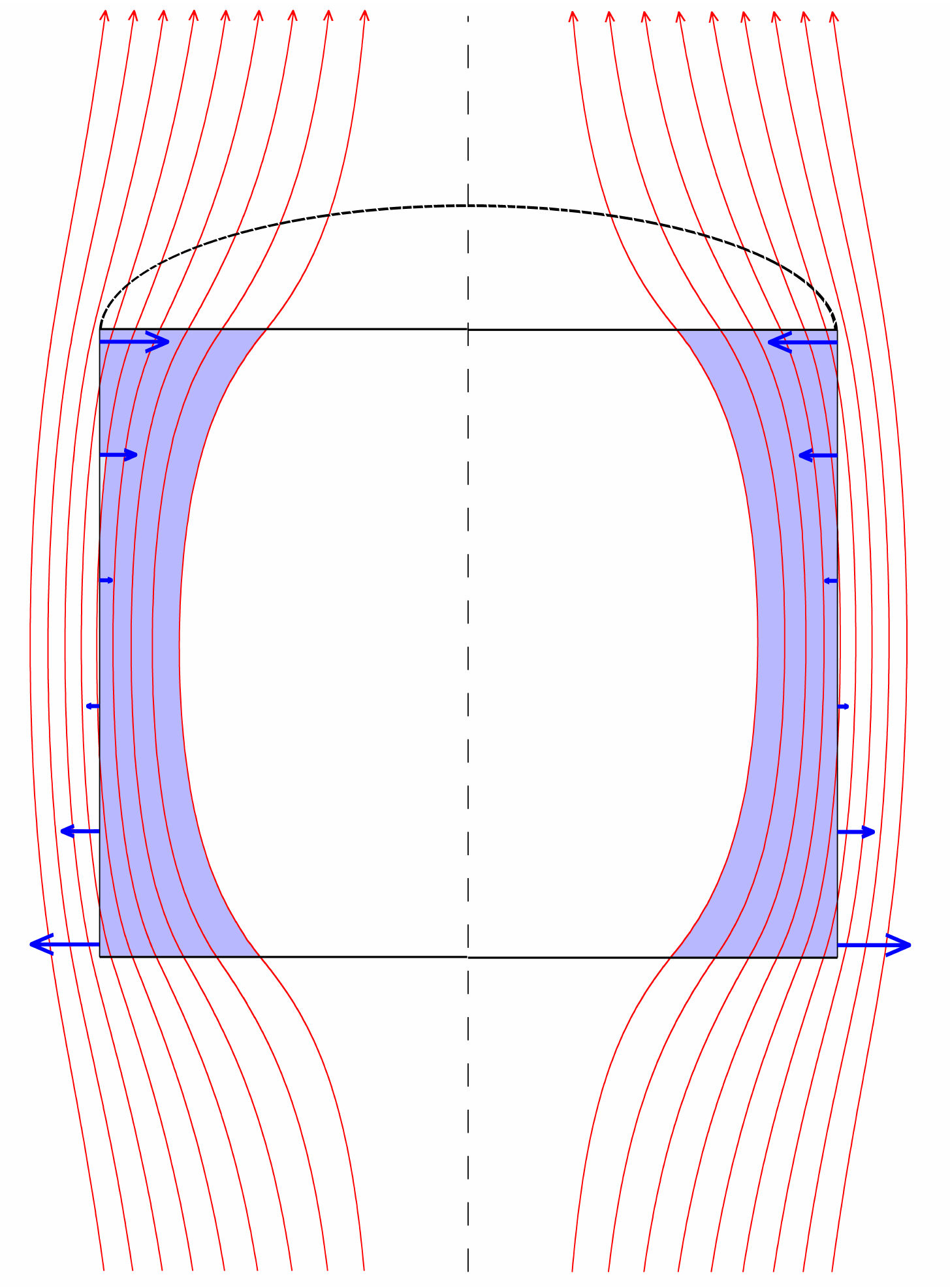}
\caption*{\label{Figure_Conc_5_1} Figure II-2: Penetration of a magnetic field
parallel to the axis of a finite superconducting cylinder (only one radial cut
for the longitudinal section is shown for symmetry reasons). The component of
the magnetic field perpendicular to the lateral surface is visualized by a set
of arrows with normalized lengths. The central dashed line represents the
symmetry axis.
}
\end{SCfigure}

\noindent Additional physical considerations can be done so as to cover the
full experimental scenario for the longitudinal transport problems. 
In particular, although our analysis has been performed within the infinite slab
geometry, one can straightforwardly argue about the extrapolation to real
experiments by means the inclusion of the third component of the magnetic field,
$H_{z0}$, which qualitatively may be related to the importance of the {\em
finite size effects} in real superconducting samples. Notice that our
numerical calculations, imply that negative currents should be more
prominent in those regions of the sample where the component of ${\bf H}$
perpendicular to the current density layers is less important. Thus, considering
that a real sample in a longitudinal configuration will be typically a rod with
field and transport current along the axis, the above idea is straightforwardly
shown by plotting the penetration of an axial field in a finite cylinder
(see Fig.~\hyperref[Figure_Conc_5_1]{II-2}). Then, the aforementioned effect
will occur at the central region of the sample, where end effects are minimal.

Just for visual purposes, Fig.~\hyperref[Figure_Conc_5_1]{II-2} shows the
distortion of the magnetic field, shielded by the induced supercurrents in a
finite superconducting cylinder, where the horizontal component of the
magnetic field along the lateral surface layer has been outlined.
It is apparent that the normal component of ${\bf H}$ will be enhanced close to
the bases and tend to zero at the central region. Then, inhomogeneous surface
current densities with negative flow at the mid part should be expected in
agreement with the experimental evidences reported in
Refs.~\cite{P2-Voloshin_2001,P2-LeBlanc_2003,P2-LeBlanc_2002,P2-Matsushita_1998,
P2-LeBlanc_1993,P2-LeBlanc_1991,P2-Voloshin_1991,P2-Matsushita_1984,
P2-Cave_1978,P2-Walmsley_1977,P2-Esaki_1976,P2-Walmsley_1972,P2-London_1968,
P2-LeBlanc_1966}.

\vspace*{0.5cm}
\paragraph{\textbf{\textit{Chapter 6 \vspace{1cm}\\}}}

In this chapter, we have presented a thorough study of the local and global
electromagnetic response of a straight, infinite, cylindrical 
type-II superconducting wire subject to diverse AC-configurations of transverse
magnetic flux density $B_{0}(t)$ and/or longitudinal transport current flow
$I_{tr}(t)$. We have assumed that the superconductor follows the celebrated
critical state model with a constant threshold for the critical current density
$J_{c}$, such that $|I_{tr}|\leq |J_{c}| \pi R^{2}$.
The problem is posed over a mesh of virtual filamentary wires each carrying a
current $I_{i}$ across a surface $s_{i}$, filling up the whole
cross section of the superconducting wire whose area is defined by $\Omega=\pi
R^{2}$. 

After a brief theoretical review that concentrates on the physical nature of the
different contributions to the AC response (Sec.~\ref{ch-6-1}.1), we
have performed extensive numerical calculations for several
amplitudes of the impressed transport current, $I_{\tt{a}}$, as well as the
amplitude of the magnetic flux density associated to the external 
excitation source, $B_{\tt{a}}$, for three different regimes of excitation: 

\noindent(\textit{i}) Isolated electromagnetic sources, Fig.~\ref{Figure_6_1}
(pag.~\pageref{Figure_6_1}) and Fig.~\ref{Figure_6_2}
(pag.~\pageref{Figure_6_2}).

\noindent(\textit{ii}) Synchronous electromagnetic sources,
Fig.~\ref{Figure_6_4} (pag.~\pageref{Figure_6_4}) and Fig.~\ref{Figure_6_9}
(pag.~\pageref{Figure_6_9}).

\noindent(\textit{iii}) Asynchronous electromagnetic sources,
Fig.~\ref{Figure_6_12} (pag.~\pageref{Figure_6_12}).

For each of the above cases, and in order to understand
the influence of the electromagnetic excitations involved in the
macroscopical physical processes found in this kind of systems, we have
presented a detailed study of the local dynamics of the distribution of
screening currents, $I_{i}=\pm J_{c} s_{i}$ or $0$, as well as the related
local density of power dissipation $\textbf{E}\cdot\textbf{J}$ along a cyclic
oscillating period. Likewise, for a wide number of experiments, we have
presented the full behavior of the magnetic flux density vector $\textbf{B}$.
Many of these results are summarized in the section of supplementary material at
the end of this chapter. On the other hand, for a closer connection with the
most common experimental observables, we have calculated
the wire's magnetic moment $\textbf{M}$ and the hysteretic losses $L$ as a
function of AC external excitations, and their comparison with classical
analytical approaches has been featured. Our main conclusions are summarized
below.

\vspace*{0.5cm}
\paragraph{(~\textit{I.}~) \textit{Isolated
excitations.}\vspace*{0.8cm}\\}

Section~\ref{ch-6-2}.2  is devoted to unveil the physics behind
the simplest configurations, where the superconducting wire under zero field
cooling is subjected to isolated external excitations. On the one hand, for
cases with pure AC transport current, we have shown that our
numerical method achieves an exact comparison with the well known analytical
solution for the AC hysteretic loss (see Fig.~\ref{Figure_6_1},
pag.~\pageref{Figure_6_1}). Moreover, local magnetization effects have been
revealed from integration of the magnetic moments of the screening currents in
half cross section of the SC wire. Thus, although the global condition
$\textbf{M}(\Omega)=0$ occurs, this does not imply the absence of
hysteretic losses, which are in fact produced by the
reassembling of the distribution of screening currents in the transient between
two consecutive steady states, and therefore of local variation of the induced
magnetic flux density in those active zones where the screening currents appear.

On the other hand, when the SC wire is under the action
of an external magnetic flux density applied perpendicular to its surface,
$B_{0,y}$, and $I_{tr}(t)=0$ (Fig.~\ref{Figure_6_2}, pag.~\pageref{Figure_6_2}),
an exact analytical solution for the dynamics of the flux front profiles is not
known. Nevertheless, we have shown that the analytical approach provided
by Gurevich for monotonic losses fits well to
our numerical calculations (Fig.~\ref{Figure_6_3}, pag.~\pageref{Figure_6_3}),
if one assumes that the dependence of the hysteretic loss density for the
periods
$\pm B_{\tt{a}}\rightarrow0$ and $0\rightarrow \mp
B_{\tt{a}}$ is the same as the loss calculated for the first monotonic
branch, i.e., the excitation branch before the cyclic
peak to peak process. Then, the AC loss may be calculated by introducing a
factor of four.

Regarding the magnetization curves, we have observed that for the first
branch of the oscillating electromagnetic excitation $(0<B_{0,y}(t)\leq
B_{\tt{a}})$, or so called monotonic branch, the saturation point of the
magnetic moment $M_{p}$ may be straightforwardly identified at the value of full
penetration field $B_{p}$. On the contrary, within the cyclic stage, if
$B_{\tt{a}}\geq B_{p}$, the magnetic moment saturates at different values
of the magnetic field satisfying the dimensionless empirical relation
$B_{p^{\dag}}=\mp(2 B_{p}- B_{\tt{a}} \mp B_{0}(t')-1/2)$. The simultaneous
choice of both signs have to be made according to the sign's rule in the time
derivative of the cyclic excitation, i.e., for each half cycle of amplitude
$B_{\tt{a}}$. Considerations of premagnetized samples are allowed by
$B_{0}(t')\neq0$. Remarkably, the set of magnetization loops displayed in
Fig.~\ref{Figure_6_2} (pag.~\pageref{Figure_6_2}) serves as a map for drawing
any magnetization loop for dealing with arbitrary relations between the
experimental parameters $B_{0}(t')$ and $B_{\tt{a}}$

\paragraph{\textleaf\hspace*{0.3cm} \textit{Nature of the hysteretic losses, and
their formulations.}\vspace*{0.8cm}\\}

As a conclusion of our analysis of the basic configurations where the
cylindrical SC wire is only subjected to isolated electromagnetic excitations,
$B_{0,y}$ or $I_{tr}$, we have compared our results to the hysteretic AC loss
obtained by different methods. Thus, we have noticed the following remarkable
aspects:

\vspace*{0.5cm}\noindent(1)\hspace*{0.5cm} 
Despite the fact that an analytical solution for Eq.~(\ref{Eq.6.8}) can only be
achieved for those cases with only transport current flow, the actual AC loss
when the isolated electromagnetic excitation corresponds to an external
magnetic field applied perpendicular to the superconducting surface, may be
straightforwardly evaluated by calculating the enclosed area by the
magnetization loop between the steady-peaks of the electromagnetic excitation, 
as a function of the density of magnetic flux provided by the external
source [Eq.~(\ref{Eq.6.16}), pag.~\pageref{Eq.6.16}]. 

\vspace*{0.5cm}\noindent(2)\hspace*{0.5cm} 
The expression of the AC losses provided by the
inductive component of the vector potential may be conveniently rewritten when
simultaneous occurrence of injected current lines or screening currents
accomplishing transport current condition $I_{tr}(t)\neq0$, and the so-called
magnetization currents,  is given by
[Eq.~(\ref{Eq.6.19}),
pag.~\pageref{Eq.6.19}]. Thus, when the superconducting wire is subjected to
both electromagnetic excitations, $B_{0}(t)$ and $I_{tr}(t)$,
significant reductions (or increases) of the hysteretic losses may be envisaged
by reducing (or increasing) the magnitude of the local density of magnetic flux 
$\textbf{B}=\textbf{B}_{0}+\textbf{B}_{ind}$. 

\vspace*{0.5cm}\noindent(3)\hspace*{0.5cm} 
Likewise, an alternative approach for calculating the
AC hysteretic loss per closed cycles has been derived from the principle of
conservation of energy and the definition of the Poynting's vector, such that
for cylindrical superconducting wires the AC hysteretic loss per unit length can
be reliably calculated by [Eq.~(\ref{Eq.6.22}), pag.~\pageref{Eq.6.22}].

\vspace*{0.5cm}
\paragraph{(~\textit{II.}~) \textit{Synchronous excitations.}\vspace*{0.8cm}\\}

In section~6.3.1 we have presented a detailed study of the
physical features associated to the local electrodynamics of superconducting
wires, subjected to the simultaneous action of oscillating synchronous
excitations $B_{0,y}(t)$ and $I_{tr}(t)$, in the geometrical conditions above
considered. Thus, a wide number of experiments based upon the combination of
different amplitudes for the density of magnetic flux related to the external
field source, $B_{\tt{a}}$, and the peak intensity of the impressed transport
current, $I_{\tt{a}}$, have been analyzed for situations in which the wire is in
the virgin state [Fig.~\ref{Figure_6_4}, pag.~\pageref{Figure_6_4}], or it has
been premagnetized [Fig.~\ref{Figure_6_9}, pag.~\pageref{Figure_6_9}] before
switching on the cyclic synchronous excitation $(B_{0,y},I_{tr})$. Our main
remarks concerning the underlying physics of these systems are detailed
below.

\vspace*{0.5cm}\noindent(1)\hspace*{0.5cm} 
The local distribution of screening currents when the simultaneous
action of the electromagnetic excitations $B_{0,y}$ and $I_{tr}$ is conceived,
may be described as the consumption of the magnetization currents (screening
currents induced by the external magnetic field) by effect of the occurrence of
injected current lines (constrained by the local condition
$\sum_{i}I_{i}=I_{tr}$). As a result of this, the flux front profile
is displaced from the geometrical center of the wire towards one of the sides
by a kind of ``Lorentz force'' effect on the injected current lines [see e.g.,
Figs.~\ref{Figure_6_5}
- \ref{Figure_6_6} (pag.~\pageref{Figure_6_5}-\pageref{Figure_6_6}), and their
corresponding supplementary material, Figs. S5 (pag.~\pageref{Figure_S_5}) and 
S8 (pag.~\pageref{Figure_S_8})].

\vspace*{0.5cm}\noindent(2)\hspace*{0.5cm} 
For low magnetic fields ($B_{\tt{a}}\leq2$, in units of the penetration field),
the distribution of screening currents is \textit{current-like} as long as the
peak intensity of the transport current is so high ($I_{\tt{a}}\geq0.5$, in
units of the critical current) that the ``magnetization lines'' in counter
direction to the injected current lines may be neglected
(Fig.~\ref{Figure_6_5}). Likewise, for higher values of $B_{\tt{a}}$ and lower
values of $I_{\tt{a}}$, \textit{field-like} profiles may be
envisaged (Fig.~\ref{Figure_6_6}). Nevertheless, for most of the possible
combinations between $B_{\tt{a}}$ and $I_{\tt{a}}$, nonsymmetric distributions
of the screening currents in a \textit{nothing-like} fashion are predicted [see
also Figs. S5 (pag.~\pageref{Figure_S_5}), S8 (pag.~\pageref{Figure_S_8}), and
S11 (pag.~\pageref{Figure_S_11})].

\vspace*{0.5cm}\noindent(3)\hspace*{0.5cm} 
Remarkable distortions of the magnetic flux density outside of
the superconducting wire are mainly observed at the instants when the
synchronous excitation tends to zero in the AC excitation.

\vspace*{0.5cm}\noindent(4)\hspace*{0.5cm} 
The maximal density of magnetic flux occurs in the side opposite to the
location of the flux free core. Thus, recalling Eq.~(\ref{Eq.6.19}), the
strongest localization of the density of magnetic flux in one side of the
active zone of the material produces a remarkable localization of the local
hysteretic losses in such manner that the heat release from the
superconducting wire is highly localized too [see e.g., Fig.~\ref{Figure_6_7}
(pag.~\pageref{Figure_6_7}) or Figs. S14-S16
(pags.~\pageref{Figure_S_14}-\pageref{Figure_S_16})].

\vspace*{0.5cm}\noindent(5)\hspace*{0.5cm} 
Having in mind that the local density of magnetic flux may
involve the concomitant response of both the magnetization currents and the
injected current lines, the \textit{B-oriented} approach for the calculation of
AC losses (pag.~\pageref{B-oriented}) allows to understand why it is not
possible to discriminate the magnetization loss from the transport loss
by using electromagnetic measurements
\cite{P2-Ogawa_2011,P2-Inada_2005,P2-Jiang_2004,P2-Rabbers_2001,P2-Reuver_1985}.
On the other hand, the \textit{S-oriented} approach (pag.~\pageref{S-oriented})
justifies that calorimetric methods which directly measure the release flux of
energy over the superconducting surface [Eq.~(\ref{Eq.6.21})] have been found
crucial for the determining of the actual AC loss when simultaneous
electromagnetic excitations act on the
superconductor~\cite{P2-Nguyen_2005a,P2-Nguyen_2005b,P2-Ramos_2010,
P2-Ashworth_2000,P2-Magnusson_2000}.

\vspace*{0.5cm}\noindent(6)\hspace*{0.5cm} 
Regarding the magnetic moment curves [Fig.~\ref{Figure_6_8},
pag.~\pageref{Figure_6_8}], we argue that only for small values of
$I_{\tt{a}}$, Bean-like loops are expected. However, as $I_{\tt{a}}$ increases,
the derived effect by the consumption of the magnetization currents is most
prominent, ending up with the symmetrization of the magnetization loop for
cyclic periods as function of $B_{0}$ or $I_{tr}$ into striking
lenticular shapes. As a consequence of this process, a distinct low-pass
filtering effect comes to the fore which, in
the case of the triangular input excitations with $I_{\tt{a}}=I_{c}$, yields a
nearly perfect sinusoidal (first-harmonic) output signal $M_{y}(t)$.

\vspace*{0.5cm}\noindent(7)\hspace*{0.5cm} 
If the superconducting wire has been magnetized before switching on
the synchronous AC excitation ($\textbf{B}_{0},I_{tr}$), say at $t=t'$ [see
Fig.~\ref{Figure_6_9} (pag.~\pageref{Figure_6_9})], the center of the
magnetization loop drifts from $\textbf{M}(0,0)$ towards
$\textbf{M}(B_{0}(t'),0)$, such that the corners of the magnetization loop
$\textbf{M}(\pm B_{\tt{a}},\pm I_{\tt{a}})$ lie on the excitation coordinates
$(B_{0}(t') \pm B_{\tt{a}},\pm I_{\tt{a}})$. Thus, as the area enclosed by the
magnetization loop remains the same, the power losses attained along the
premagnetization process plays no role in the calculation of the hysteretic AC
loss. Nevertheless, in these cases, the profiles drawn for the screening
currents have revealed highly intricate patterns regarding to the coexistence
of the magnetization currents and the injected current lines, as well as
defining the flux front profile [Fig. S11 (pag.~\pageref{Figure_S_11})]

\vspace*{0.5cm}\noindent(8)\hspace*{0.5cm} 
Our straightforward calculation of the actual hysteretic losses by means the
general definition $\delta L=\int_{\Phi}\textbf{E}\cdot\textbf{J}dr$, reveals
important differences concerning the approximate formulae customarily used.
In fact, we have
shown that the actual AC losses are always higher than those envisaged by the
linear superposition of contributions due to either type of AC excitation [see
Fig.~\ref{Figure_6_10}, pag.~\pageref{Figure_6_10}]. Comparisons reveal the
important fact that the linear approach $L(B_{\tt{a}})+L(I_{\tt{a}})$ is only
appropriate for high strengths of the magnetic flux density and low transport
currents (screening currents with distribution \textit{field-like}), as well as
for low magnetic field and high transport current (screening currents with
distribution \textit{current like}). Likewise, for high amplitudes of the
magnetic flux density ($B_{\tt{a}}\geq B_{p}$) and for any value of the
transport current, significant reductions of the actual hysteretic loss may be
envisaged if one compares it with the predicted losses for a wire carrying a DC
current instead to the AC case [Fig.~\ref{Figure_6_11},
pag.~\pageref{Figure_6_11}].

\vspace*{0.5cm}
\paragraph{(~\textit{III.}~) \textit{Asynchronous
excitations.}\vspace*{0.8cm}\\}

Our detailed analysis of the underlying physics behind the local and global
electromagnetic response of superconducting wires subjected to isolated
electromagnetic excitations $B_{0}(t)$ or $I_{tr}(t)$ in oscillating regimes,
and then of the synchronous action of both, have revealed that the total AC loss
may be controlled by locally reducing the total magnetic flux density resulting
from the addition of the external magnetic field and that induced by the
concomitant occurrence of magnetization currents and the injected current lines.
Thus, either by displacing in time the electromagnetic excitations
(phase shift), by introducing changes in frequency, or simply by
considering excitation branches with time derivatives
in counter directions, one can help to counterbalance the local density of
magnetic flux in the zone of maximum heat release, which is further translated
to the reduction of the hysteretic losses.

Thus, as long as both excitations evolve with the same oscillating frequency, it
is possible to assert that at least a minimal reduction
of the AC loss should appear for relative phase changes between the
electromagnetic excitations. This is in agreement with the experimental and
numerical evidences reported in
Refs.~\cite{P2-Nguyen_2005a,P2-Nguyen_2005b,P2-Vojenciak_2006}. For example, by
the simple overlapping between the excitation curves of the electromagnetic
sources, it is evident that the maximal losses are envisaged for the
synchronous cases, as well as for those cases with a relative change in phase of
half excitation period. Likewise, minimal losses are attained around a phase
shifting of a quarter of period, although its exact position as a function of
the electromagnetic excitations straightforwardly depends on the entangled
competition between the external magnetic field and the induced fields by the
injected current lines and the so called magnetization currents. Thus, unless
the distribution of screening currents show patterns of the kind
\textit{current-like} or \textit{field-like}, is not possible assert that the
lower AC loss is given for a change of phase of a quarter of period.

Within the above scenario, we have argued that predictions are not
straightforward for non-trivial time dependencies of the simultaneous
electromagnetic excitations and, higher hysteretic AC losses could be also
expected. Along this line, we have introduced a thorough study of the so-called
double frequency effects, which arise when one of the isolated electromagnetic
sources, $B_{0}(t)$ or $I_{tr}(t)$, is connected to a power supply with a double
oscillating frequency [see Fig.~\ref{Figure_6_12}, pag.~\pageref{Figure_6_12}].
The most outstanding observations about the local and global behavior of the
involved electromagnetic quantities are detailed below:

\vspace*{0.5cm}\noindent(1)\hspace*{0.5cm} 
For asynchronous excitations, the distribution of screening currents in the
first peak of the dominant excitation (i.e., that with a longer period), may be
strongly different as compared to the attained distributions for the subsequent
excitation peaks [Fig. S17, pag.~\pageref{Figure_S_17}]. Thus, the
first peak of excitation cannot be considered as a steady-state for integrating
the AC loss when the integral is reduced to half period of excitation. In other
words, the knowledge of the distribution of screening
currents in the time elapsed for the second half-period of the dominant
excitation is relevant. The latter fact
has been experimentally recognized in
Ref.~\cite{P2-Nguyen_2005a}.

\vspace*{0.5cm}\noindent(2)\hspace*{0.5cm} 
Whatever electromagnetic excitation carries the double frequency, we have found
that, in the AC regime, complex arrays of domains connected by boundary lines
with currents switching between $I_{c}$ and $-I_{c}$
appear [see Fig.~\ref{Figure_6_13} (pag.~\pageref{Figure_6_13}) or Fig. S17
(pag.~\pageref{Figure_S_17})].

\vspace*{0.5cm}\noindent(3)\hspace*{0.5cm} 
Contrary to the strong localization of the local density of power losses
observed in cases with synchronous excitations, for asynchronous excitations the
active zone with higher heat release is no longer focused at only one side of
the superconducting wire [see Fig.~\ref{Figure_6_13}
(pag.~\pageref{Figure_6_13}) or Fig. S20
(pag.~\pageref{Figure_S_20})]. 

\vspace*{0.5cm}\noindent(4)\hspace*{0.5cm} 
As far as concerns to the magnetic response of the superconducting wire
when double frequency effects are incorporated, \textit{exotic} magnetization
loops are predicted [see Fig.~\ref{Figure_6_14} (pag.~\pageref{Figure_6_14}) 
and Fig.~\ref{Figure_6_15} (pag.~\pageref{Figure_6_15})]. Outstandingly, for
low amplitudes of the transport current, $I_{\tt{a}}$, Bean-like loops may be
observed for any of the above mentioned cases. However, as $I_{\tt{a}}$
increases,
strongest differences between the magnetization loops arise, and the global
behavior is radically different to the synchronous cases. 

\vspace*{0.5cm}\noindent(5)\hspace*{0.5cm} 
A comprehensive analysis of the magnetization curves as a function of the
temporal evolution of the electromagnetic excitations has been carried
out. Thus, several symmetry conditions for the magnetization loop, when the
electromagnetic excitation with double frequency is either $I_{tr}$ or $B_{0}$,
concerning to the amplitudes $I_{\tt{a}}$ and $B_{\tt{a}}$ have been
illustrated.

\vspace*{0.5cm}\noindent(6)\hspace*{0.5cm} 
In order to attain the low-pass filtering effect in the temporal evolution of
the magnetization curve when $I_{\tt{a}}\rightarrow I_{c}$, it becomes
absolutely necessary to assure that both oscillating excitations evolve
synchronous in time.

\vspace*{0.5cm}\noindent(7)\hspace*{0.5cm} 
Accordingly to the different setups in panes (b) and (c)
of Fig.~\ref{Figure_6_12}, we have calculated the hysteretic AC loss for several
values of $B_{\tt{a}}$ and $I_{\tt{a}}$,  when the dominant excitation is either
the applied magnetic field, $L_{asynch}^{(b)}$, or the impressed transport
current, $L_{asynch}^{(c)}$ [Fig.~\ref{Figure_6_16},
pag.~\pageref{Figure_6_16}]. By comparing them with the hysteretic AC loss
predicted for the synchronous cases [see also Fig.~\ref{Figure_6_17},
pag.~\pageref{Figure_6_17}], we have shown the AC loss may either increase or
decrease by double frequency effects when asynchronous excitations are involved.
The relative intensities of the excitations $B_{\tt{a}}$ and $I_{\tt{a}}$ play
an additional role on this. Then, the big number of possible combinations
makes it imperative to have a previous knowledge of
the operational environment of the superconducting wire, for attaining valid
predictions of the actual AC losses.


\renewcommand\bibname{References II}

\section*{}
\break
\vspace*{0.0cm}
\subsection*{\label{ch-6-s}\Huge{Supplementary Material II}}
\vspace*{0.0cm}
\markboth{\sc \textbf{Supplementary Material II}}{\sc \textbf{Supplementary
Material II}}
\addcontentsline{toc}{chapter}{\sc \textbf{Supplementary Material II}}

\begin{itemize}
 \item \textbf{\textit{\Large{SC wire subjected to an AC transport current
\\ condition}}}
\end{itemize}

\begin{figure}[H]
\centering
\includegraphics[height=13cm,width=5.8cm]{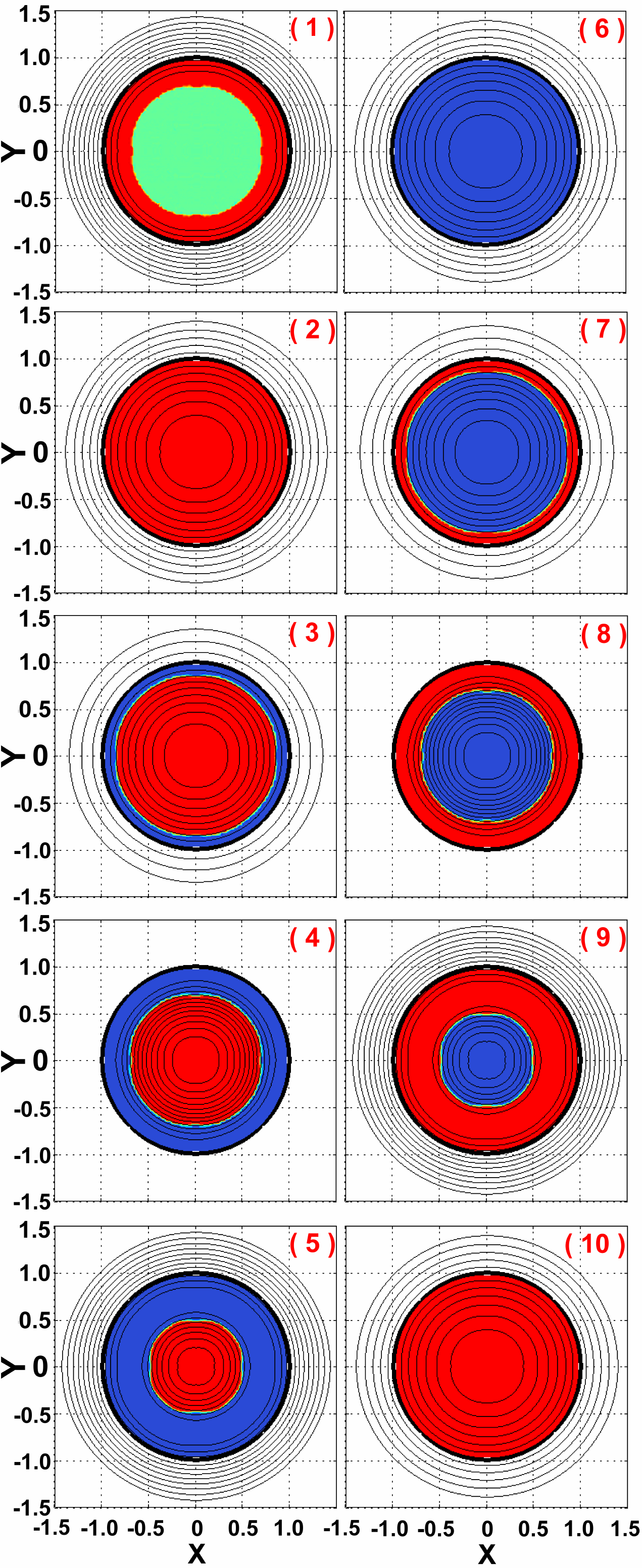}\hspace*{-0.1cm}
\includegraphics[height=13cm,width=6.1cm]{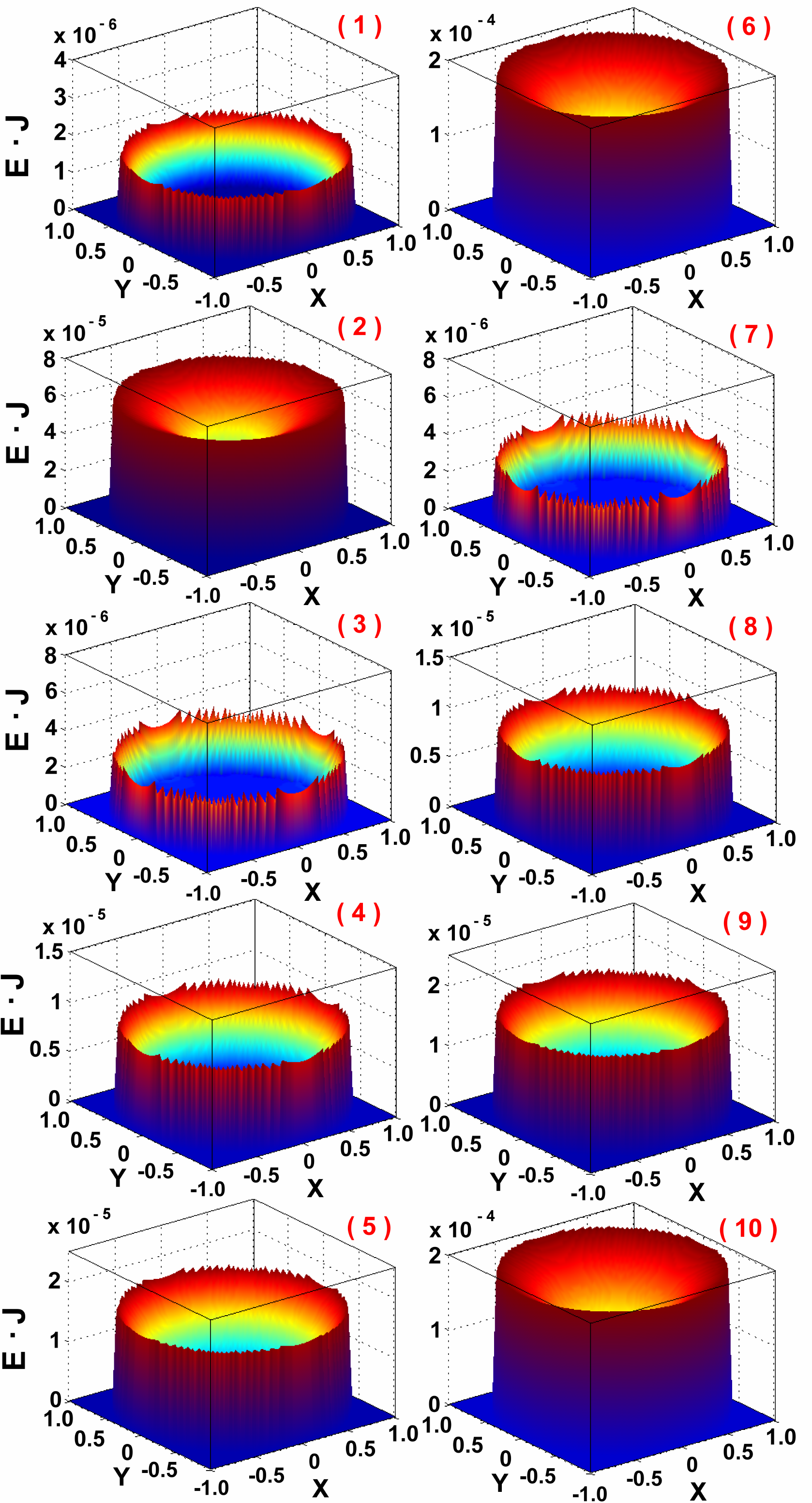}
\caption*{\label{Figure_S_1} Figure S1: For an oscillating transport current of
amplitude $I_{\tt{a}}=I_{c}$, and the temporal steps defined in
Fig.~\ref{Figure_6_1}(b) (pag.~\pageref{Figure_6_1}), we show the
dynamics of the magnetic flux lines (projected isolevels of the vector
potential), together with the profiles of current $I_{i}$ across the
superconducting wire, in the 1st and 2nd column. Next, 3rd and 4th column show
the corresponding evolution of the local density of power dissipation. The
profiles of current are displayed according
to: red ($+I_{i}$), blue ($-I_{i}$), and green (zero). The plotting
interval is $\Delta I_{tr}(t)=I_{\tt{a}}/2$, with t=0 defining the virgin state
(i.e., $I_{tr}=0$). Units are $\pi R^{2}J_{c}\equiv{I_c}$ for
$I_{\rm tr}$, and
$(\mu_{0}/4\pi)J_{c}^{2}R^{2}\delta t^{-1}$ for E$\cdotp$J.
}
\end{figure}

\vspace*{-1.0cm}

\begin{figure}[H]
\centering
\includegraphics[height=14cm,width=13cm]{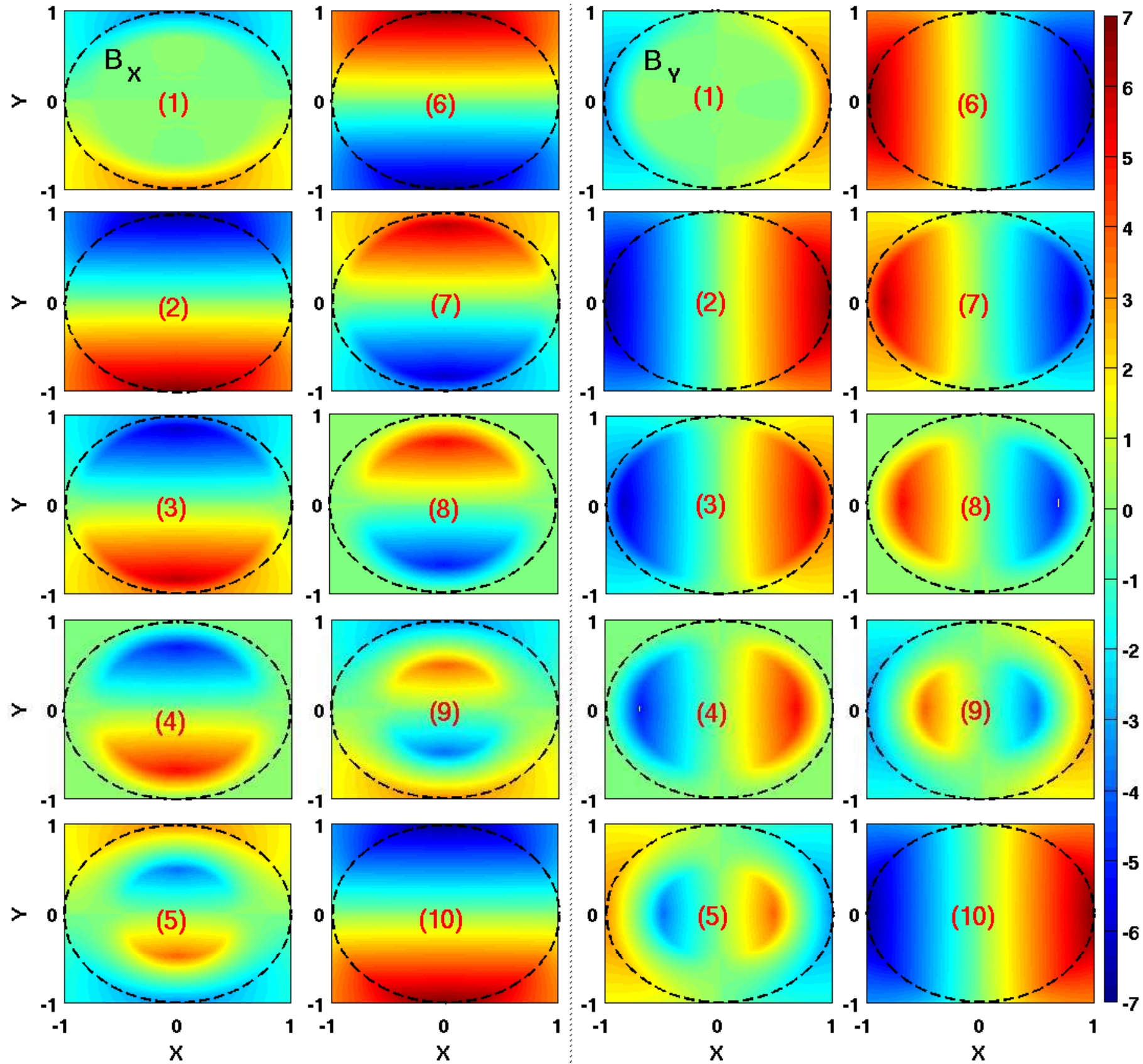}
\caption*{\label{Figure_S_2} Figure S2: Colormaps for the intensity
of the components of magnetic flux density $B_{x}$ (left pane) and $B_{y}$
(right pane) for the square section of area $4R^{2}$ enclosing the cylindrical
wire of radius $R=1$.
The profiles are plotted in according to Fig. S1, and the
superconducting surface is depicted by black dashed lines. Units are
$(\mu_{0}/4\pi)J_{c}R$ for B.
}
\end{figure}

\break
\vspace*{0.0cm}
\begin{itemize}
 \item \textbf{\textit{\Large{SC wire subjected to an AC external magnetic
flux}}}
\end{itemize}

\begin{figure}[H]
\centering
\includegraphics[height=13cm,width=5.8cm]{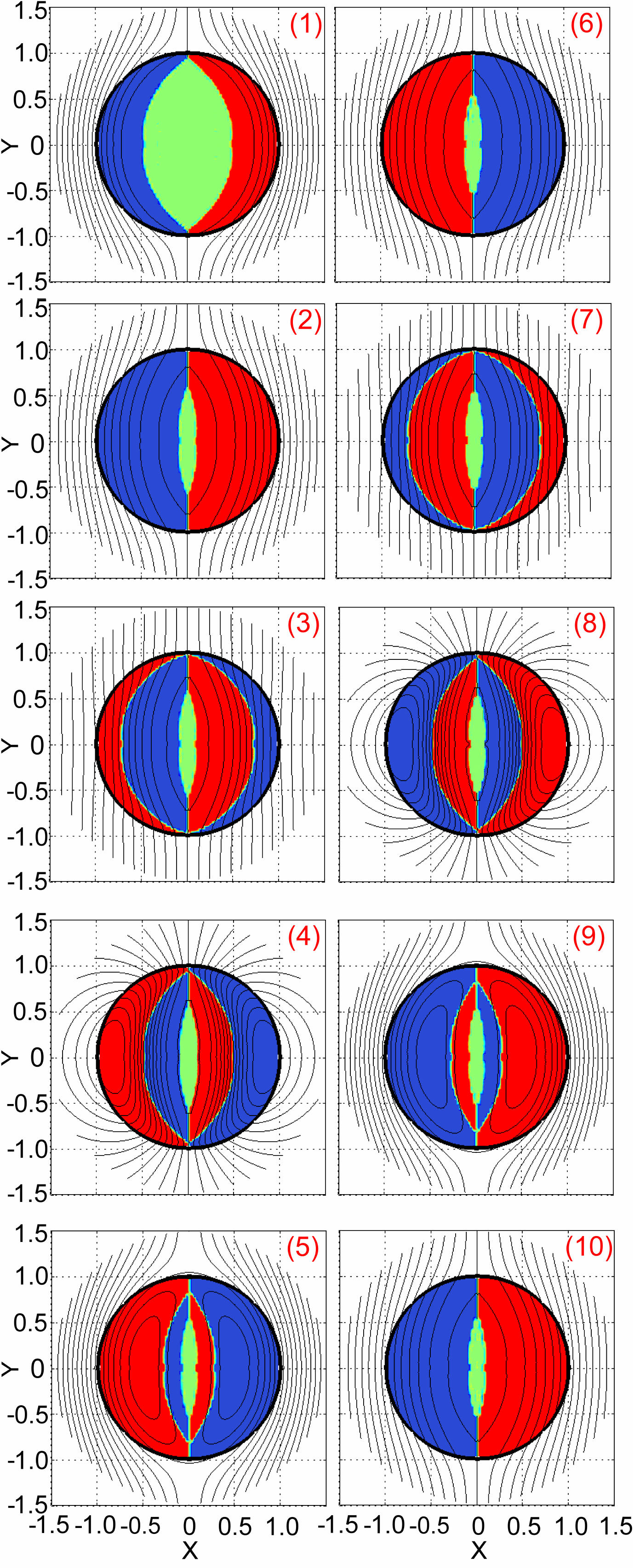}\hspace*{-0.1cm}
\includegraphics[height=13cm,width=6.1cm]{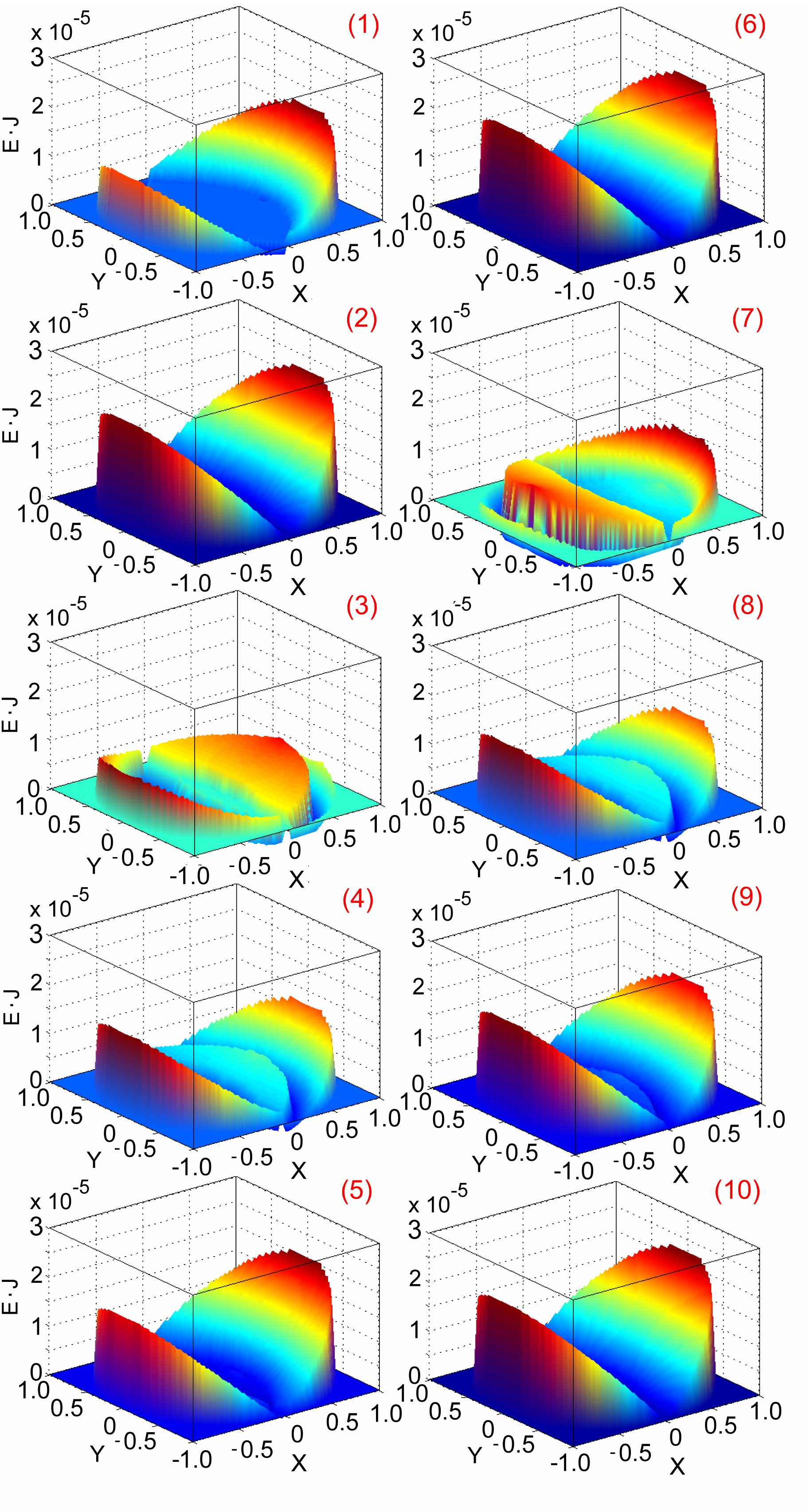}
\caption*{\label{Figure_S_3} Figure S3: For an external AC magnetic flux applied
along the $y-axis$, with amplitude $B_{\tt{a}}=6$, and for the
temporal steps defined in Fig.~\ref{Figure_6_2}(b) (pag.~\pageref{Figure_6_2});
the 1st and 2nd column show the
dynamics of the magnetic flux lines and their corresponding profiles of current
$I_{i}$. Next, 3rd and 4th column show the corresponding dynamics of the local
density of power dissipation. The profiles of current are displayed according
to: red ($+I_{i}$), blue ($-I_{i}$), and green (zero). The plotting interval is
$\Delta B_{0,y}=3$, with t=0 defining the virgin state (i.e., $B_{0}=0$).
Units are $(\mu_{0}/4\pi)J_{c}R$ for B, and
$(\mu_{0}/4\pi)J_{c}^{2}R^{2}\delta t^{-1}$ for E$\cdotp$J.
}
\end{figure}

\vspace*{-1.0cm}

\begin{figure}[H]
\centering
\includegraphics[height=14cm,width=13cm]{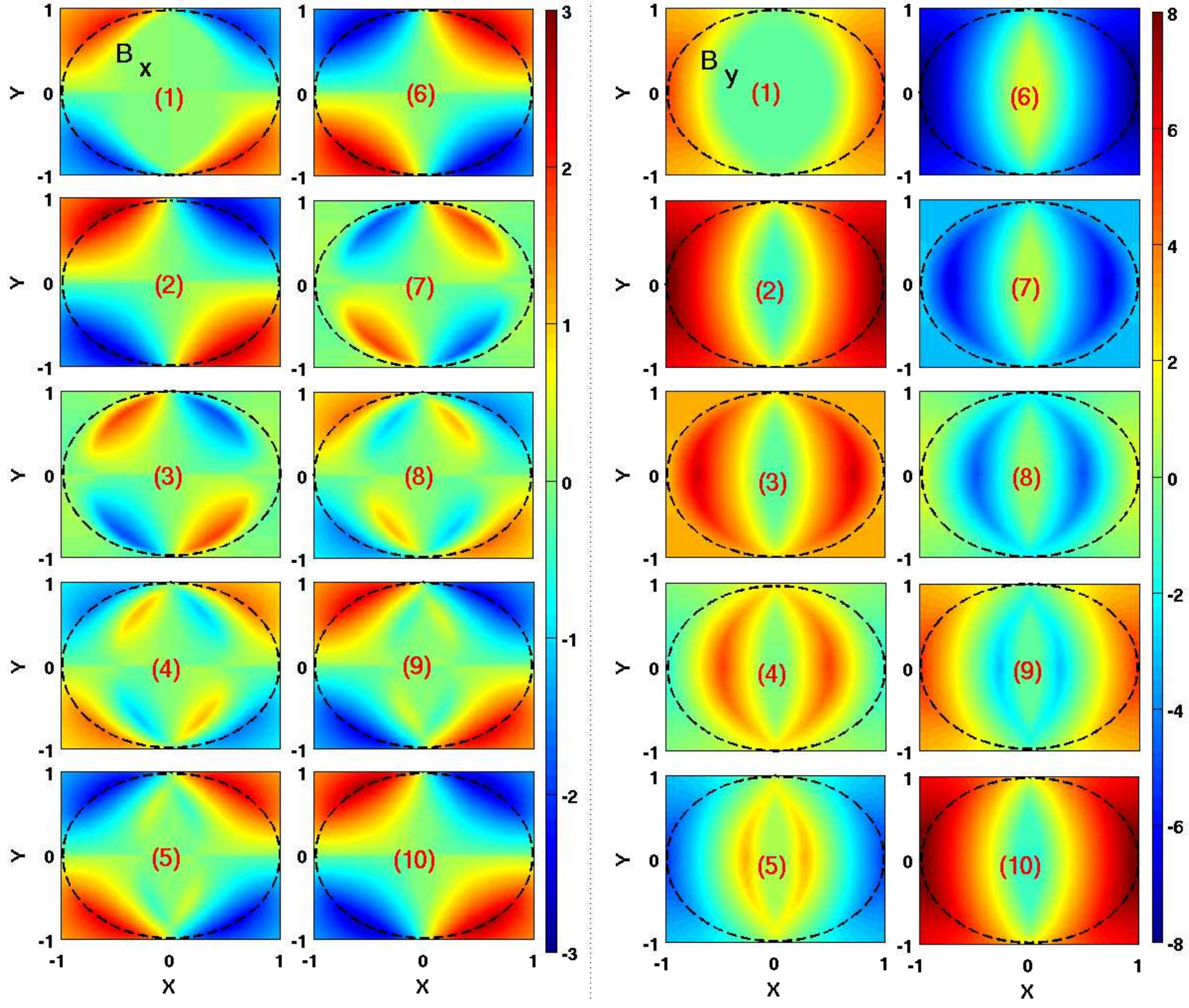}
\caption*{\label{Figure_S_4} Figure S4: Colormaps for the intensity of the
components of magnetic flux density $B_{x}$ (left pane) and $B_{y}$ (right pane)
for the square section of area $4R^{2}$ enclosing the cylindrical wire of radius
$R=1$. The profiles are plotted in according to the Fig.~S3. The
superconducting surface is depicted by the black dashed lines. Recall that units
for B are $(\mu_{0}/4\pi)J_{c}R$.
}
\end{figure}

\break
\vspace*{-0.5cm}
\begin{itemize}
 \item \textbf{\textit{\Large{SC wire subjected to synchronous oscillating
\\ excitations}}}
\end{itemize}

\begin{figure}[H]
\begin{center}
\includegraphics[height=15cm,width=3cm]{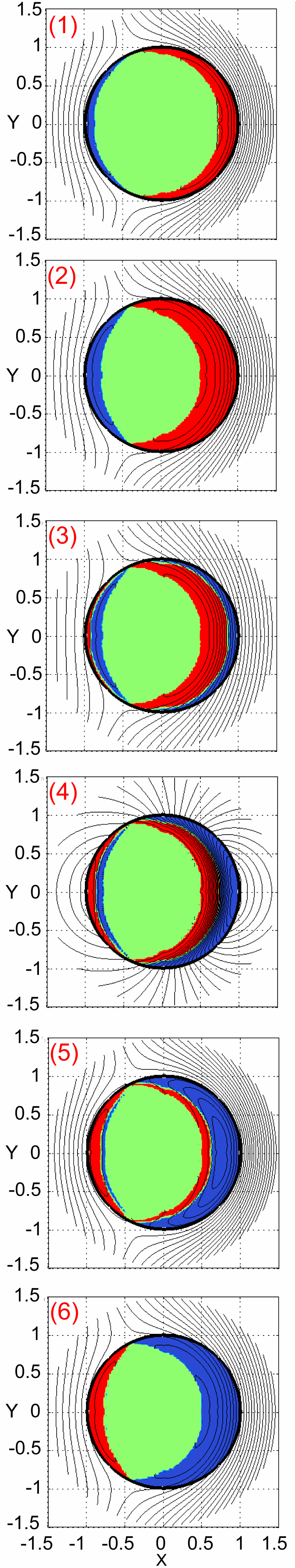} \hspace*{-0.1cm}
\includegraphics[height=15cm,width=3cm]{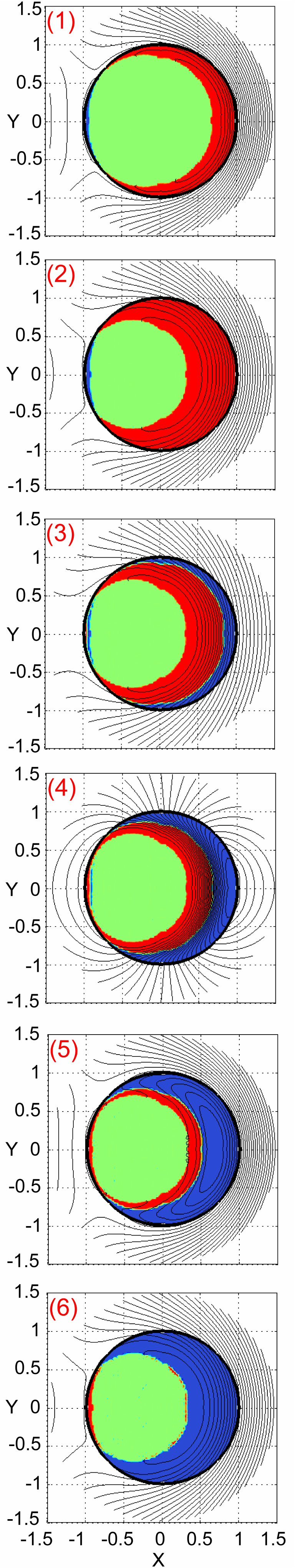} \hspace*{-0.1cm}
\includegraphics[height=15cm,width=3cm]{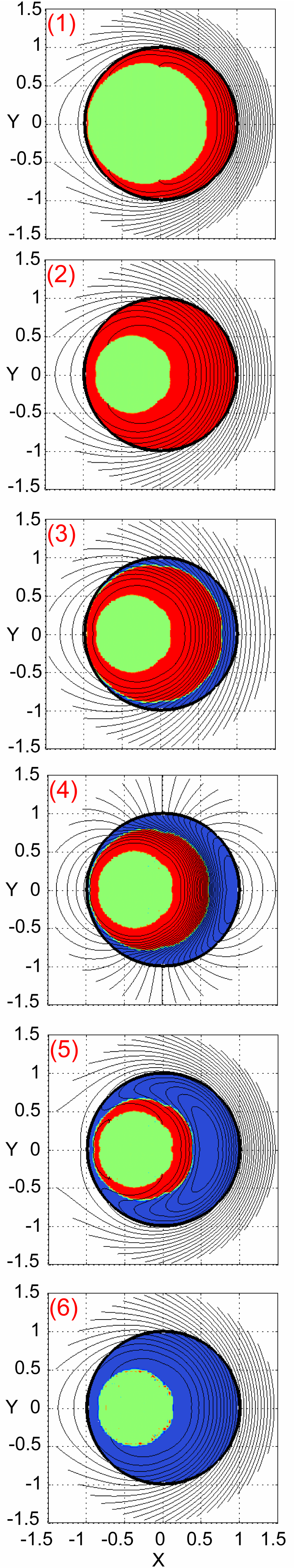} \hspace*{-0.1cm}
\includegraphics[height=15cm,width=3cm]{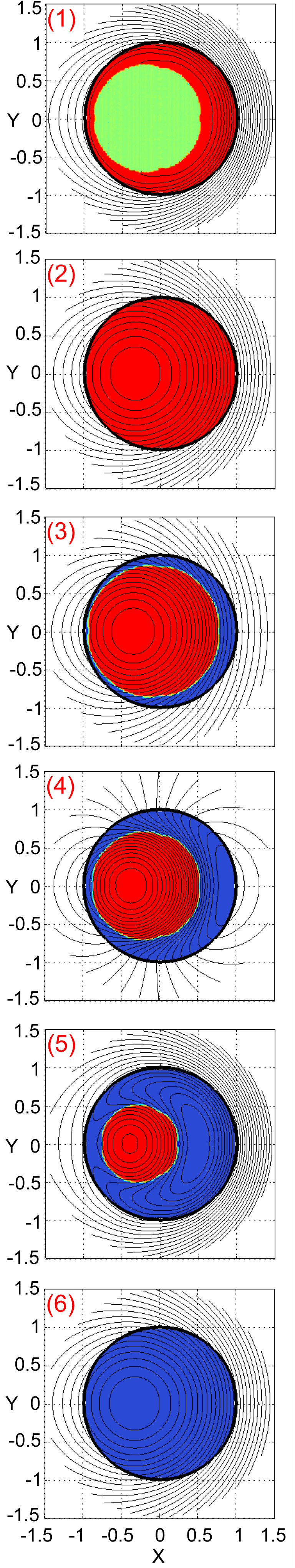} \hspace*{-0.1cm}
\end{center}
\caption*{\label{Figure_S_5} Figure S5: Evolution of the magnetic flux lines
and profiles of current with simultaneous oscillating sources $B_{0,y}$ and
$I_{tr}$, of amplitudes $B_{\tt{a}}=2$ (\textit{low field}) and:
$I_{\tt{a}}=0.25$ \textit{(1st column)}, $I_{\tt{a}}=0.5$ \textit{(2nd column)},
$I_{\tt{a}}=0.75$ \textit{(3rd column)}, and $I_{\tt{a}}=1$ \textit{(4th
column)}. Plotting interval is
$(B_{\tt{a}}/2,I_{\tt{a}}/2)$, where the time-step (1) defines the condition 
$(1,I_{\tt{a}}/2)$ [see also Fig.~\ref{Figure_6_4}
(pag.~\pageref{Figure_6_4})]. By symmetry, only the first half of the AC cycle
is shown [i.e., time-steps (2) to (6)].
}
\end{figure}

\break
\vspace*{0.0cm}
\begin{figure}[H]
\centering
\includegraphics[height=12.5cm,width=13cm]{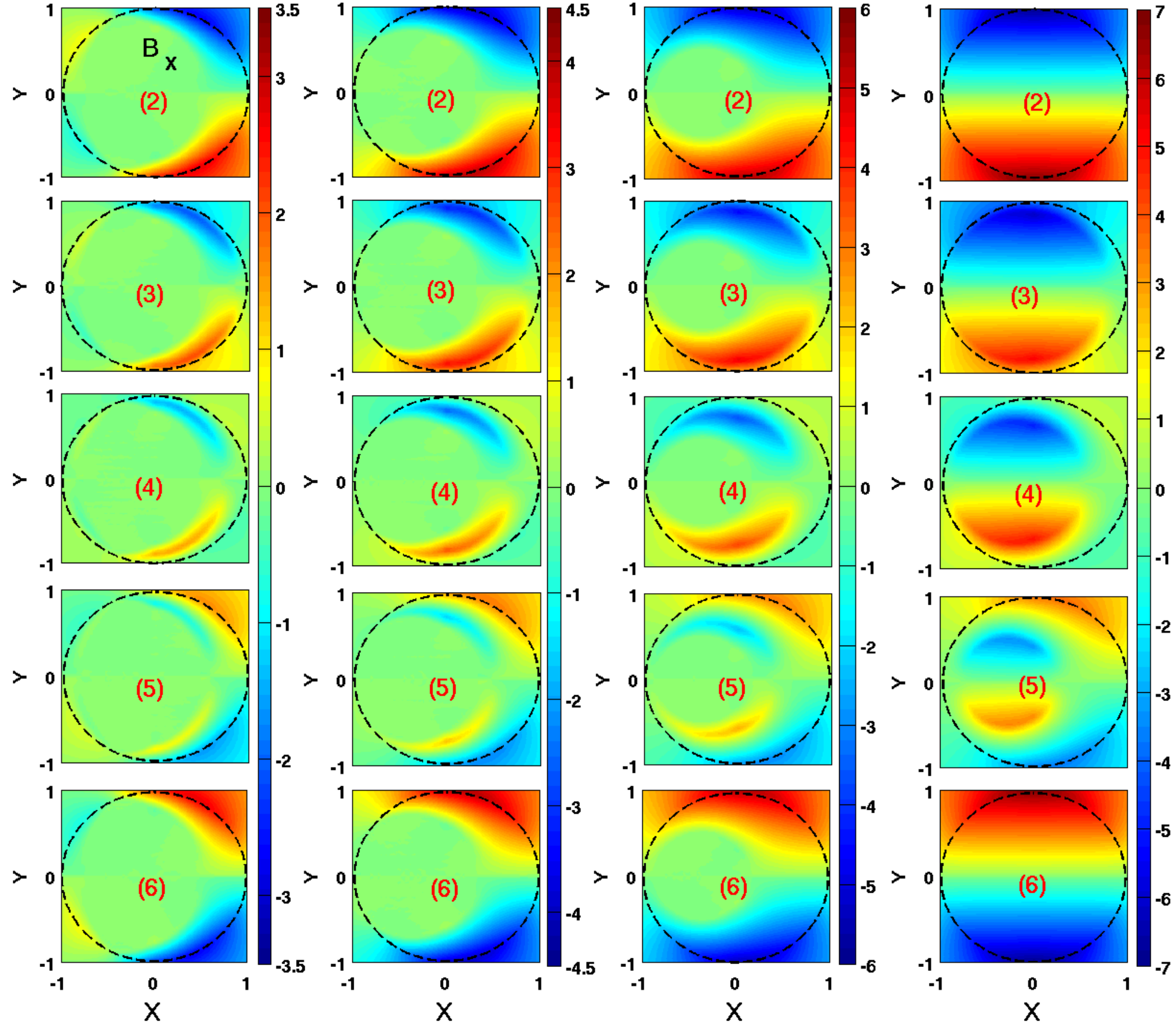}
\caption*{\label{Figure_S_6} Figure S6: Colormaps for the evolution of the
component of magnetic flux density $B_{x}$, corresponding to the current
profiles displayed in Fig.~S5.
}
\end{figure}

\break
\vspace*{0.0cm}
\begin{figure}[H]
\centering
\includegraphics[height=12.5cm,width=13cm]{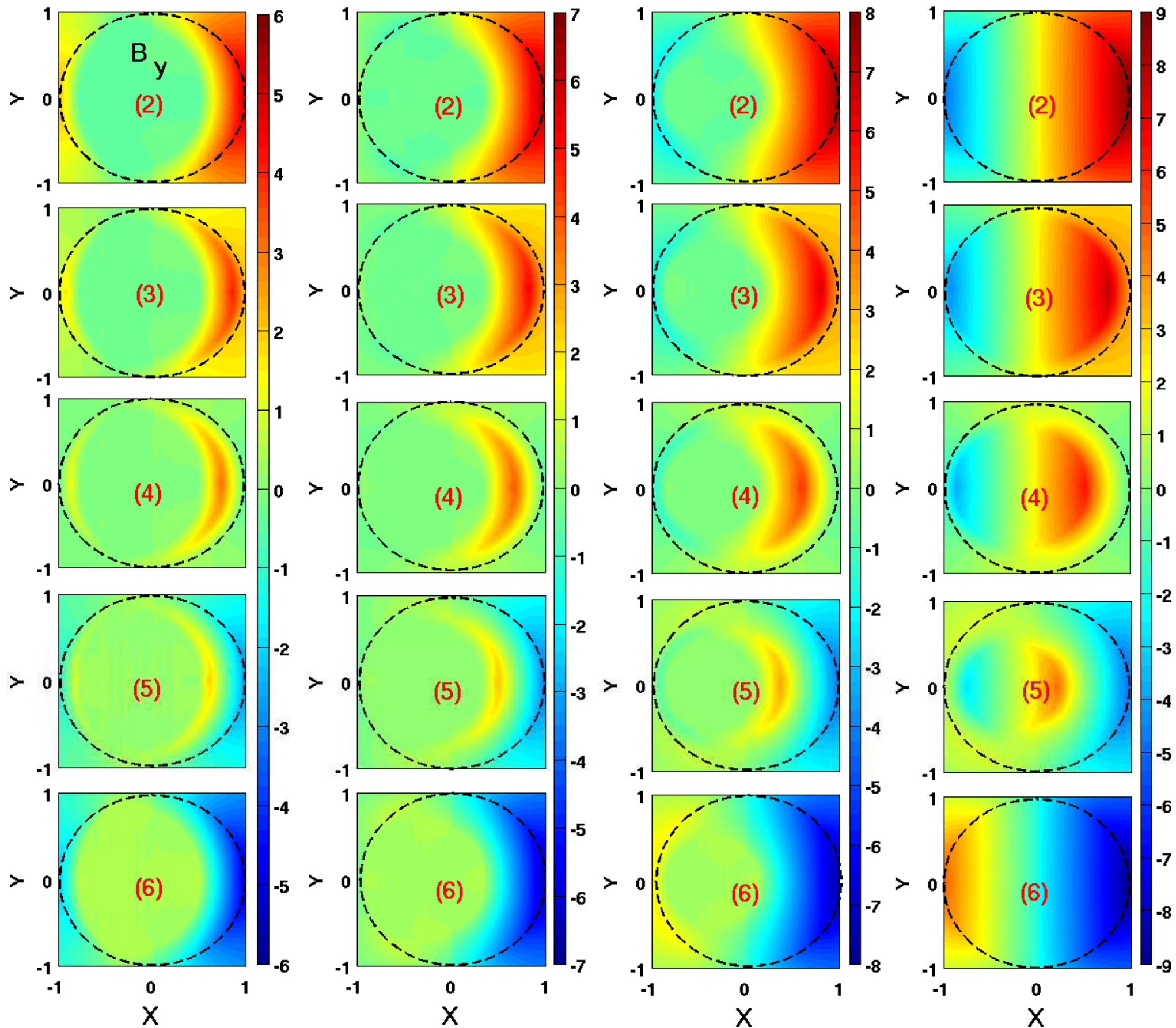}
\caption*{\label{Figure_S_7} Figure S7: Colormaps for the evolution of the
component of magnetic flux density $B_{y}$, corresponding to the current
profiles displayed in Fig.~S5.
}
\end{figure}

\break
\vspace*{0.5cm}
\begin{figure}[H]
\begin{center}
\includegraphics[height=15cm,width=3cm]{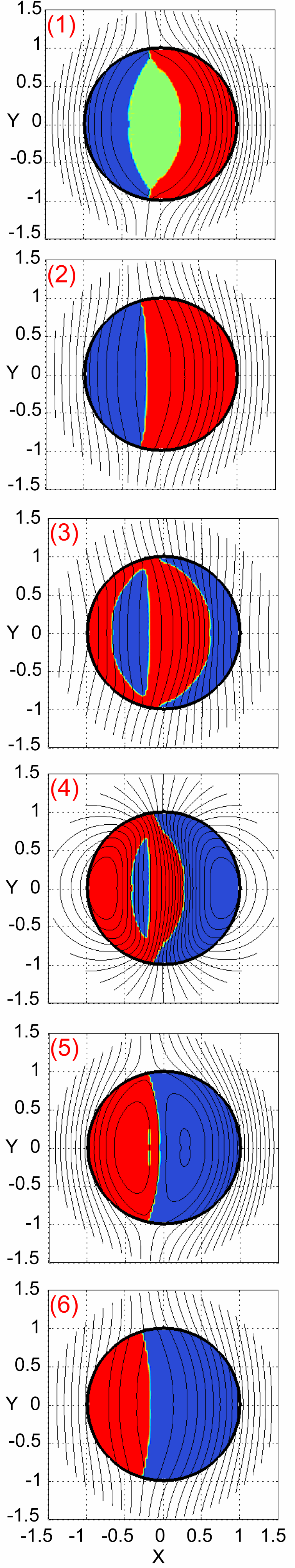} \hspace*{-0.1cm}
\includegraphics[height=15cm,width=3cm]{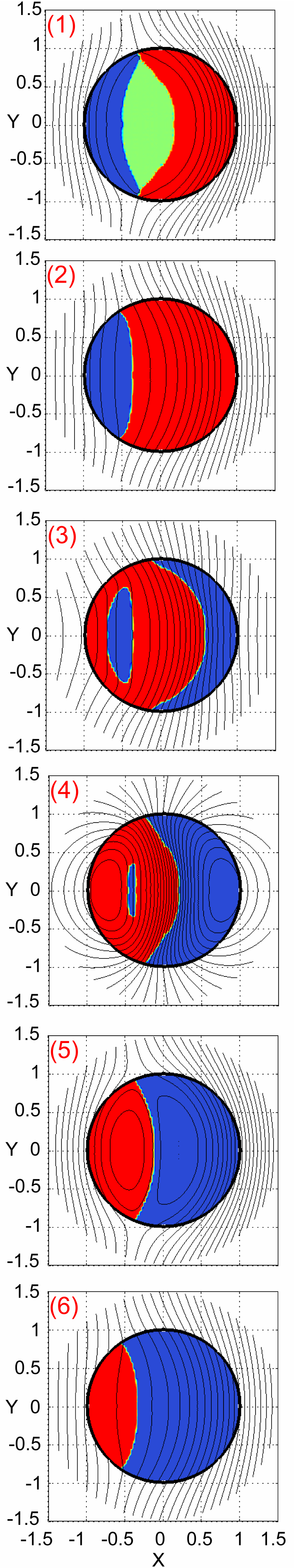} \hspace*{-0.1cm}
\includegraphics[height=15cm,width=3cm]{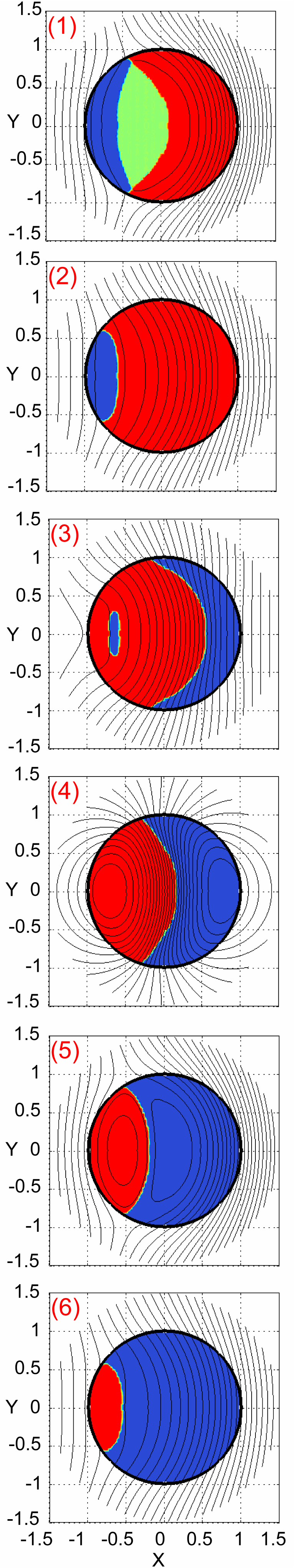} \hspace*{-0.1cm}
\includegraphics[height=15cm,width=3cm]{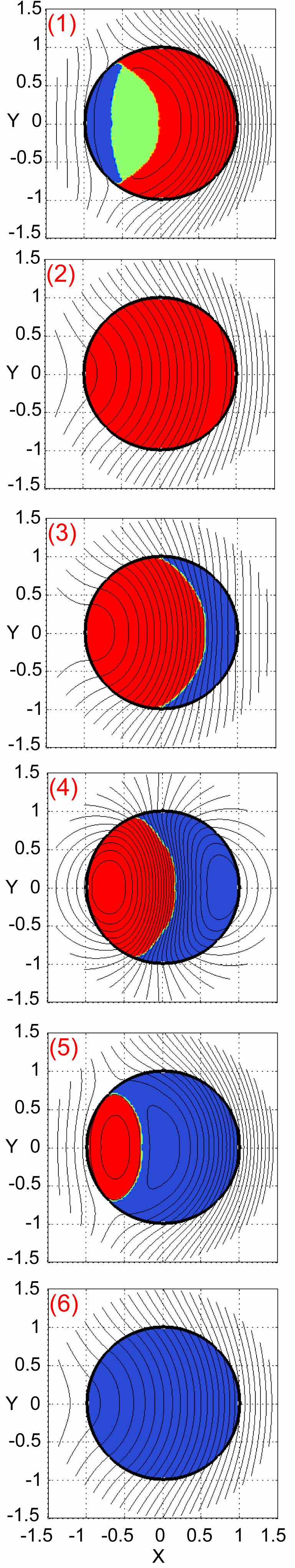} \hspace*{-0.1cm}
\end{center}
\caption*{\label{Figure_S_8} Figure S8: Evolution of the magnetic flux lines
and profiles of current with synchronous oscillating sources
$(B_{0,y},I_{tr})$
of amplitudes $B_{\tt{a}}=8$ (\textit{high field}) and, $I_{\tt{a}}=0.25$
\textit{(1st column)}, $I_{\tt{a}}=0.5$ \textit{(2nd column)},
$I_{\tt{a}}=0.75$ \textit{(3rd column)}, and $I_{\tt{a}}=1$ \textit{(4th
column)}. Plotting interval is $(B_{\tt{a}}/2,I_{\tt{a}}/2)$, where the time
step (1) defines the condition $(8,I_{\tt{a}}/2)$ [see also
Fig.~\ref{Figure_6_4} (pag.~\pageref{Figure_6_4})].
}
\end{figure}

\break
\vspace*{0.5cm}
\begin{figure}[H]
\centering
\includegraphics[height=12.5cm,width=13cm]{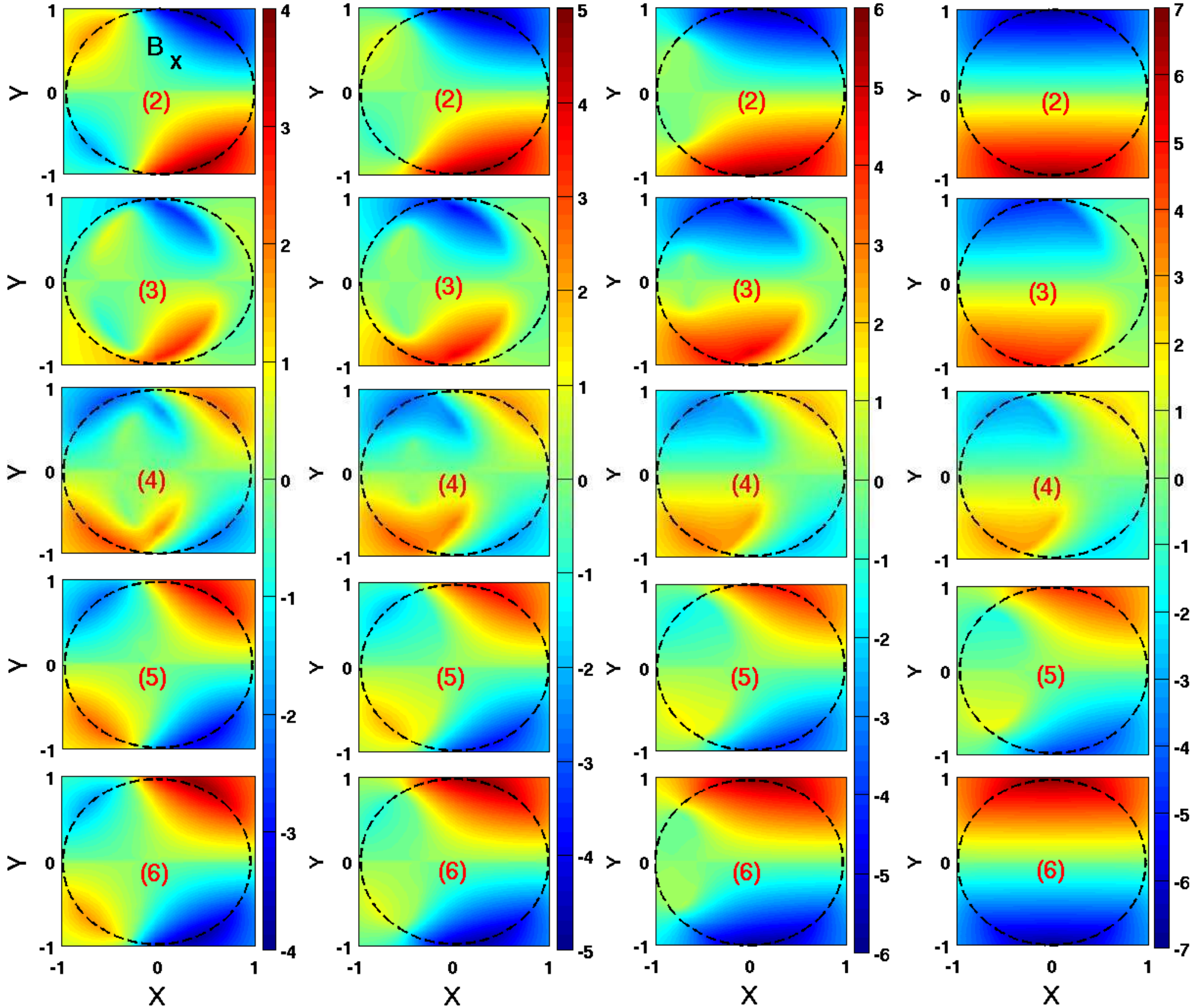}
\caption*{\label{Figure_S_9} Figure S9: Colormaps for the evolution of the
component of magnetic flux density $B_{x}$, corresponding to the current
profiles displayed in Fig.~S8.
}
\end{figure}

\break
\vspace*{0.5cm}
\begin{figure}[H]
\centering
\includegraphics[height=12.5cm,width=13cm]{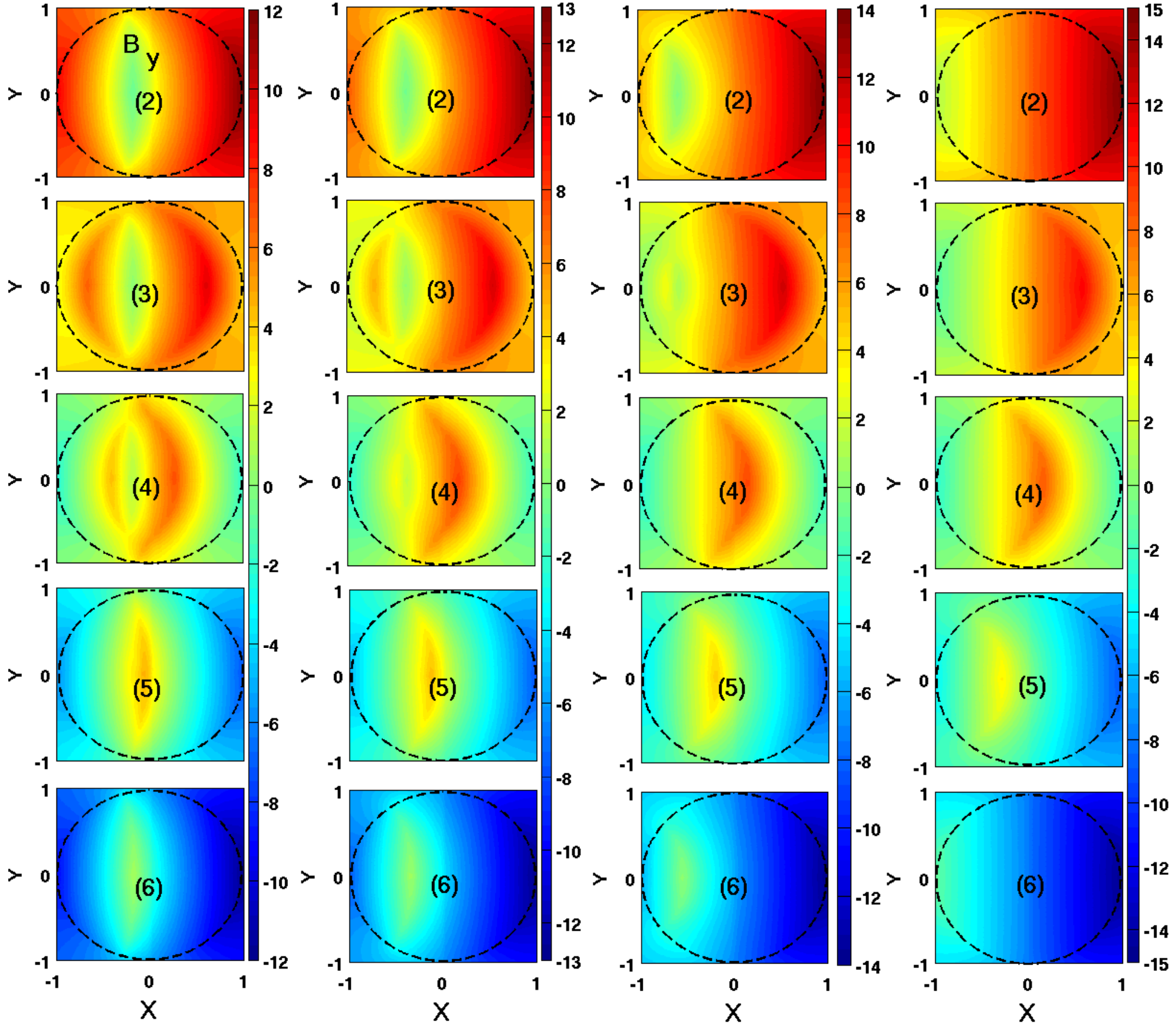}
\caption*{\label{Figure_S_10} Figure S10: Colormaps for the evolution of the
component of magnetic flux density $B_{y}$, corresponding to the current
profiles displayed in Fig.~S8.
}
\end{figure}

\break
\vspace*{-0.8cm}
\begin{figure}[H]
\begin{center}
\includegraphics[height=15cm,width=3cm]{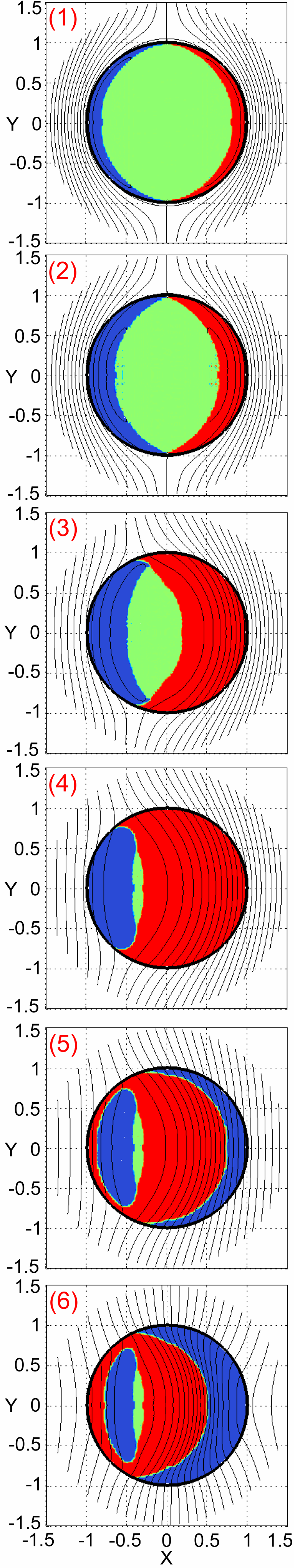} \hspace*{-0.1cm}
\includegraphics[height=15cm,width=3cm]{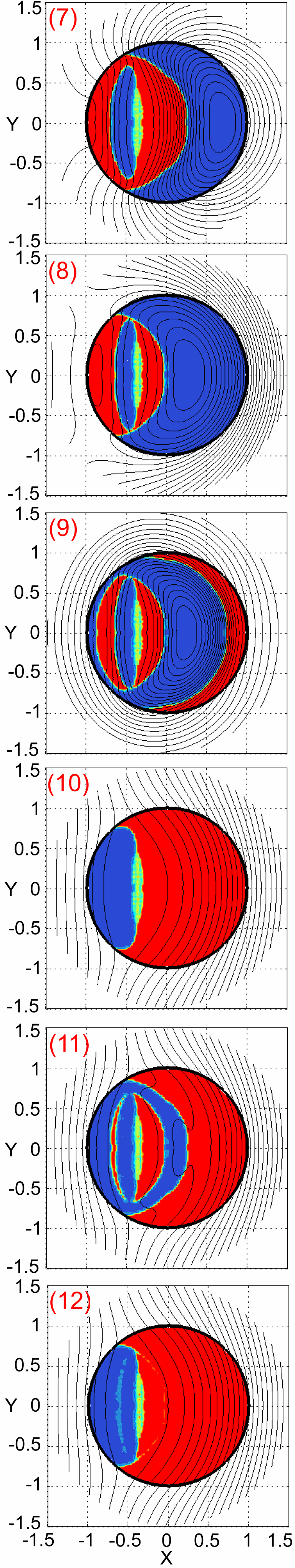}
\hspace*{-0.1cm}\vline
\includegraphics[height=15cm,width=3cm]{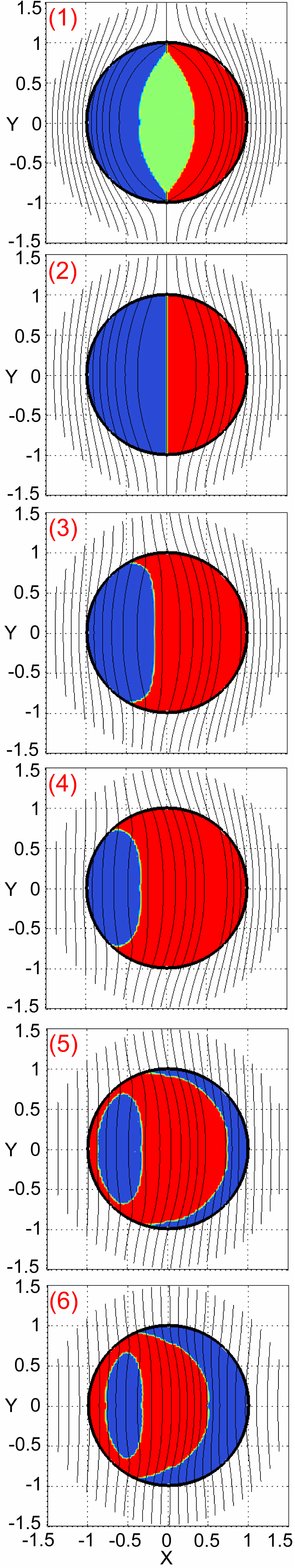} \hspace*{-0.1cm}
\includegraphics[height=15cm,width=3cm]{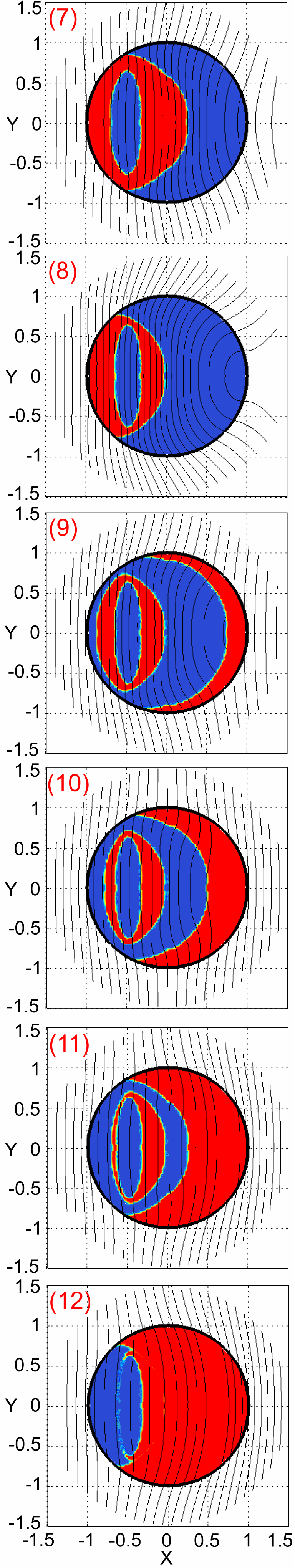} \hspace*{-0.1cm}
\end{center}
\caption*{\label{Figure_S_11} Figure S11: Evolution of the magnetic flux lines
and profiles of current with synchronous oscillating sources $B_{0,y}$ and
$I_{tr}$ of amplitudes $B_{\tt{a}}=4$ (\textit{intermediate field}) and
$I_{\tt{a}}=0.5$. Results for two premagnetized
samples with $B(t')=2$ (left pane) and $B(t')=8$ (right pane) are shown. 
Numeric tags in the upper left corner of each subplot have been
incorporated according to the following time-steps for the
experimental process depicted in Fig.~\ref{Figure_6_9}:
(1) corresponds to half of the time between $t=0$ and $t=t'$, then (2) at
$t=t'$, and for (2)
to (12) increases of $\Delta t\equiv 1/8$ per unit cycle have been considered.
Thus, the full cycle peak-to-peak corresponds to the subplots (4) to (12), 
respectively.
}
\end{figure}

\break
\vspace*{-1.0cm}
\begin{figure}[H]
\centering
\includegraphics[height=16cm,width=13cm]{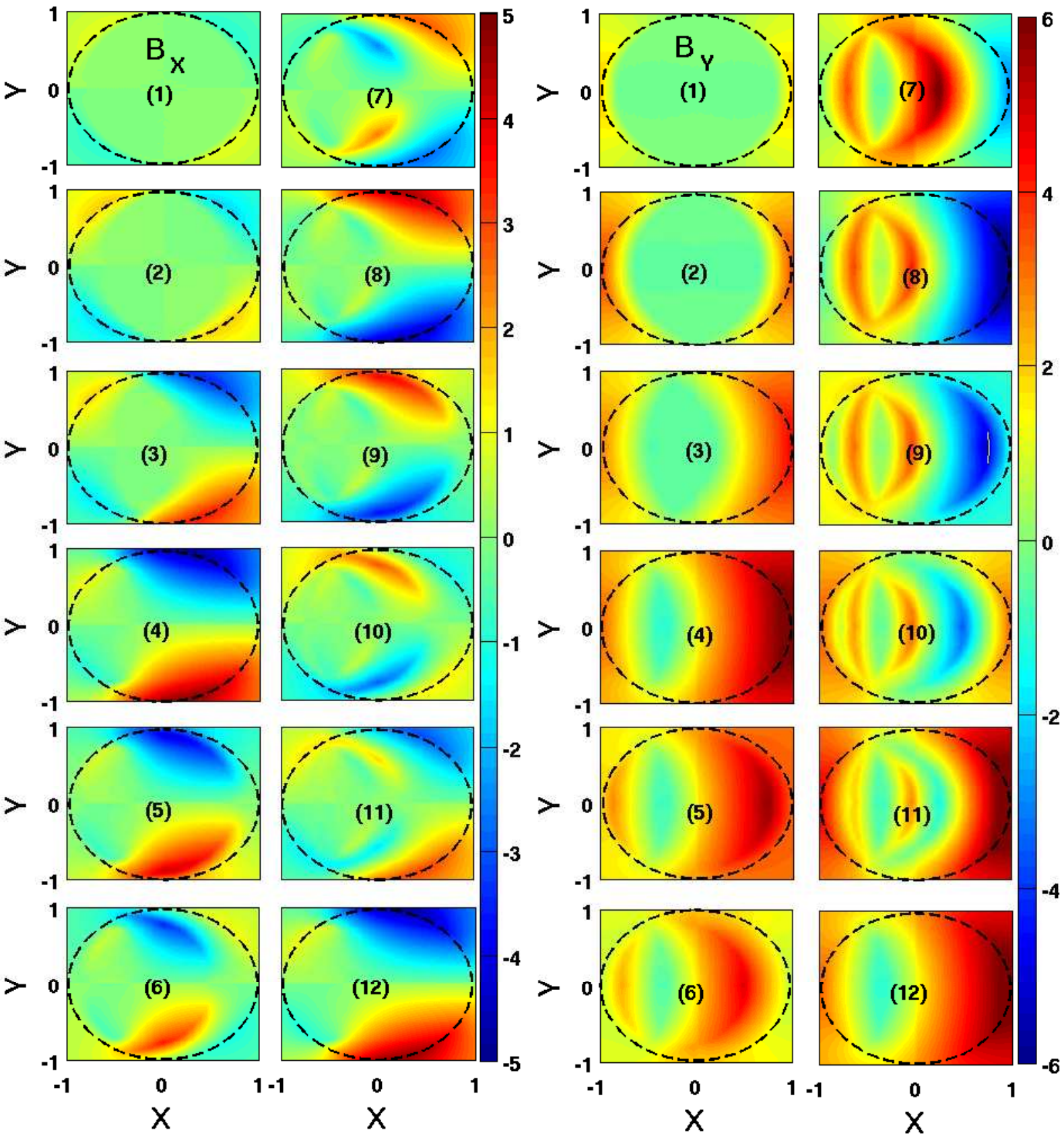}
\caption*{\label{Figure_S_12} Figure S12: Colormaps for the evolution of the
components of magnetix flux density $B_{x}$ (left) and $B_{y}$ (right)
for the current density profiles displayed at the left pane of Fig.~S11.
For $B_{y}$, subplots (3) to (6), and (12), the colormap have to be renormalized
to a linear scale of limits 10 and -10.
}
\end{figure}

\break
\vspace*{-1.0cm}
\begin{figure}[H]
\centering
\includegraphics[height=16cm,width=13cm]{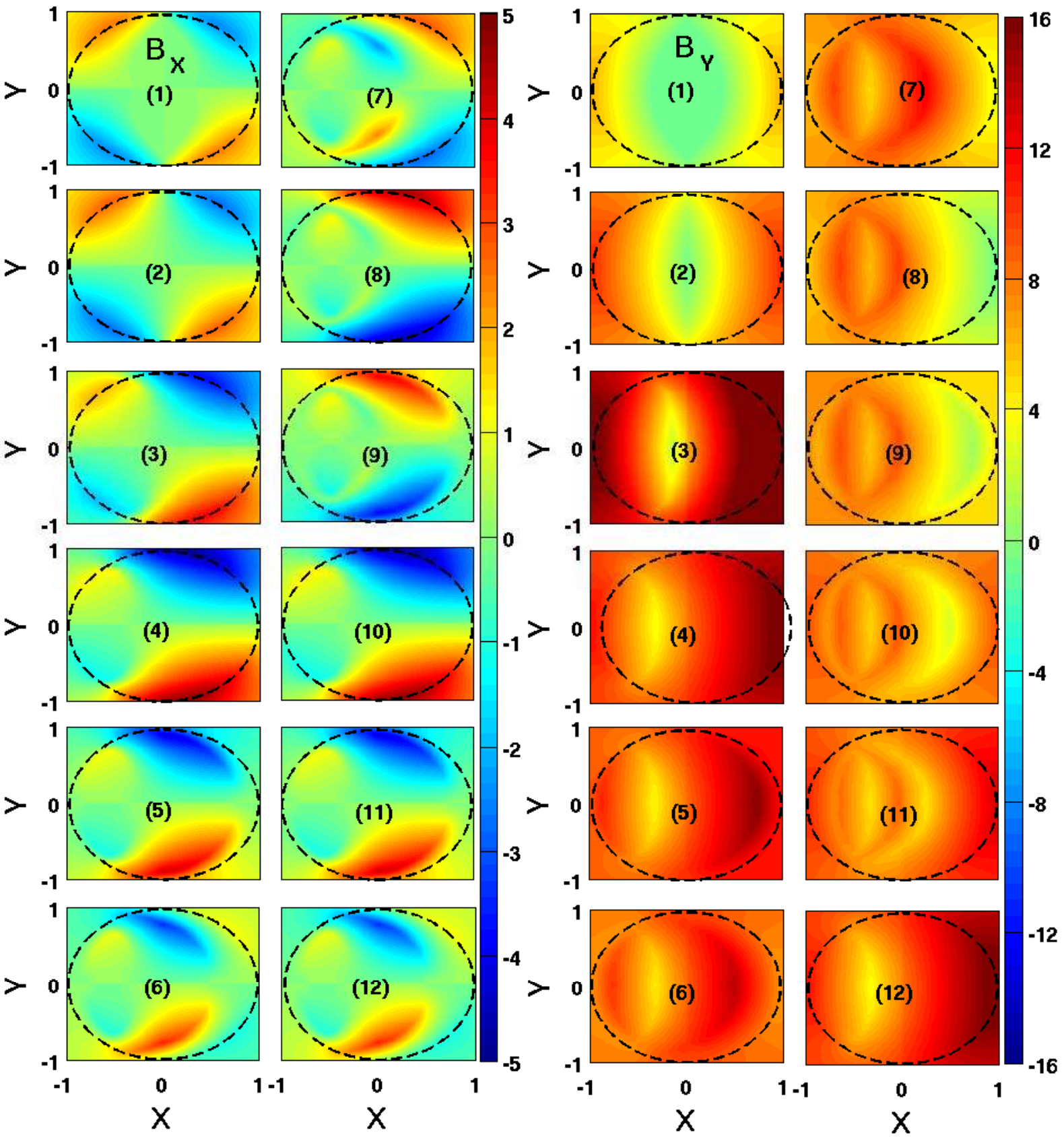}
\caption*{\label{Figure_S_13} Figure S13: Colormaps for the evolution of the
components of magnetix flux density $B_{x}$ (left) and $B_{y}$ (right)
for the current density profiles displayed at the rigth pane of Fig.~S11.
}
\end{figure}

\break
\vspace*{0.5cm}
\begin{figure}[H]
\begin{center}
\includegraphics[height=12cm,width=6cm]{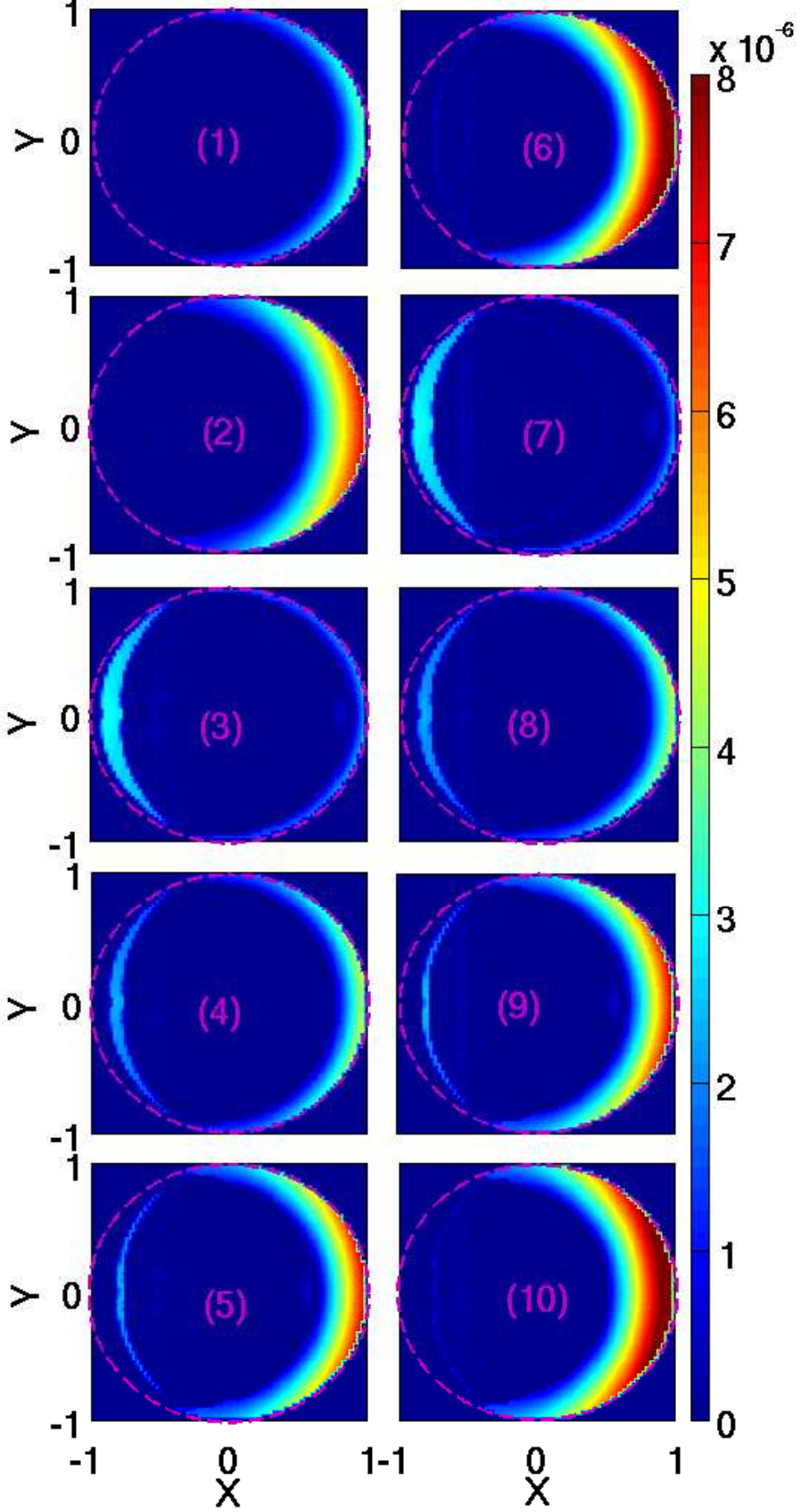} \hspace*{-0.1cm}
\includegraphics[height=12cm,width=6cm]{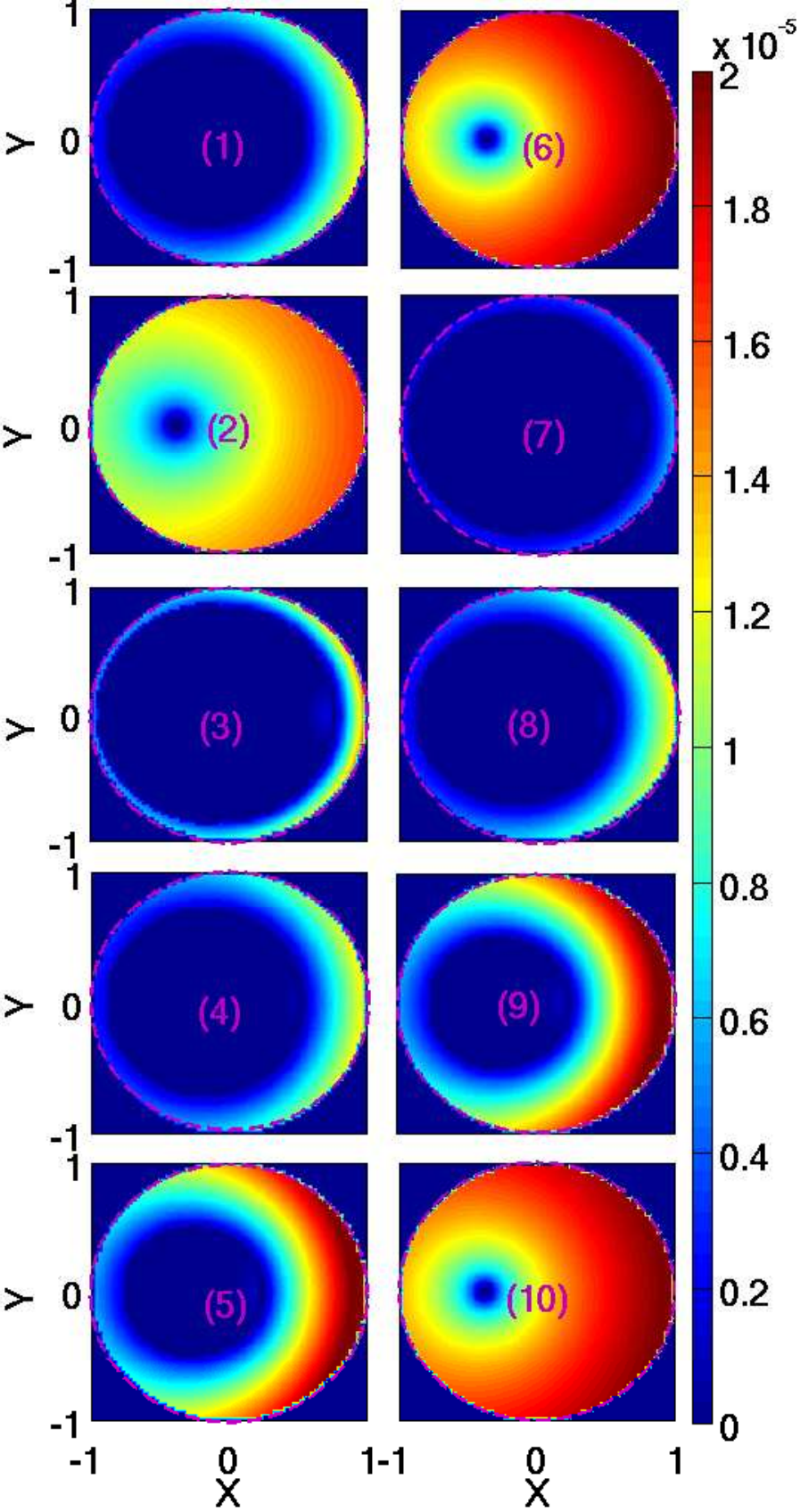} \hspace*{-0.1cm}
\end{center}
\caption*{\label{Figure_S_14} Figure S14: Evolution of the local density
of power dissipation \textbf{E}$\cdotp$\textbf{J} for oscillating sources of
amplitude $B_{\tt{a}}=2$ and: $I_{\tt{a}}=0.25$ \textit{(left pane)}, and
$I_{\tt{a}}=1$ \textit{(right pane)}. 
Each step has been plotted according to the
temporal process depicted in Fig.~\ref{Figure_6_4}. In the right pane, the
colormap for subplots
(2), (6), and (10) have to be renormalized by a factor of 5. Also, some of the
corresponding flux profiles have been displayed in the left and right columns
of Figs.~S5-~S7.
}
\end{figure}

\break
\vspace*{0.5cm}
\begin{figure}[H]
\begin{center}
\includegraphics[height=12cm,width=6cm]{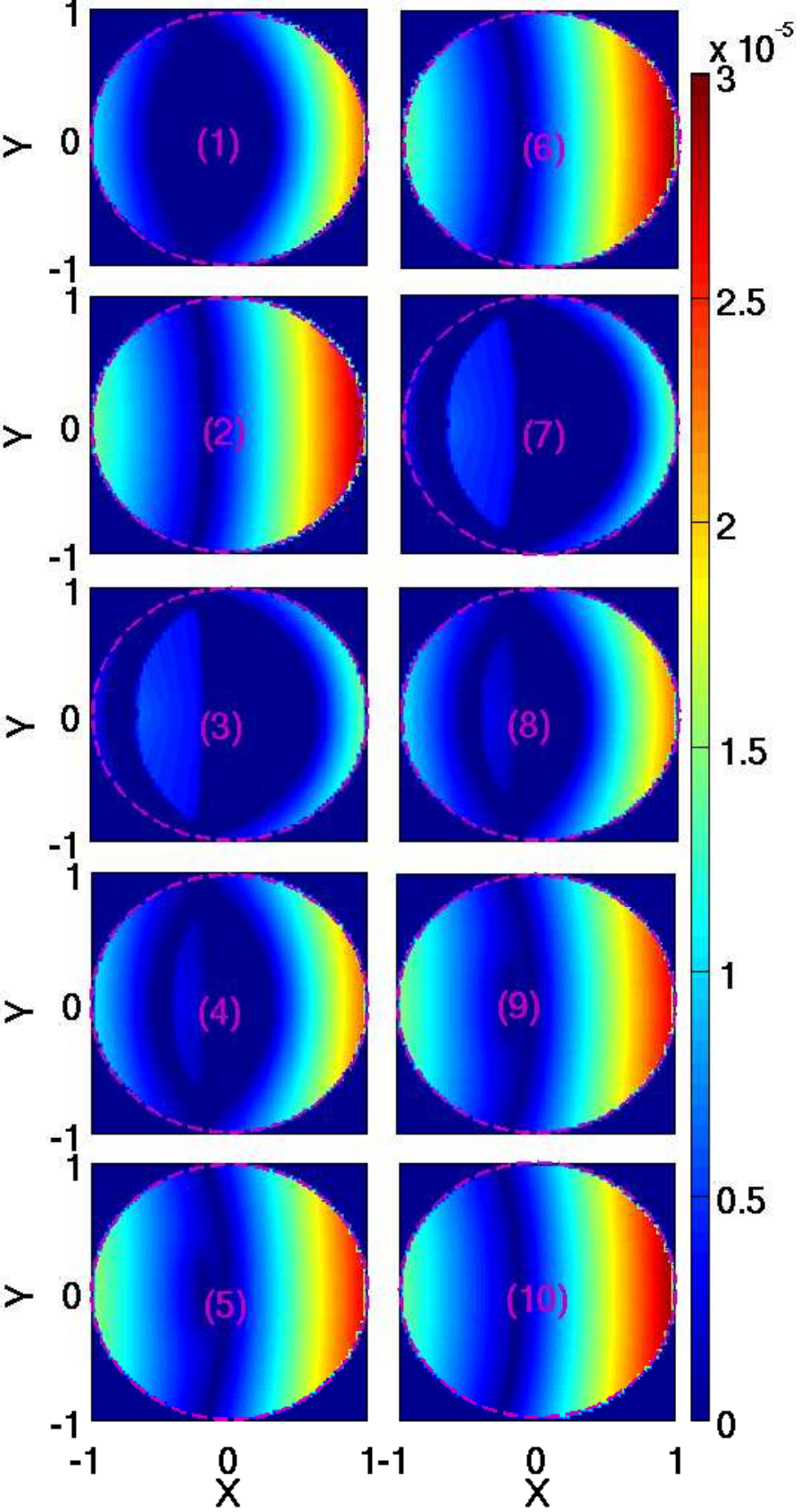} \hspace*{-0.1cm}
\includegraphics[height=12cm,width=6cm]{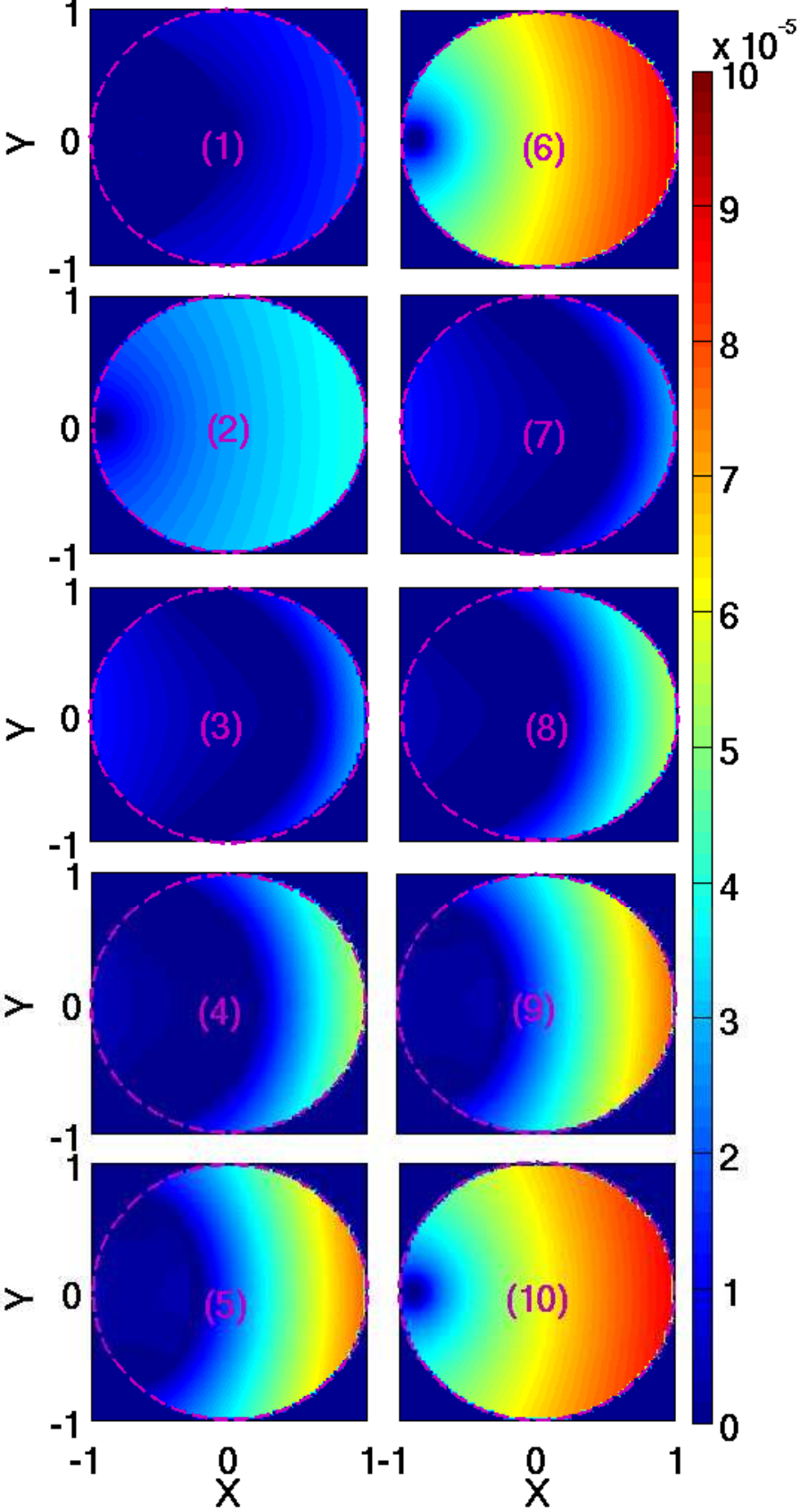} \hspace*{-0.1cm}
\end{center}
\caption*{\label{Figure_S_15} Figure S15: Evolution of the local density
of power dissipation \textbf{E}$\cdotp$\textbf{J} for oscillating sources of
amplitude $B_{\tt{a}}=8$ and, $I_{\tt{a}}=0.25$ \textit{(left pane)}, and
$I_{\tt{a}}=1$ \textit{(right pane)}. Each step has been plotted
according to the temporal process depicted in Fig.~\ref{Figure_6_4}, and some of
the corresponding flux profiles have been displayed in the left and right
columns
of Figs.~S8-~S10. In the right pane, the colormap for subplots
(6) and (10) have to be renormalized by a factor of 2. 
}
\end{figure}

\break
\vspace*{0.5cm}
\begin{figure}[H]
\begin{center}
\includegraphics[height=15cm,width=6cm]{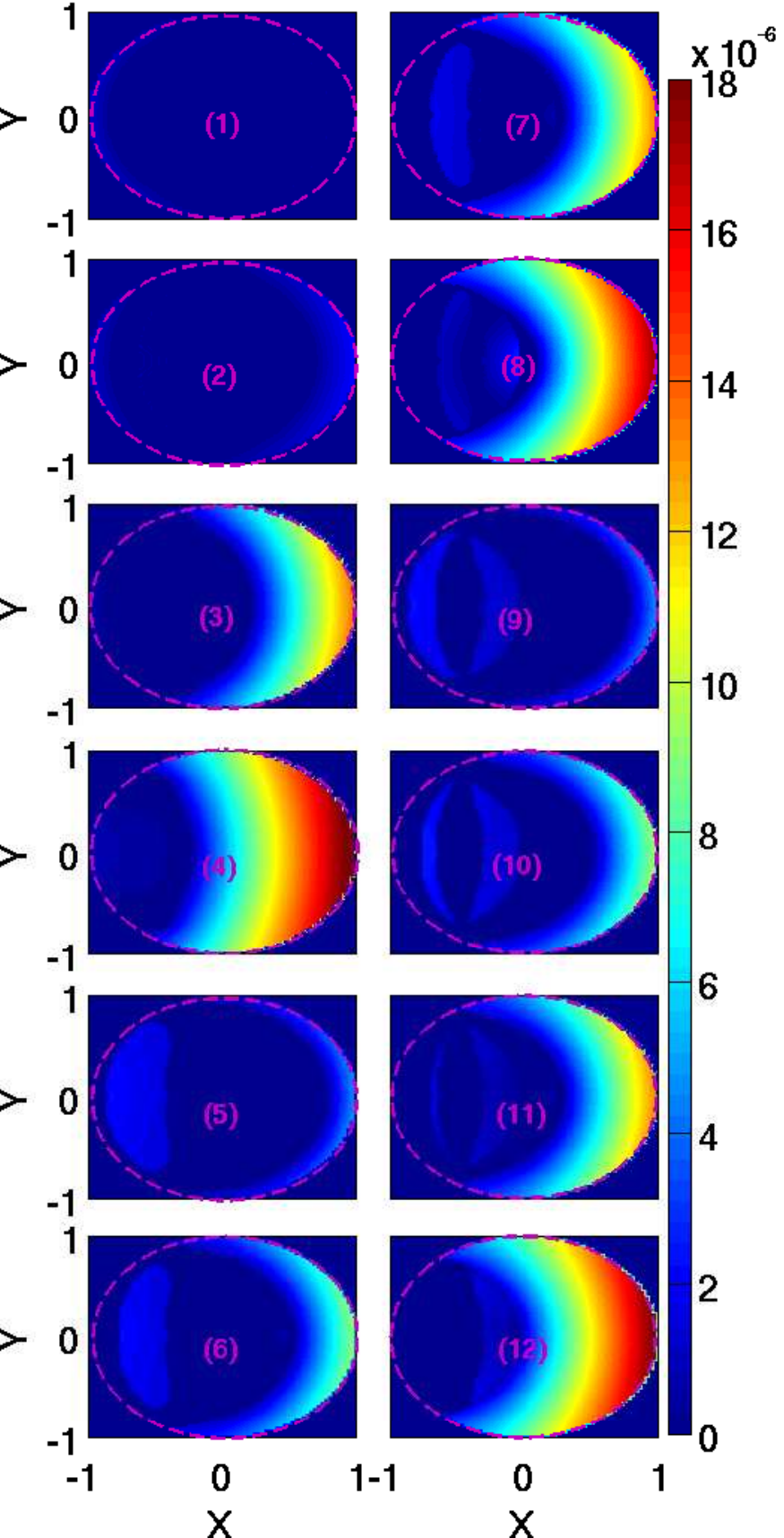} \hspace*{-0.1cm}
\includegraphics[height=15cm,width=6cm]{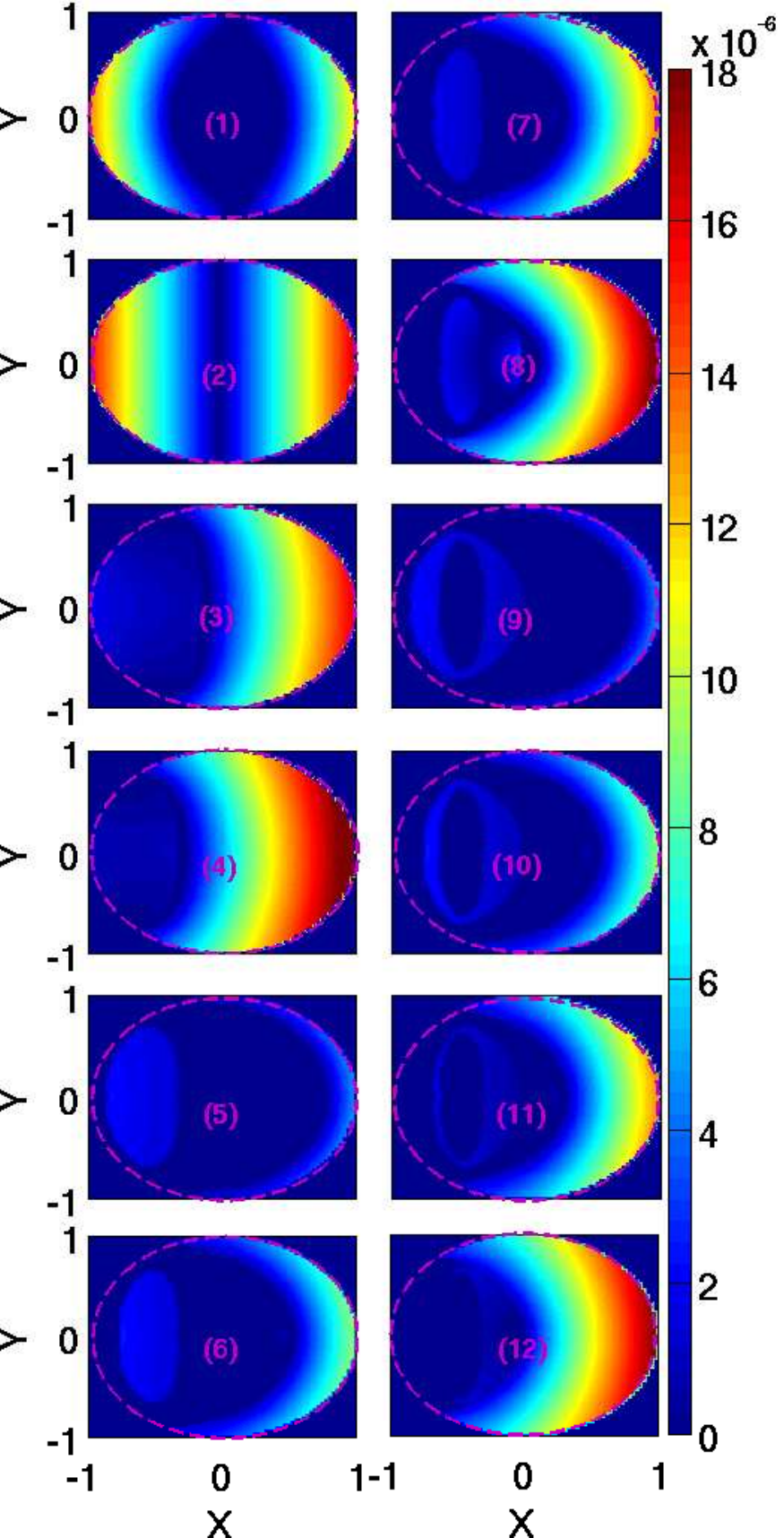} \hspace*{-0.1cm}
\end{center}
\caption*{\label{Figure_S_16} Figure S16: Evolution of the local density
of power dissipation \textbf{E}$\cdotp$\textbf{J} for cases with synchronous
sources $(B_{0,y},I_{tr})$ and premagnetized superconducting wires with:
$B(t')=2$ (\textit{left pane}) and $B(t')=8$ (\textit{right pane}),  in
correspondence to the profiles of current density displayed in Fig.~S11.
}
\end{figure}

\break
\vspace*{-0.5cm}
\begin{itemize}
 \item \textbf{\textit{\Large{SC wire subjected to asynchronous oscillating
\\ excitations}}}
\end{itemize}

\begin{figure}[H]
\begin{center}
\includegraphics[height=13cm,width=3cm]{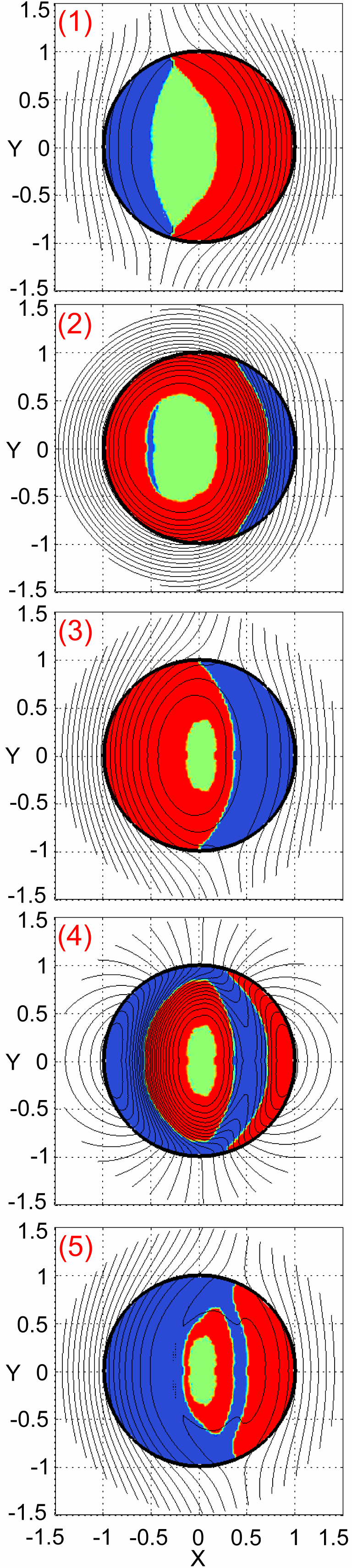} \hspace*{-0.1cm}
\includegraphics[height=13cm,width=3cm]{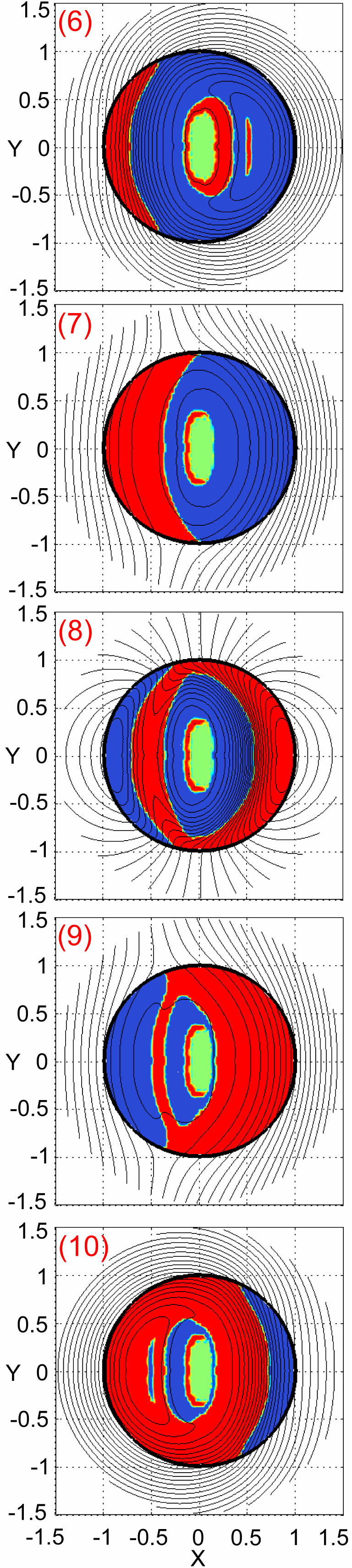}
\hspace*{-0.1cm}\vline
\includegraphics[height=13cm,width=3cm]{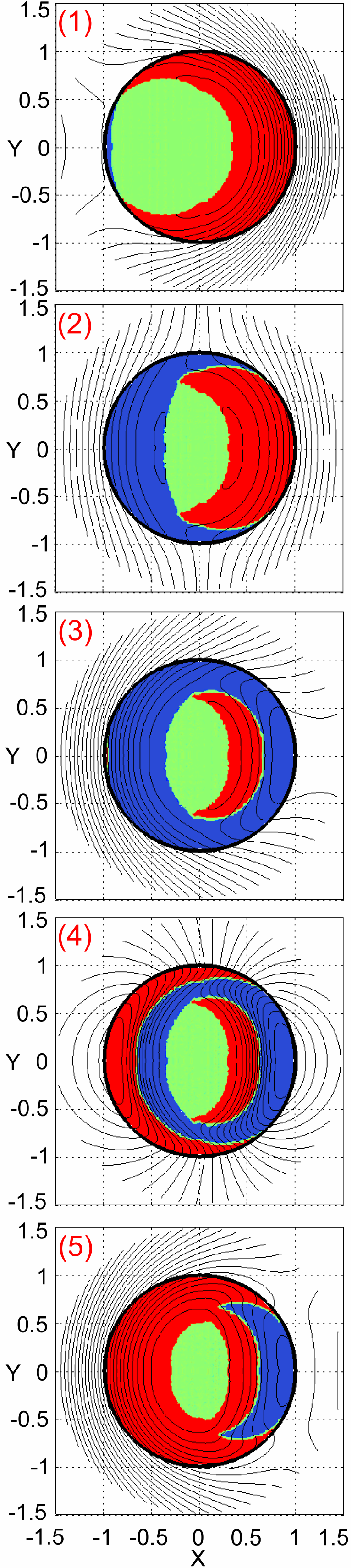} \hspace*{-0.1cm}
\includegraphics[height=13cm,width=3cm]{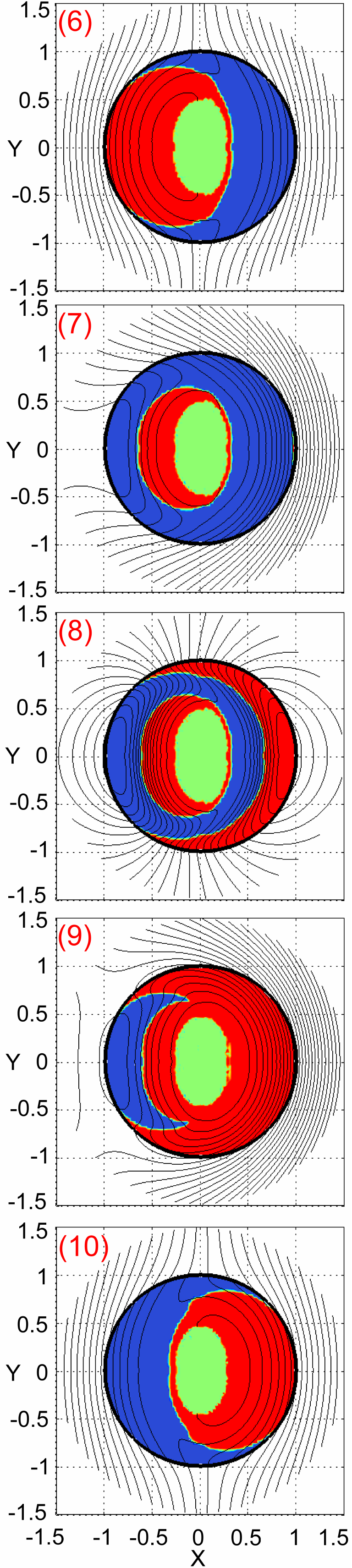} \hspace*{-0.1cm}
\end{center}
\caption*{\label{Figure_S_17} Figure S17: Evolution of the magnetic flux lines
and profiles of current with asynchronous oscillating sources $B_{0,y}$ and
$I_{tr}$, of amplitudes $B_{\tt{a}}=4$ and $I_{\tt{a}}=0.5$. In the left pane,
some of the results for the temporal process displayed in pane (b) of
Fig.~\ref{Figure_6_12} ($B_{0,y}$ has the lower
frequency) are shown. Analogously, the results for the
pane (c) of Fig.~\ref{Figure_6_12} ($I_{tr}$ has the lower frequency) are
shown at the right pane.
}
\end{figure}

\break
\vspace*{-0.0cm}
\begin{figure}[H]
\centering
\includegraphics[height=13.5cm,width=13cm]{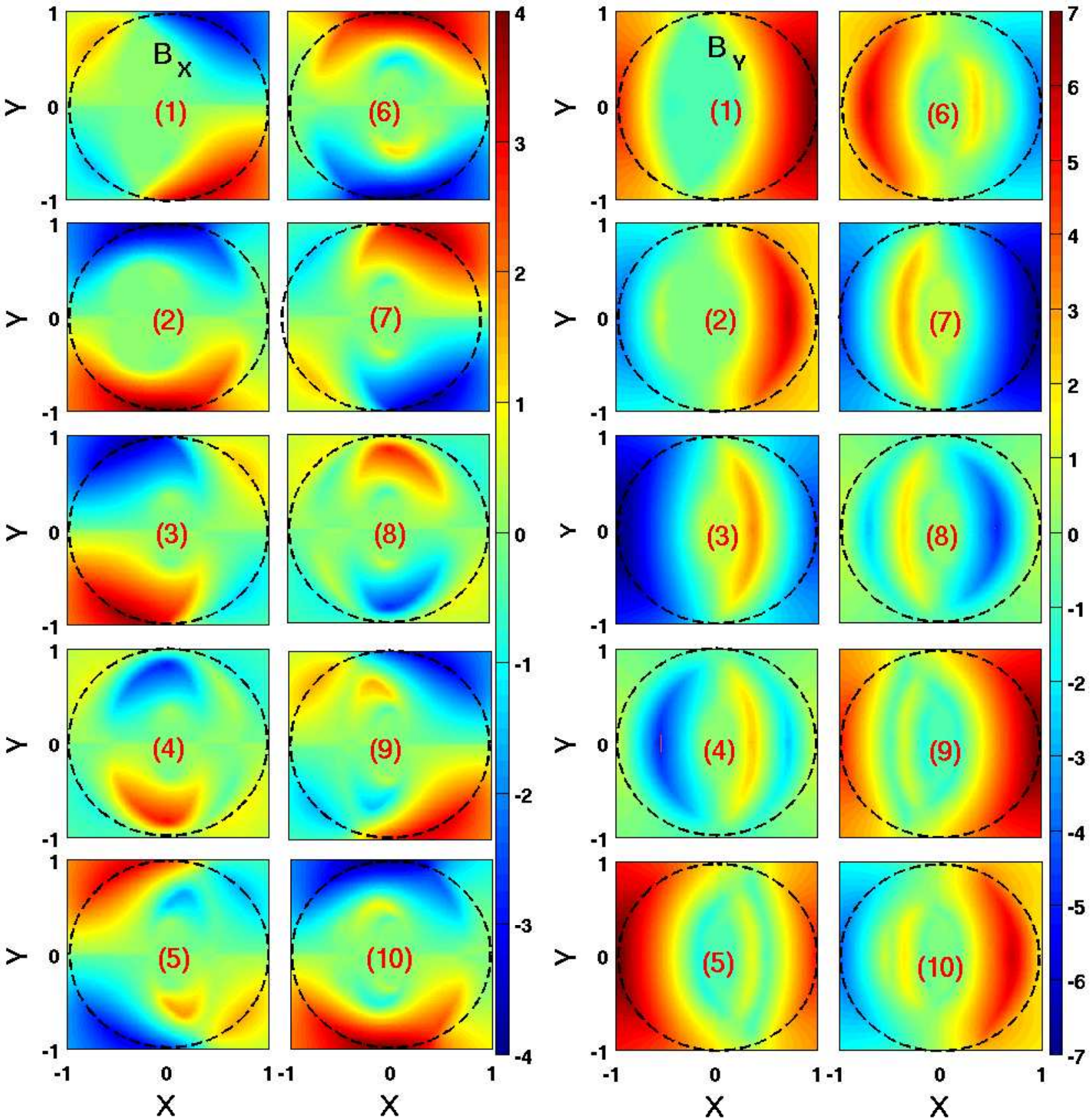}
\caption*{\label{Figure_S_18} Figure S18: Colormaps for the evolution of the
components of magnetix flux density $B_{x}$ (left) and $B_{y}$ (right)
for the current density profiles displayed at the left pane of Fig.~S17.
}
\end{figure}

\break
\vspace*{-0.0cm}
\begin{figure}[H]
\centering
\includegraphics[height=13.5cm,width=13cm]{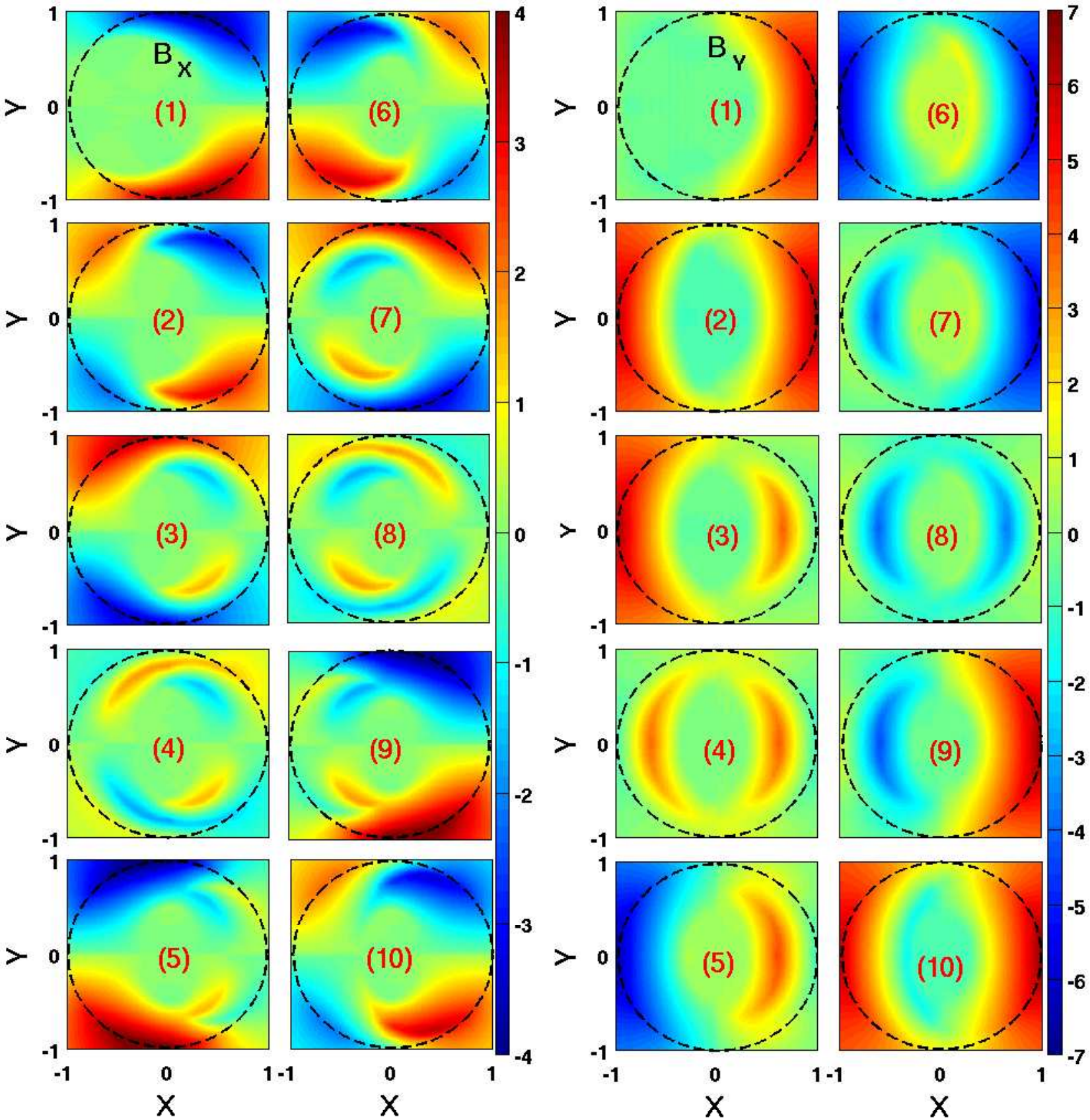}
\caption*{\label{Figure_S_19} Figure S19: Colormaps for the evolution of the
components of magnetix flux density $B_{x}$ (left) and $B_{y}$ (right)
for the current density profiles displayed at the right pane of Fig.~S17.
}
\end{figure}

\break
\vspace*{0.5cm}
\begin{figure}[H]
\begin{center}
\includegraphics[height=12.5cm,width=6cm]{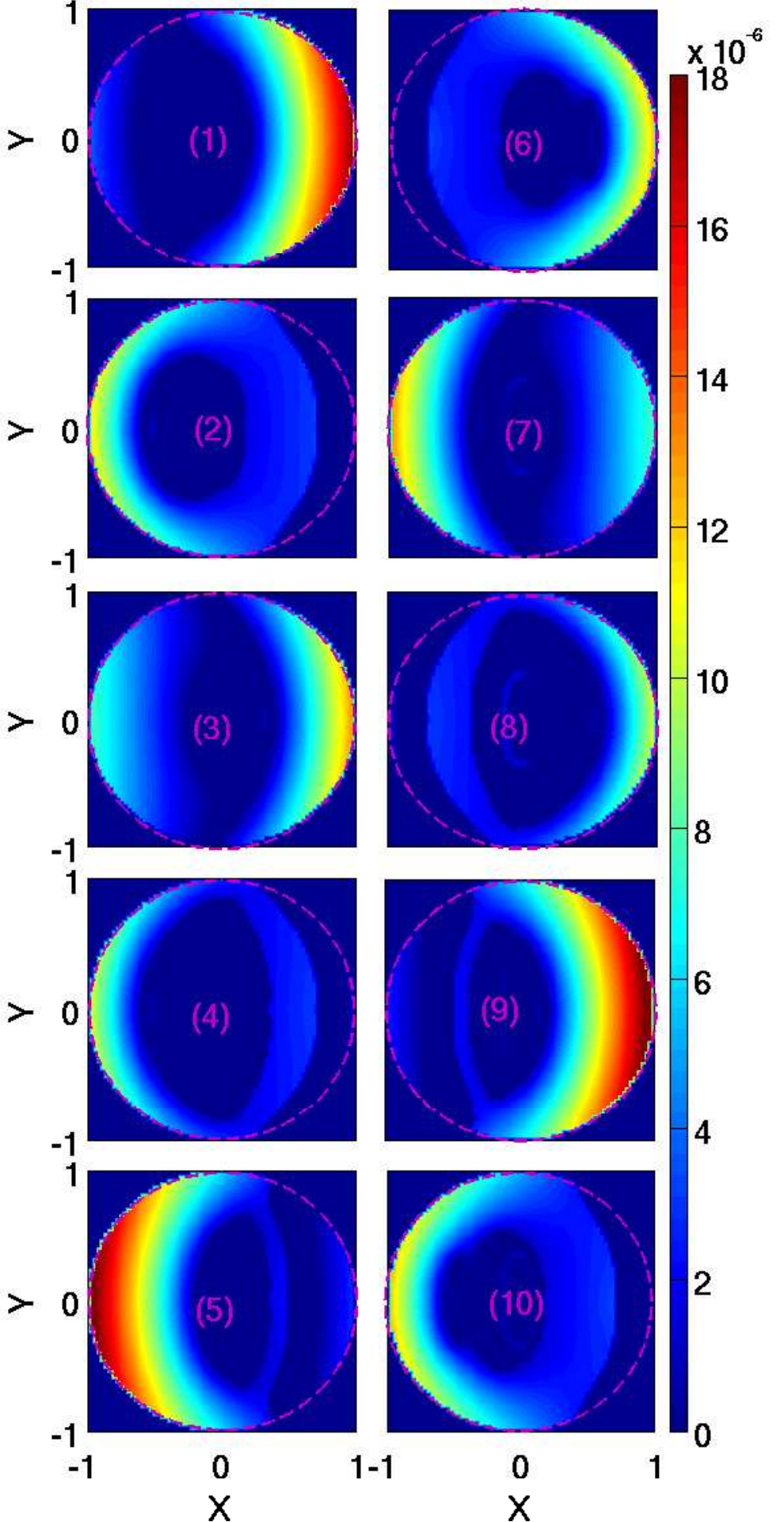} \hspace*{-0.1cm}
\includegraphics[height=12.5cm,width=6cm]{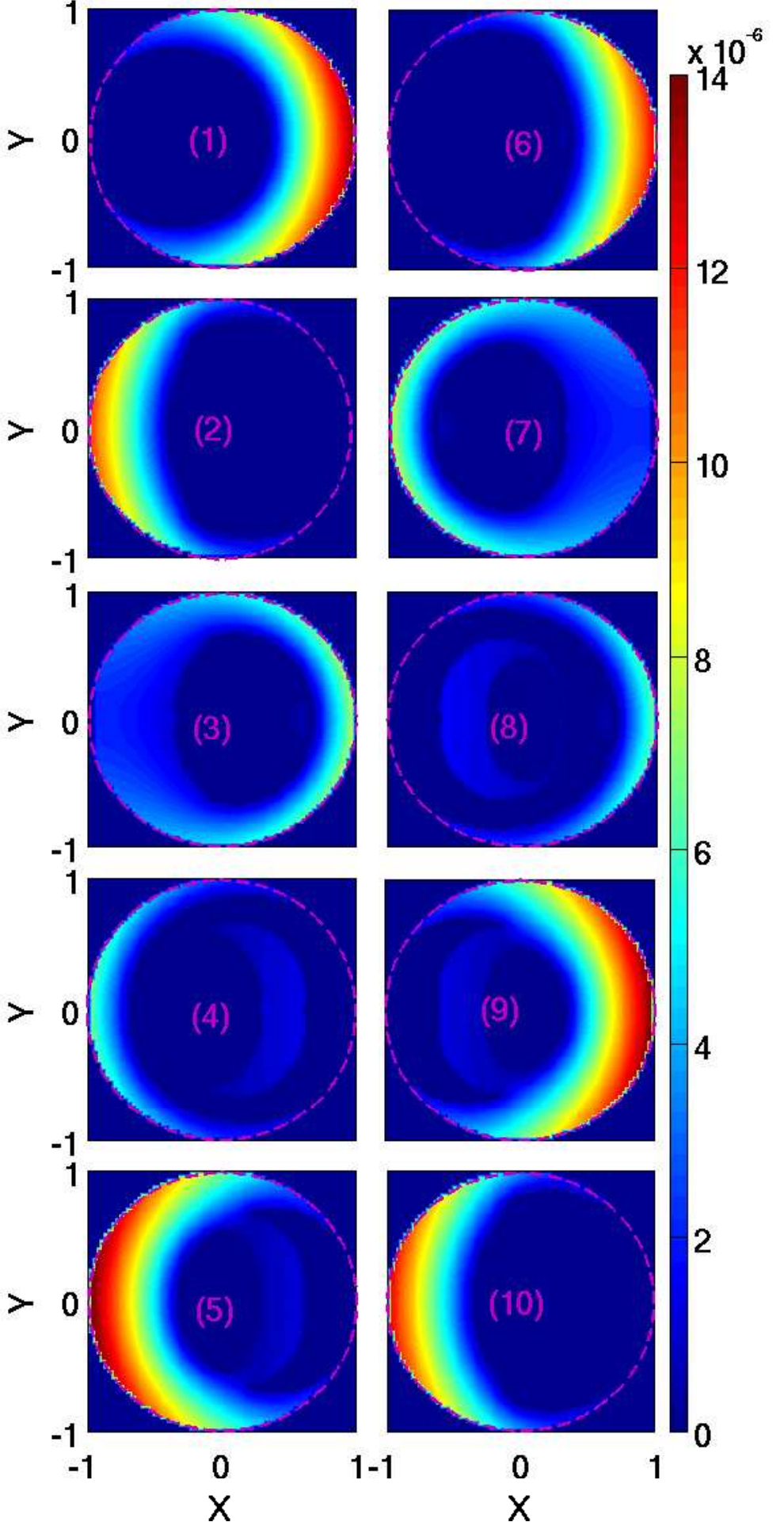} \hspace*{-0.1cm}
\end{center}
\caption*{\label{Figure_S_20} Figure S20: Evolution of the local density
of power dissipation \textbf{E}$\cdotp$\textbf{J} for cases with asynchronous AC
sources $B_{0,y}$ and $I_{tr}$, according to the current density profiles
displayed in Fig.~S17.
}
\end{figure}

%



\part{\label{Part_3}\textsc{\textbf{Microscopical aspects also analyzed}}}


\chapter*{}
\vspace*{-3cm}
\section*{\label{Intro-P2}\hspace*{0.55cm} \sc \textbf{Introduction}}
\markboth{\sc \textbf{Introduction}}{\hspace*{0.55cm} \sc \textbf{Introduction}}
\addcontentsline{toc}{chapter}{\hspace*{0.55cm} \sc \textbf{Introduction}}

After one century of the discovery of superconductivity  we are still
awaiting for a conclusive theory at least beyond normal metals, that are
described within the framework of the BCS theory~\cite{P3-BCS57}. However, this
does not mean that we lack well-established theories to explain some of the
experimental facts, despite many of the thermodynamic properties as the high
superconducting transition temperature $(T_{c})$ can not be reproduced under a
unique scheme. Interestingly, although that the theoretical background
behind the understanding of the microscopical aspects involved in the creation
of the superconducting state is rather complex, we have found that relatively
simple numerical techniques may be used in order to withdraw relevant
conclusions from specific experimental
data~\cite{P3-Ruiz2009,P3-Ruiz2011_JS,P3-Ruiz2011_CAP}.

Our point of interest is the following. In conventional metals, the
electron-phonon (E-Ph) coupling mode has long been recognized as the main
mechanism involved in the superconducting properties, because the strength of
this interaction essentially determines the value of $T_{c}$. However, in the
HTSC the experimentally determined $d-$wave pairing in their layered crystal
structure with one or more CuO$_{2}$ planes per unit cell~\cite{P3-Bednorz86},
introduces considerable theoretical complications even when other coupling modes
are considered~\cite{P3-Kulic04}. Fortunately, with the appearance of a new era
of analyzers for Angle Resolved Photoemission Spectroscopies (ARPES) with
improved resolution both in energy
and momentum~\cite{P3-Hufner07}, the controversy on the influence of the
anisotropic character of the superconducting gap in the electron properties of
HTSC can be directly avoided just by analyzing preferential directions within
the $CuO_{2}$ planes or the so-called nodal directions. These directions are
basically characterized by a negligible contribution of the superconducting gap
along the $(0,0)-(\pi,\pi)$ direction in the Brillouin zone,
providing a smart solution if one is merely interested in identifying the
energy modes of the quasiparticles involved in the superconducting pairs
formation, i.e., in the origin of the coupling mode which binds two
electrons (holes) in the formation of Cooper pairs. 

In order to  understand the relevance of the ARPES technique, we
recall that the photoemission process results in both an excited photoelectron
and a photohole in the final state. On the one hand, unlike other probes, the
ARPES technique has the advantage of momentum
resolving, which becomes a useful probe of the related scattering mechanisms
contributing to the electrical transport in different materials. We want to note
that the single-particle scattering rate measured in ARPES is not identical to
the scattering rate measured in transport studies themselves. Nonetheless,
direct proportionality between them has been
established~\cite{P3-Kulic00,P3-Smith01}. On the other hand, one of the most
telling manifestations of the E-Ph mode is a mass renormalization of the
electronic dispersion at the energy scale associated with the phonons. This
renormalization effect is directly observable in the ARPES measurements as a
low-energy excitation band in the dispersion curves of photo-emitted electrons,
known as \textit{kink}~\cite{P3-Lanzara01,P3-Zhou03}. In other words, the kink
effect can be understood as a well-defined slope change in the electronic
energy-momentum dispersion in a similar energy scale ($E_{k}-E_{F}\sim40 -
80meV$).
This feature, so far universal in the HTSC, has been regarded as a signature of
the strength of the boson mechanism which causes the pair formation in the
superconducting state. In fact, all the interactions of the electrons which are
responsible for the unusual normal and superconducting properties of cuprates
are believed to be represented in this anomaly~\cite{P3-Zhou07}. This has
prompted an intense debate about the nature of the coupling mode involved in the
density of low-energy electronic excited states in the momentum-energy space,
and its influence on the emergence of the superconducting 
state~\cite{P3-Ruiz2009,P3-Ruiz2011_JS,P3-Ruiz2011_CAP,P3-Lanzara01,P3-Zhou03,
P3-Zhou07,P3-Devereaux04,P3-Zhou02,P3-Zhou05,P3-Kordyuk06,P3-Xiao07,
P3-Takahashi07,P3-Gweon04,P3-Douglas07,P3-Johnson01,P3-Borisenko06,P3-Zhang08,
P3-Graf08,P3-Reznik08,P3-Giustino08,P3-Park08,P3-Chang08}.  

In an effort to clarify the influence of the phonon coupling mode (either weak
or strong), we have analyzed the influence of the E-Ph interaction
on the electronic dispersion relations for several cuprate compounds. A full
discussion of the ARPES technique, as well as a detailed description of
the strong correlations theory which have been used to reproduce the nodal kink
effect in HTSC is more appropriately reserved for specialized texts in
photoemission spectroscopies and many body theories [see e.g.,
Refs.~\cite{P3-Doniach98,P3-Gross86,P3-Allen76,P3-Allen82,P3-Carbotte90,
P3-Ashcroft76}]. However, we can, in a brief way, introduce the basic
concepts of the E-Ph coupling theory for photoemission spectroscopies, and give
a thorough interpretation of the influence of this boson coupling mode on the
momentum distribution-curves (MDC) measured
by ARPES (chapter~\ref{ch-7}), and finally argue about how strong can be
considered the phonon coupling mode from the
analysis of the predicted values for $T_{c}$, the ratio
gap $2\Delta_{0}/k_{B}T_{c}$, and the zero temperature gap $\Delta_{0}$
(chapter~\ref{ch-8}). 

On the one hand, in chapter~\ref{ch-7} our analysis shows a remarkable
agreement between theory and experiment for different samples and at different
doping levels. Universal effects such as the nodal kinks at low energies are
theoretically reproduced, emphasizing the necessary distinction between the
general electron mass-enhancement parameter $\lambda^{*}$and the conventional
electron-phonon coupling parameter $\lambda$. On the other hand, in
chapter~\ref{ch-8} a thorough analysis of the superconducting thermodynamic
quantities and the Coulomb effects based on different approaches
will reveal as, contrary to the predictions for LSCO samples, in more
anisotropic materials as Bi2212 and Y123 families, it seems unavoidable to
consider additional coupling modes in order to justify their high critical
temperatures.








\chapter{\label{ch-7} \sc \textbf{E-Ph Theory And The Nodal Kink Effect In
HTSC}}

The photoemission process formally measures a complicated nonlinear response
function. However, it is helpful to notice that the analysis of the optical
excitation of the electron in the bulk greatly simplifies within the ``sudden
approximation''~\cite{P3-Randeria95,P3-Hedin02}. It
means that the photoemission process is supposed to occur {\em
suddenly}, with no post-collisional interaction between the
photoelectron and the system left behind~\cite{P3-Damascelli03}. In
particular, it is assumed that the excited state of the sample
(created by the ejection of the photo-electron) does not relax in
the time it takes for the photo-electron to reach the
detector~\cite{P3-Zhou07}. It can be shown that within the sudden
approximation using Fermi's Golden Rule for the transition rate, the
measured photo-current density is basically proportional to the
spectral function of the occupied electronic states in the solid,
i.e.: $J_{\bf k}\propto A_{\bf k}(E_{k})$. Eventually, and validated by
whether or not the spectra can be understood in terms of well
defined peaks representing poles in the spectral function, one may
connect $A_{\bf k}(E_{k})$ to the quasiparticle Green's function
$G(\textbf{k},E)=1/(E_{k}-\Sigma_{k}(E_{k})-\varepsilon_{k})$, with
$\Sigma_{k}(E_{k})$ defining the electronic self-energy and $\varepsilon_{k}$
the
bare band dispersion. In fact, customarily the inversion method for the
experimental data in ARPES is based upon the so called sudden approximation
trough the relation $A(\textbf{k},\omega)=-(1/\pi)$Im$
G(\textbf{k},\omega+i0^{+})$. Beyond the sudden approximation, one
would have to take into account the screening of the photoelectron
by the rest of the system, and the photoemission process could be
described by the generalized golden rule formula, i.e, a
three-particle correlation function~\cite{P3-Hedin02}. However, for our
purposes,
it is important mention the evidence that the sudden approximation
is justified for the cuprate superconductors even at low photon
energies~\cite{P3-Randeria95,P3-Koralek06}. In the end, the suitability of
the approximations invoked, will be justified by the agreement
between the theory and the experimental observations.

In the diagrammatic language, the above approach can be reduced to calculate the
quasiparticle self energy $\Sigma_{k}$ within the framework of the
Fermi-liquid theory, where electron-like quasiparticles populate bands in the
energy-momentum space up to the cut-off at the Fermi energy. In the case of
normal metals, this sophisticated description was firstly introduced by
Migdal~\cite{P3-Migdal58} who showed that the small parameter
$N(0)\theta_{D}$ allows to consider the higher order corrections negligible,
assuming that the
density of states $N(\varepsilon)$ is approximately a constant $N(0)$ over the
interval $(-\theta_{D},\theta_{D})$ around the Fermi level $\varepsilon_{F}$.
Here, $\theta_{D}$ is the so-called Debye energy. 

However, in the case of superconductors, the theory requires to incorporate the
Cooper-pairs condensation through a bosonic coupling function assuming that both
the electronic and the bosonic spectrum are possible to obtain from inelastic
neutron scattering (INS) measurements, X-Ray scattering (XRS) experiments,
tunneling experiments, or ab-initio calculations of the electronic band
structure. This complex picture can be
understood, in general terms, from the so-called Eliashberg
theory~\cite{P3-Eliashberg60}, and the works by Nambu~\cite{P3-Nambu60},
Schrieffer~\cite{P3-Schrieffer63}, and Morel and Anderson~\cite{P3-Morel62}. In
this scenario, the boson spectrum is directly associated to the lattice
vibration (phonons) as the binding  mechanism for the Cooper-pairs formation.
Nevertheless, this theory also has been often considered as the base of another
possible mechanisms with a magnetic
origin~\cite{P3-Kulic00,Muschler2010,P3-Hwang07,Schachinger06,P3-Carbotte05,
P3-Dordevic05,P3-Schachinger03,P3-Schachinger00,P3-Carbotte99,P3-Dahm09}. As
there is not any argument which allows to validate this assumption, onwards, we
will refer only to phonons as the boson coupling mechanism.

In order to support our ``phononic choice'' we want to recall that the absence
of the magnetic-resonance mode in LSCO ($La_{2-x}Sr_{x}CuO_{4}$)
\cite{P3-Zhou05}, Bi2212 ($Bi_{2}Sr_{2}CaCu_{2}O_{8+x}$) over-doped
(x=0.23) \cite{P3-Hwang04}, and its appearance only below $T_{c}$ in some
cuprates (e.g.,
Bi2201 \cite{P3-Sato03}) are not consistent with the universality of the kink
effect. Moreover, recent studies on electron doped systems
\cite{P3-Park08,P3-Graf07,P3-Wilson06,P3-Zhao07} have shown that the intensity
of the  magnetic resonance mode is seemingly weak in comparison with the phonon
mode to be considered as the cause of the strong electron energy dispersion
measured by ARPES. 

In the forthcoming paragraphs, and relying on a solid theoretical basis that is
introduced within the supplementary material section for the reader's sake, we
present our analysis of available experimental data.

\break
\vspace*{0.0cm}
\subsection*{\label{ch-7-1}
\hspace*{0.1cm} \textsl{7.1 Basic statements for the E-Ph
coupling}}
\vspace*{1cm}
\markboth{\hspace*{0.1cm}\textsl{7.1 Basic statements for the E-Ph
coupling}} 
{\hspace*{0.1cm} \textsl{7.1 Basic statements for the E-Ph
coupling}}
\addcontentsline{toc}{chapter}
{\hspace*{0.1cm} \textsl{7.1 Basic statements for the E-Ph
coupling}}

As a manifestation of the electron-phonon coupling interaction one
can introduce the mass renormalization of the electronic dispersion at the
energy scale associated with the phonons. This may be
technically\footnotemark[2]  defined from
the real part of the electron self energy as a
mass-enhancement parameter $\lambda^{*}$~\cite{P3-Grimvall81} given
by,
\begin{eqnarray}\label{Eq-7.1}
\lambda_{k}^{*}\equiv\left.-\partial_{\omega}\Sigma_{1}\right|_{\omega=0} \, .
\end{eqnarray}

\footnotetext[2]{For more details, we invite reader's to see the supplementary
material section at the end of this part.}

Although related (as it will be later clarified), $\lambda^{*}$ must not be
interpreted as
the strength of the E-Ph interaction which is
estimated from the so-called boson coupling parameter $\lambda$. This
dimensionless parameter is commonly defined in terms of the electron-phonon
spectral density as
\begin{equation}\label{Eq-7.2}
\lambda\equiv2\int_{0}^{\infty}d\nu\,\frac{\alpha^{2}F(\nu )}{\nu}\, ,
\end{equation}
and it is customarily related to the superconducting transition temperature. In
fact, as we have argued in Ref.~\cite{P3-Ruiz2009}, equality would just be
warranted at very low temperatures, and ensuring that the spectral
density measured from the experiments is fully satisfying the ME
approach for the Feynman diagram of lowest order for the E-Ph interaction 
[see Fig.~SIII-I (pag.~\pageref{Figure_SIII_1}), in the section of supplementary
material III].

\begin{figure}[t]
\centering
\includegraphics[width=1.0\textwidth]{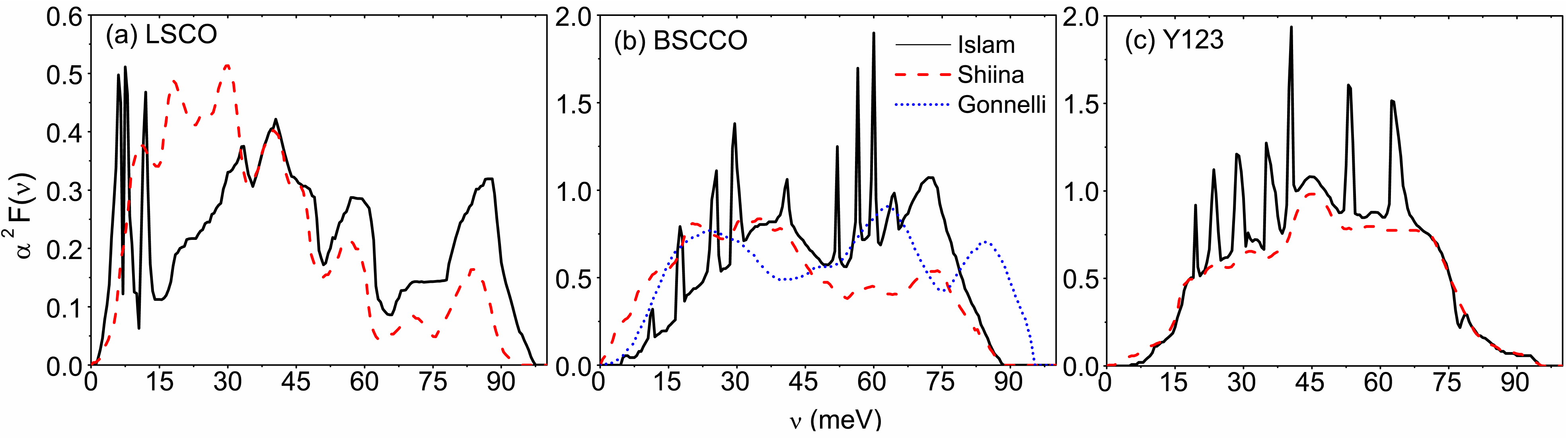}
\\
\caption{\label{Figure_7_1} 
E-Ph spectral density
$\alpha^{2}\textit{F}(\nu)$ for (a) $La_{2-x}Sr_{x}CuO_{4}$ (LSCO), (b)
$Bi_{2}Sr_{2}CaCu_{2}O_{8+x}$ (Bi2212), and (c) $YBa_{2}Cu_{3}O_{6+x}$ (Y123),
determined by different methods. The solid lines correspond to the method in
Ref.~\cite{P3-Islam00}, dashed lines to the method in Ref.~\cite{P3-Shiina90},
and the
dotted line in (b) corresponds to the method of Ref.~\cite{P3-Gonnelli98}.}
\end{figure}

Hence, taking into consideration the inherent existence of phonons in the
HTSC, we have evaluated the electron spectral densities for
$La_{2-x}Sr_{x}CuO_{4}$ (LSCO), $Bi_{2}Sr_{2}CaCu_{2}O_{8+x}$ (Bi2212), and
$YBa_{2}Cu_{3}O_{6+x}$ (Y123), in the flat model of Ref.~\cite{P3-Shiina90} as
well as by solving the isotropic
Eliashberg equations on the Matsubara frequencies. The
structure of these spectral densities is restricted to the isotropic nodal
direction under the assumption that $\alpha^{2}F(\nu)=G(\nu)\times C$, with C an
adjustable constant and $G(\nu)$ the generalized phonon density of states
extracted from the inelastic neutron scattering experiments~\cite{P3-Shiina90}.
Our results are shown in Fig.~\ref{Figure_7_1} (\textit{Shiina's} lines). In
addition, other reproducible methods to calculate the E-Ph spectral
density have been taken into consideration. Specifically, we mean the simple
method by Islam \& Islam~\cite{P3-Islam00} (\textit{Islam's} lines), and
the method by Gonnelli et al.~\cite{P3-Gonnelli98}, the last only referred for
the
Bi2212 family (\textit{Gonnelli's} line). The method of
Refs.~\cite{P3-Shiina90} \& \cite{P3-Islam00} is based on the INS experimental
data
by Renker \textit{et al.} \cite{P3-Renker87,P3-Renker88,P3-Renker89}, and the
method
in Ref.~\cite{P3-Gonnelli98} is based on the tunneling data reported in
Refs.~\cite{P3-Gonnelli97,P3-Ummarino97}.  

On the other hand, as it is shown in the supplementary material section, and
after some simple mathematical manipulations, the E-Ph self energy ($\Sigma$)
can be rewritten within the ME approach as a complex function of the form:
\begin{eqnarray}\label{Eq-7.3}
\Sigma(\textbf{k},i\omega)=\int_{0}^{\infty}d\nu\alpha^{2}F(\textbf{k},\nu)
\left\{-2\pi~i\left[N(\nu)+\frac{1}{2}\right]+
\Upsilon(\nu,\omega,T)
\right\}
\, ,
\end{eqnarray}
where the function $\Upsilon(\nu,\omega,T)$ is defined in terms of the so-called
digamma functions $\psi(z)$, 
\begin{eqnarray}\label{Eq-7.4}
\Upsilon(\nu,\omega,T)=\psi\left(\frac{1}{2}+i\frac{\nu-\omega}{2\pi
T}\right)-\psi\left(\frac{1}{2}-i\frac{\nu+\omega}{2\pi T}\right)
\, .
\end{eqnarray}

At this point, is useful recall that
the bare band energy $\varepsilon_{\textbf{k}}$
is related to the dressed band energy $E_{\textbf{k}}$ by
\begin{eqnarray}\label{Eq-7.5}
E_{\textbf{k}}=\varepsilon_{\textbf{k}}+Re\Sigma(E_{k}) \, .
\end{eqnarray}

Whereas the direct extraction of the self-energy from experiments appears to be
troublesome because the underlying band
structure of the bare electrons is \textit{a priori} unknown, a theoretical
determination of the bare band structure and
its relation to the full energy renormalization effects observed in the
experiments seems much more attractive. In this
sense, the nodal ARPES spectra are of great importance to check
the validity of the quasiparticle concept discussed above, and also for
understanding the nature of the involved interactions.

\break
\vspace*{0.0cm}
\subsection*{\label{ch-7-2}
\hspace*{0.1cm} \textsl{7.2 E-Ph model for the nodal
kink effect}}
\vspace*{1cm}
\markboth{\hspace*{0.1cm}\textsl{7.2 E-Ph model for the nodal
kink effect}} 
{\hspace*{0.1cm} \textsl{7.2 E-Ph model for the nodal
kink effect}}
\addcontentsline{toc}{chapter}
{\hspace*{0.1cm} \textsl{7.2 E-Ph model for the nodal
kink effect}}

Let me make some final remarks before moving onto the
application of the above ideas to the analysis of the nodal kink effect in the
ARPES data.
On the one hand, is to be mentioned, that the energy distribution- and momentum
distribution-curves (EDC - MDC) are the two most popular
ways for analyzing photoemission data. The dichotomy between the MDC- and
EDC-derived bands from the same data raises
critical questions about its origin and also about which one represents the
intrinsic band structure. In this sense, we
want to recall that at the larger bandwidth along the nodal direction, the MDC
method can be reliably used to extract
high quality data of dispersion in searching for fine structure. It has also
been shown theoretically that this approach
is reasonable in spite of the momentum-dependent coupling~\cite{P3-Devereaux04}.
In
a typical Fermi liquid picture, the MDC- and EDC-derived dispersions are
identical. Moreover, in an electron-boson coupling system the lower and higher
energy regions of the MDC- and EDC-derived dispersions are still
consistent, except right over the kink region \cite{P3-Zhang08}. On the other
hand, recalling that the bare electron band energy
$\varepsilon_k$ is not directly available from the experiments.
Instead, the electron momentum dispersion curve $E_{k}(k-k_{F})$ may
be measured. Below, we show that the good
agreement between our theoretical data and
the experimental facts supports this general picture.

As the kink effect is a common feature of the whole set of HTSC
families i.e., it is observable regardless of the
doping level or the temperature at which the measurement is performed, the only
possible scenario seems to be the coupling of quasiparticle with phonons. Thus,
we have noted that the relation between the dressed and bare energies is a
central property as related to the kink structure. The
key issue seems to be the existence of an energy scale (in the range of
40-80 meV). 

We have proposed an integral equation formalism based on the combination of
Eqs.~\ref{Eq-7.2}~--~\ref{Eq-7.5} from which, one
can consider that $\varepsilon_k$ implicitly depends on $E_k$
through the boson coupling parameter $\lambda$~\cite{P3-Ruiz2009}. In other
words, Eq.~\ref{Eq-7.2} acts as an integral constraint, through the definition
of $\lambda$. Thus, based on an interpolation
scheme between the numerical behavior of the dressed energy band and the
experimental data, one can introduce a universal dispersion relation
which allows to reproduce in a quite general scheme the nodal dispersions close
to the Fermi level~\cite{P3-Ruiz2009,P3-Ruiz2011_JS,P3-Ruiz2011_CAP}. As a
central result, we have encountered that the practical
totality of data are accurately reproduced by a universal linear dispersion
relation of the kind
\begin{eqnarray}\label{Eq-7.6}
{k}-{k}_{F}=\frac{\varepsilon_{k}}{v_{F_{<}}}\left(1-\delta\lambda\right)\, ,
\end{eqnarray}
with $v_{F_{<}}$ the Fermi velocity at low-energies, and $\varpi$ as the
(only) ``\textit{free}'' parameter required
for incorporating the specific renormalization for each
superconductor. From the physical point
of view, our empirical {\em ansatz}
[Eq.(\ref{Eq-7.6})] may interpreted as follows. Into the ME approach,
the involved quantities are not far from their values at the Fermi
level, and one can start with $\varepsilon_k$ replaced by
$E_{k}-\Sigma_{1}(E_{k})$, i.e.,
\begin{eqnarray}\label{Eq-7.7}
E_{k}-\Sigma_{1}(E_{k})=\left({k}-{k}_{F}\right)v_{F_{<}}
\left(1-\delta\lambda\right)^{-1} \, .
\end{eqnarray}
Then, taking derivatives with respect to $E_{k}$
and evaluating them for $E_{k}\to 0$, one gets
\begin{eqnarray}\label{Eq-7.8}
1+\lambda^{*}=\left.\frac{\partial (k-k_{F})}{\partial
E_{k}}\right|_{E_{k}=0} v_{F<}(1-\delta\lambda)^{-1}\, ,
\end{eqnarray}
and recalling that $v_{F<}$ is obtained as the slope of the
lower part of the momentum dispersion curve, this
equation leads to $1+\lambda^{*}=(1-\delta\lambda)^{-1}$. Thus, a
physical interpretation of the fit parameter $\delta$ is straightforwardly
obtained from the analytical relation
\begin{eqnarray}\label{Eq-7.9}
\delta =\frac{\lambda^{*}}{\lambda(1+\lambda^{*})} \, .
\end{eqnarray}

To the lowest order, the dimensionless parameter $\delta$ is basically the
ratio between the defined mass-enhancement and phonon-coupling
parameters $\delta\approx\lambda^{*}/\lambda$. Outstandingly, this fact
reassembles the differences obtained by
tight-binding Hamiltonian models \cite{P3-Weber87,P3-Weber88} and the
``density-functional'' band theories \cite{P3-Giustino08,P3-Heid08} appealing to
the differences between $\lambda$ and $\lambda^{*}$. To understand
these differences, recall that, in principle, the density-functional theory
\cite{P3-Kohn66} gives a correct ground-state energy, but the bands do
not necessarily fit the quasi-particle band structure used to
describe low-lying excitations. 

\break
\vspace*{0.0cm}
\subsection*{\label{ch-7-3}
\hspace*{0.1cm} \textsl{7.3 Numerical procedure and
results}}
\vspace*{1cm}
\markboth{\hspace*{0.1cm}\textsl{7.3 Numerical procedure and
results}} 
{\hspace*{0.1cm} \textsl{7.3 Numerical procedure and
results}}
\addcontentsline{toc}{chapter}
{\hspace*{0.1cm} \textsl{7.3 Numerical procedure and
results}}

In order to reproduce the experimental results, our numerical program is as
follows: 

\begin{enumerate}
 \item $v_{F_{<}}$ is determined from the momentum dispersion curve in the
ARPES measurements [$\sim $ 1.4 eV$\cdot$\AA{} - 2.2 eV$\cdot$\AA{}],

\item To determine $\delta$, we calculate the so called logarithmic frequency
$\omega_{\rm log}$ as introduced by Allen and Dynes~\cite{P3-Allen75},
\begin{eqnarray}\label{Eq-7.10}
\omega_{\rm
log}\equiv
exp\left\{\frac{2}{\lambda}\int_{0}^{\infty}\ln(\nu)[{\alpha^{2}F(\nu)}/{
\nu }]
d\nu\right\} \, .
\end{eqnarray}

This constant frequency (properly defining the corresponding spectral densities
$\alpha^{2}F(\nu)$ for each $\lambda$) is introduced only for
scaling purposes $(\delta=\varpi/\omega_{\rm log})$. Thus, we achieve a
reduction in the scattering of numerical values when dealing with different
samples.
In this sense, $\varpi$ is a mere mathematical instrument. For the spectral
densities shown in Fig.~\ref{Figure_7_1} we get $\omega_{\rm
log}^{LSCO}\simeq
16.1455\, meV$, $\omega_{\rm
log}^{Y123}\simeq 35.5900\, meV$ and $\omega_{\rm
log}^{Bi2212}\simeq 33.8984\, meV$, respectively.

\item $\varepsilon_{k}$ is numerically determined from the relation
$E_{k}=\varepsilon_{k}+Re\{\Sigma(E_{k})\}$, where the digamma functions are
simply subroutines of our code.

\item Finally, correlation between theory and experiment is established by the
application of Eq.~(\ref{Eq-7.6}). The best fit with the experimental
curves has been obtained for $\delta=0.185$ in LSCO, $\delta=0.354$ in Bi2212,
and $delta=0.365$ in Y123 families, respectively.
\end{enumerate}

\begin{figure}[t]
\centering
\includegraphics[height=11.2cm,width=13cm]{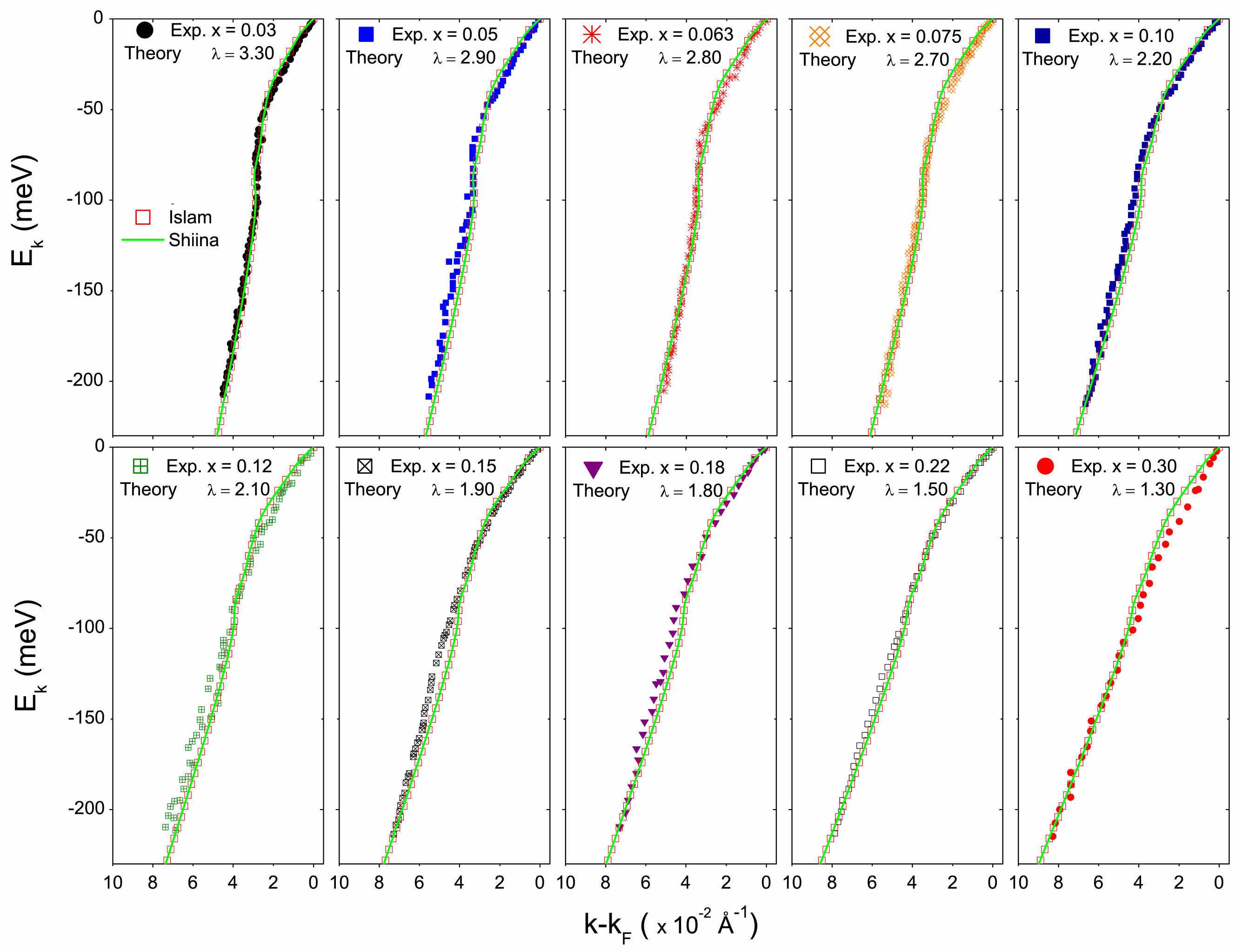}
\caption{\label{Figure_7_2} The renormalized electron quasiparticle energy
dispersion $E_{k}$ as a function of the momentum
$k-k_{F}$ for several samples of
$\texttt{La}_{2-x}\texttt{Sr}_{x}\texttt{CuO}_{4}$, measured along the
(0,0)-($\pi$,$\pi$) nodal direction at a temperature of 20 K. The doping level
$x$ ranges between 0.03 (top left) up to
0.30 (bottom right). The experimental data (symbols) are taken from
Ref.~\cite{P3-Zhou03} and the theoretical curves have
been obtained using the spectral densities of Islam \& Islam
[Ref.~\cite{P3-Islam00}] (open squares), and  Shiina \&
Nakamura (solid lines)\cite{P3-Shiina90}.}
\end{figure}

In Fig.~\ref{Figure_7_2} we show the results found in
LSCO covering the doping range ($0<x\leq0.3$).
Remarkably, within this range, the hole concentration in the $CuO_{2}$ 
plane is well controlled by the $Sr$ content and a small oxygen
non-stoichiometry, where the physical properties span over the
insulating, superconducting, and overdoped non-superconducting metal
behavior. Superconducting transition temperatures $T_{c}$
in the interval of 30-40K have been observed in
Refs.~\cite{P3-Bednorz86,P3-Uchida87,P3-Dietrich87}.
For the application of Eq.~(\ref{Eq-7.6}), we have considered
$v_{F_{<}}=2 eV\cdot$\AA{} as related to the experimental results of
Refs.~\cite{P3-Lanzara01,P3-Zhou03,P3-Zhou07,P3-Zhou05}. The best fit of the
whole set of
experimental data has been
obtained for $\delta=0.185$, and the derived $\lambda^{*}$
values are shown in Table~\ref{Table_7_1}. For comparison,
recall that values of ``$\lambda=2-2.5$'' in the range $0.2>x>0.1$
were reported in the Ref.~\cite{P3-Weber87} by Weber. In that
case, these values were obtained within the framework of the
nonorthogonal-tight-binding theory of lattice dynamics, based on the
energy band results of Mattheiss~\cite{P3-Mattheiss87}. It must be emphasized
that
in both cases (Ref.~\cite{P3-Weber87} \&
Refs~\cite{P3-Ruiz2009,P3-Ruiz2011_JS}), $\lambda$ was obtained in agreement
with
the observed $T_c$ values in LSCO. Moderate discrepancies between
the predictions of our model and the analysis of
Refs.~\cite{P3-Weber87} \& \cite{P3-Weber88} may be ascribed
to some uncertainty in the experimental spectral densities. 

\begin{figure}[t]
\centering
\includegraphics[width=0.51\textwidth]{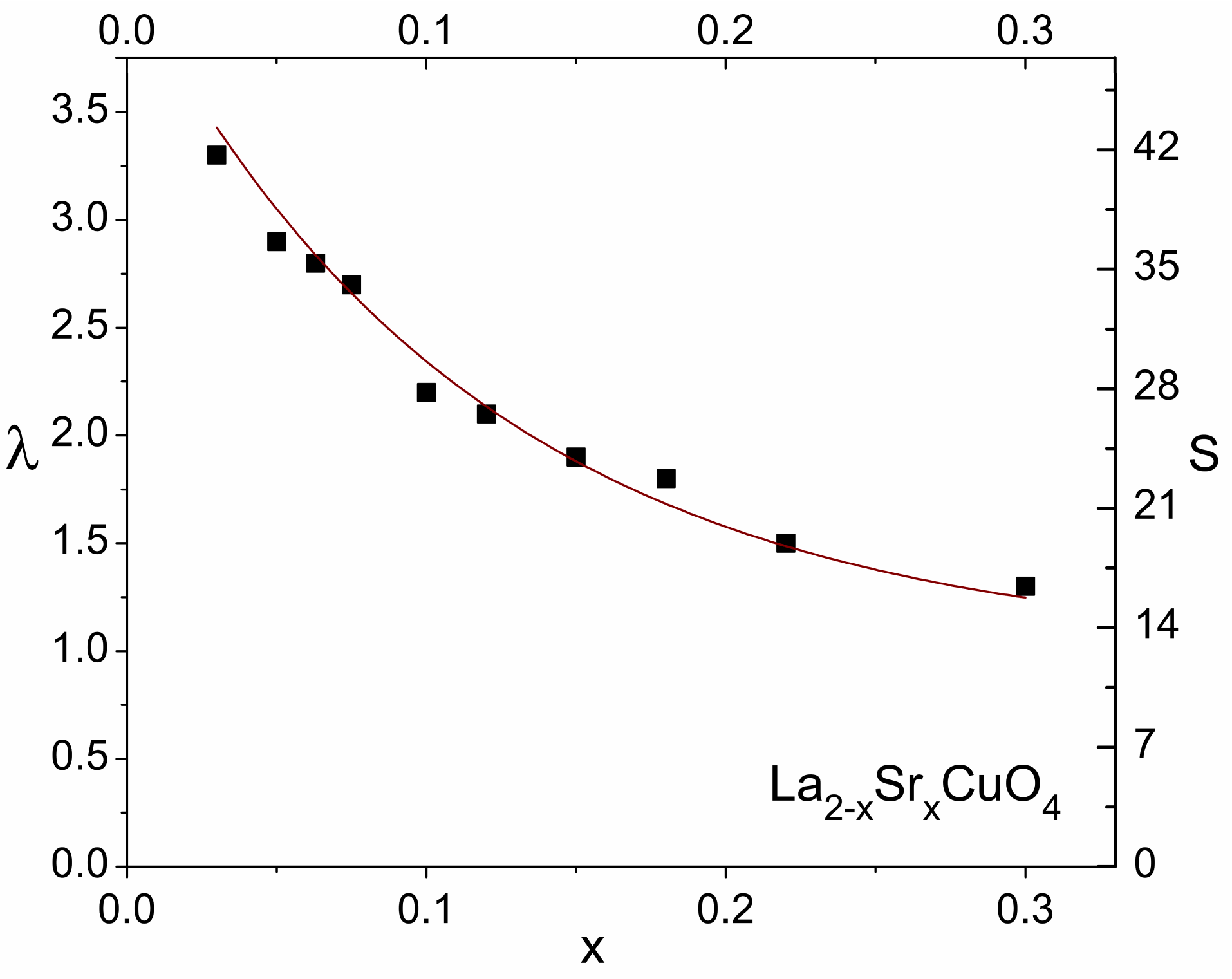}
\\
\caption{\label{Figure_7_3} Evolution of the E-Ph coupling parameter $\lambda$
as a function of the dopant content in LSCO. The
$\lambda$-values have been obtained from our best fit with the nodal kink
dispersion (black squares) shown in Fig.~\ref{Figure_7_2}.
Correspondingly, the evolution of the area $S$ as a function of the dopant
content is also shown (right scale).}
\end{figure}

On the other hand, we have taken advantage of the widespread availability of
experimental information in LSCO~\cite{P3-Lanzara01} and our numerical method,
to
describe the evolution of the E-Ph coupling parameter $\lambda$ as a function of
the doping level. It can be fitted to the simple expression
\begin{eqnarray}\label{Eq-7.11}
\lambda=2\tilde{\omega}exp(-\frac{\tilde{\omega}}{\delta} x)+1 \, ,
\end{eqnarray}
within a precision factor  of $\sim95$\% (Fig.~\ref{Figure_7_3}). Here, 
$\tilde{\omega}$ defines the ratio between the phonon characteristic energies
introduced by McMillan \cite{P3-McMillan68},
\begin{eqnarray}\label{Eq-7.12}
\omega_{\rm 1}=(2/\lambda)\int_{0}^{\infty}\alpha^{2}F(\nu)d\nu\equiv
(2/\lambda)S \, , 
\end{eqnarray}
and the logarithmic frequency $\omega_{\rm log}$ [Eq.~(\ref{Eq-7.10})]. In this
case, we get
$\omega_{\rm log}^{LSCO}\simeq 16.1455\, meV$ and
$\omega_{1}^{LSCO}\simeq 25.2627\, meV$.

\begin{figure}[t]
\centering
\includegraphics[width=0.8\textwidth]{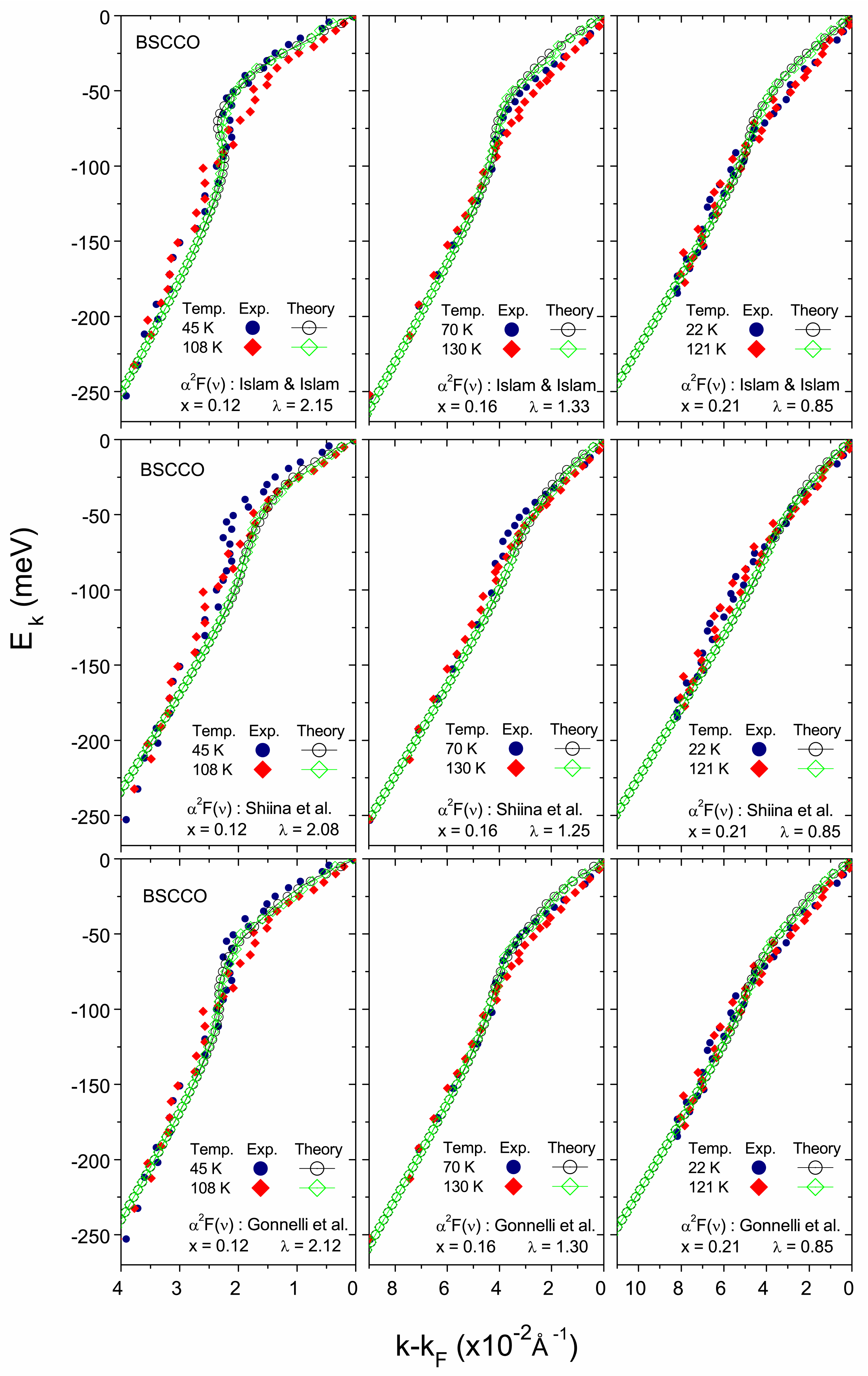}
\\
\caption{\label{Figure_7_4} Same as in Fig.~\ref{Figure_7_2} but in samples of
Bi2212. The experimental data for the normal state (full diamond) and
superconducting state (full circle) both have
been taken from the  Ref.~\cite{P3-Johnson01}. The theoretical curves (lines
-diamond, or -circles) have been obtained from
our interpolation method [Eq.~(\ref{Eq-7.6})] according to the best fit values
for the E-Ph coupling parameter $\lambda$, and
the spectral densities of Islam \& Islam~\cite{P3-Islam00} (top), Shiina \&
Nakamura~\cite{P3-Shiina90}
(middle), and Gonnelli et al.~\cite{P3-Gonnelli97} (bottom). The different
plots correspond to the doping levels of
Bi2212: under-doped ``x=0.12'' (left), optimally doped ``x=0.16'' (center), and
over-doped ``x=0.21'' (right)
respectively.}
\end{figure}

\begin{figure}[t]
\centering
\includegraphics[width=0.6\textwidth]{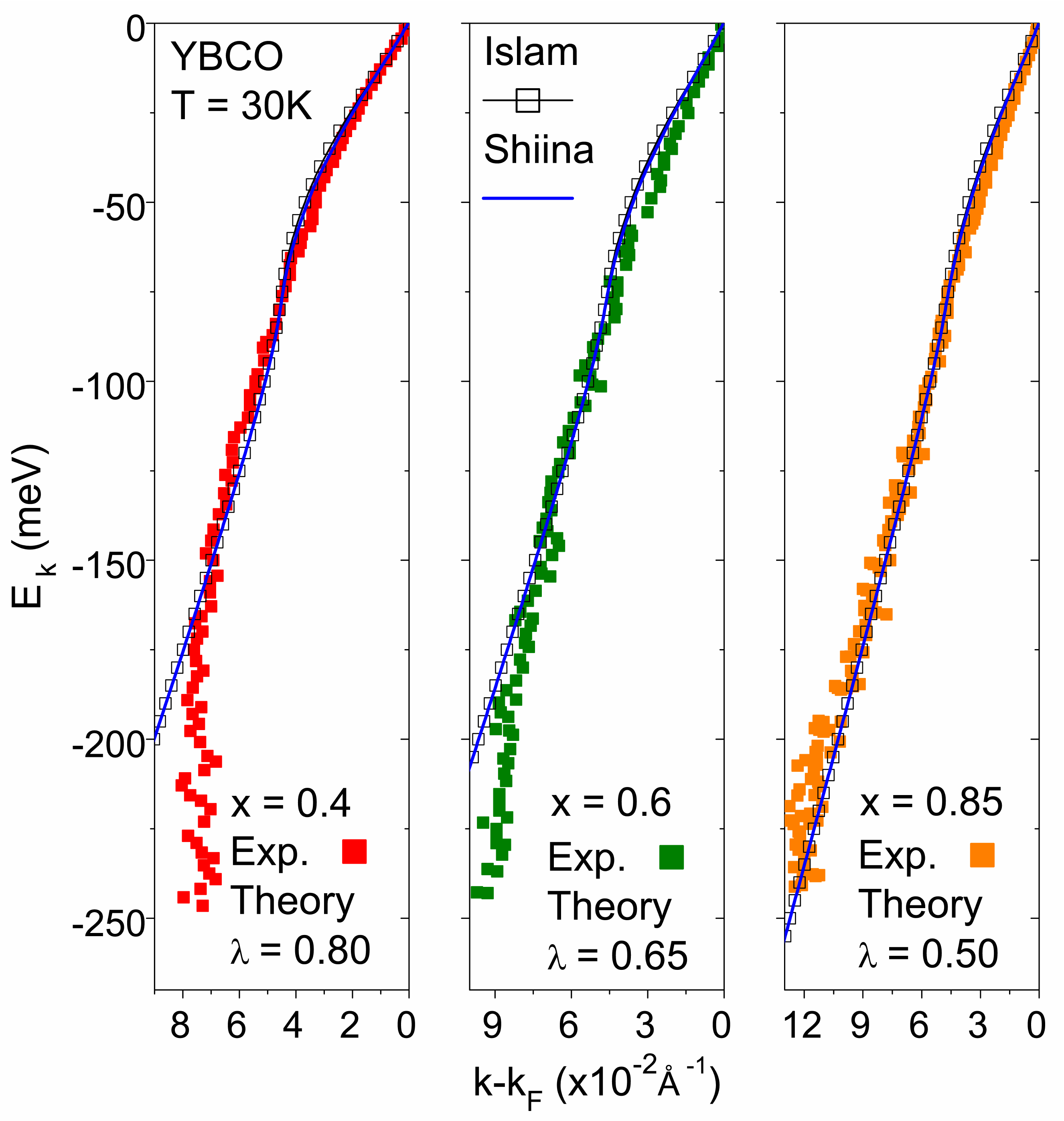}
\\
\caption{\label{Figure_7_5} Same as in Fig.~\ref{Figure_7_2} but in
samples of Y123 with (left to right): $x=0.4$
(underdoped), $x=0.6$ (underdoped), and $x=0.85$ (overdoped). The
solid squares correspond to the experimental data of
Ref.~\cite{P3-Borisenko06}. The theoretical curves have been obtained using the
spectral densities of Islam \& Islam [Ref.~\cite{P3-Islam00}] (open squares),
and
Shiina \& Nakamura [Ref.~\cite{P3-Shiina90}]
(solid line). All curves have been obtained at 30K.
}
\end{figure}

%
\begin{table}[t]
\caption{\label{Table_7_1} Values of the E-Ph coupling parameter $\lambda$ and
the corresponding mass-enhancement parameter $\lambda^{*}$ obtained from the
analysis of ARPES data at several doping levels ($x$) of
$La_{2-x}Sr_{x}CuO_{4}$, $Bi_{2}Sr_{2}CaCu_{2}O_{8+x}$. $\lambda^{*}$ has been
obtained by
means of Eq.~(\ref{Eq-7.6}) to the lowest order
$\lambda^{*}\approx\delta\lambda$. The predicted superconducting transition
temperatures $T_{c}$ are also shown. Our results are presented in contrast with
other models available in the literature.}\centering
\begin{tabular}{ccccc}
\hline\hline\rowcolor[gray]{.9}
$La_{2-x}Sr_{x}CuO_{4}$ & (LSCO). & & & $\delta=0.85$\\
\hline
$x$&Ref.&$\lambda$&$\lambda^{*}$&$T_{c}(\lambda)$  [K]\\
\hline
 0.03  & \footnotemark[1] & 3.30 & 0.61 & -- \\
 0.05  & \footnotemark[1] & 2.90 & 0.54 & -- \\
 0.063 & \footnotemark[1] & 2.80 & 0.52 & -- \\
 0.075 & \footnotemark[1] & 2.70 & 0.50 & -- \\
 0.10  & \footnotemark[1] & 2.20 & 0.41 & 42.10 \\
 0.12  & \footnotemark[1] & 2.10 & 0.39 & 40.77 \\
 0.15  & \footnotemark[1] & 1.90 & 0.35 & 37.83 \\
 0.18  & \footnotemark[1] & 1.80 & 0.33 & 36.19 \\
 0.22  & \footnotemark[1] & 1.50 & 0.28 & 30.47 \\
 0.30  & \footnotemark[1] & 1.30 & 0.24 & -- \\
 0.1-0.2 & \cite{P3-Weber87}\footnotemark[2]& 2-2.5 & -- & 30--40\\
 --      & \cite{P3-Shiina90}\footnotemark[2] & 1.78 & -- & 40.6\\
 0.15    & \cite{P3-Giustino08}\footnotemark[3] & 1--1.32 & 0.14--0.22 &-- \\
 0.22  & \cite{P3-Giustino08}\footnotemark[3] & 0.75--0.99 & 0.14--0.20 & -- \\
\hline\hline
\end{tabular}
\end{table}
\footnotetext[1]{This book and Ref.~\cite{P3-Ruiz2009}. The results corresponds
to
Figs.~\ref{Figure_7_2},~\ref{Figure_7_4}, \&~\ref{Figure_7_5}, respectively.
Here, we allow a margin of error in $\lambda$ of $\sim
\pm 0.3$ as related to the numerical interpolation procedure between
theory and experiment.}
\footnotetext[2]{The $\lambda$ values reported in that reference
were obtained so as to fit $T_{c}$ at the indicated values.}
\footnotetext[3]{In Ref.~\cite{P3-Giustino08} the electronic
structure of LSCO has been calculated employing a generalized
gradient approximation to the density functional theory and used to
determine $\lambda$.}
%

%

\begin{table}
\caption{\label{Table_7_2} Same as Table~\ref{Table_7_1} but for
$Bi_{2}Sr_{2}CaCu_{2}O_{8+x}$.}\centering
\begin{tabular}{ccccc}
\hline\hline\rowcolor[gray]{.9}
$Bi_{2}Sr_{2}CaCu_{2}O_{8+x}$ & (Bi2212). & & & $\delta=0.354$\\
\hline
$x$&Ref.&$\lambda$&$\lambda^{*}$&$T_{c}(\lambda)$  [K]\\
\hline
0.12 & \footnotemark[1] & 2.15 & 0.76 & 64.81 \\
0.16 & \footnotemark[1] & 1.33 & 0.47 & 42.45 \\
0.21 & \footnotemark[1] & 0.85 & 0.30 & 19.93 \\
--  & \cite{P3-Shiina90}\footnotemark[2] & 3.28 & -- & 85 \\
--  & \footnotemark[1] & 3.28 & 1.16 & 81.66 \\
--  & \cite{P3-Gonnelli98}\footnotemark[2] & 3.34 & 1.05 & 93 \\
--  & \footnotemark[1] & 3.34 & 1.18 & 82.40 \\
0.16 &\cite{P3-Kordyuk06}\footnotemark[4] & $\sim 1.28$ & $\sim0.43$ &--\\
\hline\hline
\end{tabular}
\end{table}
\pagebreak
\footnotetext[1]{\textit{Ibid.,} Table~\ref{Table_7_1}.}
\footnotetext[2]{\textit{Ibid.,} Table~\ref{Table_7_1}.}
\footnotetext[4]{In Ref.~\cite{P3-Kordyuk06} two channels are
defined for $\lambda$. $\lambda_{1}=0.43\pm0.02$ corresponds to the
``\emph{primary}'' channel (close to Fermi level) and is free from
normalization effects. $\lambda_{2}$ is obtained from the
Kramers-Kronig transformation and the experiment.\cite{P3-Kordyuk05} In
this sense, and within the notation employed in the current work, we
obtain, $\lambda^{*}=\lambda_{1}$, and $\lambda\simeq0.85+0.43$.}

We want to clarify that, although Eq.~(\ref{Eq-7.11}) can be considered as a
useful relation between the physical and chemical properties of LSCO, the
shaping of other HTSC families by similar expressions
cannot be guaranteed.

In Table~\ref{Table_7_2} and Fig.~\ref{Figure_7_4}, we display the results
found for the available
experimental data in Bi2212 samples. In this case, the
theoretical curves for $E_{k}(k-k_{F})$ have been predicted by our interpolation
method for the under-doped (UD70)
``$x=0.12$, $T_{c}\approx 70K$'', optimally doped (0PD90) ``$x=0.16$,
$T_{c}\approx 90K$'', and over-doped (OVD58)
``$x=0.21$, $T_{c}\approx 58K$'' samples, having use of the spectral densities
of Fig.~\ref{Figure_7_1}. The experimental data were taken from
the work by Johnson \emph{et al.}~\cite{P3-Johnson01} with
$v_{F_{<}}=1.6eV\cdot${\AA} as a value consistent with the experimental results
of Refs.~\cite{P3-Lanzara01,P3-Zhou02,P3-Kordyuk06,P3-Johnson01}. The
best fit with experimental data has been found for $\delta=0.354$. We
show that, regardless the method used for obtaining the E-Ph spectral density, 
the same conclusions can be achieved. 

\begin{table}
\caption{\label{Table_7_3} Same as Table~\ref{Table_7_1} but for
$YBa_{2}Cu_{3}O_{6+x}$.}\centering
\begin{tabular}{ccccc}
\hline\hline\rowcolor[gray]{.9}
$YBa_{2}Cu_{3}O_{6+x}$ & (Y123). & & & $\delta=0.365$\\
\hline
$x$&Ref.&$\lambda$&$\lambda^{*}$&$T_{c}(\lambda)$  [K]\\
\hline
0.4  & \footnotemark[1] & 0.80 & 0.29 & 17.51 \\
0.6  & \footnotemark[1] & 0.65 & 0.24 & 9.82 \\
0.85 & \footnotemark[1] & 0.50 & 0.18 & 3.39 \\
--  & \cite{P3-Shiina90}\footnotemark[2] & $\sim$3.4 & -- & 91 \\
--  & \footnotemark[1] & 3.45 & 1.26 & 84.19 \\
-- & \cite{P3-Weber88}\footnotemark[2] & $\sim$0.5 & -- & $\sim$3 \\
-- & \cite{P3-Weber88}\footnotemark[2] & $\sim$1.3 & -- & $\sim$30 \\
-- & \footnotemark[1] & 1.30       & 0.47 & 36.43 \\
-- & \cite{P3-Heid08}\footnotemark[5]  & -- & 0.18 - 0.22 & -- \\
-- & \footnotemark[1] & 0.49 - 0.60 & 0.18 - 0.22 &$\sim$3.0--6.6\\
\hline\hline
\end{tabular}
\end{table}
%
\footnotetext[1]{\textit{Ibid.,} Table~\ref{Table_7_1}.}
\footnotetext[2]{\textit{Ibid.,} Table~\ref{Table_7_1}.}
\footnotetext[5]{In Ref.~\cite{P3-Heid08} the parameter
``$\lambda$'' has been obtained from the spectral density
$\alpha^{2}F(\textbf{k},\nu)$ employing the local density
approximation to density functional theory (see text). This value
corresponds to the mass-enhancement parameter $\lambda^{*}$}

Finally, in Table~\ref{Table_7_3} and Fig.~\ref{Figure_7_5} we show the
results found for Y123 with the
following dopant levels: Under-Doped 35
``$x=0.4$, $T_{c}\approx 35K$'' (UD35), Under-Doped 61
``$x=0.6$, $T_{c}\approx 61K$'' (UD61), and Over-Doped 90 ``$x=0.85$,
$T_{c}\approx
90K$'' (OVD90). The experimental data were taken
from the work by Borisenko \emph{et al.}~\cite{P3-Borisenko06}. To our
knowledge,
no more
experimental evidence of kinks in the nodal direction is available
for this material. The value $v_{F_{<}}=1.63eV \cdot$\AA{} has been used
for consistency with the experimental results reported by those authors.
The best fitting between the experimental data and our model has been
found with the value $\delta=0.365$. It must be noted that the appearance of a
second kink in the underdoped case may
not be allocated with the simple assumptions of our model. Nevertheless,
although this effect could be explained either by introducing high perturbation
orders
beyond the ME approach or introducing a renormalization factor for the
electronic band structure, the range of energies for the E-Ph spectral density
does not seems provide a physically admissible explanation.

%


\chapter{\label{ch-8} \sc \textbf{Is it necessary to go beyond the E-Ph
mode?}}

According to the previous chapter, the results for optimally doped Bi2212
(OPD90, $x=0.16$) and over doped Bi2212 (OVD85, $x=0.21$) samples
(Fig.~\ref{Figure_7_4}), with measured
$T_{c}=90K$ and $85K$ respectively, have revealed that the influence of
the E-Ph coupling mechanism is seemingly weak in spite of the kink effect is
reproduced for the whole energy spectrum. On the other hand, the discrepancies
become even larger when the thermodynamic properties for Y123 samples are
analyzed. 

\vspace*{0.5cm} \textleaf \hspace*{0.5cm}\textit{To address the
question:}\vspace*{0.5cm}

\begin{figure}[t]
\centering
\includegraphics[height=11.5cm,width=11cm]{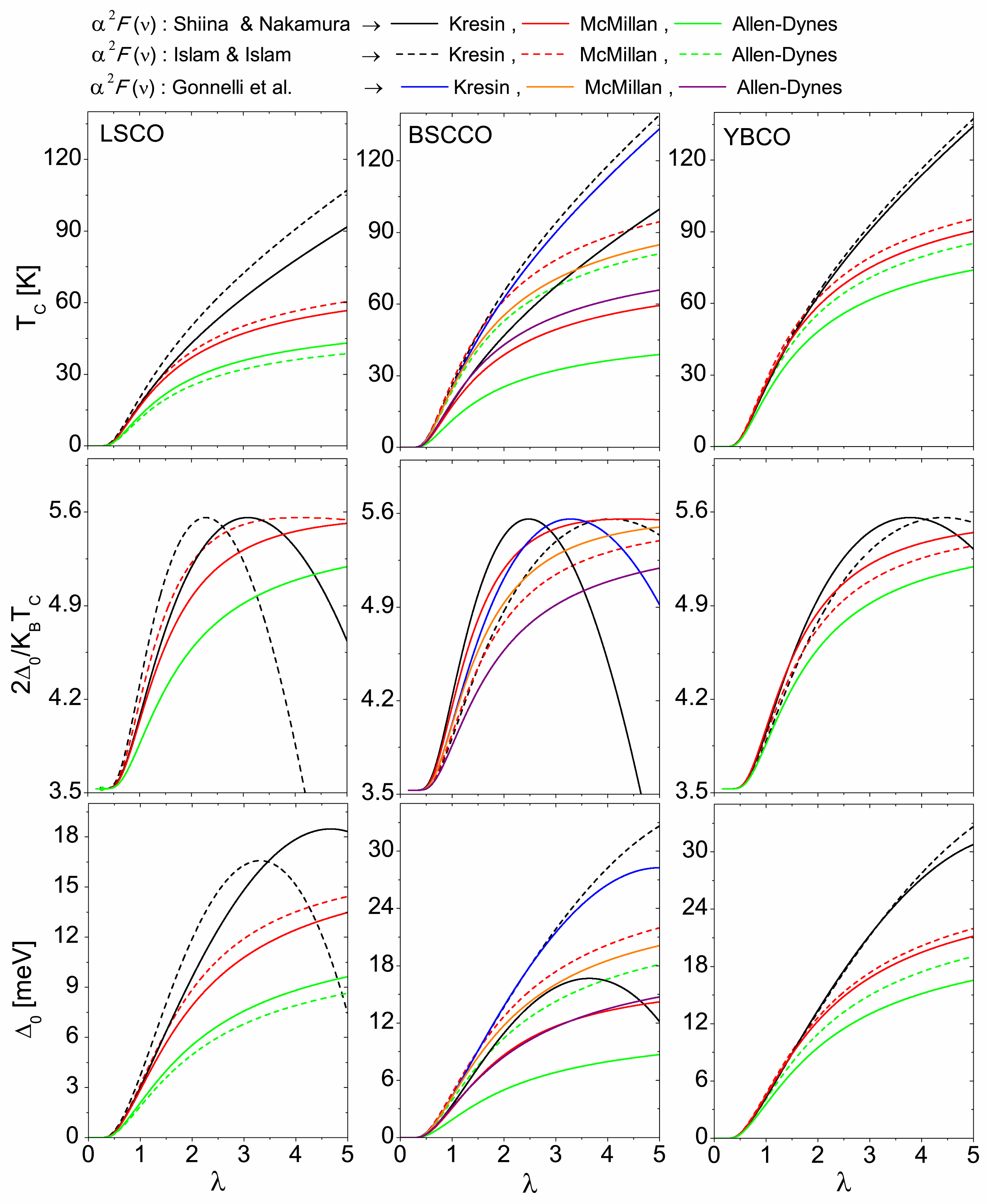}
\\
\caption{\label{Figure_8_1} Plots of the
critical temperatures $T_{c}$ (top), the ratio gap $2\Delta_{0}/k_{B}T_{c}$
(middle) and the gap $\Delta_{0}$ (bottom)
for LSCO, Bi2212 and Y123,  all of them represented as functions of the
E-Ph coupling parameter $\lambda$ obtained from the E-Ph spectral densities
shown in Figure~\ref{Figure_8_1}. We have used three different approaches: the
McMillan's formula~\cite{P3-McMillan68}, the
Allen-Dynes formula~\cite{P3-Allen75}, and the Kresin's
formula~\cite{P3-Kresin87}.
}
\end{figure}

In order to answer Is it necessary to go beyond the E-Ph coupling mode for the
superconducting pairs formation?, we can \textit{a priori} consider that some
of the E-Ph spectral densities of Fig.~\ref{Figure_7_1} will allow explain the
high $T_{c}$ values and the zero temperature gap $\Delta_{0}$ reported in the
literature. From this point of view, several approaches can be argued for each
one of the materials. For example, in Fig.~\ref{Figure_8_1} we show our results
for $T_{c}$, the ratio gap $2\Delta_{0}/k_{B}T_{c}$ and the zero temperature gap
$\Delta_{0}$ for the different HTSC families analyzed in the above chapter. The
different curves have been obtained from the point of view of three different
approaches
%
%

\begin{enumerate}
 \item The celebrated McMillan's equation~\cite{P3-McMillan68},
\begin{eqnarray}\label{Eq-8.1}
T_{c}=\frac{\omega_{1}}{1.2}exp\left[-1.04\frac{1+\lambda}{\lambda-\mu^{*}
(1+0.62\lambda)}\right] \, ,
\end{eqnarray}
with $\mu^{*}$ the so-called Coulomb pseudopotential.

 \item The refined formula by Allen and Dynes~\cite{P3-Allen75}, which is
obtained by replacing $\omega_{1}$ [Eq.~(\ref{Eq-7.12})] in Eq.~(\ref{Eq-8.1})
by the
so called logarithmic frequency $\omega_{log}$ [see
Eq.~(\ref{Eq-7.10})].

For the HTSC families considered along this section, and the different spectral
densities of Fig~\ref{Figure_8_1}, we get $\omega_{\rm log}^{LSCO}\simeq 16\,
meV$, $\omega_{1}^{LSCO}\simeq 25\, meV$, $\omega_{\rm log}^{Bi2212}\simeq 33\,
meV$, $\omega_{1}^{Bi2212}\simeq 39\, meV$, $\omega_{\rm log}^{Y123}\simeq 35\,
meV$, and $\omega_{1}^{Y123}\simeq 39\, meV$.

\item Finally, the less conventional Kresin's formula \cite{P3-Kresin87},
%
%
\begin{eqnarray}
T_{c}=0.25\; \varpi \; exp\left(\frac{2}{\lambda_{eff}}-1\right)^{-1/2}
\; ,\label{Eq-8.2}
\end{eqnarray}
where $\varpi=[(2/\lambda)\int_{0}^{\infty}\nu\alpha^{2}F(\nu)d\nu]^{1/2}$ and
$\lambda_{eff}=(\lambda-\mu^{*})[1+2\mu^{*} + (3/2) \lambda \mu^{*}
exp(-0.28\lambda)]$.
\end{enumerate}

In all calculations, the
Coulomb's pseudopotencial was given by a typical value,
$\mu^{*}=0.13$~\cite{P3-Allen82}. We have to be mention that is essential to be
aware that there is no a small parameter that enables a satisfactory
perturbation theory to be
constructed for the Coulomb interaction between electrons. Thus, Coulomb
contributions to the electron self energy $\Sigma$ introduced in
chapter~\ref{ch-7} cannot be reliably calculated~\cite{P3-Allen75}. Fortunately
this is not a serious problem in superconductivity because a reasonable
assumption is to consider that the large Coulomb effects for the normal state
are already included in the electronic bare band structure $\varepsilon_{k}$.
The remaining off-diagonal terms of the superconducting components of the
Coulomb self-energy turn out to have only a small effect on superconductivity,
which is treated phenomenologically~\cite{P3-Allen82}. 

Is to be noticed that the high values of the critical temperatures strongly
depend on the approximation invoked, and (in some cases) on the inversion
method for the boson coupling spectral density. Thus, we argue that the critical
temperature $T_{c}$ should not be considered as a fit parameter for
adjusting the theory, i.e.: one should not predict $\lambda$ from the
approximate $T_{c}$ formulas and then use it for calculating the electron
self-energy. As a proof of this, we recall that although attractive,
this idea has led to underestimates the
phonon contribution to the photoemission kink in HTSC
\cite{P3-Giustino08,P3-Heid08}.

To our knowledge, the most suitable way for determining the influence of an
interaction mechanism in the pair formation for HTSC could be (i) evaluate
the strength of the boson coupling mode from the electron renormalization
effects and then (ii) solve the Eliashberg equations for the superconducting
$T_{c}$, or have use of semiempirical approaches as the introduced before. From
such analysis, we conclude that the consideration
of the E-Ph interaction in LSCO strongly suggests that
the high $T_{c}$ values can be caused by the conventional
coupling to phonons (see table~\ref{Table_7_1}). However, for the Bi2212 and
Y123 samples the obtained results are not so
encouraging as they are for the LSCO family
(see tables~\ref{Table_7_2}~\&~\ref{Table_7_3}). 

On the one hand, we have noticed that the obtained
critical temperatures for Bi2212-UD70 in the framework of the spectral densities
of Islam \& Islam [$T_{c}^{Kresin}(\lambda=2.15)=70.0 K$ and
$T_{c}^{McMillan}(\lambda=2.15)=64.81 K$], and Gonnelli et al.
[$T_{c}^{Kresin}(\lambda=2.15)=67.06 K$ and
$T_{c}^{McMillan}(\lambda=2.15)=58.19 K$], are in some sense, good estimations
for the experimental values of $T_{c}$. However, from the method by Shiina \&
Nakamura~\cite{P3-Shiina90}, and the framework of the Allen
\& Dynes formula~\cite{P3-Allen75}, a strong reduction of $T_{c}$ ($\sim 40\%$)
can be found. This can be interpreted as a first signal about the need
of considering additional perturbation mechanisms into the matrix
elements of the Eliashberg equations, or perhaps, and in an optimistic way for
the phonon hypothesis, this fact could be revealing that the flat model used in
Ref.~\cite{P3-Shiina90} is not consistent with the experimental facts of this
kind of material.
 
On the other hand, regarding the Bi2212-OPD90 and Bi2212-OVD85 samples,
independently of the inversion method used for the E-Ph spectral density
(Fig.~\ref{Figure_7_1}), the maximal values for
$\lambda$ which are able to reproduce the kink structure ($\lambda \simeq 1.3$
and $\lambda \simeq 0.85$ respectively), both underestimate the experimental
critical temperatures in about $50\%$. The disagreement can be even higher
($\sim80\%$) if we consider the results for Y123 samples
(see Fig.~\ref{Figure_7_5} \& Table~\ref{Table_7_3}). However, before moving on
thinking
in the necessity of additional boson coupling mechanisms, is necessary to
revalidate
the influence of the Coulomb effects along the framework of the different
invoked approaches.

\begin{figure}[t]
\centering
\includegraphics[height=11cm,width=13cm]{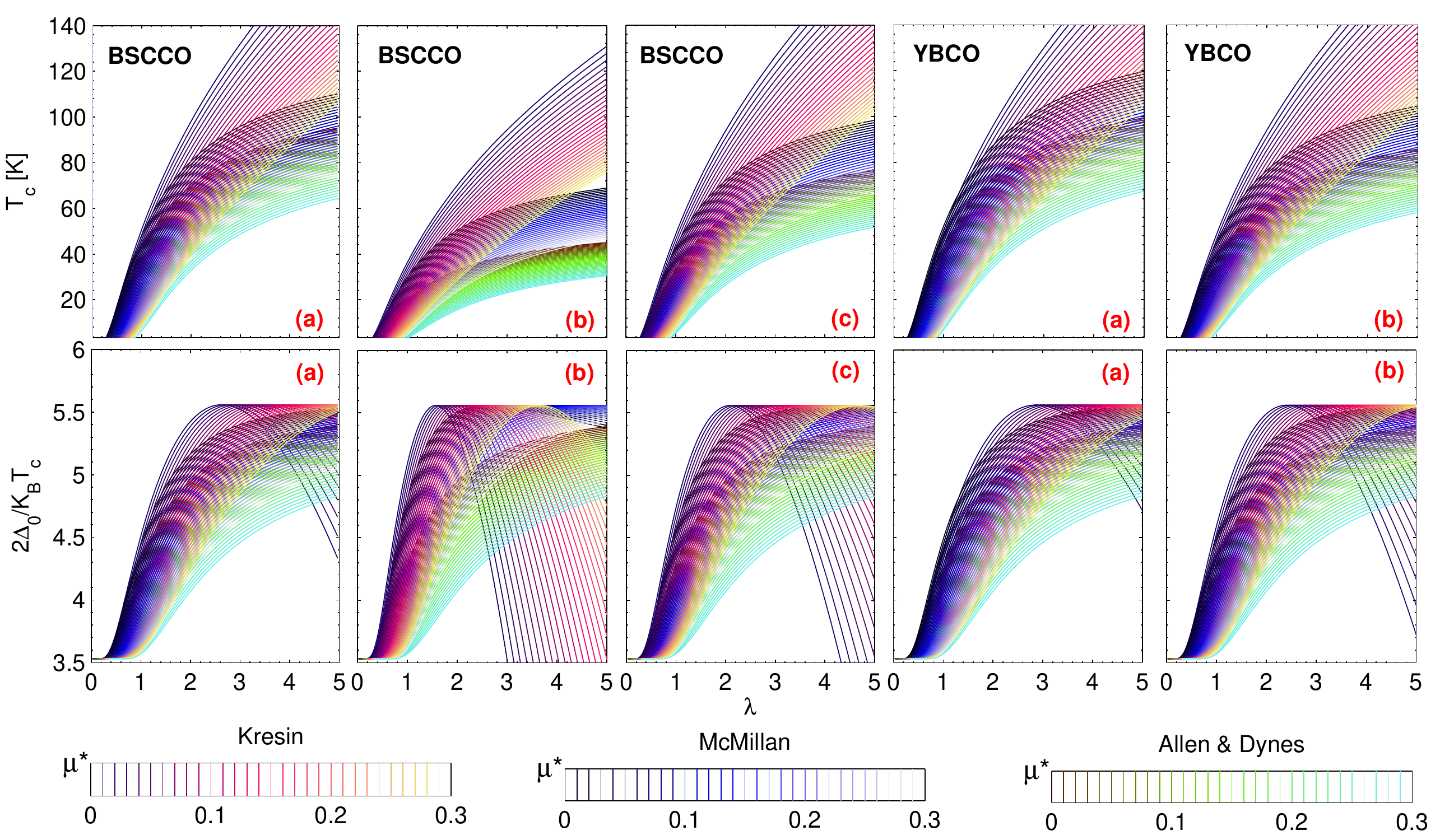}
\\
\caption{\label{Figure_8_2} Plots of the critical temperatures $T_{c}$ (top),
and the ratio gap $2\Delta_{0}/k_{B}T_{c}$
for Bi2212 and Y123 samples, as functions of the E-Ph coupling parameter
$\lambda$ and the Coulomb pseudopotential $\mu^{*}$. $\lambda$ has been
determined from the spectral densities of: (a) Islam \& Islam~\cite{P3-Islam00},
(b) Shiina \& Nakamura~\cite{P3-Shiina90}, and (c) Gonnelli et
al.~\cite{P3-Gonnelli98}. Three different approximations are shown: the Kresin's
formula~\cite{P3-Kresin87}, the McMillan's formula~\cite{P3-McMillan68}, and
the 
Allen-Dynes formula~\cite{P3-Allen75}.
}
\end{figure}

\begin{figure}[t]
\centering
\includegraphics[height=9cm,width=13cm]{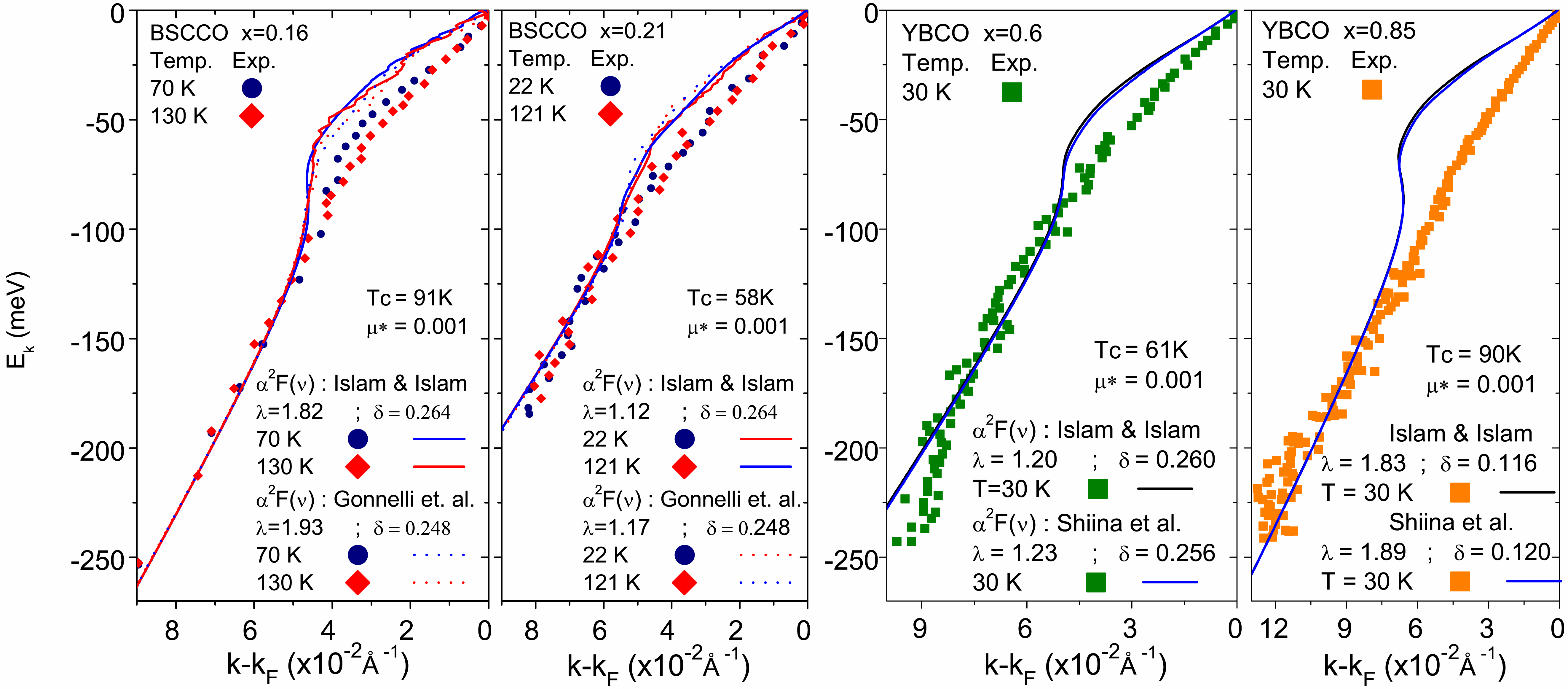}
\\
\caption{\label{Figure_8_3} The renormalized energy $E_{k}$ as a function of
the momentum $k-k_{F}$ under the assumption that the nodal dispersion is a
consequence of the E-Ph coupling mode and the Coulomb effects are
completely contained in the bare energies. Then, we assume the very weak Coulomb
pseudopotential $\mu^*=0.001$ within the more favourable approximation for the
calculus of $T_{c}$ and the phonon hypothesis, i.e. the Kresin Formula
and the E-Ph spectral density $\alpha^2\emph{F}(\omega)$
obtained by the methods of Refs.~\cite{P3-Islam00,P3-Gonnelli97} in Bi2212
cases,
and Refs.~\cite{P3-Islam00,P3-Shiina90} in Y123 cases.
}
\end{figure}

Assuming that the Coulomb effects are almost negligible for the
renormalized energy of the electronic quasiparticles participating in the
superconducting pairs formation, a remarkable enhancement of the
thermodynamic properties could be expected (Fig.~\ref{Figure_8_2}). For
example, taking $\mu^{*}=0.001$ rather than
the conventional $\mu^{*}=0.13$, and assuming the most
favorable scenario for the phonon hypothesis, i.e.: (\textit{i})
determine the E-Ph spectral density from the methods by Islam \& Islam
(Ref.~\cite{P3-Islam00}), and/or Gonnelli et al. (Ref.~\cite{P3-Gonnelli98}),
(\textit{ii}) use of the empirical Kresin's formula to determine $T_{c}$
(see Fig.~\ref{Figure_8_2}), and (\textit{iii}) check if it is possible to
reproduce the renormalization effects of the scattered quasiparticles along the
nodal direction by ARPES measurements; the maximal influence of the E-Ph
mechanism for the families of Bi2212 and Y123 can be estimated. For
Bi2212-OPD90, we have obtained $\lambda_{Islam}(T_{c}^{Kresin}=91K)\simeq1.82$
and $\lambda_{Gonnelli}(T_{c}^{Kresin}=91K)\simeq1.93$, with the renormalization
parameters for the ARPES ``bare'' dispersion $\delta=0.264$ and $\delta=0.248$,
respectively. Regarding to Bi2212-OVD58, the obtained 
E-Ph coupling parameters are: $\lambda_{Islam}(T_{c}^{Kresin}=58K)\simeq1.12$
and
$\lambda_{Gonnelli}(T_{c}^{Kresin}=58K)\simeq1.17$, for the same $\delta$
values above considered. Along this line, we have observed a widening of the
kink effect which can be only explained by the
existence of at least one additional perturbation mechanism reducing the
momentum of the dispersed quasiparticles  (see Fig.~\ref{Figure_8_3}). 

On the other hand, regarding the Y123 samples, the effect of reducing the
Coulomb pseudopotential shows that the expected values for $T_{c}$ in Y123-UD35
can be explained under any of the aforementioned formulas. In detail, for
$\lambda=0.80$ (see Fig.~\ref{Figure_7_5}) and the E-Ph spectral density
extracted from the method by Islam \& Islam (see Fig.~\ref{Figure_7_1}), we
have obtained  $T_{c}^{Kresin}\simeq37.69$, $T_{c}^{McMillan}\simeq40.04$, and
$T_{c}^{Allen-Dynes}\simeq33.25$. And from the most rigorous
method of Ref.~\cite{P3-Shiina90}, we have obtained:
$T_{c}^{Kresin}\simeq36.86$,
$T_{c}^{McMillan}\simeq35.01$, and $T_{c}^{Allen-Dynes}\simeq28.71$. Then, at
least for Y123-UD35, is possible keep the idea that the E-Ph coupling
is the the most relevant interaction mechanism for the superconducting pair
formation. However, we can not argue the same for the whole set of Y123 samples,
because the values of $T_{c}$ are still underestimated in about $62\%$ in
Y123-UD61, and about $80\%$ in Y123-OVD90. Indeed, if once again we allow assume
the most favorable scenario for the phonon hypothesis with the aim of reproduce
the $T_{c}$ value for Y123-UD61
[$\lambda_{Islam}(T_{c}^{Kresin}=61K)\simeq1.20$;
$\lambda_{Shiina}(T_{c}^{Kresin}=61K)\simeq1.23$], and Y123-OVD90
[$\lambda_{Islam}(T_{c}^{Kresin}=90K)\simeq1.83$;
$\lambda_{Shiina}(T_{c}^{Kresin}=90K)\simeq1.89$], with the ARPES ``bare''
dispersion renormalized according to the experiments, a
most notorious widening of the kink effect appears in disagreement with the
experiments (see Fig.~\ref{Figure_8_3}). 

Thus, despite the E-Ph mechanism by itself is able to explain many of the
properties of low temperature superconductors and
even some of the HTSC, it does not seem possible to avoid the
idea of the existence of some additional perturbation mechanism contributing to
the formation of pairs of Cooper. In fact, in Y123
samples, the importance of an additional mechanism seems to be more significant
that in Bi2212, and it could be related to the appearance of a second kink in
the
under-doped phase. Recent experimental results on the oxygen isotope effect in
Bi2212 also assert the need of consider additional coupling mechanism for the
formation of pairs~\cite{P3-Carbotte10}. 

Summarizing, we recall that there are
several possibilities as candidates of additional perturbation mechanisms but
any of them should modify the perturbation theory introduced along the last two
chapters, i.e., by
the superposition (adding) of different spectral functions emulating the boson
coupling mechanism, we know that the thermodynamic properties could be
reproduced. However, the calculation of the intrinsic effects in the
renormalized energy band near of the Fermi Level might be compromised, and a
satisfactory theory for both effects cannot achieved under the same scheme.


 \chapter*{}
 \vspace*{-3cm}
 \section*{\Huge{Conclusions III}}
 \markboth{\sc \textbf{Conclusions III}}{\sc \textbf{Conclusions III}}
 \addcontentsline{toc}{chapter}{\sc \textbf{Conclusions III}}
 \vspace*{2cm}

In summary, we have introduced a numerical model that allows to reproduce the
appearance of the ubiquitous nodal kink for a wide set of ARPES
experiments in cuprate superconductors. Our proposal is grounded on
the Migdal-Eliashberg approach for the self-energy of
quasi-particles within the electron-phonon coupling scenario. The
main issue is the use of a linear dispersion relation for the
bare band energy, i.e.:
$\varepsilon_{k}=(k-k_{F})v_{F<}(1-\delta\lambda)^{-1}$. $\delta$,
the only free parameter of the theory is a universal property for
each family of cuprates, which has been interpreted as the relation
between the mass-enhancement $\lambda^{*}$ and electron-phonon
coupling $\lambda$ parameters. 

On the one hand, for decades, a well-known controversy has arisen on the role
of the parameter ``$\lambda$'' whose values noticeably scatter among
different model calculations. As a central result, our proposal
re-ensembles the ``$\lambda$'' values obtained from different models
and, as a first approximation, it solves the controversy through the
relation $\lambda^{*}\cong\delta\lambda$. We emphasize that the
phenomenological parameter $\delta$ (obtained through the analysis
of a wide collection of data) has allowed to go beyond the
conventional Migdal-Eliashberg approach for restricted sets
of experiments. Our model is directly supported by the ``$\lambda$'' values
obtained in
Refs.~\cite{P3-Kordyuk06,P3-Giustino08,P3-Shiina90,P3-Gonnelli98,P3-Weber87,
P3-Weber88,P3-Heid08}.
Furthermore, an excellent agreement between the theory and the available
collection of experiments is achieved. 

On the other hand, our results support the idea
that the strong renormalization of the band structure and the so-called
universal nodal Fermi velocity, customarily related to the dressing of the
electron with excitations, can be explained in terms of the conventional
electron-phonon interaction. In fact, our
results suggest that the electron-phonon interaction strongly influences the
electron dynamics of the high-$T_{c}$ superconductors, and it is an important
mechanism linked with the Fermi surface topology. Thus, we conclude that the
electron-phonon interaction (strong or weak) must be included in any realistic
microscopic theory of superconductivity, although its effect in the appearance
of the superconducting state and the high critical temperatures is not clear
yet.

In detail, we have studied the influence of the electron-phonon coupling
mechanism through different doping levels in several families of HTSC-cuprates.
On the one hand, we have evaluated different methods for obtaining the
electron-phonon spectral densities and their influence on the electron bare band
energy, and on the other hand, several approaches have been recalled to obtain
the critical temperatures. Our results suggest that at least in the LSCO
family, and in the so-called Bi2212-UD70 and Y123-UD35 superconductors, the
electron-phonon interaction could be the most relevant mechanism involved in the
formation of Cooper's pairs. Our conclusion is supported
by the experimental evidence of a mass renormalization of the electronic
dispersion curves measured along the nodal direction in ARPES, and the reported
$T_{c}$ values in good agreement with our theoretical predictions. In addition, 
we have evaluated the consequences of assuming an
enhanced phonon mechanism, through the reduction of the Coulomb's
pseudopotential weight. When appropriate $T_c$ values are obtained by this
method, a remarkable widening of the predicted kink effect arises. This fact,
suggest that independently of the approximations invoked and even avoiding the
influence of the \textit{d}-wave superconducting gap through the nodal ARPES
measurements, it doesn't seem possible to elude the existence of additional
mechanisms that reduce the momentum of the dispersed quasiparticles in
comparison with the phonon mechanism. In this sense, despite the fact that in
LSCO the influence of the magnetic mode seems not relevant, it is not possible
to ignore its importance over the electron properties of other HTSC families.


\renewcommand\bibname{References III}

\section*{}
\break
\vspace*{0.0cm}
\subsection*{\label{PIII-s}\Huge{Supplementary Material III}}
\markboth{\sc \textbf{Supplementary Material III}}{\sc \textbf{Supplementary
Material III}}
\addcontentsline{toc}{chapter}{\sc \textbf{Supplementary Material III}}

\vspace*{2.0cm}
\textleaf \textbf{\textit{\Large{Theoretical Framework For The Electron Self
Energy}}}
\vspace*{1.0cm}

Lattice vibrations couple to electrons because the displacements
of atoms from their equilibrium positions alter the band dispersions, either
lowering or raising the total electron quasiparticle energy, where the phonon
propagator or Green's function for the phonon contribution can be defined as 
\begin{eqnarray}\label{Eq-SIII-1}
D_{\alpha,\beta}(\textbf{Q},\tau)=-\left\langle
T_{\tau}\textbf{u}_{Q,\alpha}(\tau)_{H}\textbf{u}_{-Q,\beta}^{\dag}(\tau)_{H}
\right\rangle \, .
\nonumber
\end{eqnarray}
The displacement operators $\textbf{u}_{Q}$ have been written in the
Heisenberg picture, with the exception that the time \emph{t} is replaced by an
imaginary time $-i\tau$ with $\beta=T^{-1}$. $T_{\tau}$ is the Wick operator
which
reorders the operators in such a way that $\tau$ increases
from right to left. In detail, the displacement operator $\textbf{u}_{Q}$ is a
sum of phononic operators, where the operator
$a_{Qi}\left(a_{Qi}^{\dag}\right)$ destroy (create) a phonon with energy
$\omega_{Qi}$, wave vector $\textbf{Q}=Qi$, branch index i, and
polarization vector $\hat{\epsilon}_{Qi}$, i.e.,
\begin{eqnarray}\label{Eq-SIII-2}
\textbf{u}_{Q}=\sum_{i}\left(\frac{\hbar}{2NM\omega_{Qi}}\right)^{1/2}\hat{
\epsilon}_{Qi}\left(a_{Qi}+a_{Qi}^{\dag}\right) \, .
\nonumber\end{eqnarray}
Here, by simplicity, we have considered only one kind of ion-mass \emph{M} for
the displaced atoms. On the other hand, as we are interested in determine the
Green's function for the interacting system, is  helpful to write the Dyson's
equations for electrons and phonons, i.e., 
\begin{eqnarray}\label{Eq-SIII-3}
G^{-1}(\textbf{k},i\omega_{n})=G_{0}^{-1}(\textbf{k},i\omega_{n})-\Sigma(\textbf
{k},i\omega_{n}) \, , 
\nonumber\end{eqnarray}
\begin{eqnarray}\label{Eq-SIII-4}
\left[D^{-1}(\textbf{Q},i\omega_{\nu})\right]_{ij}=&\left[D_{0}^{-1}(\textbf{Q},
i\omega_{\nu})\right]_{ij}-\Pi_{ij}(\textbf{Q},i\omega_{\nu}) \, ,
\nonumber\end{eqnarray}
in terms of their corresponding spectral representations, as follows:
\begin{eqnarray}\label{Eq-SIII-5}
G(\textbf{k},i\omega_{n})=\int_{-\infty}^{\infty}d\varepsilon
C(\textbf{k},\varepsilon)(i\omega_{n}-\varepsilon)^{-1} \, ,
\nonumber\end{eqnarray}
\begin{eqnarray}\label{Eq-SIII-6}
D_{\alpha,\beta}(\textbf{Q},i\omega_{\nu})=\int_{-\infty}^{\infty}d\nu
B_{\alpha,\beta}(\textbf{Q},\nu)\times[(i\omega_{\nu}-\nu)^{-1}-(i\omega_{\nu}
+\nu)^{-1}] \, .
\nonumber\end{eqnarray}

Notice as, the last couple of equations are defined within the framework of the
``Matsubara Frequencies'' $(i\omega_{n}=i(2n+1)\pi T)$, where the Green's
function for non-interacting electrons is defined as follows: 
\begin{eqnarray}\label{Eq-SIII-7}
G_{0}(\textbf{k},i\omega_{n})=\left(i\omega_{n}-\varepsilon_{k}\right)^{-1} \, ,
\nonumber\end{eqnarray}
and the phonon propagator in the harmonic
approximation~\cite{P3-Doniach98} can be defined as
\begin{eqnarray}\label{Eq-SIII-8}
D_{\alpha,\beta}^{0}(\textbf{Q},i\omega_{\nu})=\sum_{\textbf{Q}}
\left(\frac{\hbar}{2NM\omega_{\textbf{Q}}}\right)\epsilon_{Q\alpha}\epsilon_{
-Q\beta}
\left(\frac{1}{i\omega_{\nu}-\omega_{\textbf{Q}}}-\frac{1}{i\omega_{\nu}+\omega_
{\textbf{Q}}
}\right) \, .
\nonumber\end{eqnarray}
$C(\textbf{k},\varepsilon)$ and $B_{\alpha,\beta}(\textbf{Q},\nu)$ are the
electron and phonon spectral functions, respectively, and
$\Sigma(\textbf{k},i\omega_{n})$ and $\Pi(\textbf{Q},i\nu_{n})$ are the
corresponding electron and phonon self-energies. On the one hand, the phonon
self-energy $\Pi$ causes a large shift of the phonon
energies which is very difficult to evaluate by numerical methods. Nevertheless,
the spectral function $B_{\alpha,\beta}$ is directly measurable by INS or, more
precisely, from a Born-Von Karman interpolation \cite{P3-Doniach98}. On the
other
hand, there is not possible get a direct measure of the spectral
function \emph{C}, and evaluate it is too complicated because in the
electronic density of states $N(\varepsilon)$ there are some critical points in
the Brillouin zone (Van-Hove singularities) where
$\left|\nabla\varepsilon_{k}\right|$ becomes zero by symmetry reasons. Thus,
is easier evaluate $\Sigma$ even though a proper theory implies
the use of a Hamiltonian model that in the normal state corresponds to the
Fr\"{o}hlich Hamiltonian \cite{P3-Gross86}. Here, the key lies in resolving the
Feynman diagram of figure~SIII-I within the
Migdal-Eliashberg (ME)
approach. 

\begin{SCfigure}
\centering
\includegraphics[height=5.5cm,width=6.5cm]{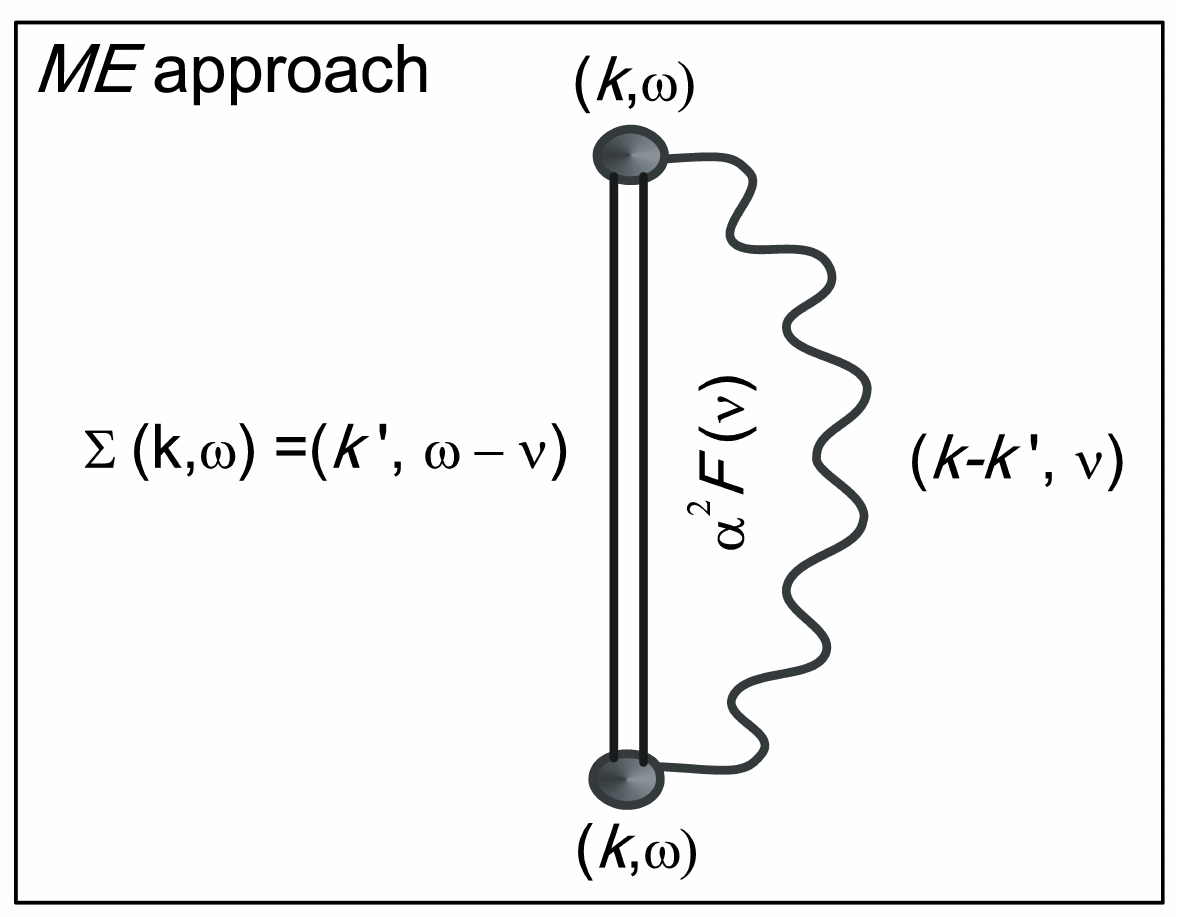}
\caption*{\label{Figure_SIII_1} Figure SIII-1: Feynman diagram for the ME
approach. The wavy
line
represents the phonon Green's function $D$, the double solid line is the
renormalized electron Green's function $G$, and the shaded circles represent the
small electron-phonon vertex corrections which have been neglected in this
chapter.}
\end{SCfigure}

Strictly speaking, the Feynman diagram of Fig.~SIII-I represents
the lowest perturbation diagram
for the self energy of electron quasiparticles $\Sigma(k,i\omega_{n})$, 
scattered from the state \textit{k} to \textit{k'} through the phonon
propagator, 

\begin{eqnarray}
\Sigma(k,i\omega_{n})=-\beta^{-1}\sum_{k',\nu} \langle
k|\nabla_{\alpha}V|k'\rangle
D_{\alpha,\beta}(k-k',i\omega_{\nu})\langle k'|\nabla_{\beta}V|k\rangle
G(k',i\omega_{n}-i\omega_{\nu}) \nonumber \, .
\nonumber\end{eqnarray}

Here, we have assumed that the electron correlations are responsible of the
formation of quasiparticles which are well defined near the Fermi level.
In addition, the so called vertex corrections
can be a \textit{priori} neglected because these can be shown to be reduced by
the ratio between the phonon frequency ($0-100~meV$) and  $\varepsilon_{F}$
($\sim1-10~eV$). Thus, the presence of strong electron correlations
mediated by the electron-phonon interaction is avoided, and the multi-phonon
excitations are reduced to the single-loop approximation or ME approach. The
advantage of this formulation is that the electron self energy can be
defined in terms of an experimental spectral density which allow estimate the
interaction between the electrons and the lattice vibrations for each one of
the materials. In this sense, is useful define the E-Ph spectral density as 
\begin{eqnarray}\label{Eq-SIII-9}
\alpha^{2}F(k,k',\nu)\equiv N(0)\langle k|\nabla_{\alpha}V|k'\rangle
B_{\alpha,\beta}(k-k',\nu)\langle k'|\nabla_{\beta}V|k\rangle\, ,
\nonumber\end{eqnarray}
where $N(0)=\sum_{k}\delta(\varepsilon_{k})$ represents the
single-spin electronic density of states at the Fermi surface. 

Is to be mentioned, that some of the photoemission experiments require consider
the complexities of the $d-$band electron structure, and phonons from the
``angular'' and ``energy'' components of the phase space ($k$). Then, the
$k$-space can be represented in terms of its harmonic components
$(J,\varepsilon)$~\cite{P3-Allen76}, and the electron self energy can be defined
 in terms of this set as,

\begin{eqnarray}\label{Eq-SIII-10}
\Sigma_{J}(k,i\omega_{n})=T\sum_{J',\nu}\int_{0}^{\infty}d\varepsilon'  &&
\frac{ N(\varepsilon')}{N(0)}\int_{0}^{\infty}d\nu
\alpha^{2}\emph{F}(J,J',\varepsilon,\varepsilon',\nu)\nonumber \\ &&
\left(\frac{2\nu}{\omega_{\nu}^{2}+\nu^{2}}\right)
G_{J'}(\varepsilon',i\omega_{n}-i\omega_{\nu}) \, . 
\nonumber\end{eqnarray}

The electron-phonon spectral
density $\alpha^{2}\emph{F}(J,J',\varepsilon,\varepsilon',\nu)$ represents a
measure of the effectiveness of the phonons of frequency $\nu$ in the scattering
of electrons from $k(J,\varepsilon)$  to $k'(J',\varepsilon')$. Here,
only the harmonic approximation to the phonon propagator
$D(k-k',\nu)=2\nu/(\omega_{\nu}^{2}+\nu^{2})$ has been considered.
Nevertheless, the above equation for the electron self energy is still
cumbersome, and
requires an accurate determination of the electron-phonon spectral density
from theoretical ab-initio calculations.
Then, to avoid this  tricky procedure our position is invoke the ME approach. 

In order, the ME approach consists in simplify $\Sigma_{J}(k,i\omega_{n})$ by
assuming that, it is possible to neglect the dependence on the energy surfaces
($\varepsilon,\varepsilon'$) of
the $N(\varepsilon')\alpha^{2}F(J,J',\varepsilon,\varepsilon',\nu)$ function.
This allow omit the processes violating the Born-Oppenheimer adiabatic
theorem contained within the high order graphs. Formally, this means that the
spectral function $\alpha^{2}F(JJ',\nu)$ is to be diagonal where its
representation for the normal state will continue to hold in the superconducting
state for the isotropic Cooper pairing or
$s$-\textit{wave} gap. However, other pairing schemes which break
rotation symmetry are, in principle, possible~\cite{P3-Allen82,P3-Carbotte90}.

Thus, taking advantage that the ARPES measurements at the nodal direction are
not influenced by the anisotropy of the superconducting gap, we will refer to a
(non-directional) isotropic quasiparticle spectral density, defined as the
double
average over the Fermi surface of the electron-phonon spectral density
$\alpha^{2}F(\textbf{k},\textbf{k}',\nu)$;
i.e., 
\begin{eqnarray}\label{Eq-SIII-11}
    \alpha^{2}F(\nu)=\frac{1}{N(0)}\sum_{\textbf{kk'},j} |
    g_{\textbf{kk'}}^{j}|^{2}\delta(\nu-\nu_{\textbf{k}-\textbf{k'}}^{j})
    \delta(\varepsilon_{k})\delta(\varepsilon_{k'}) \, ,
\nonumber\end{eqnarray}
where,
$g_{\textbf{k}\textbf{k}'}^{j}=[\hbar/2M\nu^{j}_{\textbf{k}'\textbf{k}}]^{1/2}
\langle
\textbf{k}|\hat{\epsilon}^{\; j}_{\textbf{k}'\textbf{k}}\cdot\nabla
V|\textbf{k}'\rangle$ defines the matrix elements of the E-Ph interaction
for electron
scattering from $\textbf{k}$ to $\textbf{k'}$ with a phonon of
frequency $\nu_{\textbf{k}-\textbf{k'}}^{j}$ ($j$ is a branch index).
$V$ stands for the crystal potential, $\hat{\epsilon}^{\;
j}_{\textbf{k}'\textbf{k}}$ is the polarization
vector, and $\delta(x)$ denotes the Dirac's delta function evaluated at
$x'=0$. 

Notice that, $|g_{\textbf{k}\textbf{k}'}^{j}|^{2}$ is
inversely proportional to the number of
charge carriers contributed by each atom of the crystal to the
bosonic coupling mode. Therefore, an increase in the doping level,
which causes an increment in the hole concentration of the $CuO_{2}$
plane must be reflected in the coupling parameters as we will show
in chapter~\ref{ch-8}. Moreover, recalling the
outstanding feature of the theory of metals, that
$|g_{\textbf{k}\textbf{k}'}^{j}|^{2}$ vanishes linearly with
$|\textbf{k}-\textbf{k}'|$ when $|\textbf{k}-\textbf{k}'| \ll
{k}_{F}$ \cite{P3-Ashcroft76}, one would expect a {\em linear}
disappearance of the coupling effect that gives rise to the nodal
kink in the vicinity of the Fermi surface. Thus, 
inspired by recent results on the ``universality'' of the nodal Fermi
velocity $v_{F<}$ (at low energies) in the HTSC, a prominent
role of this quantity is also expected.

Finally, in photoelectron scattering experiments the
relevant dynamical information is contained in the
analytic continuation $G(k,\omega+i0^{+})$ to points just above the
real frequency axis, known as the ``retarded'' Green's
function~\cite{P3-Doniach98}. One
is therefore led to continue the electronic self-energy
$\Sigma(k,i\omega_{n})$ analytically by $\Sigma(k,\omega+i0^{+})\equiv
\Sigma_{1}(k,\omega)+i\Sigma_{2}(k,\omega)$, where the bare electron band energy
is determined by the poles of the Green's function $G(k,\omega+i0^{+})$ or the
zeros of $G^{-1}(k,\omega+i0^{+})$. 

Assuming that a pole occurs near $\omega=0$, one gets
\begin{eqnarray}\label{Eq-SIII-12}
G^{-1}(k,\omega+i0^{+})
&& =\omega-\varepsilon_{k}-\Sigma_{1}(k,\omega)-i\Sigma_{2 } (k,\omega)
\nonumber \\ &&
\simeq\omega\left(1-\left.\frac{\partial\Sigma_{1}(k,\omega)}{\partial\omega}
\right|_{\omega=0}\right)-\left[
\varepsilon_{k}+\Sigma_{1}(k,0)\right]-i\Sigma_{2}(k,\omega) \nonumber \, .%
\nonumber\end{eqnarray}
Then, the pole of $G$ occurs at a frequency $\omega_{0}$ given by
$\omega_{0}=E_{k}-i/2\tau_{k}$, with the quasiparticle scattering time defined
by
$\tau_{k}^{-1}=-2\left(1-\partial_{\omega}\Sigma_{1}\right)^{-1}\Sigma_{2}(k,E_{
k})$, and the electron dressed band energy $E_{k}$ by
\begin{eqnarray}\label{Eq-SIII-13}
E_{k}=(1-\partial_{\omega}\Sigma_{1})^{-1}\left[\varepsilon_{k}+\Sigma_{1}(k,
0)\right] \, .
\nonumber\end{eqnarray}
%

%



\part{\label{Part_IV}\textsc{\textbf{Addenda}}}


\chapter*{}
\vspace*{-3cm}
\section*{\label{Glossary}\sc \textbf{Glossary}}
\markboth{\sc \textbf{Glossary}}{\sc \textbf{Glossary}}
\addcontentsline{toc}{chapter}{\sc \textbf{Glossary}}
\vspace*{1cm}

In order to provide an easiest reading of this book, below we introduce a list
of the most used abbreviations in text. Greek symbols are either incorporated by
their phonetic translation into Latin.

\begin{itemize}

\item \textbf{A}
\subitem A. Magnetic vector potential.
\subitem AC. Alternate Cycle.
\subitem ARPES. Angle resolved photoemission spectroscopies.

\item \textbf{B}
\subitem B. Magnetic induction field (Bold-facing means vector).
\subitem $B_{\tt{a}}$. Peak amplitude for the AC excitation $B_{0}$.
\subitem $B_{0}$. Applied magnetic flux density.
\subitem $B_{p}$. Penetration field.
\subitem $B_{ind}$. Self (induced) magnetic flux density.
\subitem Bi2212. $Bi_{2}Sr_{2}CaCu_{2}O_{8+x}$.
\subitem Bi2201. $Bi_{2}Sr_{1.65}La_{0.35}CuO_{6+x}$.
\subitem Bi-2221. $(Bi,Pb)_{2}Sr_{2}Ca_{2}Cu_{3}O_{x}$
\subitem BM. Brandt-Mikitik

\item \textbf{C}
\subitem $\chi$. Bandwidth of the critical state material law incorporated by
the SDCST, i.e., $J_{c\parallel}/J_{c\perp}$.
\subitem CS. Critical state.

\item \textbf{D}
\subitem $\Delta$. Variation (increment) of ...
\subitem $\Delta_{0}$. The zero temperature superconducting gap (Only used
within the third part of this dissertation).
\subitem $\Delta_{\textbf{r}}$. Material law for the critical state problems.
\subitem DC. Direct current.
\subitem DCSM. Double critical state model.

\item \textbf{E}
\subitem E. Induced transient electric field (In bold means vector).
\subitem $E_{c}$. Critical electrical field.
\subitem $E_{F}$. Electron energy at the Fermi level.
\subitem $E_{k}$. Electron dressed band energy.
\subitem $\varepsilon_{k}$. Electron bare band energy.
\subitem E-Ph. Electron-Phonon.
\subitem EDC. Energy Distribution Curves.
\subitem EDCSM. Elliptical double critical state models.
\subitem Eq(s). Equation(s).

\item \textbf{F}
\subitem ${\cal F}$. Minimization functional or so-called Objective function.
\subitem Fig(s). Figure(s)


\item \textbf{H}
\subitem H. Magnetic field (In bold means vector).
\subitem HTSC. High-temperature superconducting copper oxides.

\item \textbf{I}
\subitem $I_{\tt{a}}$. Peak amplitude for the AC excitation $I_{tr}$.
\subitem $I_{c}$. Critical current.
\subitem $I_{tr}$. Transport current.
\subitem $I_{\parallel}$. Parallel current.
\subitem $I_{\perp}$. Perpendicular current.
\subitem ICSM. Isotropic critical state model.
\subitem INS. Inelastic Neutron Scattering.

\item \textbf{J}
\subitem J. Electrical current density (In bold means vector).
\subitem $J_{c}$. Critical current density.
\subitem $J_{c\parallel}$. Parallel component of J.
\subitem $J_{c\perp}$. Perpendicular component of J.

\item \textbf{K}
\subitem $k_{B}$. Boltzmann's constant.
\subitem $k_{F}$. Electrom momentum at the Fermi level

\item \textbf{L}
\subitem ${\cal L}$. Lagrange density.
\subitem $L$. Hysteretic AC loss.
\subitem $\lambda$. Electron-phonon coupling parameter.
\subitem $\lambda^{*}$. Mass-enhancement parameter.
\subitem LANCELOT. A FORTRAN package for large-scale nonlinear optimization.
\subitem LEED. Low electronic energy diffraction.
\subitem LSCO. $La_{2-x}Sr_{x}CuO_{4}$.

\item \textbf{M}
\subitem $\mu^{*}$. Coulomb psudopotential.
\subitem $\mu_{0}$. Permeability of the free space.
\subitem $\mu_{r}$. Relative permeability associated to a ferromagnetic.
material.
\subitem M. Magnetization (In bold means vector).
\subitem $M_{ij}$. Mutual/Self inductance matrix.
\subitem MDC. Momentum distribution-curves.
\subitem ME. Migdal-Eliashberg.
\subitem MRI. Magnetic resonance imaging.

\item \textbf{N}
\subitem n. Smoothing index.

\item \textbf{O}
\subitem $\omega$. Electromagnetic oscillating frequency.
\subitem $\Omega$. Superconducting volume.
\subitem OPD. Optimally doped.
\subitem OVD. Over doped.

\item \textbf{P}
\subitem $\Phi$. SC volume.
\subitem $\varPhi$. Electric scalar potential.
\subitem $\textbf{p}$. Lagrange multiplier.
\subitem Pag(s). Page(s).


\item \textbf{R}
\subitem $R$. Radius of the cylinder.

\item \textbf{S}
\subitem \textbf{S}. Poynting's vector.
\subitem SC. Superconductor.
\subitem SDCST. Smooth double critical state theory.
\subitem SIF. Standard Input Format.
\subitem STM. Scanning tunneling microscopy.

\item \textbf{T}
\subitem $T_{c}$. Superconducting critical temperature.
\subitem TCSM. T critical state model.

\item \textbf{U}
\subitem UD. Under doped.



\item \textbf{X}
\subitem XRS. X-ray scattering.

\item \textbf{Y}
\subitem Y123. $YBa_{2}Cu_{3}O_{6+x}$.


\end{itemize}





 \chapter*{}
 \vspace*{-3cm}
 \section*{\sc \textbf{Publications}}
 \markboth{\sc \textbf{Publications}}{\sc \textbf{Publications}}
 \addcontentsline{toc}{chapter}{\sc \textbf{Publications}}
 \vspace*{1cm}

\begin{flushleft}
Some of the results presented in this dissertation have been published in the
following scientific communications:
\end{flushleft}

\begin{enumerate}
 \item \textcolor{blue}{H. S. Ruiz}, A. Bad\'ia-Maj\'os, Yu. A. Genenko,
S.V. Yampolskii and H. Rauh.\\
\hyperlink{http://arxiv.org/abs/1108.0659v1}{\textcolor{blue}{Applied Physics
Letters. February 24, 2012. AIP ID: 102211APL}}\\ 
Superconducting wires under simultaneous oscillating sources: Magnetic
response, dissipation of energy and low pass filtering.
 \item \textcolor{blue}{H. S. Ruiz} and A. Bad\'ia-Maj\'os.\\
\hyperlink{http://www.sciencedirect.com/science/article/pii/S1567173911004639}{
\textcolor{blue}{Current Applied Physics \textbf{12}, 550 (2012).}}\\ 
Strength of the phonon-coupling mode in $La_{2-x}Sr_{x}CuO_{4}$, \\
$Bi_{2}Sr_{2}CaCu_{2}O_{8+x}$ and $YBa_{2}Cu_{3}O_{6+x}$ composites along the
nodal direction.
\item \textcolor{blue}{H. S. Ruiz}, A. Bad\'ia-Maj\'os and C. L\'opez.\\
\hyperlink{http://iopscience.iop.org/0953-2048/24/11/115005}{\textcolor{blue}{
Superconductor Science and Technology \textbf{24}, 115005
(2011).}}\\ 
Material laws and related uncommon phenomena in the electromagnetic response of
type~\textendash~II superconductors in longitudinal geometry.
 \item \textcolor{blue}{H. S. Ruiz}, C. L\'opez and A. Bad\'ia-Maj\'os.\\
\hyperlink{http://publish.aps.org/search/field/author/Ruiz_H_S}{\textcolor{blue}
{Physical Review B \textbf{83}, 014506 (2011).}}\textcolor{red}{*}\\ 
Inversion mechanism for the transport current in type~\textendash~II
superconductors.\\
\textcolor{red}{*} \textit{Selected for the Virtual Journal of Applications of
Superconductivity, Vol 29, Issue 9 (May 1st, 2011), Properties Important for
Applications.}
 \item \textcolor{blue}{H. S. Ruiz} and A. Bad\'ia-Maj\'os.\\
\hyperlink{http://www.springerlink.com/content/u325174340t1655k/}
{\textcolor{blue}{Journal of Superconductivity and Novel Magnetism \textbf{24},
1273 (2011).}}\\ 
Relevance of the Phonon-Coupling Mode on the Superconducting Pairing Interaction
of $La_{2-x}Sr_{x}CuO_{4}$.
 \item \textcolor{blue}{H. S. Ruiz} and A. Bad\'ia-Maj\'os and.\\
\hyperlink{http://iopscience.iop.org/0953-2048/23/10/105007}{\textcolor{blue}{
Superconductor Science and Technology \textbf{23}, 105007
(2010).}}\\ 
Smooth double critical state theory for type~\textendash~II superconductors.
 \item A. Bad\'ia-Maj\'os, C. L\'opez and \textcolor{blue}{H. S. Ruiz}.\\
\hyperlink{http://publish.aps.org/search/field/author/Ruiz_H_S}{\textcolor{blue}
{Physical Review B \textbf{80}, 144509 (2009).}}\textcolor{red}{*}\\ 
General critical states in type~\textendash~II superconductors.\\
\textcolor{red}{*} \textit{Selected for the Virtual Journal of Applications of
Superconductivity, Vol 17, Issue 8 (October 15th, 2009), Properties Important
for Applications.}
 \item \textcolor{blue}{H. S. Ruiz} and A. Bad\'ia-Maj\'os.\\
\hyperlink{http://publish.aps.org/search/field/author/Ruiz_H_S}{\textcolor{blue}
{Physical Review B \textbf{79}, 054528 (2009).}}\textcolor{red}{*}\\ 
Nature of the nodal kink in angle-resolved spectra of cuprate superconductors.\\
\textcolor{red}{*} \textit{Selected for the Virtual Journal of Applications of
Superconductivity, Vol 16, Issue 5 (March 1st, 2009), Materials Important for
Applications.}

\end{enumerate}





 \chapter*{}
 \vspace*{-3cm}
 \section*{\sc \textbf{Acknowledgments}}
 \markboth{\sc \textbf{Acknowledgments}}{\sc \textbf{Acknowledgments}}
 \addcontentsline{toc}{chapter}{\sc \textbf{Acknowledgments}}
 \vspace*{1cm}

I am especially thankful for the great confidence and support provided
by my advisor Dr. Antonio Bad\'ia-Maj\'os. 

I would like also to express my gratitude to Dr. Luis Alberto Angurel and Dr.
Rafael Navarro for their undeniable support along my different stages at
University of Zaragoza. I also thank to Dr. Yuri Genenko, Dr. Sergey Yampolskii,
and Dr. Hermann Rauh, for the interesting discussions we held in Darmstadt
University of Technology, as well as for their generous hospitality.

Thanks to the Spanish National Research Council (CSIC), the Institute of
Materials Science of Ar\'agon (ICMA), and the Department of Condensed Matter
Physics of the University of Zaragoza by their academic and economical support.

Finally but not less important, a special thankful to my girlfriend Edna
Corredor, who shares the same passion for physics than me. Without her
inexhaustible patience and support, as well as that from my family, this would
have not been possible. 

It is to be mentioned that, funding of the research within this dissertation
has been sponsored by the JAE program of the Spanish National Research Council
(CSIC). Also, the attendance to events and dissemination of results was
supported
by the Spanish CICyT and FEDER program (projects No. MAT2008-05983-C03-01,
MTM2006-10531, and MAT2005-06279-C03-01), the DGA project PI049/08, and DGA
grant T12/2011.



\newpage
\thispagestyle{plain}

\textit{I authorize to the University of Zaragoza for the distribution and
reproduction
of this thesis in the digital repositories of ZAGUAN and TESEO, as well as any
other one by direct consent of the University of Zaragoza, the Institute of
Materials Science of Ar\'agon (ICMA), or the Spanish National Research Council
(CSIC).}

\begin{flushright}
\textcopyright~\textsc{Harold Steven Ruiz Rondan} 
\end{flushright}

\end{document}